\documentclass[a4paper,fleqn,longmktitle]{cas-sc}

\usepackage[numbers,sort&compress]{natbib}
\usepackage{lipsum}
\usepackage{braket}
\usepackage{mathrsfs} 
\usepackage{soul}
\usepackage{upgreek}
\usepackage{threeparttable, tablefootnote}
\begin{document}

\newpage
\makeatletter
\renewcommand\section{\@startsection{section}{1}{\z@}%
	{15pt \@plus 3\p@ \@minus 3\p@}%
	{4\p@}%
	{
		\sectionfont\raggedright\hst[13pt]}}
\renewcommand\subsection{\@startsection{subsection}{2}{\z@}%
	{10pt \@plus 3\p@ \@minus 2\p@}%
	{.1\p@}%
	{
		\ssectionfont\raggedright }}
\renewcommand\subsubsection{\@startsection{subsubsection}{3}{\z@}
	{10pt \@plus 1\p@ \@minus .3\p@}%
	{.1\p@}%
	{
		\sssectionfont\raggedright}}
\makeatother

\let\WriteBookmarks\relax
\def\floatpagepagefraction{1}
\def\textpagefraction{.001}
\shorttitle{Method of Higher-order Operators}
\shortauthors{S. Khorasani}

\setcounter{page}{1}
\title[mode=title]{Operator approach in nonlinear stochastic open quantum physics}

\author[1]{Sina Khorasani}[orcid=0000-0001-9652-0112]


\address[1]{Vienna Center for Quantum Science and Technology, University of Vienna, 1090 Vienna, Austria}


\begin{abstract}[S U M M A R Y]
The success of quantum physics in description of various physical interaction phenomena relies primarily on the accuracy of analytical methods used. In quantum mechanics, many of such interactions such as those found in quantum optomechanics and quantum computing have a highly nonlinear nature, which makes their analysis extraordinarily difficult using classical schemes. Typically, modern quantum systems of interest nowadays come with four basic properties: (i) quantumness, (ii) openness, (iii) randomness, and (iv) nonlinearity. The newly introduced method of higher-order operators targets analytical solutions to such systems, and while providing at least mathematically approximate expressions with improved accuracy over the fully linearized schemes, some cases admit exact solutions. Many different applications of this method in quantum and classically nonlinear systems are demonstrated throughout. This review is purposed to provide the reader with ease of access to this recent and well-established operator algebra, while going over a moderate amount of literature review. The reader with basic knowledge of quantum mechanics and quantum noise theory should be able to start using this scheme to his or her own problem of interest. 
\end{abstract}

\maketitle

\tableofcontents

\section{Introduction}\label{Section-1}

This review article comprises a summary of the recent studies of the author on a new analytical tool to treat nonlinear quantum physics. Originated from a fundamental research on quantum optomechanics \cite{S2-1,S2-2,S2-3,S2-4,S2-5,S2-6}, the method of higher-order operators was invented first to make an attempt in solution of fourth-order quadratic interactions, without having to go through fully linearized algebra. Any developed method useful to address such a complex system should be in principle able to satisfy four basic properties simultaneously. These are enlisted below.
\begin{enumerate}
	\item \textit{Quantumness}: The first property of quantumness corresponds to the nature of quantities, which demand operator, rather than scalar function, description. The situation is complicated by the fact that these mostly refer to bosonic baths with ladder operators. Once there are two or more different bosonic baths to consider, the expansion on brakets simply is exhausting and impossible to study analytically. Mostly, master equation approach is used to tackle such problems, which has to be implemented fully numerically.
	\item \textit{Openness}: The second property of openness occurs since most experimental setups have multiple physical input and output channels, through which reflection or transmission unto different channels could be measured. While abstract Hamiltonian description of quantum systems excludes such a property, incorporation of input-output scheme unto the Langevin equations of motion provides an elegant way to deal with this. In the case of master equation approach, Lindblad operators could be added to the governing equation, which of course, keeps the full numerical approach in effect.
	\item \textit{Randomness}: The third property of randomness is caused by letting random fluctuations of zero point bosonic fields leak into the system through corresponding input channels. This is usually depited using stochastic noise operators following the well-established theory of quantum noise. In this scheme, all random fluctuations are approximated as Gaussian white. 
	\item \textit{Nonlinearity}: The fourth property of nonlinearity mostly is caused by the nature of interaction amongst various observables in quantum mechanics, which could deviate slightly or significantly from a factorizable or simple first-order product of observables. A factorizable interaction leads to linear Langevin equations and thus is exactly solvable. This is not the case for most of the actual real physical systems, and linearization is not guaranteed to provide all details and implications of a truly nonlinear physical interactions.
\end{enumerate}

Preservation of the first property prohibits usage of expectation values and classical variables instead of operators. The second property necessitates the use of input/output formalism, which is known to be compatible with Langevin's equation approach. However, presence of higher-order operators will quite often lead to multiplicative noise, which in turn will require some simplifications or approximations. Once the system is allowed to take in input, then quantum noise shows up, of course, and the algebraic method must be amended and be capable of evaluating spectral densities. Lastly, all these must be taken care of, not at the expense of dismissing nonlinearity.  

In absence of nonlinearity, it is highly convenient to use the linear Langevin equations for bosonic operators instead of master equations, which give immediate access to the time-evolution of operators and their momenta, as well as their spectral response. This can lead to a relatively accurate and explicit description of linear quantum interactions in a noisy environment. However, the case of nonlinearity has to be considered with utmost attention. For nonlinear systems, as long as tiny fluctuations around equilibrium points are of interest, then linearization of Langevin equations around equilibrium values is expected to work well. But two situations need careful consideration: 
\begin{itemize}
	\item When either the oscillating amplitudes are not small, or the nonlinearity is too strong because of large amplitude excitations. This is, for instance, exactly the situation for quantum computing hardware, even in absence of noise. One would need to develop a convenient mathematical tool which enjoys the freedom and convenience of Langevin equations, and at the same time is not fully numerical, or at least does not rely on brute force and numerics as much  as possible. 
	\item When the nonlinearity leads to observable effects even when only fluctuations are considered. This case actually occurs frequently in the context of modern quantum mechanics, such as standard and quadratic quantum optomechanics. A very striking example of such effects is side-band inequivalence, where frequency shifts of blue and red scattered lights are surprisingly a bit different. 
\end{itemize}

The method of higher-order operators \cite{S2-Paper1,S2-Paper2,S2-Paper3,S2-Paper4,S2-Paper5,S2-Paper6,S2-Paper7} is capable to successfully address all of the above concerns. It is based on the extension of Langevin equations to higher-order combinations of ladder operators, and is able to deal with all requirements of a typical nonlinear quantum system. Its application to standard optomechanics, quadratic optomechanics, as well as cross-Kerr interactions are established, and it has been also shown that it can reproduce the expected linearized results as long as nonlinearity is not too strong. Results of higher-order analysis are not always exact, but in few cases it allows to find exact mathematical solutions, which were otherwise unknown.

While the method of higher-order operators happens to provide the right \textit{Analysis Tool} for study of nonlinearity in such quantum systems, it finds many appealing applications outside the domain of quantum mechanics. Such applications include nonlinear electrical, electronic, and quantum circuits, optoelectronics, too mention a few, which are already discussed in extensive details.

It is the intention of the author to neither summarize the bulky works into a single article, nor present a literature review. The scope of nonlinear analysis is just too huge and there are already too many exceptional reviews and books available. However, it will provide a mid-way solution between a merger of extensive research articles and a book. 

So, this paper is targeted to assist graduate students and post-graduate researchers to tackle their own nonlinear problem of interest within the domains of quantum and classical physics. For those who are involved in quantum optomechanics, there is a lot of new physical insight, some of which are results of significant mathematical analyses of respective nonlinear phenomena. Having said that, it comes with lots of fresh mathematical analysis and physical insights. Because of the focus of this article, graphs and numerical calculations are not displayed. This decision was made not only for the sake of brevity or conciseness, but also for putting emphasis on the mathematical approach and not the curves and graphs. The author believes that each of the minor contributions deserves a thorough and detailed numerical and experimental research on its own in future. Therefore, the method of higher-order operators presented here can by no means considered as complete, as there is significantly much room available for interested students and scientists to explore and contribute. 

Each section comes with a reasonably well written opening introduction, which is strongly recommended to everyone to read consecutively one by one to the end of article. These openings do not contain mathematics and in turn present a compact literature review alongside new physical insights and major new findings in each section. Nevertheless, to assist with the curious readers on how to read such a long article, he or she may take notice of either of the following recommendations:
\begin{enumerate}
	\item For a quick overview of how the mathematical method of higher-order operators works and how its general approach towards a nonlinear problem is \cite{S2-Paper6}, study of \S\ref{Section-2} should suffice. All readers who want to use the method for any problem need to go through this section first. Further details of higher-order operators later follow in \S\ref{Section-4}, but this may be skipped for convenience, too.
	\item For those who are interested in derivation of nonlinear quantum optomechanics from first principles, the appropriate discussion is properly given in \S\ref{Section-3}, followed by rigorous treatment of conventional nonlinear quantum optomechanics in \S\ref{Section-8}, and very detailed analysis of the unexpected symmetry breaking, side-band inequivalence \cite{S2-Paper7}, detailed in \S\ref{Section-9}.
	\item For those who are doing research in quantum nonlinear circuits, study of \S\ref{Section-6}, \S\ref{Section-7}, and \S\ref{Section-11} is sufficient.
	\item For those who are doing research in quadratic optomechanics \cite{S2-Paper1,S2-Paper4}, study of \S\ref{Section-3} and then \S\ref{Section-10} is a must.
	\item For newcomers from the outside of nonlinear quantum physics, such as electrical engineers who have neither been exposed to similar stuff in the past, nor are going to use it, study of \S\ref{Section-2} and \S\ref{Section-12} is appropriate. However, results of analysis on nonlinear circuits, though very promising, are preliminary, and this area calls for much further research.
\end{enumerate} 

This article is written with lots of motivation and great expectations, and the author deeply appreciates in knowing the opinion of interested readers worldwide. The author has made every reasonable effort to keep the focus and pace on the track in composing this paper, however, it would not be impossible for a very precise reader and examiner to identify and find minor errors and shortcomings. Neither, answers to all possible questions in this context are all found and known. 

The method of higher-order operators is already past its infancy stage, but definitely deserves decades of further research until its full potential is unleashed and employed in various areas of science and technology.

\section{Nonlinear Interactions}\label{Section-2}

This section presents a fundamental overview to the method of higher-order operators as applied to nonlinear quantum phenomena. Basic definitions are presented and application of the proposed method to quantum optomechanics is discussed in brief, while further details follow later in the subsequent sections of the text. We keep the mathematics as simple as possible for the purpose of this section, which is to pick-up an understanding of how higher-order operators could help solving a nonlinear system. In-depth and more rigorous discussions for the moment being are left for the upcoming sections. 

So to speak, this section also summarizes the recent progress in the theory and analytical tools of quadratic optomechanical interactions, as one of the prominent domains of contemporary nonlinear quantum optics. Emphasis has been put here first to show what types of nonlinear interactions do exist, and what physical interpretations follow each. 

The standard quadratic interactions between light and mechanical motion is expressed as the product of cavity light intensity and squared mirror position. However, there exists a non-standard quadratic optomechanical interaction as well, which assumes a mathematically different form and appears as the squared product of field and mirror momenta. This non-standard type of quadratic interaction originates from two corrections: the momentum exchange and conservation among mirror and field, as well as relativistic corrections due to different mechanisms. Both these types of non-standard interactions become relevant when the ratio of mechanical to optical frequency is no longer negligible. The derivation of non-standard quadratic optomechanical interactions from first principles will be the subject of focus in \S\ref{Section-3}.

Next, we turn to the solution technique of such interactions, and introduce a formal higher-order operator method to tackle the nonlinear evolution of quantum systems. This enables one to accurately study any type of quantum nonlinear interaction using the analysis tools of linear algebra. It sounds definitely surprising that nonlinear systems might admit explicit solutions. Well, this is not always the case, and not all nonlinear systems can be integrated. However, the method of higher-order operators come to rescue when a nonlinear interaction must be studied outside the fully linearized regime. In most cases, at least approximate explicit expressions could be found, which are otherwise unknown. Some nonlinear systems, however, actually admit fully analytical solutions, and higher-order operators could provide straightforward scheme to obtaining those solutions.

In order to employ the analytical power of higher-order operators, one first needs to identify a closed Lie algebra, which should satisfy closedness property under commutation either exactly or approximately, and is referred to as the basis. Having the basis of higher-order operators known, one may proceed to construct the corresponding Langevin equations, which can be now conveniently analyzed using the existing mathematical toolbox of linear algebra to yield the spectral densities, moments, and expectation values. Application of this operator method results in a new type of symmetry breaking in standard nonlinear quantum optomechanics, referred to as the side-band inequivalence, to be discussed in complete details in \S\ref{Section-9}.

Mathematically speaking, it should be added that existence of a closed Lie algebra is neither necessary nor sufficient for integrability of a nonlinear system. However, having such a basis of closed operators known, is quite helpful in correct construction of equation system and recognition of what terms to approximate. So, the success of approach needs a balanced combination of intuitive guess and some analytical skills.

Nonlinear interactions are at the heart of many physical phenomena, both in the classical and quantum regimes. It is the way physical quantities are described, that sets border between mathematical methods of classical and quantum physics. Classical quantities are described as real-valued scalars belonging to $\mathscr{R}$, while quantum quantities are described by operators belonging to an associated Hilbert space such as $\mathscr{C}$. Mathematically, any operator is equivalent to a finite- or infinite-dimensional matrix of complex scalar values. While a Fermionic operator such as an atomic transition and spin component operator, is normally finite-dimensional in its Heisenberg matrix representation, in which a bosonic operator such as ladder creator and annihilator becomes essentially infinite-dimensional. 

An interaction between two physical quantities $A$ and $B$ is contingent on the existence of algebraic expressions in the system Hamiltonian such as $f_{nm} A^n B^m$, where $n,m\in\mathscr{N}$ are natural numbers and $f_{nm}$ represents the interaction strength. For the interaction strength to make mathematical sense, it is necessary to be independent of $A$ and $B$. That implies $\partial f_{nm}/\partial A=\partial f_{nm}/\partial B=0$. There is furthermore no relationship whatsoever between $f_{nm}$ and $f_{mn}$ in general, and these two can be completely independent. 

When $n=m=1$ the interaction is referred to be linear, and otherwise nonlinear. In quantum mechanics, a product of two observables is only a special case of linear interactions; this we shall observe shortly below. The integer $o=n+m$ shall represent the interaction order, and in that sense, any nonlinear interaction demands a minimum order of $o\geq 3$. Any other type of nonlinear interaction not basically complying to the form $A^n B^m$ can be normally expanded using appropriate polynomial expansion methods as $\sum f_{nm} A^n B^m$ over non-zero integers $n,m\in\mathscr{Z}-\{0\}$ around some expansion point $(\bar{A},\bar{B})\in\mathscr{R}\times\mathscr{R}$. The interaction order needs then to be redefined as $|n|+|m|$, however, such interactions will assume singular forms around zero and shall be excluded from the subject of present study. 

For instance, an ideal light modulator with $n=m=1$ performs a simple product of career light field with the modulating signal, similar to $AB$. Since the governing evolution equations of light and signal will turn to be exactly integrable, then the system is actually linear. In practice, a real light-signal multiplier such as the electrooptic modulation is mostly based on a nonlinear $\chi^{2}$ effect with $n=2$ and $m=1$ similar to $A^2B$, which could only approximate a mathematically exact product. For that reason, electrooptic modulation represents an example of a nonlinear system, which only approximates an ideal multiplier. This will inevitably cause measureable non-idealities, and additional distortions introduced to the modulated career, for instance, due to side-band inequivalence.

We refer to the interaction among two quantities as bipartite. Definition of interaction between three and more physical quantities can be extended and generalized in a similar way, but a tripartite interaction such as $ABC$ can never be linear, unless decomposable into summation of bipartite linear interactions such as $f_{11}AB+g_{11}BC+h_{11}CA$ and so on. Solution of tripartite interactions is beyond the scope of present discussion, and we limit ourselves only to bipartite interactions, which also happen to be the case in optomechanics.

\subsection{Linear Interactions}

Nonlinear classical problems normally turn into the form of scalar nonlinear differential equations, which can be solved by conventional numerical methods. However, description of nonlinear quantum phenomena always turn into operator differential equations, which are basically a system of infinite-dimensional matrix equations wherever one interacting partition is a bosonic field with ladder operators involved. 

As opposed to the linear quantum interactions which involve direct products of two operator quantities such as $\hat{a}\hat{b}$ and its adjoint $\hat{b}^\dagger\hat{a}^\dagger$, nonlinear quantum interactions involve terms of at least third in the order such as $\hat{a}^2\hat{b}$ and the corresponding interaction Hamiltoninan is composed of a third-degree polynomial in terms of the most basic operators. Henceforth, a physical and Hermitian interaction having the type
\begin{equation}
\label{S2-eq1}
\mathbb{H}_\text{int}=\hbar(\alpha\hat{a}\hat{b}+\beta\hat{a}^\dagger\hat{b}+\beta^\ast\hat{a}\hat{b}^\dagger+\alpha^\ast\hat{a}^\dagger\hat{b}^\dagger),
\end{equation}
where $\alpha,\beta\in\mathscr{C}$ are complex numbers, is having the most general form of linear interactions in quantum mechanics. It is easy to show that this interaction can be always factorized if $|\beta|=|\alpha|$. A factorizable bipartite interaction can be normally simplified using definition of linearly combined new operators, as $\hbar\xi\hat{x}\hat{y}$, where $\xi\in\mathscr{R}$ is a real number, and  $\hat{x}=\chi\hat{a}+\chi^\ast\hat{a}^\dagger$ and $\hat{y}=\eta\hat{b}+\eta^\ast\hat{b}^\dagger$ are Hermitian operators. Here, $\chi,\eta\in\mathscr{C}$ need to be half-modular complex constants satisfying $|\chi|^2=|\eta|^2=\frac{1}{2}$. This will set $\xi$ as the linear interaction strength. It is all clear now that any linear interaction which is factorizable, is actually a simple classical product, and such classical product cannot raise significant entanglement or squeezing. Obviously, an interaction which cannot be factorized will cause squeezing through appropriate Bogoliubov transformation as shown later in \S\ref{Section-3} under (\ref{S3-55}) and bipartite entanglement among the two bosonic fields.

\subsection{Quantum Optomechanics}

Sometimes, the interaction is nonlinear in terms of some quantities or operators, but assumes a linear form by defining some appropriate higher-order quantity operators. These new quantities could be mixtures and cross-overs of bosonic fields or higher-powers of each of the individual field operators. Quantum optomechanics is one such example of interaction, with bipartite and third-order nonlinearity, in which there exist two bosonic partitions interacting nonlinearly. These partitions correspond to electromagnetic and mechanical bosonic fields respectively composed of photons and phonon, which get correlated nonlinearly. The optomechanical interaction Hamiltonian is here denoted as \cite{S2-1,S2-2,S2-3,S2-4,S2-5,S2-6,S2-7,S2-8,S2-9}
\begin{equation}
\label{S2-eq2}
\mathbb{H}_0=-\hbar g_0\hat{a}^\dagger\hat{a}(\hat{b}+\hat{b}^\dagger),
\end{equation}
\noindent
where $g_0\in\mathscr{R}^+$ is referred to as the single-photon interaction rate. The negative sign corresponds to the fact that the optomechanical interaction is repulsive in nature because of the radiation pressure. However, the sign is actually not mathematically relevant and can be sometimes absorbed in $g_0$. The same form of Hamiltonian appears elsewhere also in other types of physical systems, such as ion/Paul traps, electrooptic/acousto-optic modulation, and Brillouin/Raman scattering. At the lowest-order, the basic optomechanical interactions which describes the effect of radiation pressure on mirror's position are third in order. But with redefining the Hermitian operator $\hat{n}=\hat{a}^\dagger\hat{a}$ it could be seen that the optomechanical interaction is actually linear between $\hat{n}$ and $\hat{b}$. This fact can later be used to make the optomechanical Hamiltonian (\ref{S2-eq2}) integrable, to be discussed in \S\ref{S2-Fourth}

Normally, linearization of only optical field ladder operator $\hat{a}$ is sufficient to reduce the interaction degree to two. This can be done by expanding the operator $\hat{a}=\bar{a}+\delta\hat{a}$, where $\bar{a}$ is the complex valued constant representing the mean classical value $\bar{a}=\braket{\hat{a}}$ and $\delta\hat{a}$ represents fluctuations with zero-mean as $\braket{\delta\hat{a}}=0$. Since photons inside the cavity do not necessarily follow the phase of external pump, it is mostly convenient to assume $\bar{a}\in\mathscr{R}^+$ as a positive real valued constant, leaving out the phase difference to the optical pump. Normally, once the linearization is done, non-interacting terms such as $\bar{a}\delta\hat{b}$ no longer contribute to the system dynamics and can be mathematically removed by appropriate unitary transformations. What remains as $\delta\hat{a}\delta\hat{b}$ represent nonlinear interactions which are to be discarded in the linearized picture. Then replacements are made as $\delta\hat{a}\rightarrow\hat{a}$, which makes the expressions more convenient to study analytically.

Surprisingly enough, the method of higher-order operators reveals that even though fluctuations could be small in amplitude and oscillate around mean values, they actually could contribute to observable nonlinear effects, depending on the mathematical form of interaction Hamiltonian involved. Side-band inequivalence is one such example of an observable nonlinear quantity, which once recognized in quantum optomechanics can no longer be ignored.

Thus performing such linearization makes the system integrable under linearized basic optomechanical Hamiltonian, effectively expressible after appropriate redefinition of operators as 
\begin{equation}
\label{S2-eq3}
\mathbb{H}_\text{lin}=-\hbar g(\hat{a}+\hat{a}^\dagger)(\hat{b}+\hat{b}^\dagger),
\end{equation}
where $g=g_0\bar{a}\propto g_0$ is the enhanced interaction rate. Further linearization of $\hat{b}$ does not discard any physics of standard optomechanics, and only can make the analysis a bit simpler if needed. This already proves to be quite sufficient for clear understanding of many basic optomechanical pheonmena in the quantum regime. Furthermore, the linearized optomechanical interaction (\ref{S2-eq3}) is now factorizable and therefore has already assumed the form of a simple product.

Further application of the same linearization method to quadratic and higher-order optomechanical interactions \cite{S2-Paper1,S2-Paper2}, leaves also behind only simple products quite similar to the linearized basic optomechanics (\ref{S2-eq3}) though with different interaction rates. This highlights the fact that any such linearization for quadratic, quartic and higher-order quantum interactions could be problematic in study of nonlinear effects. This is because of the underlying physics of such interactions having the third-order and the above, which is completely wiped out through this way of linearization process. That is known to be the major obstacle in any large amplitude analysis of quadratic interactions, since some of the nonlinear behavior should be somehow kept within the governing remaining equations. For example, the fourth-order cross-Kerr interaction $\hbar g \hat{a}^\dagger\hat{a}\hat{b}^\dagger\hat{b}$ discussed later in \S\ref{Section-11} shall equally reduce to a simple product having the form (\ref{S2-eq3}).

The use of second-order operators in optomechanics can preserve some information, which is reminiscent of the essentially nonlinear optomechanical interaction \cite{S2-Paper3}. This has been done and effects such as zero-point induced spring effect have been found. However, appropriate redefinition of third- and fourth-order operators \cite{S2-Paper3} ultimately allows full integrability as will be demonstrated later in the text. Without this technique of higher-order operator algebra, the exact solvability of a quantum optomechanical problem using available conventional tools is out of question.

The analysis of quadratic interaction in quantum optomechanics \cite{S2-Paper4,S2-Paper5,S2-Paper6,S2-Bruschi,S2-Quad} gets more complicated by the fact that there exists two mathematically distinct and different interaction types. The first being referred to as the standard quadratic interaction is the light-position quadratic term expressed as 
\begin{equation}
\label{S2-eq4}
\mathbb{H}_1=\hbar g_1\hat{a}^\dagger\hat{a}(\hat{b}+\hat{b}^\dagger)^2,
\end{equation}
while the non-standard quadratic interaction to be discussed in \S\ref{Section-3}, results from either momentum exchange between light and mirror or relativistic corrections, and can be written as \cite{S2-Paper1}
\begin{equation}
\label{S2-eq5}
\mathbb{H}_2=-\hbar g_2(\hat{a}-\hat{a}^\dagger)^2(\hat{b}-\hat{b}^\dagger)^2.
\end{equation}
Here, the strengths of standard and non-standard quadratic interactions are actually related as
\begin{equation}
\label{S2-eq6}
g_2=\frac{1}{4}\left(\frac{\pi^2}{3}+\frac{1}{4}\right)\left(\frac{\Omega}{\omega}\right)^2 g_1,
\end{equation}
where $\omega$ and $\Omega$ respectively represent the optical and mechanical resonant angular frequencies of the cavity. It has to be mentioned that (\ref{S2-eq5}) could be rewritten in many other equivalent forms such as 
\begin{equation}
\label{S2-eq7}
\mathbb{H}_2=-\hbar g_2(\hat{a}+\hat{a}^\dagger)^2(\hat{b}-\hat{b}^\dagger)^2,
\end{equation}
since an arbitrary $\frac{\pi}{2}$ phase shift in $\hat{a}$ interchanges (\ref{S2-eq7}) and (\ref{S2-eq5}), while leaving the standard quadratic (\ref{S2-eq4}) and even the basic standard (\ref{S2-eq2}) optomechanical interactions unaltered; obviously, there exist infinitely many equivalent forms due to gauge equivalence of such phase-shifted bosonic fields.

Hence, the overall optomechanical interaction, in a system composed of an optomechanical cavity and a laser drive, and up to the quartic \cite{S2-Paper1} order can be written after noticing the appropriate sign conventions and grouping terms having the same interaction order together as
\begin{eqnarray}
\label{S2-eq8}
\mathbb{H}&=&\mathbb{H}_\text{s}+\mathbb{H}_0+(\mathbb{H}_1+\mathbb{H}_2)+(\mathbb{H}_3+\mathbb{H}_4)+\mathbb{H}_\text{d}, \\ \nonumber
\mathbb{H}_\text{s}&=&\hbar\omega(\hat{a}^\dagger\hat{a}+\frac{1}{2})+\hbar\Omega(\hat{b}^\dagger\hat{b}+\frac{1}{2}),\\ \nonumber
\mathbb{H}_0&=&-\hbar g_0\hat{a}^\dagger\hat{a}(\hat{b}+\hat{b}^\dagger), \\ \nonumber
\mathbb{H}_1&=&+\hbar g_1\hat{a}^\dagger\hat{a}(\hat{b}+\hat{b}^\dagger)^2,\\ \nonumber
\mathbb{H}_2&=&-\hbar g_2(\hat{a}+\hat{a}^\dagger)^2(\hat{b}-\hat{b}^\dagger)^2,\\ \nonumber
\mathbb{H}_3&=&+\hbar g_3\hat{a}^\dagger\hat{a}(\hat{b}+\hat{b}^\dagger)^3,\\ \nonumber
\mathbb{H}_4&=&-\hbar g_4(\hat{a}+\hat{a}^\dagger)^2\times\left[(\hat{b}-\hat{b}^\dagger)^2(\hat{b}+\hat{b}^\dagger)+(\hat{b}+\hat{b}^\dagger)(\hat{b}-\hat{b}^\dagger)^2+(\hat{b}-\hat{b}^\dagger)(\hat{b}+\hat{b}^\dagger)(\hat{b}-\hat{b}^\dagger)\right],\\ \nonumber
\mathbb{H}_\text{d}&=&i\hbar(\alpha^\ast e^{i \tilde{\omega} t}\hat{a}-\alpha e^{-i \tilde{\omega} t}\hat{a}^\dagger).
\end{eqnarray}
Here, $\tilde{\omega}$ is the frequency of drive, which is mostly set in resonance with the cavity at $\tilde{\omega}=\omega$. There could be multiple drives with different frequencies and amplitudes present, which simply add up to the number of drive terms in the Hamiltonian (\ref{S2-eq8}). For instance, pumping the cavity with a second drive at frequency $\omega-\Omega$ on red-side band causes depletion of cavity from phonons, referred to side-band cooling. Typically, there exist more than one drive in a practical setup and the drive term should be rewritten as $\mathbb{H}_\text{d}=\sum_j \mathbb{H}_{\text{d},j}$ with $\mathbb{H}_{\text{d},j}=i\hbar(\alpha_j^\ast e^{i \tilde{\omega}_j t}\hat{a}-\alpha_j e^{-i \tilde{\omega}_j t}\hat{a}^\dagger)$. In $\mathbb{H}_\text{s}$ it is customary to drop the zero-point energy $E_\text{zp}=\frac{1}{2}\hbar\omega+\frac{1}{2}\hbar\Omega$ as it has no effect on the system behavior.

It is furthermore possible to write for an idealized one-dimensional cavity with parallel mirrors \cite{S2-Paper1}
\begin{equation}
\label{S2-eq9}
g_1=\frac{x_\text{zp}}{l}g_0,
\end{equation}
where $x_\text{zp}=\sqrt{\hbar/m\Omega}$ is the zero-point displacement of the cavity and $l$ is the cavity separation, with $m$ being the motion effective mass.
Also, we shall later observe in \S\ref{Section-3} that the standard $g_3$ and non-standard $g_4$ quartic interaction rates for such an ideal cavity are connected as
\begin{equation}
\label{S2-eq10}
g_4=\frac{1}{3\sqrt{2}}g_3\frac{1}{4}\left(\frac{\pi^2}{3}+\frac{1}{4}\right)\left(\frac{\Omega}{\omega}\right)^2=\frac{1}{3\sqrt{2}}\frac{g_2}{g_1}g_3=\frac{1}{3\sqrt{2}}\left(\frac{x_\text{zp}}{l}\right) g_2.
\end{equation} 
It is thus generally correct, that quadratic interactions need to be weaker than optomechanical interactions by a factor of $x_\text{zp}/l$. Similarly, quartic interactions are weaker than quadratic interactions by the same factor of $x_\text{zp}/l$. However, this argument does not necessarily hold for non-ideal cavities, where transverse geometry can influence $g_0$ independently. However, the relationships (\ref{S2-eq9},\ref{S2-eq10}) still give a sense of why quadratic and quartic interactions get progressively weak with the order increasing. 

While $g_0$ can be engineered to be made identically zero, $g_1$ does never vanish. This follows the fact that $g_0$ is proportional to the overlap integral of light upon mechanical displacement of mirror, and then an even distribution of light together with an odd mechanical mode leaves out zero interaction for the given mechanical mode. An odd-profiled mechanical mode will have zero overlap with an even profiled incident optical field in optomechanics, or with the microwave field in superconducting electromechanics. Membrane-in-the-middle set up and tuning to the first odd-shaped mechanical frequency can theoretically establish the condition to achieve $g_0=0$, in a rather convenient way.

There are mostly more than one single mechanical mode in an optomechanical cavity or equivalent, and for instance there exist an infinitely dense spectrum of interacting phonons with various polarizations and frequencies in Raman and Brillouin scattering phenomena. The mathematical study of multi-mode mechanical systems is essentially indifferent to the present formalism, however, the resulting photonic fluctuations of each individual optomechanical interaction should be integrated or summed up as appropriate over all mechanical modes. A similar argument also goes for the mechanical displacements, which must be expressed as supervision of individually interacting eigenmodes.  

Hence, quadratic effects always survive regardless of the existence of optomechanical interactions. Quartic effects also normally are expected to vanish when $g_0=0$, and that implies the ideal condition for observation of quadratic effects is to design the optomechanical system in such a way that $g_0=0$. Since, neither of the lower- or higher-order interactions than the quadratic would effectively exist. This is actually not as difficult as it seems. 

In practice, a bit of the optomechanical interaction $g_0$ may survive because of fabrication errors and that could be a source of inconvenience in quadratic measurements. Unless a practical tuning method to completely discard $g_0$ is available, it is safe to keep track of possible contributions from a small optomechanical term $\mathbb{H}_0$ in (\ref{S2-eq8}) along with the quadratic terms $\mathbb{H}_1$ and $\mathbb{H}_2$.

For the reasons discussed in the above, quartic effects will be neglected from now on, and we will focus on two separate cases: Basic optomechanics with the quadratic and higher-order interactions dropped; this is discussed in \S\ref{S2-Opto} Quadratic optomechanics with all other interactions dropped; this is discussed in \S\ref{S2-Quad} Corrections to the quadratic interactions as a result of non-vanishing $g_0$ is discussed in \S\ref{S2-OptoQuad}.

\subsection{Langevin Equations}

The analysis of a given Hamiltonian is done here through the well-known method of Langevin equations, which provide the equation of motion for operators in an open-system with input and output channels, interacting with a bath. This will cause a constant supply of fluctuating noise from each of the input channels to the system.

The corresponding Langevin \cite{S2-Noise0,S2-Noise1,S2-Noise2,S2-Noise3} equations to a given operator $\hat{z}$ of arbitrary order in an open system with multiple inputs are
\begin{eqnarray}
\label{S2-eq11}
\dot{\hat{z}}&=&-\frac{i}{\hbar}[\hat{z},\mathbb{H}]\\ \nonumber
&+&\sum_j\left\{-[\hat{z},\hat{x}_j^\dagger]\left(\frac{\gamma_j}{2}\hat{x}_j+\sqrt{\gamma_j}\hat{x}_{j,\text{in}}\right)+\left(\frac{\gamma_j}{2}\hat{x}_j^\dagger+\sqrt{\gamma_j}\hat{x}_{j,\text{in}}^\dagger\right)[\hat{z},\hat{x}_j]\right\},
\end{eqnarray}
\noindent
where $\hat{x}_j$ is a system operator with the decay/coupling rate $\gamma_j$ and input flux $\hat{x}_{j,\text{in}}$. The first term on the right-hand-side of the above $-\frac{i}{\hbar}[\hat{z},\mathbb{H}]$ is simply the contribution of Heisenberg equation of motion. However, in an open system with input/output the system energy can be exchanged with the exterior and thus the second term corresponding to the contribution of exchange at system input/output ports arises. Obviously, a system with no input/output port conserves energy with a Hermitian Hamiltonian and the Langevin's equation reduces to the Heisenberg's equation of motion.

The drives can be assigned to and properly included in the input terms  $\hat{x}_{j,\text{in}}$ too, but normally this approach is avoided and $\hat{x}_{j,\text{in}}$ mostly denote random noise. Basic properties of noise operators within a Gaussian white approximation are already well known and are  discussed in much detail later in \S\ref{Section-4}, which include autocorrelation and commutation relationships \cite{S2-Noise0,S2-Noise1,S2-Noise2,S2-Noise3}. Some further properties of noise operator are to be discussed later in \S\ref{Section-11}.  For optomechanical problems, the summation can be run over the bath operators $\hat{a}$ and $\hat{b}$. This method also enables one to construct the noise term of any higher-order operator $\hat{z}$ in a straightforward manner. This gives
\begin{equation}
\label{S2-eq12}
\sqrt{\gamma}\hat{z}_\text{in}=\sum_j\sqrt{\gamma_j}\left\{-[\hat{z},\hat{x}_j^\dagger]\hat{x}_{j,\text{in}}+\hat{x}_{j,\text{in}}^\dagger[\hat{z},\hat{x}_j]\right\},
\end{equation}
in which $\gamma=\sum_j n_j\gamma_j$ is the effective decay rate of the higher-order operator $\hat{z}$ with $n_j\in\mathscr{N}$ being some constants which typically are natural numbers or positive integers.

\label{S2-Opto}

In standard optomechanics with only one external drive term, where no quardratic and higher-order terms exist, the Hamiltonian simply is
\begin{equation}
\label{S2-eq13}
\mathbb{H}=\hbar\omega\hat{a}^\dagger\hat{a}+\hbar\Omega\hat{b}^\dagger\hat{b}-\hbar g_0\hat{a}^\dagger\hat{a}(\hat{b}+\hat{b}^\dagger)-\hbar(\alpha e^{i \tilde{\omega} t}\hat{a}+\alpha^\ast e^{-i \tilde{\omega} t}\hat{a}^\dagger).
\end{equation}
Here, the drive makes the Hamiltonian time-dependent at the frequency of $\tilde{\omega}$. For the usual case of resonant drive with $\tilde{\omega}=\omega$, the time-dependence of the Hamiltonian can be removed by transformation to the rotating frame as $\hat{a}\to e^{-i\omega t}\hat{a}(t)$, which makes the operator $\hat{a}(t)$ explicitly time-dependent, after removing a fast oscillation part. However, for the non-resonant drives also this is equally helpful, and nonetheless will not change the normal commutation relationship $[\hat{a}(t),\hat{a}^\dagger(t)]=1$. This will however change the time-derivative of the transformed operator $\hat{a}(t)$ as
\begin{equation}
\label{S2-eq14}
\dot{\hat{a}}=\frac{d}{dt}\left[e^{-i\tilde{\omega} t}\hat{a}(t)\right]=e^{-i\tilde{\omega} t}\left[\dot{\hat{a}}(t)-i\tilde{\omega}\hat{a}(t)\right].
\end{equation}
While this transformation to the rotating frame is helpful with the standard interactions $\mathbb{H}_0$, $\mathbb{H}_1$, and $\mathbb{H}_3$, the algebraic form of non-standard interactions $\mathbb{H}_2$ and $\mathbb{H}_4$ excludes usefulness of this transformation. However, the non-standard interactions become only significant when $\omega$ is not far larger than $\Omega$, hence any further notion of fast oscillating part is point-less.

\subsection{Linear Optomechanics}

In standard optomechanics, the Langevin equations have to be set up for the first-order basis $\{A\}^\text{T}=\{\hat{a},\hat{a}^\dagger,\hat{b},\hat{b}^\dagger\}$. This chosen basis is the lowest order possible of basis and the obvious choice for analysis of optomechanics. However, the Hamiltonian $\mathbb{H}_0$ is still nonlinear and the resulting Langevin equations will remain nonlinear. Calculation of (\ref{S2-eq13}) for each of bath operators by setting $\hat{x}\in\{A\}$ gives four coupled equations. Henceforth, the exact Langevin equations while using (\ref{S2-eq14}) are
\begin{eqnarray}
\label{S2-eq15}
\dot{\hat{a}}&=&\left(i\Delta-\frac{1}{2}\kappa\right) \hat{a}+i g_0\hat{a}(\hat{b}+\hat{b}^\dagger)-\alpha-\sqrt{\kappa}\hat{a}_\text{in}, \\ \nonumber
\dot{\hat{a}}^\dagger&=&\left(-i\Delta-\frac{1}{2}\kappa\right) \hat{a}^\dagger-i g_0\hat{a}^\dagger(\hat{b}+\hat{b}^\dagger)-\alpha^\ast-\sqrt{\kappa}\hat{a}^\dagger_\text{in}, \\ \nonumber
\dot{\hat{b}}&=&\left(-i\Omega-\frac{1}{2}\Gamma\right) \hat{b}+i g_0\hat{a}^\dagger\hat{a}-\sqrt{\Gamma}\hat{b}_\text{in}, \\ \nonumber
\dot{\hat{b}}^\dagger&=&\left(i\Omega-\frac{1}{2}\Gamma\right) \hat{b}^\dagger-i g_0\hat{a}^\dagger\hat{a}-\sqrt{\Gamma}\hat{b}^\dagger_\text{in}.
\end{eqnarray}
where $\Delta=\tilde{\omega}-\omega$ is the optical detuning from drive frequency, $\kappa$ and $\Gamma$ are respectively the decay rates of $\hat{a}$ and $\hat{b}$, and the explicit time-dependence of operators after the transformation to the rotating frame has not been shown. Positively (negatively) detuned frequencies with $\Delta>0$ ($\Delta<0$) shall thus correspond to physically smaller or red-shifted (physically larger or blue-shifted) frequencies.

The terms $\pm i g_0 \hat{a}^\dagger\hat{a}$ in the third and fourth equations for $\hat{b}$ and $\hat{b}^\dagger$ are called backaction terms, and become only negligible in the limit of very weak coupling and weak optical pump. Similarly, the terms $i g_0 \hat{a}(\hat{b}+\hat{b}^\dagger)$ and $i g_0 \hat{a}^\dagger(\hat{b}+\hat{b}^\dagger)$ in the first and second equations for $\hat{a}$ and $\hat{a}^\dagger$ provide the main modulation of the optical field and formation of side-bands. The peculiar optomechanical interaction between optical field and mechanical field is maintained by simultaneous presence both of the backaction and nonlinear terms. For some purposes, the backaction is dropped but this will cause loss of significant information regarding optomechanical spring effect. Also, the modulation terms have nearly always been linearized so far in the existing reported literature, unless a numerical simulation of nonlinear effects has been carried out.

The set of equations (\ref{S2-eq15}) can be rewritten in the nonlinear matrix form
\begin{equation}
\label{S2-eq15a}
\frac{d}{dt}\{A\}=[\hat{\textbf{M}}]\{A\}-\sqrt{[\hat{\gamma}]}\{A_\text{in}\}-\{A_\text{d}\},
\end{equation}
in which the coefficients matrix $[\hat{\textbf{M}}]$ is not independent of $\{A\}$. This is why (\ref{S2-eq15a}) is not readily integrable on the first-order basis $\{A\}$, unless linearized first. Furthermore, $\{A_\text{d}\}^\text{T}=\{\alpha,\alpha^\ast,0,0\}$ is the drive input vector. It is easy to set up the elements of $[\hat{\textbf{M}}]$ by inspection from (\ref{S2-eq15}), and there is of course no unique way to write it down because of the terms such as $\hat{a}\hat{b}$, whose either of their first or second operators could be put within the coefficients matrix. 

The resulting Langevin equations for the first-order basis $\{A\}$ is nonlinear because of the presence of second-order operators $\hat{a}\hat{b}$ and $\hat{a}\hat{b}^\dagger$ in the first equation, $\hat{a}^\dagger\hat{b}$ and $\hat{a}^\dagger\hat{b}^\dagger$ in the second equation, and $\hat{a}^\dagger\hat{a}$ in the last two equations. These terms actually do have physical significance, and for instance $\hat{a}\hat{b}^\dagger$ (its conjugate) correspond to the creation (annihilation) of a phonon and annihilation (creation) of a photon. This is the red or Stokes process which is responsible for formation of first-order side-band at the expected detuned frequency $\Delta+\Omega$, as well as higher-order side-bands at the expected frequencies $\Delta+j\Omega$. Similarly, $\hat{a}\hat{b}$ (its conjugate) correspond to the annihilation (creation) of a phonon and annihilation (creation) of a photon. This is the blue or anti-Stokes process which is responsible for formation of first-order side-band at the expected detuned frequency $\Delta-\Omega$, as well as higher-order side-bands at the expected frequencies $\Delta-j\Omega$. Obviously, the second- and higher-order red/blue side-bands disappear in the fully linearized approximation. A careful nonlinear analysis of this Hamiltonian shall reveal that these side-bands are not equally spaced from the pump, meaning that there exists a non-zero side-band inequivalence. 

The first step to linearize (\ref{S2-eq15}) is to do the replacement $\hat{a}\to\bar{a}+\hat{a}$, where $\bar{a}$ is the average value of the field operator. We here may suppose that  $\bar{a}\in\mathscr{R}$ is taken to be real-valued, as its corresponding phase can be adjusted in the pump $\alpha$. Then, we can linearize the photon number $\hat{a}^\dagger\hat{a}\to\bar{a}^2+\bar{a}(\hat{a}+\hat{a}^\dagger)$, which drives the radiation pressure term with the time average $\bar{n}=\bar{a}^2$. The presence of an average radiation pressure term, puts a constant displacement upon mirror $\bar{b}$ around which the mirror movements fluctuate. Hence, we arrive at a similar linearization of $\hat{b}\to\bar{b}+\hat{b}$ and thus $\hat{a}\hat{b}\to\bar{a}\hat{b}+\bar{b}\hat{a}$.

Once these replacements are plugged in (\ref{S2-eq15}), the Langevin equations are linearized and the static expressions in terms of average values can be separated to yield
\begin{eqnarray}
\label{S2-eq16}
\left(i\Delta-\frac{1}{2}\kappa\right) \bar{a}+i 2g_0\Re[\bar{b}]\bar{a}=\alpha, \\ \nonumber\left(-i\Omega-\frac{1}{2}\Gamma\right) \bar{b}=-i g_0\bar{n}.
\end{eqnarray}
These equations can be combined to obtain the real-valued $\bar{n}=\bar{a}^2$ in terms of $|\alpha|$ through solution of a third-order algebraic equation, given as
\begin{equation}
\label{S2-eq17}
\left(\Delta+ \frac{4g_0^2\Omega}{\Omega^2+\frac{1}{4}\Gamma^2}\bar{n}\right)^2\bar{n} +\frac{\kappa^2}{4}\bar{n}=|\alpha|^2.
\end{equation}
The linearized $4\times 4$ Langevin equations around the average values now take the form
\begin{eqnarray}
\label{S2-eq18}
\dot{\hat{a}}&=&\left[i(\Delta+f)-\frac{1}{2}\kappa\right] \hat{a}+i g(\hat{b}+\hat{b}^\dagger)-\sqrt{\kappa}\hat{a}_\text{in}, \\ \nonumber
\dot{\hat{a}}^\dagger&=&\left[-i(\Delta+f)-\frac{1}{2}\kappa\right] \hat{a}^\dagger-i g(\hat{b}+\hat{b}^\dagger)-\sqrt{\kappa}\hat{a}^\dagger_\text{in}, \\ \nonumber
\dot{\hat{b}}&=&\left(-i\Omega-\frac{1}{2}\Gamma\right) \hat{b}+i g(\hat{a}+\hat{a}^\dagger)-\sqrt{\Gamma}\hat{b}_\text{in}, \\ \nonumber
\dot{\hat{b}}^\dagger&=&\left(i\Omega-\frac{1}{2}\Gamma\right) \hat{b}^\dagger-i g(\hat{a}+\hat{a}^\dagger)-\sqrt{\Gamma}\hat{b}^\dagger_\text{in}.
\end{eqnarray}
in which $g=g_0\bar{a}$ and $f=2g_0\Re[\bar{b}]=2g_0^2\Omega\bar{n}/(\Omega^2+\frac{1}{4}\Gamma^2)$. The analysis of the system of equations (\ref{S2-eq18}) is extensively discussed elsewhere  \cite{S2-1,S2-2,S2-3,S2-4,S2-6}. It is convenient, nevertheless, to put the integrable system (\ref{S2-eq18}) in the generic matrix form and rewrite the linearized Langevin equation as
\begin{equation}
\label{S2-eq19}
\frac{d}{dt}\{A\}=[\textbf{M}]\{A\}-\sqrt{[\gamma]}\{A_\text{in}\},
\end{equation}
with definitions
\begin{eqnarray}
\label{S2-eq20}
[\textbf{M}]&=&\left[
\begin{array}{cccc}
i(\Delta+f)-\frac{1}{2}\kappa & 0 & ig & ig \\ 
0 & -i(\Delta+f)-\frac{1}{2}\kappa & -ig & -ig \\
ig & ig & -(i\Omega+\frac{1}{2}\Gamma) & 0 \\
-ig & -ig & 0 & i\Omega-\frac{1}{2}\Gamma 
\end{array}
\right], \nonumber \\ \nonumber
[\gamma]&=&\text{Diag}\{\kappa,\kappa,\Gamma,\Gamma\}, \\ 
\{A_\text{in}\}^\text{T}&=&\{\hat{a}_\text{in},\hat{a}_\text{in}^\dagger,\hat{b}_\text{in},\hat{b}_\text{in}^\dagger \}.
\end{eqnarray}
Now, input-output relations \cite{S2-Noise1,S2-Noise2} can be used with the definition of the scattering matrices in Fourier domain with probe detuning $w=\omega_\text{p}-\omega$ to yield
\begin{eqnarray}
\label{S2-eq21}
\{A_\text{out}(w)\}&=&\{A_\text{in}(w)\}-\sqrt{[\gamma]}\{A(w)\}, \\ \nonumber
\{A_\text{out}(w)\}&=&[\textbf{S}(w)]\{A_\text{in}(w)\}, \\ \nonumber
[\textbf{S}(w)]&=&[\textbf{I}]-\sqrt{[\gamma]}\left(iw[\textbf{I}]-[\textbf{M}]\right)^{-1}\sqrt{[\gamma]}.
\end{eqnarray}
It is to be kept in mind that it is mostly the spectral density of $\hat{a}$ denoted by $\mathcal{S}_{AA}(w)$ which can be measured in an either heterodyne or homodyne experiment. However, the spectral density of input is known, which enables one to obtain the spectral density of output in terms of the input using the relation
\begin{equation}
\label{S2-eq22}
\mathcal{S}_{AA}(w)=\sum_{j=1}^{4}|S_{1j}(w)|^2 \mathcal{S}_{j,\text{in}}(w),
\end{equation}
where $S_{1j}(w)$ are elements of the first row of the scattering matrix $[\textbf{S}(w)]$ and 
\begin{eqnarray}
\label{S2-eq23}
\{\mathcal{S}_{\text{in}}(w>0)\}^\text{T}&=&\{1,0,m_{\rm th}+1,m_{\rm th}\}, \\ \nonumber
\{\mathcal{S}_{\text{in}}(w<0)\}^\text{T}&=&\{0,1,m_{\rm th},m_{\rm th}+1\},
\end{eqnarray}
is the spectral density vector of noise channel due to photonic and phononic baths. We here notice that $\mathscr{F}\{\hat{a}(t)\}(w)=\mathscr{F}\{\hat{a}^\dagger(t)\}^\ast(-w)$, and so on. Being different from the cavity phonon number or coherent phonon population $\bar{m}$, the quantity $m_{\rm th}$ represents the thermal population of phonons, which could be either determined by Bose-Einstein statistics of the given mechanical mode under thermodynamic equilibrium as $m_{\rm th}=1/[\exp(\hbar\Omega/k_B T)-1]$ with $k_B$ and $T$ respectively being the Boltzmann's constant and absolute temperature, or constrained by some cooling process. The coherent phonon population $\bar{m}=\braket{\hat{b}^\dagger\hat{b}}$ corresponds to the average number of phonons driven coherently by the optomechanical interaction. The correct expression for $\bar{m}$ can be determined by nonlinear higher-operator analysis, which shall be presented later in \ref{Section-8}. It shall be noticed that the so-called mean-field approximation $\bar{m}\approx\braket{\hat{b}}^\ast\braket{\hat{b}}=|\bar{b}|^2$ could be actually off the actual value of $\bar{m}=\braket{\hat{b}^\dagger\hat{b}}$ up to a factor of 2.

\subsection{Second-Order Optomechanics}
The second-order equations of optomechanics are obtained by using the mixed first- and second-order basis $\{A\}^\text{T}=\{\hat{a},\hat{b},\hat{a}\hat{b},\hat{a}\hat{b}^\dagger,\hat{n},\hat{c}\}$ \cite{S2-Paper3}, in which we have the definition $\hat{c}=\frac{1}{2}\hat{a}^2$ \cite{S2-Paper1,S2-Paper2}. It is easy to observe that this set forms a closed basis, since the commutator of every pair of operators belonging to this basis results from a linear combination of the operators therein. The non-zero commutators related to this basis are
\begin{eqnarray}
\label{S2-eq24}
[\hat{c},\hat{n}]&=&2\hat{c}, \\ \nonumber
[\hat{a},\hat{n}]&=&\hat{a}, \\ \nonumber
[\hat{b},\hat{a}\hat{b}^\dagger]&=&\hat{a},\\ \nonumber
[\hat{a}\hat{b},\hat{n}]&=&\hat{a}\hat{b},\\ \nonumber
[\hat{a}\hat{b}^\dagger,\hat{n}]&=&\hat{a}\hat{b}^\dagger,\\ \nonumber
[\hat{a}\hat{b},\hat{a}\hat{b}^\dagger]&=&2\hat{c}.
\end{eqnarray}
The resulting $6\times 6$ Langevin equations should be constructed based on the Hamiltonian (\ref{S2-eq13}) and this has been extensively discussed later in \S\ref{Section-8}. 

While these are nonlinearly coupled, further linearization of the system shall decouple three equations out of six as it will be shown shortly, leaving the optomechanical system expressible in terms of the reduced basis $\{A\}^\text{T}=\{\hat{a},\hat{a}\hat{b},\hat{a}\hat{b}^\dagger\}$, which is now not any longer closed under commutation. Putting into the standard form (\ref{S2-eq19}) and before linearization, the $3\times 3$ coefficients operator matrix $[\hat{\textbf{M}}]$ for the set of Langevin equations of the reduced basis looks like
\begin{equation}
\label{S2-eq25}
[\hat{\textbf{M}}]=
\left[ 
\begin{array}{ccc}
i\Delta-\frac{\kappa }{2} & ig_0 & ig_0 \\ 
ig_0\left(\hat{m}+\hat{n}+1\right)  & -i\left(\Omega-\Delta-g_0 \hat{b}\right)-\frac{\gamma }{2} & 0  \\ 
ig_0\left(\hat{m}-\hat{n}\right) &  0 & i\left(\Omega+\Delta+g_0\hat{b}^\dagger\right)-\frac{\gamma }{2} 
\end{array}
\right],
\end{equation}
where $\gamma=\kappa+\Gamma$ is the total optomechanical decay rate, with the decay rate matrix and input noise and drive vectors defined as
\begin{eqnarray}
\label{S2-eq26}
[\hat{\gamma}]&=&\left[
\begin{array}{ccc}
\kappa & 0 & 0\\
\kappa\hat{b} & \Gamma\hat{a} & 0\\
\kappa\hat{b} & 0 & \Gamma\hat{a}
\end{array}
\right],\\ \nonumber
\{A_\text{in}\}^\text{T}&=&\{\hat{a}_\text{in},\hat{b}_\text{in},\hat{b}^\dagger_\text{in}\}, \\ \nonumber
\{A_\text{d}\}^\text{T}&=&\{\alpha,\alpha\hat{b},\alpha\hat{b}^\dagger\}.
\end{eqnarray}
Here, the noise terms are assumed to be approximated as linearizations from the exact multiplicative forms, which appear naturally in the higher-order operator formalism. Wherever multiplicative noises appear in formalisms, analytical solutions are extremely rare and extraordinarily difficult to construct. It has been observed through numerous numerical experiments that linearized noise terms provide sufficiently accurate descriptions of most studied phenomena. This will be elaborated in further details later in \S\ref{Section-8}. Increased precision can be still obtained by using convolution integrals in the frequency domain. Moreover, the coefficients matrix (\ref{S2-eq25}) is still a function of photon $\hat{n}$ and phonon $\hat{m}$ number operators, which makes the exact integration impossible at this stage. Approximation of these operators by their respective mean values as the intracavity photon number $\bar{n}$ and coherent phonon population $\bar{m}$ is actually equivalent to linearization of optomechanical equations of motion at the second-order, instead of the first-order in the full linearization scheme. Furthermore, it allows us to obtain a fairly accurate expression for the coherent phonon population $\bar{m}$, which is otherwise unknown. 

After linearization of (\ref{S2-eq25}) and (\ref{S2-eq26}), it is possible to show that the steady state values of $\bar{b}$ and $\bar{a}=\sqrt{\bar{n}}$ must exactly satisfy (\ref{S2-eq16}) and (\ref{S2-eq17}) again. The first Langevin equation is already linear. However, linearization of second and third equations at this stage gives
\begin{eqnarray}
\label{S2-eq27}
[\textbf{M}]&=&
\left[ 
\begin{array}{ccc}
i\Delta-\frac{\kappa }{2} & ig_0 & ig_0 \\ 
ig_0\left(\bar{m}+\bar{n}+1\right)  & -i\left(\Omega-\Delta\right)-\frac{\gamma }{2} & 0  \\ 
ig_0\left(\bar{m}-\bar{n}\right) &  0 & i\left(\Omega+\Delta\right)-\frac{\gamma }{2}  
\end{array}
\right],\\ \nonumber
[\gamma]&=&\left[
\begin{array}{ccc}
\kappa & 0 & 0\\
\kappa|\bar{b}| & \Gamma\sqrt{\bar{n}} & 0\\
\kappa|\bar{b}| & 0 & \Gamma\sqrt{\bar{n}}
\end{array}
\right], \\ \nonumber
\{A_\text{d}\}^\text{T}&=&\{\alpha,0,0\}.
\end{eqnarray}

\subsection{Fourth-Order Optomechanics}\label{S2-Fourth}

It is possible to select a third-order closed basis of operators \cite{S2-Paper3}, which can surprisingly lead to a fully linear and therefore integrable system of equations. The particular choice of $\{A\}^\text{T}=\{\hat{n}^2,\hat{n}\hat{b},\hat{n}\hat{b}^\dagger\}$ here is composed of a fourth-order operator $\hat{N}=\hat{n}^2$ and a third-order operator $\hat{B}=\hat{n}\hat{b}$ and its conjugate $\hat{B}^\dagger$. This system will lead to a linearized system of Langevin equations which is actually decoupled for $\hat{B}$ and $\hat{B}^\dagger$. This particular choice, referred to as the minimal basis, reduces to a simple $2\times 2$ Langevin system (\ref{S2-eq15}) with
\begin{eqnarray}
\label{S2-eq29}
\{A\}^\text{T}&=&\{\hat{N},\hat{B}\}\\ \nonumber
[\textbf{M}]&=&\left[
\begin{array}{cc}
-2\kappa & 0 \\
ig_0 & -i\Omega-\frac{\gamma}{2}
\end{array}
\right], \\ \nonumber
[\gamma]&=&\text{Diag}\{2\kappa,\gamma\}, \\ \nonumber
\{A_\text{in}\}^\text{T}&=&\{\hat{N}_\text{in},\hat{B}_\text{in}\} = \{\hat{n}\hat{a}^\dagger\hat{a}_\text{in}+\hat{a}_\text{in}^\dagger\hat{a}\hat{n},\sqrt{2\kappa}\hat{b}\hat{n}_\text{in}+\sqrt{\Gamma}\hat{n}\hat{b}_\text{in}\},\\ \nonumber
\{A_\text{d}\}^\text{T}&=&\{\alpha\hat{n}\hat{a}^\dagger+\alpha^\ast\hat{a}\hat{n},\alpha\hat{b}\hat{a}^\dagger+\alpha^\ast\hat{b}\hat{a}\}.
\end{eqnarray}
One should notice  that the noise terms of higher-order operators are no longer Gaussian white. As a matter of fact, the new noise operators such as $\hat{n}\hat{a}^\dagger\hat{a}_\text{in}$ and $\hat{a}_\text{in}^\dagger\hat{a}\hat{n}$ and so on do not necessarily satisfy Gaussian white properties of original noise operators such as  $\hat{a}_\text{in}$ and $\hat{a}_\text{in}^\dagger$. Meanwhile in practice as it has been demonstrated for the highly nonlinear case of cross-Kerr interaction in \S\ref{Section-11}, there exist practical approximations to the resulting spectral densities of these noise terms.

It is now fairly easy and quite straightforward to construct the exact and explicit solution to the system (\ref{S2-eq29}) as
\begin{eqnarray}
\label{S2-eq30}
\hat{N}(t)&=&\hat{N}(0)e^{-2\kappa t}-2\sqrt{\kappa}\int_{0}^{t}e^{-2\kappa (t-\tau)}\hat{N}_\text{in}(\tau)d\tau, \\ \nonumber
\hat{B}(t)&=&\hat{B}(0)e^{-(i\Omega+\frac{\gamma}{2}) t}-\int_{0}^{t}e^{-(i\Omega+\frac{\gamma}{2}) (t-\tau)}\left[ig_0\hat{N}(\tau)+\sqrt{\gamma}\hat{B}_\text{in}(\tau)\right]d\tau.
\end{eqnarray}
If the multiplicative noise terms \cite{S2-Noise4} and drive vector in (\ref{S2-eq29}) can be approximated in the lowest order as Gaussian white noises, then we obtain
\begin{eqnarray}
\label{S2-eq31}
\{A_\text{in}\}^\text{T}&=&\left\{\frac{\sqrt{\bar{n}}\bar{n}}{2}\left(\hat{a}_\text{in}+\hat{a}_\text{in}^\dagger\right),\sqrt{\frac{\kappa}{\gamma}\bar{n}}\bar{b}\left(\hat{a}_\text{in}+\hat{a}_\text{in}^\dagger\right)+\sqrt{\frac{\Gamma}{\gamma}}\bar{n}\hat{b}_\text{in}\right\}, \\ \nonumber
\{A_\text{d}\}^\text{T}&=&2\sqrt{\bar{n}}\{\bar{n},\bar{b}\}\Re[\alpha].
\end{eqnarray}
While thus the optomechanical Hamiltonian can be exactly integrated, the peculiar form of (\ref{S2-eq30}) might not be very useful in practice since it presents the time evolution of higher-order operators, which does not directly relate to any of measureable quantities such as optical spectrum. Furthermore, the noise terms are multiplicative and require either approximations or recursive calculations. Anyhow, for a time-domain numerical experiment, the solution offered in (\ref{S2-eq30}) is quite practical.

\label{S2-Quad}

Solution of quadratic optomechanics needs a different basis of higher-order operators \cite{S2-Paper2}. The basis with minimum dimension required to deal with the Hamiltonain (\ref{S2-eq8}) is \cite{S2-Paper4}
\begin{equation}
\label{S2-eq32}
\{A\}^\text{T}=\{\hat{c},\hat{c}^\dagger,\hat{n},\hat{d},\hat{d}^\dagger,\hat{m}\},
\end{equation}
in which $\hat{d}=\frac{1}{2}\hat{b}^2$ and $\hat{m}=\hat{b}^\dagger\hat{b}$. This basis is also closed under commutation and satisfies a Lie algebra. To see this, it is sufficient to verify the commutators $[\hat{c},\hat{n}]=2\hat{c}$ from (\ref{S2-eq24}), which gives $[\hat{n},\hat{c}^\dagger]=2\hat{c}^\dagger$, and $[\hat{c},\hat{c}^\dagger]=\hat{n}+\frac{1}{2}$ for photons, and in a similar manner $[\hat{d},\hat{d}^\dagger]=\hat{m}+\frac{1}{2}$,
$[\hat{d},\hat{m}]=2\hat{d}$, and $[\hat{m},\hat{d}^\dagger]=2\hat{d}^\dagger$ for phonons. All other commutators within the above basis are zero.

First, the coefficients of Langevin equations (\ref{S2-eq16}) can be partitioned as
\begin{equation}
\label{S2-eq33}
[\hat{\textbf{M}}]=\left[
\begin{array}{c|c}
\hat{\textbf{M}}_\text{aa} & \hat{\textbf{M}}_\text{ab} \\ \hline
\hat{\textbf{M}}_\text{ba} & \hat{\textbf{M}}_\text{bb}
\end{array}\right],
\end{equation}
with indices a and b referring to photons and phonons. The $3\times 3$ partitions are now given by 
\begin{eqnarray}
\label{S2-eq34}
[\hat{\textbf{M}}_\text{aa}]&=&\left[
\begin{array}{ccc}
-2i\omega-\kappa & 0 & i\frac{1}{2} g_2(\hat{f}-\hat{m})\\
0 & 2i\omega-\kappa & -i\frac{1}{2} g_2(\hat{f}-\hat{m})\\
-ig_2(\hat{f}-\hat{m}) & ig_2(\hat{f}-\hat{m}) & -\kappa
\end{array}
\right] , \\ \nonumber
[\hat{\textbf{M}}_\text{ab}]&=&ig_2 \left[
\begin{array}{ccc}
-\beta_-(\hat{c}-\hat{c}^\dagger) & -\beta_-(\hat{c}-\hat{c}^\dagger) & \beta_+(\hat{c}-\hat{c}^\dagger) \\
\beta_-(\hat{c}-\hat{c}^\dagger) & \beta_-(\hat{c}-\hat{c}^\dagger) & -\beta_+(\hat{c}-\hat{c}^\dagger) \\
0 & 0 & 0
\end{array}
\right], \\ \nonumber
[\hat{\textbf{M}}_\text{ba}]&=&\frac{ig_2}{2}\left[
\begin{array}{ccc}
\hat{m}-2\hat{d}+\frac{1}{2} & \hat{m}-2\hat{d}+\frac{1}{2} & \hat{m}-2\hat{d}+\frac{1}{2} \\
-\hat{m}+2\hat{d}^\dagger-\frac{1}{2} & -\hat{m}+2\hat{d}^\dagger-\frac{1}{2} & -\hat{m}+2\hat{d}^\dagger-\frac{1}{2} \\
0 & 0 & 0
\end{array}
\right], \\ \nonumber
[\hat{\textbf{M}}_\text{bb}]&=&\left[
\begin{array}{ccc}
-2i\Omega-\Gamma & 0 & -i\frac{1}{2}g_2\hat{n}\\
0 & 2i\Omega-\Gamma & i\frac{1}{2}g_2\hat{n}\\
-ig_2(\hat{e}-\beta_-\hat{n})& ig_2(\hat{e}-\beta_-\hat{n}) & -\Gamma
\end{array}
\right].
\end{eqnarray}
Here, we have adopted the notions  $\beta_\pm=(g_1/g_2)\pm 1$ as well as $\hat{e}=\hat{c}+\hat{c}^\dagger$  and $\hat{f}=\hat{d}+\hat{d}^\dagger$. It is to be noticed that there is no unique way to partition the above nonlinear system. Meanwhile, the decay matrix, multiplicative input noise terms, and drive vector are
\begin{eqnarray}
\label{S2-eq35}
[\gamma]&=&\text{Diag}\{\kappa,\kappa,\kappa,\Gamma,\Gamma,\Gamma \}, \\ \nonumber
\{A_\text{in}\}^\text{T}&=&\{\hat{a}\hat{a}_\text{in},\hat{a}_\text{in}^\dagger\hat{a}^\dagger,\hat{a}^\dagger\hat{a}_\text{in}+\hat{a}_\text{in}^\dagger\hat{a},\hat{b}\hat{b}_\text{in},\hat{b}_\text{in}^\dagger\hat{b}^\dagger,\hat{b}^\dagger\hat{b}_\text{in}+\hat{b}_\text{in}^\dagger\hat{b} \}, \\ \nonumber
\{A_\text{d}\}^\text{T}&=&\{i\alpha e^{-i\omega t}\hat{a},i\alpha^\ast e^{i\omega t}\hat{a}^\dagger, \alpha e^{-i\omega t}\hat{a}+\alpha^\ast e^{i\omega t}\hat{a}^\dagger,0,0,0\}.
\end{eqnarray}
We also here have to notice that the particular form of quadratic Hamiltonian $\mathbb{H}_2$ excludes any usefulness of transformation to rotating frame of coordinate, and the optical frequency here is measured absolutely. Linearization of the corresponding Langevin equations around the mean values gives rise to the partitions \cite{S2-Paper4}
\begin{eqnarray}
\label{S2-eq36}
[\textbf{M}_\text{aa}]&=&\left[
\begin{array}{ccc}
i2(\omega-g_2\beta_+\bar{m})-\kappa & 0 & -i\frac{1}{2}g_2\bar{m} \\ 
0 & -i2(\omega-g_2\beta_+\bar{m})-\kappa & i\frac{1}{2}g_2\bar{m} \\ 
ig_2\bar{m} & -ig_2\bar{m} & -\kappa 
\end{array}
\right] , \\ \nonumber
[\textbf{M}_\text{ab}]&=&i\frac{g_2}{2}\left(\bar{n}+\frac{1}{2}\right)\left[
\begin{array}{ccc}
1 & 1 & -1 \\ 
-1 & -1 & 1 \\ 
0 & 0 & 0 \end{array}
\right], \\ \nonumber
[\textbf{M}_\text{ba}]&=&ig_2\left(\bar{m}+\frac{1}{2}\right)\left[
\begin{array}{ccc}
1 & 1 & -\beta_- \\
-1 & -1 & \beta_- \\
0 & 0 & 0 
\end{array}
\right], \\ \nonumber
[\textbf{M}_\text{bb}]&=&\left[
\begin{array}{ccc}
-i2(\Omega+g_2\beta_+\bar{n})-\Gamma & 0 & -ig_2\beta_-\bar{n} \\
0 & i2(\Omega+g_2\beta_+\bar{n})-\Gamma & ig_2\beta_-\bar{n} \\
i2g_2\beta_-\bar{n} & -i2g_2\beta_-\bar{n} & -\Gamma
\end{array}
\right].
\end{eqnarray}
Here, photon and phonon frequencies already have shifted to the new values $\omega+g_1-2g_2\to\omega$, and $\Omega-2g_2\to\Omega$, and we have assumed that the optical drive is actually resonant with the newly shifted value. The input vector and decay matrix have to be reformatted as \cite{S2-Paper4}
\begin{eqnarray}
\label{S2-eq37}
\{A_\text{in}\}^\text{T}&=&\left\{
\hat{a}_\text{in},
\hat{a}_\text{in}^\dagger, 
\frac{1}{2}(\hat{a}_\text{in}+\hat{a}_\text{in}^\dagger), 
\hat{b}_\text{in},
\hat{b}_\text{in}^\dagger,
\frac{1}{2}(\hat{b}_\text{in}+\hat{b}_\text{in}^\dagger) 
\right\}, \\ \nonumber
[\gamma]&=&\text{Diag}\left[\bar{n}\kappa,\bar{n}\kappa,4\bar{n}\kappa,2|\bar{d}|\Gamma,2|\bar{d}|\Gamma,4|\bar{d}|\Gamma\right].
\end{eqnarray}
Finally, the mean photon and phonon values are to be found from the nonlinearly coupled algebraic equations \cite{S2-Paper4}
\begin{eqnarray}
\label{S2-eq38}
4|\alpha|^2&=&(g_1+g_2)^2\bar{m}^2\bar{n}, \\ \nonumber
4\bar{n}|\alpha|^2&=&g_1^2(\bar{m}^2-\bar{m}),
\end{eqnarray}
together with $4|\bar{d}|^2=\bar{m}^2-\bar{m}$, required to complete the set of equations (\ref{S2-eq37}). Combining these two equations leads for the near resonant case to the fourth-order algebraic equation in terms of $\sqrt{\bar{n}}$ as 
\begin{equation}
\label{S2-eq37a}
2|\alpha|\bar{n}^2+\frac{g_1^2}{g_1+g_2}\sqrt{\bar{n}}=g_1^2\frac{2|\alpha|}{(g_1+g_2)^2}.
\end{equation}
This equation reveals that in the limit of very strong near-resonant pumping, the cavity photon population gets saturated to the value $\bar{n}_\text{max}=(1+\rho)^{-1}$ where $\rho=g_2/g_1$. This particular behavior of cavity photon number has already been observed under numerical simulations \cite{S2-Paper4}. However, cavity phonon number $\bar{m}$ can still continue to increase almost proportional to the input photon flux as $\bar{m}\propto|\alpha|$. If the pump frequency is not exactly tuned to the shifted cavity frequency as described under (\ref{S2-eq36}), therefore we always have $\bar{n}_\text{max}<1$, with no apparent limit on $\bar{m}$.  In absence of momentum interactions with $g_2=0$, the cavity photon number will be clamped to $\bar{n}=1$ via non-resonant quadratic interactions. 

If the pump is however accurately tuned to the altered cavity optical frequency $\omega+g_1-2g_2$ to obtain the on-resonance criteria, it has been shown \cite{S2-Paper4} that achieving this condition can result in a virtually unlimited increase in steady-state cavity photon population $\bar{n}$ while phonon population gets saturated around unity with $\bar{m}\approx 1$. This surprising behavior at resonance and off-resonance or near-resonance reveals that an observable quadratic effect could be expected even if the cavity is not pumped resonantly, since anyhow the cross population $\psi=\bar{m}\bar{n}$ tends to increase with the pump $|\alpha|$. If tuned to the shifted resonance, this product increases as $\psi\propto|\alpha|^2$. 

This mysterious behavior can explain the fact that quadratic interactions do not need resonant pumping, and once they are strong enough and the pumping level is sufficiently high, they should be observable. This fact has also been demonstrated in the numerical studies of this phenomenon \cite{S2-Paper4}. One may therefore take the cross population $\psi$ as a measure of strength of quadratic interactions. At high pump levels we have for an on-resonance relationship $\psi\propto|\alpha|^2$, while the off-resonance relationship reads $\psi\propto|\alpha|$.

\label{S2-OptoQuad}

It is useful from a practical point of view to know to what extent and how a quadratic optomechanical interaction $\mathbb{H}_1-\mathbb{H}_2$ could be influenced and even possibly masked by a non-vanishing basic optomechanical interaction $\mathbb{H}_0$. This will enable one to rigorously differentiate between the type and strength of standard basic optomechanical and quadratic optomechanical responses to an excitation. Since the non-vanishing $g_0$ is expected to be randomly non-zero from one fabricated device to the other, the possible range of non-zero $g_0$ could be scanned and the responses should be averaged to come up with a practically meaningful expectation for the possible response. In order for this to be done, one would need to include the optomechanical interaction $\mathbb{H}_0$ in the formulation as well. Proceeding with the same basis as (\ref{S2-eq32}) is possible but with the minor correction to the coefficients matrix (\ref{S2-eq34}) and (\ref{S2-eq36}) as well as the equations for steady-state populations (\ref{S2-eq38}). 

It is straightforward to calculate the change in each of the four partitions as follows before and after linearization. The exact change in (\ref{S2-eq34}) before linearization is given by
\begin{eqnarray}
\label{S2-eq39}
[\Delta\hat{\textbf{M}}_\text{aa}]&=&2ig_0\left[
\begin{array}{ccc}
\hat{b}+\hat{b}^\dagger & 0 & 0 \\
0 & -\hat{b}-\hat{b}^\dagger & 0 \\
0 & 0 & 0
\end{array}
\right], \\ \nonumber
[\Delta\hat{\textbf{M}}_\text{ba}]&=&ig_0\left[
\begin{array}{ccc}
0 & 0 & \hat{b} \\
0 & 0 & -\hat{b}^\dagger \\
0 & 0 & -\hat{b}+\hat{b}^\dagger
\end{array}
\right], 
\end{eqnarray}
while $[\Delta\hat{\textbf{M}}_\text{ab}]=[\Delta\hat{\textbf{M}}_\text{bb}]=[\textbf{0}]$. It is seen that $[\Delta\hat{\textbf{M}}]$ is unsurprisingly proportional to the non-vanishing optomechanical interaction rate $g_0$. These relations can be further linearized as 
\begin{eqnarray}
\label{S2-eq40}
[\Delta\hat{\textbf{M}}_\text{aa}]&=&4i\Re[\bar{b}]g_0\left[
\begin{array}{ccc}
1 & 0 & 0 \\
0 & -1 & 0 \\
0 & 0 & 0
\end{array}
\right], \\ \nonumber
[\Delta\hat{\textbf{M}}_\text{ba}]&=&ig_0\left[
\begin{array}{ccc}
0 & 0 & \bar{b} \\
0 & 0 & -\bar{b} \\
0 & 0 & -2i\Im[\bar{b}]
\end{array}
\right].
\end{eqnarray}
Here, $\bar{b}$ shall be determined from the second relation of (\ref{S2-eq17}) as
\begin{equation}
\label{S2-eq41}
\bar{b}=\frac{ig_0\bar{n}}{i\Omega+\frac{1}{2}\Gamma},
\end{equation}
together with (\ref{S2-eq38}) for $\bar{n}$, $\bar{m}$, and $\bar{d}$. It should be noticed that in the ideal mathematical sense, one would expect $\bar{b}=\sqrt{\bar{d}}$. However, for a nearly vanishing $g_0$ with indeterminate and uncertain value close to zero, this approximation should be enough to give a sense of a basis optomechanical interaction entering into the picture. The reason is simply that any actual error in this calculation of $\bar{b}$ could be directly attributed to magnitude $g_0$. 

When pumping strongly on resonance with $g_0$ not vanishing, the optomechanical tri-state equation (\ref{S2-eq17}) and the quadratic state equation (\ref{S2-eq38}) compete to determine the photon population $\bar{n}$. One from (\ref{S2-eq17}) may expect that $\bar{n}\propto\sqrt[3]{|\alpha|^2}$, while it can be shown by inspection of governing state equation \cite{S2-Paper4} that for resonantly pumped quadratic interaction requires $\bar{n}\propto|\alpha|^{2}$ which grows more rapidly than the other effect. Surprisingly, quadratic interaction on resonance does not heat up the cavity, since phonon population would anyhow remains locked to a value very close to unity $\bar{m}\approx 1$. This behavior is similar to the weakly coupled optomechanics with $g=g_0\sqrt{\bar{n}}<<\sqrt{\Omega|\Delta|}$, which infers the same type of behavior according to (\ref{S2-eq17}).

Hence, for a non-vanishing optomechanical interaction rate $g_0$, we can expect that at a sufficiently high pump level, and given that it is exactly tuned to the shifted cavity resonance, the optomechanical interaction will be perfectly masked by the quadratic interaction and therefore could be neglected and completely ignored. 

\subsection{Basic Understanding of Side-band Inequivalence}

As a remarkable conclusions to this section, which could be drawn from higher-order analysis of optomechanical systems is non-zero difference in frequency shifts of red- and blue-side-bands. While it is a well-known fact that the amplitudes of these side-bands are a bit different due to finite thermal occupation number of phonons, it is almost unnoticed that the corresponding frequency shifts are also not exactly equal. Termed as side-band inequivalence \cite{S2-Paper7}, this fact has been discussed in much details later \S\ref{Section-9}. The side-band inequivalence can be estimated as  $\frac{1}{2}\delta=\frac{1}{2}(\Delta_r+\Delta_b)-\Delta$, which represents the difference of the non-zero average of red-shifted $\Delta_r$ and blue-shifted $\Delta_b$ detunings from the pump at $\Delta$. It is expected that if $\Delta_r=\Delta+\Omega$ and $\Delta_b=\Delta-\Omega$, then $\delta$ must be zero. Even, the existence of optomechanical spring effect, which causes symmetric shift of effective mechanical frequency as $\Delta_r=\Delta+(\Omega+\delta\Omega)$ and $\Delta_b=\Delta-(\Omega+\delta\Omega)$ should again lead to the same. But this is not the case, even for resonant pump which the optomechanical spring effect identically vanishes $\delta\Omega(\Delta=0)=0$.

In order to estimate $\delta$, we may first notice that the second-order system of optomechancial equations with the $3\times 3$ coefficients matrix (\ref{S2-eq27}) gives three distinct complex-valued eigenvalues corresponding to the frequencies and decay rates associated with the central (Rayleigh) peak, red, and blue side-bands. The real parts of eigenvalues corresponding to the red and blue side-bands turn out not to be equally spaced from the pump. Following a $3\times 3$ formalism leads to third-order algebraic equations for the eigenvalues and solutions need to be expanded as power series in terms of the fraction $g_0/\Omega$. When $g_0<<\Omega$, the approximation leads to the above expression as
\begin{equation}
\label{S2-eq42}
\delta\approx \frac{2g_0^2\bar{n}}{\Omega}.
\end{equation}
Obviously, this above expression is valid and makes practical sense only for weakly-coupled and side-band resolved cavities at weak coupling, where $\Omega>>g,\kappa$ where $g=g_0\sqrt{\bar{n}}$. While one would expect that a large side-band inequivalence could be observed by increased intracavity photon number and/or pumping rate, this is certainly not the case for Doppler cavities with $\kappa>\Omega$, where very large intracavity photon numbers $\bar{n}$ are attainable. However, a thorough study of this effect and various experiments in \S\ref{Section-9}shall provide further insight and a more accurate expression to replace (\ref{S2-eq42}).

Basically, a cavity which is not side-band resolved disallows clear observation of side-bands, as they are strongly suppressed by fluctuating noise. On the one hand, numerical tests reveal that side-band inequivalence does actually exist and should be mathematically strong in Doppler cavities. On the other hand, side-bands are strongly damped and suppressed, which implies that they cannot be resolved and probably are physically unimportant altogether. This combination of opposing facts shows that side-band resolved cavities must be used.

A side-band resolved cavity also sets a stringent limit on the maximum practically attainable intracavity photon number $\bar{n}$, and thus the expected amount of inequivalence $\delta\Omega$. However, the available enhanced interaction rate $G=g_0 \bar{n}$ is still limited, so that strongly coupled cavities with large $g_0$ can only accommodate very few intracavity photons $\bar{n}$. Again, there is a maximum practical limit on how large a practically observable side-band inequivalence could be. For all practical purposes, the normalized dimensionless side-band inequivalence defined as $\bar{\delta}=\delta/\Omega$  for various types of solid-state optomechanical and superconductive electromechanical systems seem to be bounded to $10^{-6}$ to $10^{-4}$. Larger values up to a few percent on the order of $10^{-2}$ are yet observable in other types of optomechanical systems and equivalent experiments, such as Raman/Brillouin scattering. 

Intuitively, the side-band inequivalence is a combined result of both modulation nonlinearity and backaction working together. Absence of either of these two mechanisms makes the effect to fade away. To this end, we may consider four different scenarios as follows:
\begin{itemize}
	\item If we ignore the backaction and keep only the nonlinearity, the side-bands will appear exactly at $\Delta\pm j \Omega$, with obviously no side-band inequivalence even for the first-order side-bands at $j=1$. 
	\item With no modulation nonlinearity and keeping only the backaction term, there will be only first-order sidebands with the optical spring effect appearing as $\Delta\pm (\Omega+\delta\Omega)$. 
	\item With no backaction and no modulation nonlinearity, there will be only two first-order red and blue side-bands exactly at $\Delta\pm \Omega$ with no optical spring effect. 
	\item With both of the backaction and modulation nonlinearity terms kept, one may expect to have both of the optical spring and side-band inequivalence terms appearing.
\end{itemize}
However, a resonant pump $\Delta=0$ demands zero spring effect $\delta\Omega=0$, while the side-band inequivalence $\delta$ is essentially independent of the pump detuning $\Delta$. Obviously, this holds true as long as the intracavity photon number $\bar{n}$ could be kept constant since $\bar{n}$ is also respectively a function of detuning $\Delta$ and pump power.

To wrap up this section, we described different approaches to solve the standard basic, and standard quadratic, and non-standard quadratic optomechanical problems. The conventional method of linearizing Langevin equations in terms of ladder operators may not be useful for study of quadratic and higher-order effects, however, definition of higher-order alternative bases could provide a means to conveniently analyze and understand higher-order nonlinear effects in quantum physics. It was shown that the cross population of photons and phonons could be taken as a measure of the strength of quadratic interactions, which increases porportional to or quadratically with respectively off-resonance and on-resonance strong pumping. We also argued that in the presence of a non-vanishing basic optomechanical interaction, on-resonance quadratic pumping could easily mask out the basic optomechanical interaction. More importantly, while the intensity of quadratic interactions increase with the square of input pump power at resonance, the cavity phonon number gets saturated normally to a sub-unity value. This highlights the fact that it is possible to observe quadratic interactions at high illumination power, without heating the cavity up. We also provided a preliminary analysis of the unexpected symmetry breaking, side-band inequivalence.

\section{Nonlinear Optomechanical Hamiltonian} \label{Section-3}
In this section, we revisit the basic theory of quantum optomechanics, which is here reconstructed from first principles. This has been accomplished by fist finding a Lagrangian from light's equation of motion, which is subsequently proceeded to the Hamiltonian. The interaction terms expressing optomechanical phenomena are are nonlinear and include quadratic and higher-order interactions as well. These quadratic interactions seem not to vanish under any possible choice of canonical parameters, and this fact lead to new coupling terms between mechanical momentum and field. While the first quadrature of mechanical motion, being the mechanical displacement is already linked to the optics through radiation pressure, there has been otherwise no known evidence of simultaneous interaction with the second quadrature. The existence of quadratic mechanical parametric interaction is then demonstrated rigorously, which has been so far assumed phenomenologically in previous studies. Corrections to the quadratic terms are shown to be particularly significant when the mechanical frequency is of the same order or larger than the electromagnetic frequency. We show that quadratic interactions fall into two major category, which we refer to as the standard and non-standard quadratic terms, respectively providing access to the first and second quadratures of mechanical field.

Further discussions on the squeezing as well as relativistic corrections are presented, and it has been demonstrated that relativistic terms also take form of non-standard quadratic optomechanics. Since any relativistic effect has to deal with the speed at which the mechanical motion takes place, it must deal with the second quadrature as well. So, the relativistic corrections to optomechanical interactions are also quadratic at least, and can be effectively combined with the non-standard optomechanical interaction terms.

The general field of quantum optomechanics is based on the standard optomechanical Hamiltonian, which is expressed as the simple product of photon number $\hat{n}$ and the position $\hat{x}$ operators, having the form ${\mathbb{H}}_{\rm OM}=-\hbar g_0\hat{n}\hat{x}$ \cite{S2-1,S2-2,S3-3,S2-4} with $g_0\ $being the single-photon coupling rate. This is mostly referred to a classical paper by Law \cite{S2-5}, where the non-relativistic Hamiltonian is obtained through Lagrangian dynamics of the system. This basic interaction is behind numerous exciting theoretical and experimental studies, which demonstrate a wide range of applications. The optomechanical interaction ${\mathbb{H}}_{\rm OM}$ is inherently nonlinear by its nature, which is quite analogous to the third-order Kerr optical effect in nonlinear optics \cite{S3-6,S3-7}. These for instance include the optomechanical arrays \cite{S3-8,S3-9,S3-10,S3-11,S3-12,S3-13,S3-14,S3-15,S3-16,S3-17}, squeezing of phonon states \cite{S3-18,S3-19,S3-20}, Heisenberg's limited measurements \cite{S3-21}, non-reciprocal optomechanical systems \cite{S3-22,S3-23,S3-24,S3-26,S3-27,S3-28}, sensing \cite{S3-29,S3-30,S3-31}, engineered dissipation \cite{S3-32}, engineered states \cite{S3-33}, and non-reciprocal acousto-optical effects in optomechanical crystals \cite{S3-34,S3-35,S3-36}. 

Recent ideas in this field such as microwave-optical conversion \cite{S3-37,S3-38,S3-39,S3-40}, cavity electrooptics \cite{S3-41,S3-42,S3-43}, optomechanical induced transparency \cite{S3-44,S3-45,S3-46,S3-57}, and optomechanical verification of Bell's inequality \cite{S3-48} all emerge as closely related duals of quantum optomechanical systems, which are described within a completely identical framework. Furthermore, quantum chaos which had been predicted in quantum optomechanics \cite{S3-49,S2-3} and cavity quantum electro-dynamical systems \cite{S3-51,S3-52}, has been recently observed in optomechanics \cite{S3-53,S3-54}. 

Usually, the analysis in all these above works is ultimately done within the linearized approximation of ladder operators. The difficulty arising with linearization is that when quadratic effects are involved, then the fourth-order quadratic and third-order optomechanical terms both simplify as products. This approximation in addition to discarding nonlinear behavior may cause additional loss of accuracy, particularly where ${\mathbb{H}}_{\rm OM}$ interaction is non-existent or vanishingly small. For applications where strongly nonlinear quadratic or even quartic effects are primarily pursued, $g_0$ may be designed to be identically zero \cite{S3-55,S3-56,S3-57,S3-58,S3-59,S3-60,S3-61}, which urges need for accurate knowledge of higher-order interaction terms. Similar situation also could arise in trapped ultracold atomic gases \cite{S3-62}, where linear interactions identically vanish. Relevant Physical phenomena in optomechanics such as four-wave mixing, also is suitably described by higher order interaction terms \cite{S3-63}. Moreover, significance and prominent role of such nonlinear interactions has been observed in few recent experiments \cite{S3-64,S3-65}.

It is to be noticed here that the method of higher-order operators actually was originally born to study the quadratic effects arising in these types of strongly nonlinear interactions. Since any attempt made to full linearization of quadratic terms will put back these interactions into the form of a simple product, and thus eliminating the underlying physics. To be discussed later in \ref{Section-10}, it has been already shown that not only non-standard interactions can be well studied using the toolbox of higher-order operators, but also, existence of non-standard quadratic terms may imply a finite and observable effect in the limit of large mechanical frequency. The successful application and general extension of higher-order operators to conventional optomechanics \S\ref{Section-8}, side-band inequivalence \S\ref{Section-9}, cross-Kerr interactions \S\ref{Section-11}, and other areas of nonlinear quantum and classical physics \S\ref{Section-12} had been neither planned nor foreseen at first.

We here show that a careful review of the underlying theory \cite{S2-5} normally involves a number of physical approximations in formulation of the problem such as the non-relativistic limit \cite{S3-66,S3-67}, which makes the unified description of relativistic photon momenta and non-relativistic mirror motion inaccurate. As it is being shown here, a full treatment of the latter will yield higher-order multi-particle interactions. Such quadratic interactions have been recently used phenomenologically \cite{S3-58} without a rigorous theoretical basis. Similar yet weaker interactions may be also drawn from relativistic corrections \cite{S3-68} as discussed here. In that sense, the quadratic interactions are shown to receive contributions from both non-relativistic and relativistic terms, which become quite significant when the mechanical frequency is comparable or larger than the electromagnetic frequency.

The basic theory to be discussed is based on two important and basic assumptions. We first assume that the cavity mode decomposition is valid independent of the mirror motion. Second, the electric fields vanish at the mirror surface in the frame that the Lagrangian is constructed. These assumptions seem reasonable in the usual discussions with much lower mechanical frequencies when the cavity mode change can be treated adiabatically. However, when the end mirror oscillates at a very high frequency, it dresses the cavity modes so that the mode frequency might become undefined invalidating the assumption of mode decomposition. At higher frequencies, it has also been known and demonstrated that the mirror undergoing relativistic motion could produce photons from vacuum, known as the dynamical Casimir effect \cite{S3-69,S3-70}. This is of course beyond the regime being explored in this section. 

The study presented here in this section is not without its own approximation. A major approximation applied is the one-dimensional description of optomechanical structure. While this is known to be quite a reasonably accurate description of optomechanics, no one has made an attempt to rederive the nonlinear full optomechanical Hamiltonian in cavities with finite size and arbitrary shapes. It can be correctly expected that as long as only linearized interactions having the form of products are concerned, the actual shape and volume of cavity is mathematically irrelevant. Certainly, this is not the case for nonlinear interactions and specially quadratic forms which are strongly nonlinear. It is not obvious either that what approach should be taken to improve the accuracy of such descriptions.

Another very interesting but not directly relevant result of this analysis is a set of new expressions for the irrational number $\pi$, which are found through Fourier analysis, one example of which is to be revealed in (\ref{pi}). 

The focus of the first two subsections \S\ref{MotionEquations} and \S\ref{Lagrangian} is to assert the claim, and demonstrate what term is missing and why it happens. As it will be shown and rigorously proven, even for the simplest case of interaction with a single-optical mode, a new term of the type ${\dot{q}}^2Q^2$ representing quadratic momentum $\dot{q}$ and optical field $Q$ interactions is found, the origin of which is also identified. For the more general case of multi-mode optical fields, the situation is even much more complex and there are a few more missing terms to consider. Once the Lagrangian is known, the Hamiltonian is subsequently constructed in \S\ref{Hamiltonian}.

Ultimately, it shall be demonstrated that the nature of optomechanical interaction under large electromagnetic frequency is remarkably different from large mechanical frequency. For the first case, only the instantaneous position of mirror is important and it is the displacement of mirror which takes part; this has been the widely known and largely investigated case up to now. However, in the regime of large mechanical frequency it is the momentum of mirror which becomes important and dominant. Intuitively, this effect enters the optomechanical interaction through momentum conservation which has to be preserved.

\subsection{The Equations of Motion}\label{MotionEquations}

The one-dimensional wave equation for transverse component of the magnetic potential $A\left(x,t\right)$ in the dimensional form is expressed as \cite{S2-5}
\begin{equation} \label{S3-1} 
c^2A_{xx}\left(x,t\right)=A_{tt}\left(x,t\right), 
\end{equation} 
where $x$ and $t$ are respectively the position and time coordinates, and $c$ is the speed of light in free space. Suppose the instantaneous Fourier series relations for the magnetic vector potential are defined as \cite{S2-5}
\begin{eqnarray}
\nonumber
Q_k\left(t\right)&=&\frac{1}{c}\sqrt{\frac{2S}{{\mu }_0q\left(t\right)}}\int^{q\left(t\right)}_0{A\left(x,t\right)\sin\left[{\kappa }_k\left(t\right)x\right]dx},\\ \nonumber
A\left(x,t\right)&=&c\sqrt{\frac{2{\mu }_0}{Sq\left(t\right)}}\sum^{\infty }_{k=1}{Q_k\left(t\right)\sin\left[{\kappa }_k\left(t\right)x\right]},\\
\label{S3-2} 
{\omega }_k\left(t\right)&=&{c\pi k}/{q\left(t\right)}=c{\kappa }_k\left(t\right), 
\end{eqnarray} 
where $S$ is the cross-sectional area, ${\mu }_0$ is the permeability of vacuum. This arrangement ensures that the definition of canonical variables can be used later, so that ${\dot{Q_k}}^2$ simply takes on the dimension of energy. By the term instantaneous we imply that the Fourier series are composed at a given time $t$, at which a momentarily snapshot of the system is taken. Hence, the coefficients of such Fourier series needs to be updated in time and become time-dependent. This is in contrast to the general picture of Fourier series where the entire spectrum fits into the expansion and coefficients take on time-independent. It would have been possible to equally expand the magnetic potential a bit different, such as $A(x,t)=\int \mathcal{A}(k)\sin[kx-\omega(k)t] dk$, using an appropriate redefinition of dispersion $\omega(k)$. While the latter is mathematically equivalent and more common in Fourier analysis, the use of assumed form (\ref{S3-2}) actually provides a much more straightforward way to study the optomechanical interaction phenomena, from which we deduce identities and obtain all sorts of nonlinear interactions.

One may furthermore define the functions $f_k=\sqrt{{2}/{q}}\sin\left({\kappa }_kx\right)$ and $g_k=\sqrt{{2}/{q}}\cos\left({\kappa }_kx\right)$, and hence $A\left(x,t\right)=s\sum{Q_kf_k}$ where $s=c\sqrt{{{\mu }_0}/{S}}$. Here, the inner product is also defined as $\left(a|b\right)={{\int^q_0{abdx}}}$ such that the following relations may be found
\begin{eqnarray} \nonumber 
\left(f_k|f_j\right)&=&{\delta }_{kj},\\ \nonumber 
\left(f_k\left|\kappa_j x\right|g_j\right)&=&{\alpha }_{kj},\\ \label{S3-3}
\left(f_k\left|\kappa_j^2 x^2\right|g_j\right)&=&{\beta }_{kj}. 
\end{eqnarray} 
After straightforward calculations one obtains
\begin{eqnarray} \nonumber
{\alpha }_{kj}&=&-\frac{1}{2}{\delta }_{kj}+g_{kj},\\
\label{S3-4} 
{\beta }_{kj}&=&\left(k^2\frac{{\pi }^2}{3}-\frac{1}{2}\right){\delta }_{kj}+h_{kj}=\left\{ \begin{array}{cc}
k^2\frac{{\pi }^2}{3}-\frac{1}{2}, & k=j, \\ 
8\frac{{\left(-1\right)}^{k+j}kj^3}{{\left(k^2-j^2\right)}^2}, & k\neq j. \end{array}\right.
\end{eqnarray} 
Here, the anti-symmetric coefficients $g_{kj}$ and $h_{kj}$ are
\begin{eqnarray}\nonumber
g_{kj}&=&-g_{jk}=\left\{ \begin{array}{cc}
0, & k=j, \\ 
2\frac{{\left(-1\right)}^{k+j}kj}{j^2-k^2}, & k\neq j, \end{array}\right.\\ 
\label{S3-5} 
h_{kj}&=&\left\{ \begin{array}{cc}
0, & k=j, \\ 
8\frac{{\left(-1\right)}^{k+j}kj^3}{{\left(k^2-j^2\right)}^2}, & k\neq j. \end{array}\right.
\end{eqnarray} 
Differentiating $f_k$ and $g_k$ with respect to $t$, noting that ${\dot{\kappa }}_k=-{\dot{q}{\kappa }_k}/{q}$, gives
\begin{eqnarray}\nonumber
{\dot{f}}_k&=&-\frac{\dot{q}}{q}\left(\frac{1}{2}f_k+x{\kappa }_kg_k\right),\\ \nonumber
{\dot{g}}_k&=&-\frac{\dot{q}}{q}\left(\frac{1}{2}g_k-x{\kappa }_kf_k\right),\\ \label{S3-6} 
{\ddot{f}}_k&=&-\frac{\ddot{q}q-{\dot{q}}^2}{q^2}\left(\frac{1}{2}f_k+x{\kappa }_kg_k\right)+\frac{{\dot{q}}^2}{q^2}\left(\frac{1}{4}f_k+2x{\kappa }_kg_k-x^2{{\kappa }_k}^2f_k\right). 
\end{eqnarray} 
It is possible now to differentiate $A\left(x,t\right)$ with respect to position and time to obtain the relations
\begin{eqnarray}\nonumber
\frac{c^2}{s}A_{xx}\left(x,t\right)&=&-\sum{{{\omega }_k}^2Q_kf_k},\\ \nonumber
\frac{1}{s}A_t\left(x,t\right)&=&\sum{{\dot{Q}}_kf_k}-\frac{\dot{q}}{q}\sum{Q_k\left(\frac{1}{2}f_k+x{\kappa }_kg_k\right)},\\ \nonumber
\frac{1}{s}A_{tt}\left(x,t\right)&=&\sum{{\ddot{Q}}_kf_k}-2\frac{\dot{q}}{q}\sum{{\dot{Q}}_k\left(\frac{1}{2}f_k+x{\kappa }_kg_k\right)}-\frac{\ddot{q}q-{\dot{q}}^2}{q^2}\sum{Q_k\left(\frac{1}{2}f_k+x{\kappa }_kg_k\right)}, \\
\label{S3-7} 
&+&\frac{{\dot{q}}^2}{q^2}\sum{Q_k\left(\frac{1}{4}f_k+2x{\kappa }_kg_k-x^2{{\kappa }_k}^2f_k\right)}. 
\end{eqnarray} 
Therefore, the wave equation reads
\begin{eqnarray}\nonumber
-\sum{{{\omega }_k}^2Q_kf_k}&=&\sum{{\ddot{Q}}_kf_k}-\frac{\dot{q}}{q}\sum{{\dot{Q}}_k\left(f_k+2x{\kappa }_kg_k\right)}-\frac{\ddot{q}}{q}\sum{Q_k\left(\frac{1}{2}f_k+x{\kappa }_kg_k\right)}\\
\label{S3-8} 
&+&\frac{{\dot{q}}^2}{q^2}\sum{Q_k\left(\frac{3}{4}f_k+3x{\kappa }_kg_k-x^2{{\kappa }_k}^2f_k\right).} 
\end{eqnarray} 
Now, the inner product relationships (\ref{S3-3}) and some further simplification to be discussed later in \S\ref{AppendixA} help us to obtain the equation of motion as
\begin{equation} \label{S3-9} 
{\ddot{Q}}_k=-{{\omega }_k}^2Q_k+r_k\frac{{\dot{q}}^2}{q^2}Q_k+2\frac{\dot{q}}{q}\sum{g_{kj}{\dot{Q}}_j}+\frac{\ddot{q}}{q}\sum{g_{kj}Q_j}+\frac{{\dot{q}}^2}{q^2}\sum{\left(h_{kj}-3g_{kj}\right)Q_j}, 
\end{equation} 
where all summations are nonzero only for $j\neq k$, and also we adopt the definition 
\begin{equation} \label{S3-10} 
r_k=k^2\frac{{\pi }^2}{3}+\frac{1}{4}. 
\end{equation} 
The related equation in Law's paper \cite{S2-5}
\begin{equation} \label{S3-11} 
{\ddot{Q}}_k=-{{\omega }_k}^2Q_k+2\frac{\dot{q}}{q}\sum{g_{kj}{\dot{Q}}_j}+\frac{\ddot{q}}{q}\sum{g_{kj}Q_j}+\frac{{\dot{q}}^2}{q^2}\sum{g_{kj}g_{lj}Q_l}-\frac{{\dot{q}}^2}{q^2}\sum{{g_{kj}Q}_j}, 
\end{equation} 
is completely equivalent, but apparently different in presentation. Similarly, one may directly deduce from the Newton's equation of motion
\begin{equation} \label{S3-12} 
m\ddot{q}=-\frac{\partial }{\partial x}V\left(x\right)+\frac{S}{2{\mu }_0}B^2\left(q,t\right)=-\frac{\partial }{\partial x}V\left(x\right)+\frac{1}{q}\sum{{\left(-1\right)}^{k+j}{\omega }_k{\omega }_jQ_kQ_j}. 
\end{equation} 

\subsection{Lagrangian}\label{Lagrangian}

The associated Lagrangian which leads to the above set of Euler equations is given by
\begin{eqnarray}\nonumber
\mathcal{L}&=&\frac{1}{2}m{\dot{q}}^2-V\left(q\right)+\frac{1}{2}\sum{\left({{\dot{Q}}_k}^2-{\omega }^2_k{Q_k}^2\right)}-\frac{\dot{q}}{q}\sum{g_{kj}Q_j{\dot{Q}}_k}\\
\label{S3-13} 
&+&\frac{{\dot{q}}^2}{2q^2}\sum{\left(h_{kj}-2g_{kj}+r_k{\delta }_{kj}\right)Q_kQ_j}. 
\end{eqnarray} 
It should be again noticed that $h_{kk}=0$ by (\ref{S3-5}), which together with the antisymmetry of coefficients $g_{kj}$ from (\ref{S3-5}), allows further simplification to reach 
\begin{equation} \label{S3-14} 
\mathcal{L}=\frac{1}{2}m{\dot{q}}^2-V\left(q\right)+\frac{1}{2}\sum{\left({{\dot{Q}}_k}^2-{\omega }^2_k{Q_k}^2\right)}+\frac{{\dot{q}}^2}{2q^2}\sum{d_{kj}Q_kQ_j}-\frac{\dot{q}}{q}\sum{g_{kj}Q_j}{\dot{Q}}_k, 
\end{equation} 
where $d_{kj}={{\frac{1}{2}}}\left(h_{kj}+h_{jk}\right)+r_k{\delta }_{kj}$ is related to the symmetric part of $h_{kj}$, given by 
\begin{equation} \label{S3-15} 
d_{kj}=d_{jk}=\left\{ \begin{array}{cc}
r_k, & k=j, \\ 
4\frac{{\left(-1\right)}^{k+j}kj\left(k^2+j^2\right)}{{\left(k^2-j^2\right)}^2}, & k\neq j. \end{array}\right. 
\end{equation} 
This is to be compared with the quite different expression by Law \cite{S2-5} given by
\begin{equation} \label{S3-16} 
\mathcal{L}=\frac{1}{2}m{\dot{q}}^2-V\left(q\right)+\frac{1}{2}\sum{\left({{\dot{Q}}_k}^2-{\omega }^2_k{Q_k}^2\right)}+\frac{{\dot{q}}^2}{2q^2}\sum{g_{lk}g_{lj}Q_kQ_j-\frac{\dot{q}}{q}\sum{g_{kj}Q_j}{\dot{Q}}_k}, 
\end{equation} 
which indeed consistently satisfies the Euler's pair of equations. 

Let us here take a deeper look into why (\ref{S3-14}) and (\ref{S3-16}) are not identical. Firstly, the mathematical discrepancy between (\ref{S3-14}) and (\ref{S3-16}) shows up in the quadratic terms as non-standard optomechanical momentum-field interactions, and both ways of formulating Lagrangians (\ref{S3-14}) and (\ref{S3-16}) lead to identical terms in third-order optomechanics $\mathbb{H}_{\rm OM}$, as well as higher-order nonlinear terms up to the standard quadratic process. The lowest order mathematical difference between (\ref{S3-14}) and (\ref{S3-16}) is therefore the fourth- and higher-order non-standard quadratic terms, and otherwise (\ref{S3-14}) and (\ref{S3-16}) are actually equivalent. And the reason for the apparent difference originates from the way modes are allowed to interact. In the single-mode 1:1 regime, where only one mechanical and one optical modes are allowed to interact, the double summations of the Lagrangian (\ref{S3-16}) involve one single term on the outermost summation. If the innermost summation is to be taken also with only one single term, then the results of \cite{S2-5} based on Lagrangian (\ref{S3-16}) retain their validity. However, physically speaking, there is no reason to cutoff both of the inner and outer summations. Hence, it seems to the author that the inner summation should be still taken over infinite terms, from which fourth-order quadratic momentum-field non-standard interactions evolve and come into existence. Ultimately, these non-standard quadratic terms are important only in the limit of large mechanical frequency $\Omega$ where the electromagnetic frequency $\omega$ is on the same order of magnitude or smaller than $\Omega$. 

Under a rather idealistic case of when an infinite number of optical modes are involved in optomechanical interaction, and both summations are let to extend into infinity, it would be possible to demonstrate that (\ref{S3-14}) and (\ref{S3-16}) are actually the same. For this to occur, we need first to show (\ref{S3-9}) and (\ref{S3-11}) are the same representations of the same identities, too. This is not too difficult to prove, indeed. Firstly, the diagonal terms are equal if 
\begin{equation} \label{S3-17} 
\sum_k{g_{kj}g_{kj}}=r_j=d_{jj}, 
\end{equation} 
which holds true, only if the identity
\begin{equation} \label{S3-18} 
\sum^{\infty }_{j=1,j\neq k}{\frac{4k^2j^2}{{\left(j^2-k^2\right)}^2}}=k^2\frac{{\pi }^2}{3}+\frac{1}{4}, 
\end{equation} 
holds, which surprisingly does. It is possible to verify the correctness of ({S3-18}) through numerical summation over the index $j$ for various choices of $k\geq 1$. Hence, we have already obtained new representations for the irrational number $\pi$ as byproduct of optomechanical analysis. For instance, we can take $k=1$ or $k=2$ from which we respectively get
\begin{eqnarray}
\nonumber
\pi&=&\sqrt{12\left[\sum^{\infty }_{j=2}\frac{j^2}{\left(j^2-1\right)^2}\right]-\frac{3}{4}}, \\ \label{pi}
\pi&=&\sqrt{12\left[\sum^{\infty }_{j=3}\frac{j^2}{\left(j^2-4\right)^2}\right]+\frac{55}{48}}.
\end{eqnarray} 
Similar weird expressions can be deduced following larger values of $k\geq3$.

For a single-mode system with only one radiation mode, we may easily notice that (\ref{S3-9}) and (\ref{S3-11}) become readily identical. This becomes more clear by noticing that $g_{kk}=h_{kk}=0$, and hence for a system with only one electromagnetic radiation mode, (\ref{S3-9}) becomes $\ddot{Q}=-{\omega }^2Q+\left(rq^{-2}{\dot{q}}^2\right)Q$, with $r=r_1$. This is while Law's expression (\ref{S3-11}) is $\ddot{Q}=-{\omega }^2Q+q^{-2}{\dot{q}}^2\left(\sum{g_{1j}g_{1j}}\right)Q$. But (\ref{S3-17}) requires that $r=\sum{g_{1j}g_{1j}}$, which confirms the equivalency for single-mode systems on the condition that both the inner and outer summations are cutoff. 

As for a multimode system, and by comparing (\ref{S3-13}) and (\ref{S3-16}) one would need
\begin{equation} \label{S3-19} 
\sum{g_{kl}g_{jl}}=d_{kj}=r_k{\delta }_{kj}-2g_{kj}+h_{kj}, 
\end{equation} 
in order to (\ref{S3-9}) and (\ref{S3-11}) be identical. This can be put to numerical tests, and is in fact accurately satisfied, too. Hence, the single-mode Lagrangian in the non-relativistic limit can be written in the form
\begin{equation} \label{S3-20} 
\mathcal{L}=\frac{1}{2}m{\dot{q}}^2-V\left(q\right)+\frac{1}{2}\left[{\dot{Q}}^2-{\omega }^2\left(q\right)Q^2\right]+\frac{{\dot{q}}^2}{2q^2}rQ^2. 
\end{equation} 
The last term has been usually ignored so far in the literature, and will result in momentum-field coupling. In the remainder of this section, we focus on nonlinear terms arising from this interaction, and then also add up the relativistic corrections in the end.

\subsection{Hamiltonian}\label{Hamiltonian}

The definition of canonical momenta taken here is
\begin{eqnarray} \nonumber 
P_k&=&{\dot{Q}}_k-\frac{\dot{q}}{q}\sum{A_{kj}Q_j},\\ \label{S3-21} p&=&m\dot{q}-\frac{1}{q}\sum{B_{kj}P_kQ_j}, 
\end{eqnarray} 
with $A_{kj}$ and $B_{kj}$ being some transformation coefficients to be determined later. One here may take advantage of the degree of freedom in choice of $A_{kj}$ and $B_{kj}$ to get rid of unwanted summation terms in the Hamiltonian. It has to be here noticed again that the existence of the last term of (\ref{S3-20}), being completely new even under the single-mode operation, has nothing to do with the choice of canonical momenta. That implies the final resulting single-mode (and therefore multi-mode) Hamiltonian will be inevitably different, incorporating a few new terms. The Hamiltonian may be now derived from the Lagrangian by iterated use of (\ref{S3-21}) and through the relationship
\begin{equation}
\mathcal{H}=p\dot{q}+\sum{P_k{\dot{Q}}_k}-\mathcal{L},
\end{equation}
as
\begin{equation} \label{S3-22} 
\mathcal{H}=\left[m\dot{q}-\frac{1}{q}\sum{B_{kj}\left({\dot{Q}}_k-\frac{\dot{q}}{q}\sum{A_{kl}Q_l}\right)Q_j}\right]\dot{q}+\sum{\left({\dot{Q}}_k-\frac{\dot{q}}{q}\sum{A_{kj}Q_j}\right){\dot{Q}}_k}-\ \mathcal{L}. 
\end{equation}

Law \cite{S2-5} arbitrates the choice $A_{kj}=B_{kj}=g_{kj}$. But by going further with this choice for a single-mode optical field, it will be evident that  $P=\dot{Q}$ and $p=m\dot{q}$, hence, still resulting in an extra nonlinear term proportional to ${p^2Q^2}/{q^2}$ in the Hamiltonian, leading to a fourth-order momentum-field interaction. This will be discussed shortly in the following. Hence, the Lagrangian $\mathcal{L}$ found in the above yields the Hamiltonian below after some algebra
\begin{eqnarray} \nonumber 
\mathcal{H}&=&\frac{1}{2}m{\dot{q}}^2+V\left(q\right)+\frac{1}{2}\sum{\left({{\dot{Q}}_k}^2+{\omega }^2_k{Q_k}^2\right)}-\frac{\dot{q}}{q}\sum{\left(A_{kj}+B_{kj}-g_{kj}\right){\dot{Q}}_kQ_j}\\
\label{S3-23}
&+&\frac{{\dot{q}}^2}{{2q}^2}\sum{{\left(2\sum{B_{kj}A_{kl}}-d_{jl}\right)Q_lQ}_j}. 
\end{eqnarray} 
Here is readily evident now by (\ref{S3-19}), that the last term can be made identically zero, only if $d_{jl}=\sum{g_{kj}g_{kl}}=2\sum{B_{kj}A_{kl}}$. This not only cannot be satisfied by Law's choice $A_{kj}=B_{kj}=g_{kj}$, but also the second summation term nonlinear in $\dot{q}{\dot{Q}}_k$ will also survive, further complicating the Hamiltonian formulation.

Now, further elimination of $\dot{q}$ and ${\dot{Q}}_k$ from (\ref{S3-21}) gives
\begin{eqnarray}\nonumber
\mathcal{H}&=&\frac{1}{2m}{\left(p+\frac{1}{q}\sum{B_{kj}P_kQ_j}\right)}^2+V\left(q\right)+\frac{1}{2}\sum{{\omega }^2_k{Q_k}^2}\\ \nonumber
&+&\frac{1}{2}\sum{{\left[P_k+\frac{1}{mq}\left(p+\frac{1}{q}\sum{B_{lm}P_lQ_m}\right)\sum{A_{kj}Q_j}\right]}^2}\\ \nonumber
&-&\frac{1}{mq}\left(p+\frac{1}{q}\sum{B_{kj}P_kQ_j}\right)\sum{C_{kj}\left[P_k+\frac{1}{mq}\left(p+\frac{1}{q}\sum{B_{kj}P_kQ_j}\right)\sum{A_{kj}Q_j}\right]Q_j}\\ 
\label{S3-24} 
&+&\frac{1}{{2m^2q}^2}{\left(p+\frac{1}{q}\sum{B_{kj}P_kQ_j}\right)}^2\sum{D_{jl}{Q_lQ}_j}. 
\end{eqnarray} 
Here, $C_{kj}=A_{kj}+B_{kj}-g_{kj}$ and $D_{jl}=2\sum{B_{kj}A_{kl}}-d_{jl}$. Dealing directly with such an intractable and long expression is without doubt too tough. Instead, we may tweak Law's choice slightly as $A_{kj}=B_{kj}={{\frac{1}{2}}}g_{kj}$, which allows (\ref{S3-23}) be greatly simplified as
\begin{equation} \label{S3-25} 
\mathcal{H}=\frac{1}{2}m{\dot{q}}^2+V\left(q\right)+\frac{1}{2}\sum{\left({{\dot{Q}}_k}^2+{\omega }^2_k{Q_k}^2\right)}-\frac{{\dot{q}}^2}{{4q}^2}\sum{{d_{kj}Q_kQ}_j}. 
\end{equation} 
This can be further eventually expanded and simplified as
\begin{eqnarray}\nonumber
\mathcal{H}&=&\frac{1}{2m}{\left(p+\frac{1}{2q}\sum{g_{kj}P_kQ_j}\right)}^2+V\left(q\right)+\frac{1}{2}\sum{\left({P_k}^2+{\omega }^2_k{Q_k}^2\right)}
\\ \label{S3-26} 
&+&\frac{1}{4mq}\left(p+\frac{1}{2q}\sum{g_{kj}P_kQ_j}\right)\sum{g_{kj}P_kQ_j}-\frac{1}{8m^2q^2}{\left(p+\frac{1}{2q}\sum{g_{kj}P_kQ_j}\right)}^2\sum{d_{kj}Q_kQ_j}. 
\end{eqnarray} 
The last two terms of this Hamiltonian can be expanded to obtain multiple orders of interactions. These include higher-order tripartite phonon/two-photon, and quadpartite two-phonon/two-photon interactions, which does not exist in the Law's Hamiltonian \cite{S2-5}, given by
\begin{equation} \label{S3-27} 
\mathcal{H}=\frac{1}{2m}{\left(p+\frac{1}{q}\sum{g_{kj}P_kQ_j}\right)}^2+V\left(q\right)+\frac{1}{2}\sum{\left({P_k}^2+{\omega }^2_k{Q_k}^2\right)}. 
\end{equation} 
As it appears, (\ref{S3-27}) is missing two very different types of momentum field interaction as the last two summation terms of (\ref{S3-26}). This fact becomes evident in below.

\subsection{Single Optical Mode}

The interesting difference between these two Hamiltonians becomes quite clear with consideration of only one optical mode in the cavity. This simplifies our derived Hamiltonian to
\begin{equation} \label{S3-28} 
\mathcal{H}=\frac{1}{2m}p^2+V\left(q\right)+\frac{1}{2}\left[P^2+\frac{{\pi }^2}{q^2}Q^2\right]-\frac{r}{8m^2q^2}p^2Q^2, 
\end{equation} 
where $\omega \left(q\right)={\pi }/{q}$ and$\ r\cong 3.8$, while the Law's Hamiltonian \cite{S2-5} gives rise to the significantly different form
\begin{equation} \label{S3-29} 
\mathcal{H}=\frac{1}{2m}p^2+V\left(q\right)+\frac{1}{2}\left[P^2+\frac{{\pi }^2}{q^2}Q^2\right]. 
\end{equation} 
As it will be shown below, (\ref{S3-28}) and (\ref{S3-29}) agree only to the first order, and hence up to the standard optomechanical Hamiltonian.

\subsection{Field Quantization}

When the obtained Hamiltonian is moved to the realm of quantum mechanics, it is first needed to define the non-commutation rules $\left[\hat{q},\hat{p}\right]=i\hbar $ and $[{\hat{Q}}_k,{\hat{P}}_j]=i\hbar {\delta }_{kj}$, with the commutation rules $[\hat{q},{\hat{Q}}_k]=[{\hat{Q}}_k,\hat{p}]=[\hat{q},{\hat{P}}_j]=[{\hat{P}}_j,\hat{p}]=0$. This allows us to introduce the field creation and annihilation operators according to
\begin{eqnarray}\nonumber
{\hat{Q}}_k&=&\sqrt{\frac{\hbar }{2{\omega }_k\left(\hat{q}\right)}}\left({{\hat{a}}_k}^{\dagger }+{\hat{a}}_k\right)=\sqrt{\frac{\hbar }{{\omega }_k\left(\hat{q}\right)}}{\mathbb{Q}}_k,
\\ \label{S3-30} 
{\hat{P}}_k&=&i\sqrt{\frac{\hbar {\omega }_k\left(\hat{q}\right)}{2}}\left({{\hat{a}}_k}^{\dagger }-{\hat{a}}_k\right)=\sqrt{\hbar {\omega }_k\left(\hat{q}\right)}{\mathbb{P}}_k, 
\end{eqnarray} 
where ${\omega }_k\left(\hat{q}\right)=c\pi k/\hat{q}$ is defined according to (\ref{S3-2}). Also for a mechanical resonant frequency $\Omega$ and a spring restoring potential
\begin{equation} \label{S3-31} 
V\left(\hat{q}\right)=\frac{1}{2}m{\Omega}^2{\left(\hat{q}-l\right)}^2, 
\end{equation} 
the displacement operator may be defined as $\hat{q}=l+\hat{x}$, where $l$ is the reference position of mirror, and hence the phonon ladder operators as
\begin{eqnarray} \nonumber 
\hat{x}&=&\sqrt{\frac{\hbar }{2m\Omega}}\left({\hat{b}}^{\dagger }+\hat{b}\right)=\sqrt{\frac{\hbar }{m\Omega}}\mathfrak{X},\\
\label{S3-32}
\hat{p}&=&i\sqrt{\frac{\hbar m\Omega}{2}}\left({\hat{b}}^{\dagger }-\hat{b}\right)=\sqrt{\hbar m\Omega}\mathfrak{P}, 
\end{eqnarray} 
with $[\hat{b},{\hat{b}}^{\dagger }]=1$ and $\left[\mathfrak{X},\mathfrak{P}\right]=i$.

Now, it is necessary first to symmetrize \cite{S3-67} the classical Hamiltonian prior to insertion of operators, to ensure correct quantization of parameters. The process of symmetrization is done according to \cite{S3-71,S3-72,S3-73}
\begin{eqnarray} \nonumber
\mathcal{S}\left\{\mathbb{A}\mathbb{B}\right\}&=&\frac{1}{2}\left(\mathbb{A}\mathbb{B}+\mathbb{B}\mathbb{A}\right),\\ \label{S3-33}  \mathcal{S}\left\{\mathbb{A}\mathbb{B}\mathbb{C}\right\}&=&\frac{1}{3}\left(\mathbb{A}\mathcal{S}\left\{\mathbb{B}\mathbb{C}\right\}+\mathbb{B}\mathcal{S}\left\{\mathbb{A}\mathbb{C}\right\}\mathrm{+}\mathbb{C}\mathcal{S}\left\{\mathbb{A}\mathbb{B}\right\}\right), 
\end{eqnarray} 
etc. Therefore, after symmetrization the final form of the Hamiltonian is given by
\begin{eqnarray}\nonumber
\mathbb{H}&=&\frac{1}{2m}{\hat{p}}^2+V\left(\hat{q}\right)+\frac{1}{2}\sum{\left({{\hat{P}}_k}^2+{\omega }^2_k{{\hat{Q}}_k}^2\right)}+\frac{1}{4m}\mathcal{S}\left\{\frac{1}{{\hat{q}}^2}\left(\hat{p}\hat{q}+\frac{1}{2}\sum{g_{kj}{\hat{P}}_k{\hat{Q}}_j}\right)\sum{g_{kj}{\hat{P}}_k{\hat{Q}}_j}\right\}
\\ \label{S3-34} 
&-&\frac{r}{8m^2}\mathcal{S}\left\{\frac{1}{{\hat{q}}^4}{\left(\hat{p}\hat{q}+\frac{1}{2}\sum{g_{kj}{\hat{P}}_k{\hat{Q}}_j}\right)}^2\sum{d_{kj}{\hat{Q}}_k{\hat{Q}}_j}\right\}. 
\end{eqnarray} 
For a single-optical mode, (\ref{S3-34}) greatly simplifies and one gets
\begin{equation} \label{S3-35} 
\mathbb{H}=\frac{1}{2m}{\hat{p}}^2+V\left(\hat{q}\right)+\frac{1}{2}\left({\hat{P}}^2+\frac{{\pi }^2}{{\hat{q}}^2}{\hat{Q}}^2\right)-\frac{r}{8m^2}\mathcal{S}\left\{\frac{1}{{\hat{q}}^2}{\hat{p}}^2\right\}{\hat{Q}}^2. 
\end{equation} 
This has to be applied to the last interacting term, which involves
\begin{equation} \label{S3-36} 
\mathcal{S}\left\{\frac{1}{{\hat{q}}^2}{\hat{p}}^2\right\}=\mathcal{S}\left\{\frac{{\hat{p}}^2}{{\hat{q}}^2}\right\}. 
\end{equation} 
But symmetrization of a term which contains $n$ non-commuting terms, results in $n!$ terms, which for this case sum up to a total of $4!=24$ different expressions. The direct way to get around this situation is to first make an estimate of which terms are the strongest in the limit of linearized interaction and ignore the rest. It is possible furthermore to use the approximate replacement
\begin{equation} \label{S3-37} 
\frac{1}{{\hat{q}}^n}\cong \frac{1}{l^n}\left(1-n\frac{\hat{x}}{l}\right), 
\end{equation} 
to obtain
\begin{equation} \label{S3-38} 
\mathbb{H}=\frac{1}{2m}{\hat{p}}^2+V\left(\hat{q}\right)+\frac{1}{2}\left[{\hat{P}}^2+{\omega }^2\left(1-2\frac{\hat{x}}{l}+\frac{4}{l^2}{\hat{x}}^2+\cdots \right){\hat{Q}}^2\right]-\frac{r}{8m^2l^2}\mathcal{S}\left\{{\hat{p}}^2\left(1-2\frac{\hat{x}}{l}\right)\right\}{\hat{Q}}^2, 
\end{equation} 
where further substitutions should be taken as
\begin{eqnarray} \nonumber
\hat{P}&\cong& \sqrt{\hbar \omega }\left(1-\frac{1}{2l}\hat{x}+\frac{3}{8l^2}{\hat{x}}^2+\cdots \right)\mathbb{P},\ \\ \label{S3-39} 
\hat{Q}&\cong& \sqrt{\frac{\hbar }{\omega }}\left(1+\frac{1}{2l}\hat{x}-\frac{1}{8l^2}{\hat{x}}^2+\cdots \right)\mathbb{Q}. 
\end{eqnarray} 
This can be decomposed to the terms in consistency with (\ref{S2-eq8}) as
\begin{eqnarray}\nonumber
\mathbb{H}&=&{\mathbb{H}}_{s}+{\mathbb{H}}_0+{\mathbb{H}}_{12}+{\mathbb{H}}_{34}+\cdots ,
\\ \label{S3-40} 
{\mathbb{H}}_{s}&=&\frac{1}{2}\hbar \Omega{\mathfrak{P}}^2+U\left(\mathfrak{X}\right)+\frac{1}{2}\hbar \omega \left({\mathbb{P}}^2+{\mathbb{Q}}^2\right), 
\end{eqnarray} 
where ${\mathbb{H}}_{s}$ contains non-interacting terms, there is no linear or product interaction, ${\mathbb{H}}_0$ is the conventional optomechanical interaction and is nonlinear having the third-order nonlinear, quadratic terms are  ${\mathbb{H}}_{12}$ and can be decomposed into the position ${\mathbb{H}}_1$ and momentum interaction ${\mathbb{H}}_2$ terms. Similarly, quintic or fifth-order nonlinear terms are ${\mathbb{H}}_{34}$ and can be decomposed into the position ${\mathbb{H}}_3$ and momentum interaction ${\mathbb{H}}_4$ terms. 
Hence, there are several distinct types of nonlinear optomechanical multi-phonon/multi-photon interactions, which by defining $R=r/4\approx 0.95$ are respectively given by
\begin{eqnarray}\nonumber
{\mathbb{H}}_0&=&-\frac{\hbar \omega }{2l}\sqrt{\frac{\hbar }{m\Omega}}\mathfrak{X}\left({\mathbb{P}}^2+{\mathbb{Q}}^2\right), \\ \nonumber {\mathbb{H}}_{12}&=&\frac{{\hbar }^2}{2l^2m}\left[\frac{\omega }{\Omega}{\mathfrak{X}}^2\left({\mathbb{P}}^2+{\mathbb{Q}}^2\right)-R\frac{\Omega}{\omega }{\mathfrak{P}}^2{\mathbb{Q}}^2\right] = {\mathbb{H}}_{1}+{\mathbb{H}}_{2}
\\ \nonumber
&=& \left[\frac{{\hbar}^2}{2l^2m}\left( \frac{\omega}{\Omega}\right) {\mathfrak{X}}^2\left({\mathbb{P}}^2 +{\mathbb{Q}}^2\right)\right]+\left[-R\frac{\hbar^2}{2l^2m}\left( \frac{\Omega}{\omega}\right){\mathfrak{P}}^2{\mathbb{Q}}^2\right],\\ \nonumber
\\ \nonumber 
{\mathbb{H}}_{34}&=&-\frac{{\hbar }^{\frac{5}{2}}}{2m^{\frac{3}{2}}l^3\sqrt{\Omega}}\left[\frac{\omega }{\Omega}{\mathfrak{X}}^3\left({\mathbb{P}}^2+{\mathbb{Q}}^2\right)-2R\frac{\Omega}{\omega }\mathcal{S}\left\{{\mathfrak{P}}^2\mathfrak{X}\right\}{\mathbb{Q}}^2\right]={\mathbb{H}}_{3}+{\mathbb{H}}_{4} \\ \label{S3-41}
&=&\left[-\frac{{\hbar }^{\frac{5}{2}}}{2m^{\frac{3}{2}}l^3\sqrt{\Omega}}\left(\frac{\omega }{\Omega}\right){\mathfrak{X}}^3\left({\mathbb{P}}^2+{\mathbb{Q}}^2\right)\right]
+\left[R\frac{{\hbar }^{\frac{5}{2}}}{m^{\frac{3}{2}}l^3\sqrt{\Omega}}\left(\frac{\Omega}{\omega }\right)\mathcal{S}\left\{{\mathfrak{P}}^2\mathfrak{X}\right\}{\mathbb{Q}}^2\right],
\end{eqnarray} 
and so on for higher order interactions. Here, the expansion of symmetrized terms, for instance, gives
\begin{equation} \label{S3-42} 
\mathcal{S}\left\{{\mathfrak{P}}^2\mathfrak{X}\right\}=\frac{1}{3}\left({\mathfrak{P}}^2\mathfrak{X}+\mathfrak{X}{\mathfrak{P}}^2+\mathfrak{P}\mathfrak{X}\mathfrak{P}\right). 
\end{equation} 
The very important conclusion of this calculation is that the behavior of optomechanical interaction under large electromagnetic frequency $\Omega\gg\omega$ should be markedly different from large mechanical frequency $\omega\gg\Omega$. On the one hand, in the regime of large electromagnetic frequency, all momentum interaction terms vanish and one may disregard $\mathbb{H}_2$ and $\mathbb{H}_4$. On the other hand, in the regime of large mechanical frequency, all momentum interaction terms dominate and one may disregard $\mathbb{H}_1$ and $\mathbb{H}_3$.

Now, it is noted that since usually $\omega \gg \Omega$, it may observed that the first terms are much weaker than the second terms. Hence, using the identity ${\mathbb{P}}^2+{\mathbb{Q}}^2={{\frac{1}{2}}}\hat{n}+{{\frac{1}{4}}}$  where $\hat{n}$ is photon number operator, the following is obtained
\begin{eqnarray}\nonumber
{\mathbb{H}}_0&=&-\frac{\hbar \omega }{l}\sqrt{\frac{\hbar }{m\Omega}}\mathfrak{X}\left(\hat{n}+\frac{1}{2}\right)=-\hbar \alpha \mathfrak{X}\left(\hat{n}+\frac{1}{2}\right),
\\ \nonumber
{\mathbb{H}}_1&=&\frac{{\hbar }^2}{l^2m}\frac{\omega }{\Omega}{\mathfrak{X}}^2\left(\hat{n}+\frac{1}{2}\right)=+\hbar \beta {\mathfrak{X}}^2\left(\hat{n}+\frac{1}{2}\right),
\\ \label{S3-43} 
{\mathbb{H}}_3&=&-\frac{\omega }{\Omega}\frac{{\hbar }^{\frac{5}{2}}}{m^{\frac{3}{2}}l^3\sqrt{\Omega}}{\mathfrak{X}}^3\left(\hat{n}+\frac{1}{2}\right)=-\hbar \gamma {\mathfrak{X}}^3\left(\hat{n}+\frac{1}{2}\right), 
\end{eqnarray} 
where ${\mathbb{H}}_0\equiv {\mathbb{H}}_{\rm OM}$ is the simple optomechanical interaction, and ${\mathbb{H}}_1$ is known as the quadratic interaction. It has to be emphasized that while ${\mathbb{H}}_0\equiv {\mathbb{H}}_{\rm OM}$ is actually nonlinear in the exact mathematical sense, it is the quadratic interaction ${\mathbb{H}}_1$ which is mostly referred to as the nonlinear interaction in the literature \cite{S3-55,S3-56,S3-57}. Since it is possible to make $g_0$ and therefore ${\mathbb{H}}_0$ identically vanish by appropriate optomechanical design \cite{S3-55,S3-56,S3-57,S3-58} in which the overlap integral of optical and mechanical modes sums up to zero, hence the quadratic interactions ${\mathbb{H}}_1$, and of course ${\mathbb{H}}_2$,  can then find physical significance. 

The quadratic interaction has been a subject of growing importance in the recent years in optomechanical systems \cite{S3-59,S3-60,S3-61} and beyond \cite{S3-62}. In \cite{S3-59} the photon statistics and blockade under ${\mathbb{H}}_1$ interactions has been studied and analytical expressions were derived. The quantum dissipative master function has been numerically solved and the corresponding correlation functions were obtained. Interestingly, quadratic optomechanical interactions can arise at the single-photon level, too, where rigorous analytical solutions have been devised \cite{S3-60}. Such type of interactions can be also well described using equivalent nonlinear electrical circuits, where a Josephson junction brings in the desired nonlinearity of quadratic interactions and terminate a pair of lumped transmission lines \cite{S3-61}. Finally, ultracold atoms also can exhibit interactions of a comparable type which is mathematically equivalent to the quadratic interaction \cite{S3-62}.

The single-photon multi-particle rates are given by
\begin{eqnarray}\nonumber
\alpha& =&\frac{\omega }{l}\sqrt{\frac{\hbar }{m\Omega}}\equiv \sqrt{2}g_0,
\\ \nonumber 
\beta& =&\frac{\hbar }{l^2m}\frac{\omega }{\Omega},
\\ \label{S3-44} 
\gamma& =&\frac{\omega }{\Omega}\frac{{\hbar }^{\frac{3}{2}}}{m^{\frac{3}{2}}l^3\sqrt{\Omega}}. 
\end{eqnarray} 
This summarizes the Hamiltonian as
\begin{equation} \label{S3-45} 
\mathbb{H}={\mathbb{H}}_{\rm non}+{\mathbb{H}}_{\rm int}. 
\end{equation} 
in which ${\mathbb{H}}_{\rm non}$ and ${\mathbb{H}}_{\rm int}$ are respectively the non-interacting and interacting Hamiltonians when $\omega \gg \Omega$ as
\begin{eqnarray} \nonumber
{\mathbb{H}}_{\rm non}&=&{\mathbb{H}}_{s}-\frac{1}{2}\hbar \left(\alpha \mathfrak{X}-\beta {\mathfrak{X}}^2+\gamma {\mathfrak{X}}^3+\cdots \right),
\\ \label{S3-46} 
{\mathbb{H}}_{\rm int}&=&\hbar \left(\alpha \mathfrak{X}-\beta {\mathfrak{X}}^2+\gamma {\mathfrak{X}}^3+\cdots \right)\hat{n}, 
\end{eqnarray} 
which implies the absence of momentum-field interactions under the assumption of large electromagnetic frequency. Now, the dimensionless constant $\theta$ is defined as
\begin{equation} \label{S3-47} 
\theta =\frac{1}{l}\sqrt{\frac{\hbar }{m\Omega}}=\frac{x_{\rm zp}}{l}, 
\end{equation} 
with $x_{\rm zp}$ being the r.m.s. value of zero-point fluctuations, by which the following is deduced
\begin{eqnarray}\nonumber
\beta &=&\theta \alpha \ll \alpha ,
\\ \label{S3-48} 
\gamma &=&\theta \beta ={\theta }^2\alpha \ll \beta . 
\end{eqnarray} 
This implies that every kind of higher-order interaction is typically $\theta $ times weaker than the interaction of the preceding-order. It should be noted that while such interactions are normally expected to rapidly vanish with the order increasing, is a well-known fact that certain physical phenomena such as magnetism in solid ${}_{3}$He cannot be understood without inclusion of four-particle interaction terms \cite{S3-74,S3-75}. It is worth here to mention that a detailed theory of optomechanics in superfluid ${}_{4}$He has been developed \cite{S3-76}, but no expression for the nonlinear terms has been reported. Interestingly, optomechanical \cite{S3-76a} and Brillouin lasing experiments \cite{S3-76b} on superfluid He a droplets show consistency with side-band inequivalence, although such systems are mechanically more complex and less understood.

In general, the interaction of mechanical and optical modes is not strictly one-dimensional, implying that the overlap integral of normalized modes should also be taken into account. For instance, odd mechanical modes with even optical modes have zero interaction. In that sense, tuning the interaction to an odd mode and then shining an even optical mode, or vice versa, makes the optomechanical interaction identically zero by setting $\alpha \equiv \sqrt{2}g_0=0$. Then the lowest order surviving interaction would be the ${\mathbb{H}}_4$ term. This method has been used in \cite{S3-55,S3-56,S3-57} to highlight the quadratic interaction and make its measurement much easier. It has been shown that these quadratic terms may be exploited for direct observation of mechanical eigenmode jumps \cite{S3-55,S3-56}, as well as two-phonon cooling and squeezing \cite{S3-57}, while the coupling strength $\beta $ could be increased by three orders of magnitude \cite{S3-56}.

Moreover, the origin of mechanical parametric coupling which has recently been phenomenologically hypothesized \cite{S3-58} for the associated physical interactions cannot be understood without the presented analysis, although based on some earlier experimental evidence \cite{S3-77}.

It must be added that the condition $\omega \gg \Omega$ may be violated in carefully designed superconducting microwave circuits and also the recently demonstrated molecular optomechanics \cite{S3-78}, which signifies the importance of the ${\mathfrak{P}}^2{\mathbb{Q}}^2$ term in ${\mathbb{H}}_4.$ It is furthermore worthwhile to point out that the regime $\omega =\Omega$ can be indeed be accessed and investigated, as it has been shown experimentally for superconducting circuit optomechanics \cite{S3-79}. The proposal of light propagation in a cylinder with rotating walls \cite{S3-80} also requires accessing regimes where $\omega $ and $\Omega$ fall within the same order of magnitude. Alternatively, in situations where $\omega \ll \Omega$, the scaling will be then given as
\begin{equation} \label{S3-49} 
\theta =\frac{R x_{\rm zp}}{l}\left(\frac{{\Omega}^2}{{\omega }^2}\right), 
\end{equation} 
which shows a significant enhancement in this type of interactions.

\subsection{Conditions for Observation of Momentum-Field Quadratic Interactions}

In summary, two general criteria should be satisfied in a carefully designed experiment to allow investigation of momentum-field quadratic interactions:

\begin{enumerate}
	\item  The optomechanical interaction ${\mathbb{H}}_3$ must vanish to allow easier study of quadratic interaction ${\mathbb{H}}_4$. This is quite possible by design as extensively has been discussed in the above and literature \cite{S3-55,S3-56,S3-57,S3-58,S3-59,S3-60,S3-61,S3-62}.
	
	\item  The mechanical frequency $\Omega$ must be of the same order of magnitude or exceeding the electromagnetic frequency $\omega $. This is also possible and at least one experiment using superconducting optomechanics \cite{S3-79} has accessed this regime. Other possibilities are molecular optomechanics \cite{S3-78} as well as a rotating cylinder \cite{S3-80}.
\end{enumerate}

Evidently, such momentum-field quadratic interactions might be more difficult to observe under normal experimental conditions compared to the regular optomechanical setups. However, progressive developments in the precision and accuracy of optomechanics experiments, such as what happened for the case of Laser Interferometric Gravitational Observatory (LIGO) \cite{S3-81}, could make it eventually possible to realize and probe such unexplored domains.

\subsection{Linearized Quantization}

The standard method to linearize the interaction Hamiltonian can be now used by making the substitutions $\hat{a}\to \bar{a}+\hat{a}$ where the new $\hat{a}$ operator from now on stands for the non-classical perturbations and $\left\langle \hat{a}\right\rangle $ is a measure of optical field amplitude. Then ignoring higher-order terms and retaining only the lowest-order interacting terms, we get
\begin{equation} \label{S3-50} 
{\mathbb{H}}_3=-\hbar g_3\left({\hat{b}}^{\dagger }+\hat{b}\right)\left(e^{i\varphi }{\hat{a}}^{\dagger }+e^{-i\varphi }\hat{a}\right), 
\end{equation} 
as well as 
\begin{eqnarray} \nonumber
{\mathbb{H}}_4&=&+\hbar g^+_4{\left({\hat{b}}^{\dagger }+\hat{b}\right)}^2\left(e^{i\varphi }{\hat{a}}^{\dagger }+e^{-i\varphi }\hat{a}\right),\ \ \omega \gg \Omega, 
\\ \label{S3-51} 
{\mathbb{H}}_4&=&+\hbar g^-_4{\left({\hat{b}}^{\dagger }-\hat{b}\right)}^2\left({\hat{a}}^{\dagger }+\hat{a}\right),\ \ \omega \ll \Omega. 
\end{eqnarray} 
Here, $\varphi =\measuredangle \bar{a}$ and the coupling frequency rates are defined as
\begin{eqnarray} \nonumber
g_3&=&\frac{\alpha }{\sqrt{2}}\left|\bar{a}\right|\equiv g_0\left|\bar{a}\right|\equiv G,
\\ \nonumber
g^+_4&=&\frac{\beta }{2}\left|\bar{a}\right|=\theta G,
\\ \label{S3-52} 
g^-_4&=&\ R\frac{{\Omega}^2}{{\omega }^2}g^+_4. 
\end{eqnarray} 

Following the same method to linearize the mechanical motions, with the replacement $\hat{b}\to \bar{b}+\hat{b}$ where the new $\hat{b}$ operator denotes the perturbations, gives rise to the expressions
\begin{eqnarray}\nonumber
{\mathbb{H}}_4&=&+\hbar G^+_4\left({\hat{b}}^{\dagger }+\hat{b}\right)\left(e^{i\varphi }{\hat{a}}^{\dagger }+e^{-i\varphi }\hat{a}\right),\ \ \omega \gg \Omega,
\\ \label{S3-53} 
{\mathbb{H}}_4&=&+\hbar G^-_4\left({\hat{b}}^{\dagger }-\hat{b}\right)\left({\hat{a}}^{\dagger }+\hat{a}\right),\ \ \omega \ll \Omega, 
\end{eqnarray} 
where $\vartheta =\measuredangle \bar{b}$ is set to zero without loss of generality, $G^+_4=2\left|\bar{b}\right|g^+_4\cos\vartheta $, and $G^-_4=2\left|\bar{b}\right|g^-_4\sin\vartheta $. In general, when $\omega \gg \Omega$ is violated, one would expect the momentum of mirror be coupled to the first quadrature of the radiation field. This type of interaction can be compared to the normal optomechanical interaction (\ref{S3-50}), in which the position is coupled to the first quadrature of the field. When the optical and mechanical frequencies do not differ by orders of magnitude so that neither $\omega \gg \Omega$ nor $\omega \ll \Omega$ hold, then the linearized Hamiltonian could be recast as
\begin{equation} \label{S3-54} 
{\mathbb{H}}_4=\hbar G^+_4\left({\hat{b}}^{\dagger }+\hat{b}\right)\left(e^{i\varphi }{\hat{a}}^{\dagger }+e^{-i\varphi }\hat{a}\right)+\hbar G^-_4\left({\hat{b}}^{\dagger }-\hat{b}\right)\left({\hat{a}}^{\dagger }+\hat{a}\right).
\end{equation} 

\subsection{Squeezing in Quadratic Optomechanical Interaction}\label{S3-Squeeze}
The linearized relationship (\ref{S3-54}) for the quadratic Hamiltonian can be written as
\begin{equation} \label{S3-55} 
{\mathbb{H}}_4=\hbar G\left(\hat{a}{{\mathbb{B}}}^{\dagger }+{\hat{a}}^{\dagger }{\mathbb{B}}\right)=\hbar G\left({{\mathbb{A}}}^{\dagger }\hat{b}+{\mathbb{A}}{\hat{b}}^{\dagger }\right), 
\end{equation} 
where
\begin{eqnarray} \nonumber
G&=&\sqrt{G^+_4G^-_4},
\\ \nonumber
{\mathbb{A}}&=&{\hat{a}}^{\dagger }\sinh\rho +\hat{a}\cosh\rho ,
\\ \nonumber
{\mathbb{B}}&=&{\hat{b}}^{\dagger }\cosh\rho +\hat{b}\sinh\rho ,
\\  \label{S3-56} 
\rho &=&{\tanh}^{-1}\left(\frac{G^+_4-G^-_4e^{i\varphi }}{G^+_4+G^-_4e^{i\varphi }}\right). 
\end{eqnarray} 
The expression (\ref{S3-55}) contains linearized terms of both of the standard and non-standard quadratic interactions. It is here to be noticed that ${\mathbb{B}}$ and ${\mathbb{A}}$ are in the standard form of Bogoliubov squeezing operator \cite{S3-54,S3-55}. It may be noted that the equation
\begin{equation} \label{S3-57} 
G^-_4=R\frac{{\Omega}^2}{{\omega }^2}G^+_4, 
\end{equation} 
is actually a function of $\vartheta $ by definition of $G^+_4$ and $G^-_4$. Simplifying the above gives the expression for squeeze ratio as
\begin{equation} \label{S3-58} 
\rho =\ln\left(\frac{\omega }{\sqrt{R}\Omega}\right)-i\frac{\varphi }{2}. 
\end{equation} 
This shows that quadratic interactions give rise to squeezed mechanical or optical states unless $\omega =\sqrt{R}\Omega$ and of course $\varphi =0$.

Let us here make a mention of the fact that based on (\ref{S3-54}), any interaction Hamiltonian which can be put into a simple product form, such as the factorizable fully linearized optomechanics (\ref{S2-eq3}), may not cause squeezing. By inspection of (\ref{S3-56}) we easily can see that setting either of $G_4^+$ or $G_4^-$ to zero will remove the possibility of Bogoliubov transformation and makes the interaction (\ref{S3-55}) singular. This explains that a fully linearized single mode 1:1 optomechanical interaction does not yield squeezing, since the only remaining interaction term is linearized down to a simple product. In the exact mathematical sense, however, quantum optomechanics is nonlinear and the possibility of squeezing only through single-mode 1:1 optomechanical interaction cannot be immediately decided using this argument. However, it will be demonstrated in \S\ref{Section-10} that non-standaard momentum-field interactions together with the standard quadratic terms may cause a tiny squeezing even in a single-mode 1:1 optomechanical cavity. Given all these facts, squeezing of optomechanical interaction even at the fully linearized level is still possible using more than one electromagnetic or mechanical interacting modes.

As discussed above, the Hamiltonian ${\mathbb{H}}_3$ can be made identically zero \cite{S3-56,S3-57,S3-82,S3-83} to access the quadratic interaction terms ${\mathbb{H}}_4$ directly. There is an interesting condition on the ratio of optical to mechanical frequencies, which could be sought here. Let
\begin{equation} \label{S3-59} 
\omega =\sqrt{\eta R}\Omega, 
\end{equation} 
in which $\eta $ is a constant to be determined later. This allows the ${\mathbb{H}}_4$ to be written as
\begin{equation} \label{S3-60} 
{\mathbb{H}}_4=2\hbar \beta \left[-\frac{1}{\eta }{\mathfrak{P}}^2{\mathbb{Q}}^2+{\mathfrak{X}}^2\left({\mathbb{P}}^2+{\mathbb{Q}}^2\right)\right]=\hbar \beta \left[\frac{1}{2\eta }{\left({\hat{b}}^{\dagger }-\hat{b}\right)}^2{\left({\hat{a}}^{\dagger }+\hat{a}\right)}^2+{\left({\hat{b}}^{\dagger }+\hat{b}\right)}^2\left({\hat{a}}^{\dagger }\hat{a}+\hat{a}{\hat{a}}^{\dagger }\right)\right]. 
\end{equation} 
Further expansion of results as shown in \S\ref{AppendixB} gives  
\begin{equation} \label{S3-61} 
{\mathbb{H}}_{4,{\rm int}}=2\hbar \beta \left|\bar{a}\right|\left[\left(1+\frac{1}{2\eta }\right)\left({{\hat{b}}^{\dagger 2}}+{\hat{b}}^2\right)+\left(1-\frac{1}{2\eta }\right)\hat{m}\right]\times \left({\hat{a}}^{\dagger }+\hat{a}\right). 
\end{equation} 
Then, for the choice of $\eta ={{\frac{1}{2}}}$, that is $\omega \cong 0.69\Omega$, one may reach the desired interaction quadratic Hamiltonian, linearized in the electromagnetic operators
\begin{equation} \label{S3-62} 
{\mathbb{H}}_{4,{\rm int}}=\hbar 2J\left({{\hat{b}}^{\dagger 2}}+{\hat{b}}^2\right)\left({\hat{a}}^{\dagger }+\hat{a}\right), 
\end{equation} 
where the interaction rate is $J=\lambda =2\beta \left|\bar{a}\right|$. When expanded in its four terms and after the replacement $\hat{c}={{\frac{1}{2}}}{\hat{b}}^2$ to be shown in \S\ref{AppendixC}, one may immediately recognize the Hamiltonian of the type
\begin{equation} \label{S3-63} 
{\mathbb{H}}_{4,{\rm int}}=\hbar J\left(\hat{c}{\hat{a}}^{\dagger }+{\hat{c}}^{\dagger }\hat{a}\right)+\hbar \lambda \left({\hat{c}}^{\dagger }{\hat{a}}^{\dagger }+\hat{c}\hat{a}\right). 
\end{equation} 
The first parenthesis represents the Hopping or Beam-Splitter term, while the second is normally referred to as the dissipation. Interestingly, the above could have been further linearized in mechanical operators to obtain
\begin{equation} \label{S3-64} 
{\mathbb{H}}_{4,{\rm int}}=i\hbar \mathcal{J}\left(e^{-i\vartheta }{\hat{b}}^{\dagger }+e^{i\vartheta }\hat{b}\right)\left({\hat{a}}^{\dagger }-\hat{a}\right)=i\hbar \mathcal{J}\left(e^{i\vartheta }\hat{b}{\hat{a}}^{\dagger }-e^{-i\vartheta }{\hat{b}}^{\dagger }\hat{a}\right)+i\hbar \mathcal{J}\left(e^{-i\vartheta }{\hat{b}}^{\dagger }{\hat{a}}^{\dagger }-e^{i\vartheta }\hat{b}\hat{a}\right). 
\end{equation} 
where $\mathcal{J}=2J\left|\bar{b}\right|.$ This latter form, may find application in non-reciprocal optomechanics \cite{S3-84}.

\subsection{Relativistic Considerations}

As a final remark, the approximate nature of the Lagranian formulation by Law \cite{S2-5} has not been left unnoticed. It could be attributed first to the non-relativistic description of mirror's motion which ultimately ignores higher-order interactions, and then to the relativistic nature of radiation friction force and the associated Doppler shift \cite{S3-66}. As a result, in a subsequent paper by Cheung and Law \cite{S3-67}, it has been made clear that the non-relativistic optomechanical Hamiltonian is correct only to the first-order in $\dot{q}$. The relativistic corrections can be however quite different in nature, and can be group into three different categories:
\begin{enumerate}
	\item  The relativistic Doppler shift \cite{S3-66}, which causes corrections in ${\dot{q}}/{c}$,
	\item  The relativistic correction in radiation pressure term \cite{S3-67}, the lowest-order of which is proportional to ${\dot{q}}/{c}$,
	\item  The length contraction \cite{S3-68} due to the moving mirror boundary, which results again in corrections as ${\dot{q}}/{c}$. 
\end{enumerate}
Not surprisingly, all these relativistic terms vanish in the limit of infinite light speed $c$. These altogether could be taken into account in a fully relativistic formulation of the Lagrangian and equations of motions for the mirror and optical field \cite{S3-85}, which has been recently carried out in an extensive research by Casta\~{n}os \& Weder \cite{S3-68}.

As shown in \S\ref{AppendixD}, the total relativistic correction terms added to the Hamiltonian takes the form
\begin{equation} \label{S3-65} 
\Delta \mathcal{H}=-\hbar {\left({\hat{b}}^{\dagger }-\hat{b}\right)}^2\sum_k{w_{kj}\left({\hat{a}}^{\dagger }_k+{\hat{a}}_k\right)\left({\hat{a}}^{\dagger }_j+{\hat{a}}_j\right)}, 
\end{equation} 
which is obviously quadratic and also assumes the general form of momentum-field interaction. For the single-mode cavity, $w={{\chi }_0\pi \hbar d\Omega}/{4mcl^2}\propto \Omega/\omega$, to be compared with $\gamma=\theta\beta \propto \Omega/\omega$ in (\ref{S3-48}). Curiously, when the mechanical frequency is large and $\Omega\ll\omega$ is violated, both of the quadratic interaction rates of the non-standard $\beta$ and relativistic correction $w$ get large proportionally. Hence, the relativistic correction to the quadratic inteeraction rate of the quadratic Hamiltonian ${\mathbb{H}}_4$  is expressible as the dimensionless fraction
\begin{equation} \label{S3-66} 
-\frac{w}{\beta }=-\frac{{\chi }_0\pi d{\Omega}^2}{4c\omega }. 
\end{equation} 
Again, it is seen that when $\omega \gg \Omega$ is violated, the relativistic corrections might be quite significant. In any case, there is no relativistic correction to ${\mathbb{H}}_3$. Furthermore, all relativistic corrections (\ref{S3-65}) vanish in the limit of $c\to\infty$ as expected.

\subsection{Derivation of (\ref{S3-9})}\label{AppendixA}

In this section, we present the step-by-step details of the derivation of (\ref{S3-9}) from the previous equations, as it constitutes the most critical part of this part. Starting from (\ref{S3-8}), one has first to rename the dummy index from $k$ to $j$, multiply both sides by $f_k$, and then take the inner product. This will yield the expression
\begin{eqnarray} \nonumber
-\sum{\omega_j^2Q_j\left(f_k|f_j\right)}&=&\sum{\ddot{Q}_j\left(f_k|f_j\right)}-\frac{\dot{q}}{q}\sum{\dot{Q}_j\left[\left(f_k|f_j\right)+2\left(f_k\left|\kappa_j x\right|g_j\right)\right]},
\\ \nonumber
&-&\frac{\ddot{q}}{q}\sum{Q_j\left[\frac{1}{2}\left(f_k|f_j\right)+\left(f_k\left|{\kappa }_jx\right|g_j\right)\right]}\\
\label{S3-A1}
&+&\frac{\dot{q}^2}{q^2}\sum{Q_j\left[\frac{3}{4}\left(f_k|f_j\right)+3\left(f_k\left|{\kappa }_jx\right|g_j\right)-\left(f_k\left|{\kappa }_j^2x^2\right|g_j\right)\right]}. 
\end{eqnarray}
Using (\ref{S3-3}), we trivially get
\begin{eqnarray}\nonumber
-\sum{{{\omega }_j}^2Q_j{\delta }_{kj}}&=&\sum{{\ddot{Q}}_j{\delta }_{kj}}-\frac{\dot{q}}{q}\sum{{\dot{Q}}_j\left[{\delta }_{kj}+2{\alpha }_{kj}\right]}-\frac{\ddot{q}}{q}\sum{Q_j\left[\frac{1}{2}{\delta }_{kj}+{\alpha }_{kj}\right]}\\
\label{S3-A2}
&+&\frac{{\dot{q}}^2}{q^2}\sum{Q_j\left[\frac{3}{4}{\delta }_{kj}+3{\alpha }_{kj}-{\beta }_{kj}\right]},
\end{eqnarray}
which after rearrangement takes the form
\begin{eqnarray} \nonumber
{\ddot{Q}}_k&=&-{{\omega }_k}^2Q_k+\frac{\dot{q}}{q}{\dot{Q}}_k+\frac{\ddot{q}}{2q}Q_k-\frac{3{\dot{q}}^2}{4q^2}Q_k+2\frac{\dot{q}}{q}\sum{{\alpha }_{kj}{\dot{Q}}_j}+\frac{\ddot{q}}{q}\sum{{{\alpha }_{kj}Q}_j}
\\ \label{S3-A3}
&-&\frac{{\dot{q}}^2}{q^2}\sum{\left(3{\alpha }_{kj}-{\beta }_{kj}\right)Q_j}.
\end{eqnarray}
We may furthermore use ${\alpha }_{kj}=-{{\frac{1}{2}}}{\delta }_{kj}+g_{kj}$ and ${\beta }_{kj}=\left({{\frac{1}{3}}}k^2{\pi }^2-{{\frac{1}{2}}}\right){\delta }_{kj}+h_{kj}$ from (\ref{S3-4}) to simplify and rewrite (\ref{S3-A3}) as
\begin{eqnarray}\nonumber
{\ddot{Q}}_k&=&-{{\omega }_k}^2Q_k-\frac{3{\dot{q}}^2}{4q^2}Q_k+\frac{3{\dot{q}}^2}{2q^2}Q_k+\frac{{\dot{q}}^2}{q^2}\left(\frac{k^2{\pi }^2}{3}-\frac{1}{2}\right)Q_k+2\frac{\dot{q}}{q}\sum{g_{kj}{\dot{Q}}_j}+\frac{\ddot{q}}{q}\sum{{g_{kj}Q}_j}\\ 
\label{S3-A4}
&-&\frac{{\dot{q}}^2}{q^2}\sum{\left(3g_{kj}-h_{kj}\right)Q_j},
\end{eqnarray}
which by plugging in the definition for $r_k$ from (\ref{S3-10}) takes the form
\begin{equation}\label{S3-A5}
{\ddot{Q}}_k=-{{\omega }_k}^2Q_k+r_k\frac{{\dot{q}}^2}{q^2}Q_k+2\frac{\dot{q}}{q}\sum{g_{kj}{\dot{Q}}_j}+\frac{\ddot{q}}{q}\sum{g_{kj}Q_j}+\frac{{\dot{q}}^2}{q^2}\sum{\left(h_{kj}-3g_{kj}\right)Q_j}.
\end{equation}
This is exactly the equation (\ref{S3-9}).

\subsection{Special Case}\label{AppendixB}

Expansion of (\ref{S3-60}) results in
\begin{eqnarray} \nonumber
{\mathbb{H}}_4&=&\ \hbar \beta \left[\frac{1}{2\eta }{\left({\hat{b}}^{\dagger }-\hat{b}\right)}^2\left({{\hat{a}}^{\dagger 2}}+{\hat{a}}^2\right)+\left(1+\frac{1}{2\eta }\right)\left(\hat{m}+\frac{1}{2}\right)\left(\hat{n}+\frac{1}{2}\right)+\left({{\hat{b}}^{\dagger 2}}+{\hat{b}}^2\right)\left(\hat{n}+\frac{1}{2}\right)\right]
\\
\nonumber
&=&\ \hbar \beta \left[\frac{1}{2\eta }\left({{\hat{b}}^{\dagger 2}}+{\hat{b}}^2\right)\left({{\hat{a}}^{\dagger 2}}+{\hat{a}}^2\right)+\left(1+\frac{1}{2\eta }\right)\left(\hat{m}+\frac{1}{2}\right)\left(\hat{n}+\frac{1}{2}\right)\right.
\\
\label{S3-B1}
&+&\left.\left({{\hat{b}}^{\dagger 2}}+{\hat{b}}^2\right)\left(\hat{n}+\frac{1}{2}\right)-\frac{1}{2\eta }\left(\hat{m}+\frac{1}{2}\right)\left({{\hat{a}}^{\dagger 2}}+{\hat{a}}^2\right)\right],
\end{eqnarray}
where $\hat{n}={\hat{a}}^{\dagger }\hat{a}$ and $\hat{m}={\hat{b}}^{\dagger }\hat{b}$ are respectively photon and photon number operators. Retaining only the interacting terms, gives the expression
\begin{equation}\label{S3-B2}
{\mathbb{H}}_{4,{\rm int}}=\ \hbar \beta \left[\frac{1}{2\eta }\left({{\hat{b}}^{\dagger 2}}+{\hat{b}}^2\right)\left({{\hat{a}}^{\dagger 2}}+{\hat{a}}^2\right)+\left(1+\frac{1}{2\eta }\right)\hat{m}\hat{n}+\left({{\hat{b}}^{\dagger 2}}+{\hat{b}}^2\right)\hat{n}-\frac{1}{\eta }\hat{m}\left({{\hat{a}}^{\dagger 2}}+{\hat{a}}^2\right)\right].
\end{equation}
In the limit $\omega \gg \Omega$ with $\eta \to \infty $, ${\mathbb{H}}_4$ as in (\ref{S3-53}) is recovered. With linearization of the electromagnetic field operators, and removing the non-interacting terms the following is found
\begin{eqnarray}\nonumber
{\mathbb{H}}_{4,{\rm int}}&=&2\hbar \beta \left[\frac{1}{2\eta }\left({{\hat{b}}^{\dagger 2}}+{\hat{b}}^2\right)\left({{\bar{a}}^*\hat{a}}^{\dagger }+\bar{a}\hat{a}\right)+\left(1+\frac{1}{2\eta }\right)\hat{m}\left(\bar{a}{\hat{a}}^{\dagger }+{\bar{a}}^*\hat{a}\right)\right.
\\ \label{S3-B3}
&+&\left.\left({{\hat{b}}^{\dagger 2}}+{\hat{b}}^2\right)\left(\bar{a}{\hat{a}}^{\dagger }+{\bar{a}}^*\hat{a}\right)-\frac{1}{\eta }\hat{m}\left({{\bar{a}}^*\hat{a}}^{\dagger }+\bar{a}\hat{a}\right)\right],
\end{eqnarray}
By continuing the work on the linearized quadratic interaction one obtains the expression
\begin{eqnarray}\nonumber
{\mathbb{H}}_{4,{\rm int}}&=&2\hbar \beta \left|\bar{a}\right|\left[\frac{1}{2\eta }\left({{\hat{b}}^{\dagger 2}}+{\hat{b}}^2\right)\left({e^{-i\varphi }\hat{a}}^{\dagger }+e^{i\varphi }\hat{a}\right)+\left(1+\frac{1}{2\eta }\right)\hat{m}\left(e^{i\varphi }{\hat{a}}^{\dagger }+e^{-i\varphi }\hat{a}\right)\right.
\\ \label{S3-B4}
&+&\left.\left({{\hat{b}}^{\dagger 2}}+{\hat{b}}^2\right)\left(e^{i\varphi }{\hat{a}}^{\dagger }+e^{-i\varphi }\hat{a}\right)-\frac{1}{\eta }\hat{m}\left({e^{-i\varphi }\hat{a}}^{\dagger }+e^{i\varphi }\hat{a}\right)\right],
\end{eqnarray}
which for $\varphi =0$ simplifies to
\begin{equation}\label{S3-B5}
{\mathbb{H}}_{4,{\rm int}}=2\hbar \beta \left|\bar{a}\right|\left[\left(1+\frac{1}{2\eta }\right)\left({{\hat{b}}^{\dagger 2}}+{\hat{b}}^2\right)+\left(1-\frac{1}{2\eta }\right)\hat{m}\right]\times \left({\hat{a}}^{\dagger }+\hat{a}\right).
\end{equation}

\subsection{Squared Annihilator}\label{AppendixC}

The operator $\hat{c}$ has clearly a simple solution for its eigenkets, which is the same as coherent states such as $\left|z\right\rangle \ $where $\hat{c}\left|z\right\rangle ={{\frac{1}{2}}}z^2\left|z\right\rangle $. Hence, the eigenvalue is simply the complex number ${{\frac{1}{2}}}z^2$. Meanwhile, one has 
\begin{equation}\label{S3-C1}
\left|z\right\rangle =e^{-\frac{1}{2}{\left|z\right|}^2}\sum^{\infty }_{m=0}{\frac{z^m}{\sqrt{m!}}\left|m\right\rangle }.
\end{equation}
It is furthermore easy to check that $\left[\hat{c},{\hat{c}}^{\dagger }\right]=\hat{m}+{{\frac{1}{2}}}={\hat{b}}^{\dagger }\hat{b}+{{\frac{1}{2}}}$ \cite{S2-Paper2}. When, the mean phonon number is $\left\langle \hat{m}\right\rangle ={{\frac{1}{2}}}$, then $\left\langle \left[\hat{c},{\hat{c}}^{\dagger }\right]\right\rangle =1$, which is quite similar to the commutator $\left[\hat{b},{\hat{b}}^{\dagger }\right]$=1.

\subsection{Derivation of (\ref{S3-65})}\label{AppendixD}

The relativistic Lagrangian density for a light field with normal incidence to a fully reflective and non-compressible moving mirror, correct to the first-order in $\dot{q}$ and $\ddot{q}$, reads \cite{S3-68}

\begin{equation}\label{S3-D1}
\frac{\partial \mathcal{L}}{\partial V}=\frac{1}{2}\left(\textbf{E}\cdot \textbf{D}-\textbf{H}\cdot\textbf{B}\right)+\frac{{\mathcal{G}}^2{\epsilon }_0}{2}\chi {\left|\textbf{E}\cdot \hat{z}-c\mathcal{B}\textbf{B}\cdot \hat{y}\right|}^2,
\end{equation}
where $\textbf{E}=-\hat{z}{{\frac{\partial }{\partial t}}}A$, $\textbf{B}=\nabla \times \left(A\hat{z}\right)=-\hat{y}{{\frac{\partial }{\partial x}}}A$, $\textbf{D}={\epsilon }_0\textbf{E}$, and $\textbf{B}={\mu }_0\textbf{H}$. Furthermore,
\begin{eqnarray}\nonumber
\mathcal{B}&=&\frac{v}{c}=\frac{\dot{q}}{c}=\frac{p}{mc},\ 
\\ \label{S3-D2}
\mathcal{G}&=&\frac{1}{\sqrt{1-\mathcal{B}^2}},
\end{eqnarray}
and $\chi $ is a dimensionless shape function independent of $v$, being zero outside mirror and relative susceptibility of the mirror's dielectric ${\chi }_0$ inside, and ${\epsilon }_0$ is the permittivity of vacuum. By expanding in the powers of $\mathcal{B}$, this Lagrangian gives the first- and second-order corrections to the quadratic Hamiltonian density as
\begin{equation}\label{S3-D3}
\frac{\partial }{\partial V}\Delta \mathcal{H}=\frac{\partial }{\partial V}{\Delta \mathcal{H}}^{\left(1\right)}+\frac{\partial }{\partial V}{\Delta \mathcal{H}}^{\left(2\right)}=-\left.\frac{\partial }{\partial \mathcal{B}}\frac{\partial \mathcal{L}}{\partial V}\right|_{\mathcal{B}=0}\mathcal{B}-\left.\frac{1}{2}\frac{{\partial }^2}{{\partial \mathcal{B}}^2}\frac{\partial \mathcal{L}}{\partial V}\right|_{\mathcal{B}=0}\mathcal{B}^2.
\end{equation}
Hence, one may obtain 
\begin{equation}\label{S3-D4}
{\Delta \mathcal{H}}^{\left(1\right)}=-{\epsilon }_0S\int^q_0{{\left.\frac{\partial }{\partial \mathcal{B}}\left[\frac{1}{2}{\left(\frac{1}{\sqrt{1-\mathcal{B}^2}}\right)}^2\chi {\left|A_t+c\mathcal{B}A_x\right|}^2\right]\right|}_{\mathcal{B}=0}\mathcal{B}dx},
\end{equation}
which further simplifies as
\begin{equation}\label{S3-D5}
{\Delta \mathcal{H}}^{\left(1\right)}=-{\epsilon }_0Sc\int^q_0{\chi A_tA_x\mathcal{B}dx}.
\end{equation}
It is appropriate to assume the approximation of conducting interface \cite{S3-87,S3-88,S3-89} for the mirror, such as the thickness is let to approach zero, while it susceptibility increases proportionally. In that limit, one may set 
\begin{equation}\label{S3-D6}
\chi \left(x,t\right)\approx {\chi }_0d\delta \left[x-q\left(t\right)\right],
\end{equation}
where $d$ is the mirror's thickness. This is similar to the assumption of the locality of interaction by Gardiner \& Zoller \cite{S2-Noise1}, too. Hence, one gets
\begin{equation}\label{S3-D7}
{\Delta \mathcal{H}}^{\left(1\right)}\approx -{\epsilon }_0\mathcal{V}c{\chi }_0A_t\left(q,t\right)A_x\left(q,t\right)\mathcal{B},
\end{equation}
with $\mathcal{V}=Sd$ is the cavity volume. Now, one has from (\ref{S3-2}), $A_x\left(q,t\right)=s\pi \sqrt{{2}/{q^3}}\sum{kQ_k},$ $A_t\left(q,t\right)=-\dot{q}A_x\left(q,t\right)$, and thus
\begin{equation}\label{S3-D8}
{\Delta \mathcal{H}}^{\left(1\right)}\approx 2{\pi }^2d{\chi }_0\frac{{\dot{q}}^2}{q^3}\sum{kjQ_kQ_j}.
\end{equation}
It is quite remarkable that (\ref{S3-D8}) is purely relativistic, and vanishes in the limit of infinite $c$, as shown below. Here, the dependence on $t$ is hidden for convenience. This term translates after symmetrization into 
\begin{equation}\label{S3-D9}
{\Delta \mathcal{H}}^{\left(1\right)}=-2\hbar {\left({\hat{b}}^{\dagger }-\hat{b}\right)}^2\sum_k{w_{kj}\left({\hat{a}}^{\dagger }_k+{\hat{a}}_k\right)\left({\hat{a}}^{\dagger }_j+{\hat{a}}_j\right)},
\end{equation}
where $w_{kj}=\sqrt{jk}{{\chi }_0\pi \hbar d\Omega}/{4mcl^2}$ are the coupling rates. Now, the quadratic correction ${\Delta \mathcal{H}}^{\left(2\right)}$ is given by
\begin{equation}\label{S3-D10}
{\Delta \mathcal{H}}^{\left(2\right)}=-\frac{1}{2}{\epsilon }_0S{\left.\int^q_0{\frac{{\partial }^2}{{\partial \mathcal{B}}^2}\left\{\frac{1}{2}{\left(\frac{1}{\sqrt{1-\mathcal{B}^2}}\right)}^2\chi {\left|A_t+c\mathcal{B}A_x\right|}^2\right\}dx}\right|}_{\mathcal{B}=0}\mathcal{B}^2.
\end{equation}
Simplifying and using the conducting interface approximation gives
\begin{eqnarray}\nonumber
{\Delta \mathcal{H}}^{\left(2\right)}&=&-\frac{1}{2}S{\epsilon }_0\int^q_0{\chi \left[{A_t}^2+c^2{A_x}^2\right]\mathcal{B}^2dx}\\ \nonumber
&\approx& -\frac{1}{2}\mathcal{V}{\epsilon }_0{\chi }_0\left[{\dot{q}}^2+c^2\right]{A_x}^2\left(q,t\right)\mathcal{B}^2\\
\label{S3-D11}
&\approx& -{\pi }^2d{\chi }_0\frac{{\dot{q}}^2}{q^3}\sum{kjQ_kQ_j}.
\end{eqnarray}
This one after insertion of operators gives$\ {\Delta \mathcal{H}}^{\left(2\right)}=-{{\frac{1}{2}}}{\Delta \mathcal{H}}^{\left(1\right)}$ and thus the total relativistic correction is found as
\begin{equation}\label{S3-D12}
{\Delta \mathcal{H}}^{\left(1\right)}+{\Delta \mathcal{H}}^{\left(2\right)}=-\hbar {\left({\hat{b}}^{\dagger }-\hat{b}\right)}^2\sum_k{w_{kj}\left({\hat{a}}^{\dagger }_k+{\hat{a}}_k\right)\left({\hat{a}}^{\dagger }_j+{\hat{a}}_j\right)}.
\end{equation}

In the end, the derivation of the optomechanical Hamiltonian has been carefully examined from the modal expansions, equations of motion, all the way to the Lagrangian, and ultimately the Hamiltonian and relativistic considerations. A set of correction terms to the nonlinear terms have been identified, which do not eliminate under any choice of canonical momenta. With the careful system design which allows $g_0=0$, these type of interactions are particularly interesting and now being actively pursued. It was shown that under these conditions one may expect coupling of mechanical momentum to the field position. Other sorts of interactions emerge under various conditions. In general, when the optical frequency is not much larger than the mechanical frequency, novel nonlinear interactions may appear. 

We furthermore showed for single-mode 1:1 optomechanical cavities where there exist one optical and one mechanical mode, that squeezing via the basic optomechanical interaction $\mathbb{H}_{\rm OM}$ is not possible. The reason is that the linearized form is factorizable and a simple product. However, quadratic interactions with non-standard momentum-field interactions give access to both of the quadratures of the mechanical/optical fields. As a result, squeezing via quadratic interactions even in a single-mode 1:1 optomechanical cavity is possible, but the non-standard quadratic corrections must survive.

The very important conclusion is that the behavior of optomechanical interaction under large electromagnetic frequency $\Omega\gg\omega$ should be markedly different from large mechanical frequency $\omega\gg\Omega$. While for $\omega\gg\Omega$, which has been the usual studied case so far, all momentum interaction terms vanish, for $\Omega\gg\omega$ momentum interaction terms dominate and this leads to an unexplored and new realm in nonlinear quantum optomechanics.

\section{Open Systems \& Langevin Equations} \label{Section-4}
This section describes the fundamental approach to solve the nonlinear Langevin equations arising from quadratic interactions in quantum mechanics using operator algebra. While, the zeroth order linearization approximation to the operators is normally used, here first and second order truncation perturbation schemes are introduced and employed. These schemes employ higher-order system operators, and then approximate number operators with their corresponding mean boson numbers, only where needed. Spectral densities of squared operators are derived later in \S\ref{Section-5}, and an expression for the second-order correlation function at zero time-delay has been found, which reveals that the cavity photon occupation of an ideal laser at threshold reaches $\sqrt{6}-2$, in good agreement with other extensive numerical calculations. As further applications, analysis of the quantum anharmonic oscillator, calculation of $Q-$functions, and also analysis of quantum limited amplifiers, and nondemoliton measurements will be later based on this scheme and discussed in \S\ref{Section-6}. An extensive application of higher-order operators to quantum and quadratic optomechanics as well as cross-Kerr interaction are to be discussed in the subsequent sections \S\ref{Section-8} through \S\ref{Section-11}, which also includes a dedicated and in depth study of an unprecedented nonlinear symmetry breaking phenomenon referred to as the side-band inequivalence \S\ref{Section-9}.

In quantum optomechanics the standard interaction Hamiltonian is simply the product of photon number $\hat{n}=\hat{a}^\dagger\hat{a}$ and the position $x_{\textrm{zp}}(\hat{b}+\hat{b}^\dagger)$ operators \cite{S2-1,S2-2,S2-3,S2-6,S2-4,S2-5}, where $x_{\textrm{zp}}$ is the zero-point motion, and $\hat{a}$ and $\hat{b}$ are respectively the photon and phonon annihilators. This type of interaction can successfully describe a vast range of phenomena, including optomechanical arrays \cite{S3-8,S4-7,S3-10,S3-11,S3-12,S3-13,S3-14}, squeezing of phonon states \cite{S3-18,S4-14a,S3-20}, non-reciprocal optomechanics \cite{S3-22,S3-23,S4-17,S4-17a}, Heisenberg’s limited measurements \cite{S3-21}, sensing \cite{S3-29,S3-30,S3-31}, engineered dissipation and states \cite{S3-32,S3-33}, and non-reciprocal acousto-optics \cite{S4-23}. In all these applications, the mathematical toolbox to estimate the measured spectrum is Langevin equations \cite{S2-Noise1,S2-Noise0,S2-Noise2,S2-Noise3}.

Usually, the analysis of quantum optomechanics is done within the linearized approximation of photon ladder operators, normally done as $\hat{a}\rightarrow\bar{a}+\delta\hat{a}$ with $|\bar{a}|^2=\bar{n}$ being the mean cavity photon number, while nonlinear terms in $\delta\hat{a}$ are ignored. But this suffers from limited accuracy wherever the basic optomechanical interaction $\mathbb{H}_{\rm OM}=-\hbar g_0\hat{n}(\hat{b}+\hat{b}^\dagger)$ is either vanishingly small or non-existent. In fact, the single-photon interaction rate $g_0$ can be identically made zero by appropriate design \cite{S3-55,S3-56,S3-57,S3-58}, when quadratic or even quartic effects are primarily pursued. This urges need for accurate knowledge of higher-order interaction terms. 

Some other optomechanical phenomena such as four-wave mixing, also can be suitably understood by incorporation of higher-order interaction terms \cite{S3-63}. Recent experiments \cite{S3-64,S3-65} have already established the significance and prominent role of such type of nonlinear interactions. In fact, quadratic nonlinear optomechanics \cite{S4-33.0,S2-9,S3-59,S4-33.3,S4-33.4,S4-33.5,S4-33.6,S4-33.6.0,S4-33.6.1,S2-7,S4-33.6.3,S4-33.6.4,S4-33.6.5,S3-53,S4-33.6.6,S4-33.6.7} is now a well recognized subject of study even down to the single-photon level \cite{S3-60}, for which circuit analogues have been constructed \cite{S3-17,S3-61} and may be regarded as fairly convenient simulators \cite{S4-33.8a,S4-33.8b,S4-33.8c} of much more complicated experimental optomechanical analogues. Dual formalisms of quadratic optomechanics are also found in ultracold atom traps \cite{S3-62,S4-33.9a} as well as optical levitation \cite{S4-33.10}. Such types of nonlinear interactions also appear elsewhere in anharmonic quantum circuits \cite{S4-33.10.0}. Quadratic interactions are in particular important for energy and non-demolition measurements of mechanical states \cite{S2-1,S2-2,S2-6,S4-33.11,S4-33.12,S4-33.13}. While the simple linearization of operators could be still good enough to explain some of the observations, there remains a need for an exact and relatively simple mathematical treatment. Method of Langevin equations also normally fails, and other known methods such as expansion unto number states and master equation, require lots of computation while giving little insight to the problem. 

Perturbative expansions and higher-order operators have been used by other researchers to study noise spectra of lasers \cite{S4-33.15,S4-33.16,S4-33.19,S4-33.20}. Also, the master equation approach \cite{S4-33a,S4-33b} can be used in combination with the quasi-probablity Wigner functions \cite{S4-33c,S4-33d} to yield integrable classical Langevin equations. Nevertheless, a method recently has been proposed \cite{S4-34}, which offers a truncation correlation scheme for solution of driven-dissipative multi-mode systems. While being general, it deals with the time evolution of expectation values instead of operators within the truncation accuracy, so the corresponding Langevin equations cannot be analytically integrated. 

Alternatively, a first-order perturbation has been proposed to tackle the nonlinear quadratic optomechanics \cite{S4-34a}. This method perturbatively expands the unknown parameters of classical Langevin equations for the nonlinear system, and proceeds to the truncation at first order. However, the expansion is accurate only where the ratio of photon loss rate to mechanical frequency $\kappa/\Omega$ is large. This condition is strongly violated for instance in superconductive electromechanical systems. 

The treatment of quadratic interactions using quantum Langevin equations is first made possible recently using higher-order operators. This section presents a perturbative mathematical treatment within the first and second order approximations to the nonlinear system of Langevin equations, which ultimately result in an integrable system of quantum mechanical operators. The trick here is to introduce operators of higher dimensionality into the solution space of the problem. Having their commutators calculated, it would be possible to set up an extended system of Langevin equations which could be conveniently solved by truncation at the desirable order. 

To understand how it intuitively works, one may consider the infamous first order quadratic nonlinear Riccati differential equation \cite{S4-34b,S4-34c}, which is exactly integrable if appropriately transformed as a system of two coupled linear first order differential equations. Alternatively, Riccati equation could be exactly transformed into a linear second order differential equation, too. But this is not what we consider here, since it will result in a much more complicated second-order system of Langevin equations involving derivatives of noise terms. 

The method introduced here is useful in other areas of quantum physics \cite{S3-62,S4-33.10} than optomechanics, where nonlinearities such as anharmonic or Kerr interactions are involved. We also describe how the $Q-$functions could be obtained for the anharmonic oscillator. Further applications of nonlinear stochastic differential equations \cite{S4-34.8,S4-34.9,S4-34.10} beyond stochastic optomechanics \cite{S4-33.6.5,S3-53} includes finance and stock-market analysis \cite{S4-34.11}, turbulence \cite{S4-34.12,S4-34.12a}, hydrology and flood prediction \cite{S4-34.13}, and solar energy \cite{S4-34.14}. Also, the Fokker-Planck equation \cite{S4-33.20,S4-34.15,S4-34.16,S4-34.16a,S4-34.16b} is actually equivalent to the nonlinear Schr\"{o}dinger equation with bosonic operator algebra, and its moments \cite{S4-34.17} translate into nonlinear Langevin equations. Similarly, this method can deal with side-band generation in optomechanics \cite{S4-35}, superconducting circuits \cite{S4-35a}, as well as spontaneous emission in open systems \cite{S4-36,S4-36.0}. Applications in estimation of other parameters such as the second order correlation $g^{(2)}(0)$ \cite{S4-36a,S4-36b,S4-36c,S4-36d}, quantum limited amplifiers \cite{S4-36e,S4-36f} and quantum nondemolition measurements \cite{S4-36f,S4-36g,S4-36h,S4-36i} are demonstrated respectively next in \S\ref{Section-5} and \S\ref{Section-6}, too, and furthermore it is found that an unsqueezed ideal laser reaches $\sqrt{6}-2$ cavity photons at threshold.

For the moment being, let us postpone the details treatment of the third-order nonlinear optomechanics with the Hamiltonian $\mathbb{H}_{\rm OM}$ to \S\ref{Section-8} and start right away now at the fourth-order nonlinear quadratic interactions. This will be revisited again with large expansions in the mathematics and practical considerations in \S\ref{Section-10} separately, where momentum-field interactions are investigated further.

\subsection{Hamiltonian}

A nonlinear quadratic optomechanical interaction in the most general form \cite{S2-Paper1} following (\ref{S3-41}) and the analysis carried out in \S\ref{Section-3} is here defined as 
\begin{equation}
\label{T4-eq1}
\mathbb{H}=\hbar \gamma (\hat{b}\pm\hat{b}^\dagger)^2(\hat{a}\pm\hat{a}^\dagger)^2,
\end{equation}
\noindent
where $\gamma$ is the corresponding interaction rate, and various combinations of signs are allowed to give access to either of the momentum- and position-field interactions. Furthermore, bosonic photon $\hat{a}$ and phonon $\hat{b}$ ladder operators satisfy $[\hat{b},\hat{b}^\dagger]=[\hat{a},\hat{a}^\dagger]=1$ as well as $[\hat{b},\hat{a}]=[\hat{b},\hat{a}^\dagger]=0$. Meanwhile, quadratic interactions normally are \cite{S2-1,S2-2,S2-6}
\begin{equation}
\label{T4-eq2}
\mathbb{H}=\hbar \gamma \hat{a}^\dagger\hat{a}(\hat{b}\pm\hat{b}^\dagger)^2,
\end{equation}
\noindent
which by defining the photon number operator $\hat{n}=\hat{a}^\dagger\hat{a}$ takes essentially the same algebraic form.

Direct expansion of (\ref{T4-eq1}) shows that it essentially brings in a different interaction type compared to (\ref{T4-eq2}). Doing so, we obtain $\mathbb{H}=\hbar \gamma (\hat{b}^2+\hat{b}^{\dagger2}\pm 2\hat{m}\pm 1)^2(\hat{a}^2+\hat{a}^{\dagger2}\pm 2\hat{n}\pm 1)$ where $\hat{m}=\hat{b}^\dagger\hat{b}$. Hence, (\ref{T4-eq1}) includes interactions of type $\hat{a}^2\hat{b}^2$, $\hat{a}^2\hat{b}^{\dagger2}$, and so on, which are absent in (\ref{T4-eq2}). It should be noticed that the widely used standard optomechanical interaction $\mathbb{H}_{\rm OM}$ results in nonlinear and linear Langevin equations when expressed respectively in the terms of $\{\hat{a},\hat{b}\}$ and $\{\hat{n},\hat{x}\}$. Hence, this type of interaction is not addressed here. In addition to the above Hamiltonians (\ref{T4-eq1},\ref{T4-eq2}), there exist still other types of nonlinear optomechanical interactions \cite{S3-20,S4-38} such as $\mathbb{H}=\hbar g (\hat{b}\pm\hat{b}^\dagger)(\hat{a}^2\pm\hat{a}^{\dagger2})$, which is also not considered explicitly here, but can be well treated using the scheme presented here. This latter Hamiltonian for instance can describes the photon pair generation, which can be addressed in a similar way and is considered separately later in \S\ref{Section-9} where side-band inequivalence is discussed.

\subsection{Linear Perturbation}

The perturbation approach at the lowest order is what is being widely used by researchers to solve the systems based on either (\ref{T4-eq1}) or (\ref{T4-eq2}). To this end, ladder field operators are replaced with their perturbations, while product terms beyond are neglected and truncated. Obviously, this will give rise to interactions of the type $\hbar(\hat{b}\pm\hat{b}^\dagger)^2(q\delta\hat{a}+q^\ast\delta\hat{a}^\dagger)$, where $q=2\gamma(\bar{a}\pm\bar{a}^\ast)$ for (\ref{T4-eq1}) and $q=\gamma\bar{a}$ for (\ref{T4-eq2}) is some complex constant in general, and $\delta\hat{a}$ now represents the perturbation term around the steady state average $|\bar{a}|=\sqrt{\bar{n}}$. This technique is mostly being referred to as the linearization of operators, and directly leads to an integrable set of Langevin equations if also applied to the mechanical displacement as well.

\subsection{Square Field Operators}

Here, we define the square field operators \cite{S2-Paper1}
\begin{eqnarray}
\label{T4-eq3}
\hat{c}&=&\frac{1}{2}\hat{a}^2,\\ \nonumber
\hat{d}&=&\frac{1}{2}\hat{b}^2.
\end{eqnarray}
\noindent
for photons, which obviously satisfy $[\hat{c},\hat{a}]=[\hat{c},\hat{b}]=[\hat{d},\hat{a}]=[\hat{d},\hat{b}]=[\hat{c},\hat{d}]=0$. Now, it is not difficult to verify that these operators furthermore satisfy the commutation relationships
\begin{eqnarray}
\label{T4-eq4}
[\hat{c},\hat{c}^\dagger]&=&\hat{n}+\frac{1}{2},\\ \nonumber
[\hat{c},\hat{n}]&=&2\hat{c},\\ \nonumber
[\hat{c}^\dagger,\hat{n}]&=&-2\hat{c}^\dagger,\\ \nonumber
[\hat{c},\hat{a}^\dagger]&=&\hat{a}.
\end{eqnarray}
Defining the phonon number operator as $\hat{m}=\hat{b}^\dagger\hat{b}$, in a similar manner we could write
\begin{eqnarray}
\label{T4-eq5}
[\hat{d},\hat{d}^\dagger]&=&\hat{m}+\frac{1}{2},\\ \nonumber
[\hat{d},\hat{m}]&=&2\hat{d},\\ \nonumber
[\hat{d}^\dagger,\hat{m}]&=&-2\hat{d}^\dagger,\\ \nonumber
[\hat{d},\hat{b}^\dagger]&=&\hat{b}.
\end{eqnarray}

The set of commutator equations (\ref{T4-eq4}) and (\ref{T4-eq5}) enables us to treat the quadratic nonlinear interaction perturbatively to the desirable accuracy, as is described in the following.

\subsection{Langevin Equations}

The input/output formalism \cite{S2-Noise1,S2-Noise0,S2-Noise2,S2-Noise3} can be used to assign decay channels to each of the quantum variables of the system. This will result in the set of Langevin equations  
\begin{equation}
\label{T4-eq6}
\frac{d}{dt}\{A\}=[\textbf{M}]\{A\}-\sqrt{[\Gamma]}\{A_{\rm in}\},
\end{equation}
\noindent
where $\{A\}$ is the system vector, $[\textbf{M}]$ is the coefficients matrix whose eigenvalues need to have negative or vanishing real parts to guarantee stability, and $[\Gamma]$ is a real-valued matrix which is diagonal if all noise terms corresponding to the members of $\{A\}$ are mutually independent. When $[\textbf{M}]$ is independent of $\{A\}$, (\ref{T4-eq6}) is linear and integrable and otherwise nonlinear and non-integrable. If  $[\textbf{M}(t)]$ is a function of time, then (\ref{T4-eq6}) is said to be time-dependent. Furthermore, $\{A_{\rm in}\}$ represents the input fields to the system at the respective ports, and $\{A_{\rm out}\}$ is the output fields, which are related together as \cite{S2-4,S2-5,S3-8}
\begin{equation}
\label{T4-eq7}
\{A_{\rm out}\}=\{A_{\rm in}\}+\sqrt{[\Gamma]}\{A\}.
\end{equation}
\noindent
Here, $[\Gamma]$ is supposed to be diagonal for simplicity. From the scattering matrix formalism we also have
\begin{equation}
\label{T4-eq8}
\{A_{\rm out}\}=[\textbf{S}]\{A_{\rm in}\}.
\end{equation}
\noindent
Hence, taking $w$ as the angular frequency and performing a Fourier transform on (\ref{T4-eq6}), the scattering matrix is found by using (\ref{T4-eq7}) and (\ref{T4-eq8}) as
\begin{equation}
\label{T4-eq9}
[\textbf{S}(w)]=[\textbf{I}]-\sqrt{[\Gamma]}\left(iw[\textbf{I}]+[\textbf{M}]\right)^{-1}\sqrt{[\Gamma]}.
\end{equation}
\noindent
Hence,  $[\textbf{S}]$ is well-defined if $[\textbf{M}]$ is known. This can be obtained by using the Langevin equations 
\begin{equation}
\label{T4-eq10}
\dot{\hat{z}}=\frac{d}{dt}\hat{z}=-\frac{i}{\hbar}[\hat{z},\mathbb{H}]-[\hat{z},\hat{x}^\dagger](\frac{1}{2}\Gamma\hat{x}+\sqrt{\Gamma}\hat{z}_{\rm in})+(\frac{1}{2}\Gamma\hat{x}^\dagger+\sqrt{\Gamma}\hat{z}_{\rm in}^\dagger)[\hat{z},\hat{x}],
\end{equation}
\noindent
where $\hat{x}$ is any system operator, which is here taken to be the same as $\hat{z}$ to comply with (\ref{T4-eq8}).

By setting either $\hat{z}=\hat{c}$ or $\hat{z}=\hat{d}$ the commutators in (\ref{T4-eq10}) by (\ref{T4-eq4}) or (\ref{T4-eq5}) always lead back to the same linear combination of these forms. Thus, the new set of Langevin equations is actually linear in terms of the square or higher-order operators, if perturbatively truncated at a finite order. So, instead of solving the nonlinear system in linearized $2\times 2$ space $\{A\}^{\rm T}=\{\hat{a},\hat{b}\}$, one may employ an expanded dimensional space with increased accuracy. There, truncation and sometimes mean field approximations are necessary to restrict the dimension, since commutators of new operators mostly lead to even higher-orders and are thus not closed under commutation. As examples, a $4\times 4$ space $\{A\}^{\rm T}=\{\hat{a},\hat{d},\hat{d}^\dagger,\hat{m}\}$ truncated at the first-order, or a $6\times 6$ space $\{A\}^{\rm T}=\{\hat{c},\hat{c}^\dagger,\hat{n},\hat{d},\hat{d}^\dagger,\hat{m}\}$ truncated at the second-order could be used for (\ref{T4-eq1},\ref{T4-eq2}). To illustrate the application of this method, we describe two examples next. It could be extended to the accuracy of the second-order perturbation, too, by defining appropriate cross product operator terms between photonic and phononic partitions.   

\subsection{Examples}

Here, we describe two examples from the nonlinear interactions of having type (\ref{T4-eq1}) or (\ref{T4-eq2}).

\subsubsection{Standard Quadratic Interaction (\ref{T4-eq2})}

Analysis of such systems requires analysis in a 4-dimensional space, spanned by $\{A\}^{\rm T}=\{\hat{a},\hat{d},\hat{d}^\dagger,\hat{m}\}$. Taking the plus sign here without loss of generality and after dropping a trivial non-interacting term $\mathbb{H}_{\rm non}=\hbar\gamma\hat{n}$, the nonlinear interaction  is
\begin{equation}
\label{T4-eq11}
\mathbb{H}=2\hbar\gamma\hat{n}(\hat{d}+\hat{d}^\dagger+\hat{m}).
\end{equation}

This can be found by expansion of (\ref{T4-eq2}), plugging in (\ref{T4-eq3}) and $[\hat{b},\hat{b}^\dagger]=1$, and dropping a trivial term $\hbar\gamma\hat{n}$. Using (\ref{T4-eq5}), $[\hat{a},\hat{n}]=\hat{a}$ and $[\hat{a}^\dagger,\hat{n}]=-\hat{a}^\dagger$ in the non-rotating frame of operators, and ignoring the self-energy Hamiltonian $\mathbb{H}_{\rm self}=\hbar(\omega+\gamma)\hat{n}+\hbar\Omega\hat{m}$ for the moment, Langevin equations become
\begin{eqnarray}
\nonumber
\dot{\hat{a}}&=&-2i\gamma\hat{a}(\hat{d}+\hat{d}^\dagger+\hat{m})-\frac{1}{2}\Gamma_1\hat{a}-\sqrt{\Gamma_1}\hat{a}_{\textrm{in}},\\ \nonumber
\dot{\hat{d}}&=&-2i\gamma\hat{n}(2\hat{d}+\hat{m}+\frac{1}{2})-(\hat{m}+\frac{1}{2})(\frac{1}{2}\Gamma_2\hat{d}+\sqrt{\Gamma_2}\hat{d}_{\textrm{in}}), \\ \nonumber
\dot{\hat{d}}^\dagger&=&2i\gamma\hat{n}(2\hat{d}^\dagger+\hat{m}+\frac{1}{2})-(\hat{m}+\frac{1}{2})(\frac{1}{2}\Gamma_2\hat{d}^\dagger+\sqrt{\Gamma_2}\hat{d}_{\textrm{in}}^\dagger),\\ \label{T4-eq12}
\dot{\hat{m}}&=&4i\gamma\hat{n}(\hat{d}-\hat{d}^\dagger).
\end{eqnarray}

So far, the set of equations (\ref{T4-eq12}) is exact. However, integration of (\ref{T4-eq12}) is still not possible at this stage, and taking Fourier transformation must be done later when arriving at a linear operator system. We present a first-order and second-order perturbative method to deal with this difficulty. 

It should be furthermore noticed that using a non-rotating frame with the self-energy Hamiltonian $\mathbb{H}_{\rm self}$ not ignored, would have resulted in identical equations, except with the addition of the trivial terms $-i\Delta\hat{a}$, $-i2\Omega\hat{d}$, and $+i2\Omega\hat{d}^\dagger$ respectively to the first three equations, where $\Delta=\omega+\gamma-\nu$ is the optical detuning with $\nu$ being the cavity optical resonance frequency, and $\omega$ and $\Omega$ are respectively the optical and mechanical frequencies. Also, the damping coefficient in high mechanical quality factor $Q_{\rm m}$ limit could be estimated as $\Gamma_2=2\Gamma_{\rm m}$, where $\Gamma_{\rm m}$ is the damping rate of the $\hat{b}$ phononic field. Here, it is preferable not to use the rotating frames since the coefficients matrix $[\textbf{M}]$ becomes time-dependent.

\subsubsection{First-order Perturbation to (\ref{T4-eq12})}

Now, if the photon and phonon baths each have a mean boson number respectively as $\left<\hat{n}\right>=\bar{n}$ and $\left<\hat{m}\right>=\bar{m}$, we could immediately write down the linear system of equations in the non-rotating frame of operators and neglection of self-energies $\mathbb{H}_{\rm self}$ as
\begin{eqnarray}
\nonumber
\dot{\hat{a}}&=&-3i\gamma\bar{m}\hat{a}-i\gamma\bar{a}\hat{d}-i\gamma\bar{a}\hat{d}^\dagger-\frac{1}{2}\Gamma_1\hat{a}-\sqrt{\Gamma_1}\hat{a}_{\textrm{in}},\\ \nonumber
\dot{\hat{d}}&=&-2i\gamma\bar{n}\left(2\hat{d}+\hat{m}+\frac{1}{2}\right)-\left(\bar{m}+\frac{1}{2}\right)\left(\frac{1}{2}\Gamma_2\hat{d}+\sqrt{\Gamma_2}\hat{d}_{\textrm{in}}\right), \\ \nonumber
\dot{\hat{d}}^\dagger&=&2i\gamma\bar{n}\left(2\hat{d}^\dagger+\hat{m}+\frac{1}{2}\right)-\left(\bar{m}+\frac{1}{2}\right)\left(\frac{1}{2}\Gamma_2\hat{d}^\dagger+\sqrt{\Gamma_2}\hat{d}_{\textrm{in}}^\dagger\right),\\ \label{T4-eq13}
\dot{\hat{m}}&=&4i\gamma\bar{n}\left(\hat{d}-\hat{d}^\dagger\right),
\end{eqnarray}
\noindent
which is now exactly integrable. Here, we use the linearization $2\hat{a}\hat{d}=(\bar{a}+\delta\hat{a})\hat{d}+\hat{a}(\bar{d}+\delta\hat{d})\rightarrow\bar{a}\hat{d}+\bar{d}\hat{a}$, where $\bar{d}=\frac{1}{2}\bar{a}^2$ and higher-order terms of the form $\delta\hat{a}\delta\hat{d}$ are dropped, and so on. But this cannot be applied to $\hat{n}\hat{m}=\hat{a}^\dagger\hat{a}\hat{m}$ since $\hat{n}$ and $\hat{a}^\dagger$ are absent from the basis. Furthermore, any linearization of this expansion would generate terms $\hat{a}\hat{m}$ and $\hat{a}^\dagger\hat{m}$ which are still nonlinear. Both of these issues can be resolved by a second-order perturbation as follows next. This results in the operator equations
\begin{eqnarray}
\raggedleft
\nonumber
\frac{d}{dt}\begin{Bmatrix}
\hat{a}\\
\hat{d}\\
\hat{d}^\dagger\\
\hat{m}
\end{Bmatrix}&=&\begin{bmatrix}
-i3\gamma\bar{m}-\frac{1}{2}\Gamma_1 & -i\gamma\bar{a} & -i\gamma\bar{a} & 0\\
0 & -i4\gamma\bar{n}-\frac{1}{2}\left(\bar{m}+\frac{1}{2}\right)\Gamma_2 & 0 & -i2\gamma\bar{n}\\
0 & 0 & +i4\gamma\bar{n}-\frac{1}{2}\left(\bar{m}+\frac{1}{2}\right)\Gamma_2 & i2\gamma\bar{n} \\
0 & i4\gamma\bar{n} & -i4\gamma\bar{n} & 0
\end{bmatrix}\begin{Bmatrix}
\hat{a}\\
\hat{d}\\
\hat{d}^\dagger\\
\hat{m}
\end{Bmatrix}\\ \label{T4-eq14}
&-&\begin{Bmatrix}
\sqrt{\Delta_1}\hat{a}_{\textrm{in}}\\
\sqrt{\Delta_2}\hat{d}_{\textrm{in}}\\
\sqrt{\Delta_2}\hat{d}^\dagger_{\textrm{in}}\\
0
\end{Bmatrix},
\end{eqnarray}	
\noindent
where $\sqrt{\Delta_1}=\sqrt{\Gamma_1}$ and $\sqrt{\Delta_2}=\left(\bar{m}+\frac{1}{2}\right)\sqrt{\Gamma_2}$. The set of equations (\ref{T4-eq14}) is linear and can be easily addressed by standard methods of stochastic Langevin equations used in optomechanics \cite{S2-1,S2-2,S2-6,S2-Noise1,S2-Noise0} and elsewhere. More specifically, one may employ analytical Fourier methods in frequency domain as an matrix algebraic problem to obtain spectra of variables, or integrate the system numerically by stochastic numerical methods in time domain to obtain time dependent behavior of expectation values.  

All that remains is to find the average cavity boson numbers for photons $\bar{n}$ and phonons $\bar{m}$. In order to do this, one may first arbitrate $d/dt=0$ in (\ref{T4-eq13}) at steady state, and then use the equality of real parts in first equation to find the expression for $\bar{n}$. Doing this, results in $\bar{n}=4|\bar{a}_{\rm in}|^2/\Gamma_1$ where $|\bar{a}_{\rm in}|$ represents the amplitude of coherent laser input. Also, the initial cavity phonon occupation number at $t=0$ could be estimated simply as  $\bar{m}=1/\left[\exp(\hbar\Omega/k_{\rm B}T)-1\right]$ \cite{S2-Noise1,S2-Noise0}, where $k_{\rm B}T$ is the thermal energy with $k_{\rm B}$ and $T$ being respectively the Boltzmann's constant and absolute temperature. Detailed numerical examinations reveal that the system of equations (\ref{T4-eq14}) is generally very well stable with $\Re\{{\rm eig}[{\bf M}]\}<0$ at sufficiently low optical intensities.

\subsubsection{Full Quadratic Interaction (\ref{T4-eq1})}

Analysis of a fully quadratic system requires analysis in a $6\times 6$ dimensional space, spanned by $\{A\}^{\rm T}=\{\hat{c},\hat{c}^\dagger,\hat{n},\hat{d}, \hat{d}^\dagger,\hat{m}\}$. Taking both of the plus signs here, the Hamiltonian could be written as
\begin{equation}
\label{T4-eq15}
\mathbb{H}=4\hbar\gamma(\hat{d}+\hat{d}^\dagger+\hat{m})(\hat{c}+\hat{c}^\dagger+\hat{n}),
\end{equation}
\noindent
where a trivial non-interacting term $\mathbb{H}_{\rm non}=2\hbar\gamma(1+\hat{n}+\hat{m}+\hat{d}+\hat{c}+\hat{d}^\dagger+\hat{c}^\dagger)$ is dropped. The set of Langevin equations can be obtained in a similar manner, and in non-rotating frame of operators with neglection of self-energies $\mathbb{H}_{\rm self}=\hbar(\omega+2\gamma)\hat{n}+\hbar(\Omega+2\gamma)\hat{m}$ for the moment, results in
\begin{eqnarray}
\raggedleft
\label{T4-eq16}
\dot{\hat{c}}&=&-i4\gamma(\hat{d}+\hat{d}^\dagger+\hat{m})\left(2\hat{c}+\hat{n}+\frac{1}{2}\right)-\left(\hat{n}+\frac{1}{2}\right)\left(\frac{1}{2}\Gamma_1\hat{c}+\sqrt{\Gamma_1}\hat{c}_{\textrm{in}}\right),\\ \nonumber
\dot{\hat{c}}^\dagger&=&i4\gamma(\hat{d}+\hat{d}^\dagger+\hat{m})\left(2\hat{c}^\dagger+\hat{n}+\frac{1}{2}\right)-\left(\hat{n}+\frac{1}{2}\right)\left(\frac{1}{2}\Gamma_1\hat{c}^\dagger+\sqrt{\Gamma_1}\hat{c}_{\textrm{in}}^\dagger\right),\\ \nonumber
\dot{\hat{n}}&=&i8\hbar\gamma(\hat{d}+\hat{d}^\dagger+\hat{m})(\hat{c}-\hat{c}^\dagger),\\ \nonumber
\dot{\hat{d}}&=&-i4\gamma(\hat{c}+\hat{c}^\dagger+\hat{n})\left(2\hat{d}+\hat{m}+\frac{1}{2}\right)-\left(\hat{m}+\frac{1}{2}\right)\left(\frac{1}{2}\Gamma_2\hat{d}+\sqrt{\Gamma_2}\hat{d}_{\textrm{in}}\right),\\ \nonumber
\dot{\hat{d}}^\dagger&=&i4\gamma(\hat{c}+\hat{c}^\dagger+\hat{n})\left(2\hat{d}^\dagger+\hat{m}+\frac{1}{2}\right)-\left(\hat{m}+\frac{1}{2}\right)\left(\frac{1}{2}\Gamma_2\hat{d}^\dagger+\sqrt{\Gamma_2}\hat{d}_{\textrm{in}}^\dagger\right),\\ \nonumber
\dot{\hat{m}}&=&i8\hbar\gamma(\hat{c}+\hat{c}^\dagger+\hat{n})(\hat{d}-\hat{d}^\dagger).
\end{eqnarray}
\noindent
Similar to (\ref{T4-eq12}), the damping rate for sufficiently high optical quality factors $Q$ could be estimated as $\Gamma_1=2\kappa$, where $\kappa$ is the damping rate of the $\hat{a}$ photonic field.

Quite clearly, should we have not ignored the self-energy Hamiltonian $\mathbb{H}_{\rm self}$, then addition of the diagonal terms $-i2\Delta\hat{c}$, $+i2\Delta\hat{c}^\dagger$ to the first two where $\Delta=\omega+2\gamma-\nu$ with $\nu$ being the optical cavity resonance frequency, and similarly $-i2\Omega\hat{d}$ and $+i2\Omega\hat{d}^\dagger$ to the fourth and fifth equations would have been necessary. These are not shown here only for the sake of convenience. Again, it is emphasized that transformation to the rotating frame of operators here would make the coefficients time-dependent in an oscillating manner, and it is far better to be avoided for these classes of nonlinear problems. 

\subsubsection{First-order Perturbation to (\ref{T4-eq16})}

In a similar manner to (\ref{T4-eq13}), we may assume photon and phonon baths each have a mean boson number respectively as $\left<\hat{n}\right>=\bar{n}$ and $\left<\hat{m}\right>=\bar{m}$, which gives
\begin{eqnarray}
\label{T4-eq17}
\dot{\hat{c}}&=&-i4\gamma\bar{m}\left(2\hat{c}+\hat{n}\right)-i4\gamma\left(\bar{n}+\frac{1}{2}\right)(\hat{d}+\hat{d}^\dagger+\hat{m})-\left(\bar{n}+\frac{1}{2}\right)\left(\frac{1}{2}\Gamma_1\hat{c}+\sqrt{\Gamma_1}\hat{c}_{\textrm{in}}\right),\\ \nonumber
\dot{\hat{c}}^\dagger&=&i4\gamma\bar{m}\left(2\hat{c}^\dagger+\hat{n}\right)+i4\gamma\left(\bar{n}+\frac{1}{2}\right)(\hat{d}+\hat{d}^\dagger+\hat{m})-\left(\bar{n}+\frac{1}{2}\right)\left(\frac{1}{2}\Gamma_1\hat{c}^\dagger+\sqrt{\Gamma_1}\hat{c}_{\textrm{in}}^\dagger\right),\\ \nonumber
\dot{\hat{n}}&=&i8\hbar\bar{m}(\hat{c}-\hat{c}^\dagger),\\ \nonumber
\dot{\hat{d}}&=&-i4\gamma\bar{n}\left(2\hat{d}+\hat{m}\right)-i4\gamma\left(\bar{m}+\frac{1}{2}\right)(\hat{c}+\hat{c}^\dagger+\hat{n})-\left(\bar{m}+\frac{1}{2}\right)\left(\frac{1}{2}\Gamma_2\hat{d}+\sqrt{\Gamma_2}\hat{d}_{\textrm{in}}\right),\\ \nonumber
\dot{\hat{d}}^\dagger&=&i4\gamma\bar{n}\left(2\hat{d}^\dagger+\hat{m}\right)+i4\gamma\left(\bar{m}+\frac{1}{2}\right)(\hat{c}+\hat{c}^\dagger+\hat{n})-\left(\bar{m}+\frac{1}{2}\right)\left(\frac{1}{2}\Gamma_2\hat{d}^\dagger+\sqrt{\Gamma_2}\hat{d}_{\textrm{in}}^\dagger\right),\\ \nonumber
\dot{\hat{m}}&=&i8\hbar\gamma\bar{n}(\hat{d}-\hat{d}^\dagger).
\end{eqnarray}

We here need to assume the redefinition $\sqrt{\Delta_1}=(\bar{n}+\frac{1}{2})\sqrt{\Gamma_1}$. Now, without taking $\mathbb{H}_{\rm self}$ into account, this will lead to the linear system of matrix Langevin equations

\begin{flalign}
\nonumber
\begin{bmatrix}
-i8\gamma\bar{m}-\frac{\bar{N}}{4}\Gamma_1 & 0 & -i4\gamma\bar{m} & -i2\gamma(\bar{N}) & -i2\gamma(\bar{N}) & -i2\gamma(\bar{N}) \\
0 & i8\gamma\bar{m}-\frac{\bar{N}}{4}\Gamma_1 & i4\gamma\bar{m} & i2\gamma(\bar{N}) & i2\gamma(\bar{N}) & i2\gamma(\bar{N}) \\
i8\gamma\bar{m} & -i8\gamma\bar{m} & 0 & 0 & 0 & 0 \\ 
-i2\gamma(\bar{M}) & -i2\gamma(\bar{M}) & -i2\gamma(\bar{M}) & -i8\gamma\bar{n}-\frac{\bar{M}}{4}\Gamma_2 & 0 & -4i\gamma\bar{n}\\
i2\gamma(\bar{M}) & i2\gamma(\bar{M}) & i2\gamma(\bar{M}) & 0 & i8\gamma\bar{n}-\frac{\bar{M}}{4}\Gamma_2 & i4\gamma\bar{n}\\
0 & 0 & 0 & i8\gamma\bar{n} & -i8\gamma\bar{n} & 0
\end{bmatrix}
\\ \label{T4-eq18}
\times
\begin{Bmatrix}
\hat{c}\\
\hat{c}^\dagger\\
\hat{n}\\
\hat{d}\\
\hat{d}^\dagger\\
\hat{m}
\end{Bmatrix}
-\begin{Bmatrix}
\sqrt{\Delta_1}\hat{c}_{\textrm{in}}\\
\sqrt{\Delta_1}\hat{c}^\dagger_{\textrm{in}}\\
0\\
\sqrt{\Delta_2}\hat{d}_{\textrm{in}}\\
\sqrt{\Delta_2}\hat{d}^\dagger_{\textrm{in}}\\
0
\end{Bmatrix}=\frac{d}{dt}\begin{Bmatrix}
\hat{c}\\
\hat{c}^\dagger\\
\hat{n}\\
\hat{d}\\
\hat{d}^\dagger\\
\hat{m}
\end{Bmatrix},
\end{flalign}
\noindent
which is, of course, integrable now. Here, we make use of the adoptions $\bar{N}=2\bar{n}+1$ and $\bar{M}=2\bar{m}+1$ to simplify the appearance. The initial cavity boson numbers $\bar{n}$ and $\bar{m}$ can be set in the same manner which was done for the system of equations (\ref{T4-eq14}). Numerical tests reveal that (\ref{T4-eq18}) is conditionally stable if the optical intensity is kept below a certain limit on the red detuning, and is otherwise unstable.  

\subsubsection{Second order Perturbation to (\ref{T4-eq12},\ref{T4-eq16})}

The set of Langevin equations (\ref{T4-eq12},\ref{T4-eq16}) can be integrated with much more accuracy, if we first identify and sort out the cross terms as individual operators. For instance, (\ref{T4-eq16}) contains the cross operators $\hat{c}\hat{d}$, $\hat{c}\hat{d}^\dagger$, $\hat{c}\hat{m}$, $\hat{c}^\dagger\hat{d}$, $\hat{c}^\dagger\hat{d}^\dagger$, $\hat{c}^\dagger\hat{m}$, $\hat{n}\hat{d}$, $\hat{n}\hat{d}^\dagger$, as well as $\hat{n}\hat{m}$ which is self-adjoint. These constitute an extra set of nine cross operators to be included in the treatment. All these cross operators are formed by multiplication of photonic and phononic single operators, whose notation order, such as $\hat{c}\hat{d}=\hat{d}\hat{c}$ and so on, is obviously immaterial. 

Now, one may proceed first to determine the commutators between these terms where relevant, which always result in linear combinations of the other existing terms. This will clearly enable a more accurate formulation of (\ref{T4-eq16}) but in a $6+9=15$ dimensional space, which is given by the array of operators $\{A\}^{\rm T}=\{\hat{c},\hat{c}^\dagger,\hat{n},\hat{d},\hat{d}^\dagger,\hat{m},\hat{c}\hat{d},\hat{c}\hat{d}^\dagger,\hat{c}\hat{m},\hat{c}^\dagger\hat{d},\hat{c}^\dagger\hat{d}^\dagger,\hat{c}^\dagger\hat{m},\hat{n}\hat{d},\hat{n}\hat{d}^\dagger,\hat{n}\hat{m}\}$. 

The independent non-trivial quadratic commutator equations among cross operators here are found after tedious but straightforward algebra as
\begin{eqnarray}
\nonumber
[\hat{c}\hat{d},\hat{c}^\dagger\hat{d}^\dagger]&=&\frac{1}{8}[(2\hat{n}\hat{m}+3)(\hat{m}+\hat{n}+2)+\hat{n}^2+\hat{m}^2-4],\\ \nonumber
[\hat{c}\hat{d},\hat{c}^\dagger\hat{m}]&=&\frac{1}{2}(\hat{n}^2+2\hat{n}\hat{m}+2\hat{n}+\hat{m}+2)\hat{d},\\ \nonumber
[\hat{c}\hat{d},\hat{n}\hat{d}^\dagger]&=&\frac{1}{2}(\hat{m}^2+3\hat{m}+2\hat{m}\hat{n}+\hat{n}+2)\hat{c},\\ \nonumber
[\hat{c}\hat{d},\hat{n}\hat{m}]&=&(\hat{n}+\hat{m}+4)\hat{c}\hat{d},\\ \nonumber
[\hat{c}\hat{d}^\dagger,\hat{c}^\dagger\hat{d}]&=&\frac{1}{8}(2\hat{n}\hat{m}+\hat{m}+\hat{n}-1)(\hat{m}-\hat{n}),\\ \nonumber
[\hat{c}\hat{d}^\dagger,\hat{c}^\dagger\hat{m}]&=&\frac{1}{2}\left[(2\hat{n}+1)\hat{m}-(\hat{n}+1)(\hat{n}+2)\right]\hat{d}^\dagger,\\ \nonumber
[\hat{c}\hat{d}^\dagger,\hat{n}\hat{d}]&=&\frac{1}{2}\left[\hat{m}(\hat{m}-2\hat{n})-(\hat{m}+\hat{n})\right]\hat{c},\\ \nonumber
[\hat{c}\hat{d}^\dagger,\hat{n}\hat{m}]&=&2(\hat{m}-\hat{n}-2)\hat{c}\hat{d}^\dagger,\\ \nonumber
[\hat{c}\hat{m},\hat{n}\hat{d}]&=&2(\hat{m}+\hat{n}+2)\hat{c}\hat{d},\\ \label{T4-eq19}
[\hat{c}\hat{m},\hat{n}\hat{d}^\dagger]&=&2(\hat{m}+\hat{n})\hat{c}\hat{d}^\dagger.
\end{eqnarray}
\noindent
The rest of commutators among cross operators are either adjoints of the above, or have a common term which makes their evaluation possible using either (\ref{T4-eq4}) or (\ref{T4-eq5}). Commutators among cross operators and single operators can be always factored, such as $[\hat{c}\hat{d},\hat{n}]=[\hat{c},\hat{n}]\hat{d}$. Commutators among single operators are already known (\ref{T4-eq4},\ref{T4-eq5}). It can be therefore seen that commutators (\ref{T4-eq19}) always lead to operators of higher orders yet, so that they do not terminate at any finite order of interest by merely expansion of operators basis. This fact puts the perturbative method put into work. There are, however, nonlinear systems such as semiconductor optical cavities \cite{S4-33.19,S4-33b} in which higher-order operators yield an exact closed algebra and satisfy a closedness property within the original space by appropriate definition. 

The set of ten commutators now can be perturbatively linearized as a second-order approximation, by replacing the number operators with their mean values, wherever needed to reduce the set of operators back to the available 15 dimensional space. This will give rise to the similar set of equations after some algebra 
\begin{eqnarray}
\nonumber
[\hat{c}\hat{d},\hat{c}^\dagger\hat{d}^\dagger]&=&\frac{1}{16}(\bar{m}+\bar{n}+8)\hat{n}\hat{m}+\frac{1}{8}\left[\bar{m}(\bar{n}+1)+\frac{1}{2}\bar{n}^2+3\right]\hat{m}+\frac{1}{8}\left[\bar{n}(\bar{m}+1)+\frac{1}{2}\bar{m}^2+3\right]\hat{n}+\frac{1}{4},\\ \nonumber
[\hat{c}\hat{d},\hat{c}^\dagger\hat{m}]&=&\frac{1}{2}(\bar{n}+2\bar{m}+2)\hat{n}\hat{d}+\frac{1}{2}(\bar{m}+2)\hat{d},\\ \nonumber
[\hat{c}\hat{d},\hat{n}\hat{d}^\dagger]&=&\frac{1}{2}(\bar{m}+3+2\bar{n})\hat{c}\hat{m}+\frac{1}{2}(\bar{n}+2)\hat{c},\\ \nonumber
[\hat{c}\hat{d},\hat{n}\hat{m}]&=&(\bar{n}+\bar{m}+4)\hat{c}\hat{d},\\ \nonumber
[\hat{c}\hat{d}^\dagger,\hat{c}^\dagger\hat{d}]&=&\frac{1}{16}(\bar{m}-\bar{n})\hat{n}\hat{m}+\frac{1}{8}\left[\bar{m}(\bar{n}+1)-1-\frac{1}{2}\bar{n}^2\right]\hat{m}-\frac{1}{8}\left[\bar{n}(\bar{m}+1)-1-\frac{1}{2}\bar{m}^2\right]\hat{n},\\ \nonumber
[\hat{c}\hat{d}^\dagger,\hat{c}^\dagger\hat{m}]&=&\frac{1}{2}(2\bar{m}-\bar{n}-3)\hat{n}\hat{d}^\dagger+\frac{1}{2}(\bar{m}-2)\hat{d}^\dagger,\\ \nonumber
[\hat{c}\hat{d}^\dagger,\hat{n}\hat{d}]&=&\frac{1}{2}(\bar{m}-2\bar{n}-1)\hat{c}\hat{m}-\frac{1}{2}\bar{n}\hat{c},\\ \nonumber
[\hat{c}\hat{d}^\dagger,\hat{n}\hat{m}]&=&2(\bar{m}-\bar{n}-2)\hat{c}\hat{d}^\dagger,\\ \nonumber
[\hat{c}\hat{m},\hat{n}\hat{d}]&=&2(\bar{m}+\bar{n}+2)\hat{c}\hat{d},\\ \label{T4-eq20}
[\hat{c}\hat{m},\hat{n}\hat{d}^\dagger]&=&2(\bar{m}+\bar{n})\hat{c}\hat{d}^\dagger.
\end{eqnarray}
\noindent
where the reduction of triple operator products among single and cross operators as  $4\hat{x}\hat{y}\hat{z}\rightarrow\bar{x}\hat{y}\hat{z}+\bar{x}\bar{y}\hat{z}+\bar{y}\bar{z}\hat{x}+\bar{z}\bar{x}\hat{y}$ is used where appropriate. For instance, the term $4\hat{n}\hat{m}^2$ is replaced as $\bar{m}\hat{m}\hat{n}+2\bar{m}\bar{n}\hat{m}+\bar{m}^2\hat{n}$ and so on. Also, similar to (\ref{T4-eq13}), products among single operators are reduced as $2\hat{x}\hat{y}\rightarrow\bar{x}\hat{y}+\bar{y}\hat{x}$. This is somewhat comparable to the mean field approach in cross Kerr optomechanics \cite{S4-38a}.

There are two basic reasons why we have adopted this particular approach to the linearization and cuting off the diverging operators of higher orders. The first reason is that number operators vary slowly in time as opposed to their bosonic counterparts which oscillate rapidly in time, given the fact that the use of rotating frames is disallowed here. Secondly, number operators are both positive-definite and self-adjoint, and thus can be approximated by a positive real number. These properties makes the replacements $\hat{n}\rightarrow\bar{n}$ and $\hat{m}\rightarrow\bar{m}$ reasonable approximations, and the replacement with mean values needs only to be restricted to the number operators, to yield a closed algebra necessary for construction of Langevin equations. Hence, the correct application of replacements only to the triple operator products appearing in (\ref{T4-eq19}) will make sure that no operator having an order beyond than that of cross operators will appear in the formulation.

Anyhow, it can be seen now that all approximate commutators in (\ref{T4-eq20}) allow the set of operators $\{A\}^{\rm T}\cup\{\hat{1}\}=\{\hat{1},\hat{c},\hat{c}^\dagger,\hat{n},\hat{d},\hat{d}^\dagger,\hat{m},\hat{c}\hat{d},\hat{c}\hat{d}^\dagger,\hat{c}\hat{m},\hat{c}^\dagger\hat{d},\hat{c}^\dagger\hat{d}^\dagger,\hat{c}^\dagger\hat{m},\hat{n}\hat{d},\hat{n}\hat{d}^\dagger,\hat{n}\hat{m}\}$ to take on linear combinations of its members among every pair of commutations possible, where $\hat{1}$ is the identity operator. Obviously, this approximate closedness property now makes the full construction of Langevin equations for the operators belonging to $\{A\}$ possible. It is noted that $\hat{1}$ is not an identity element for the commutation.

We can now define the set $\{S\}={\rm span}(\{A\}\cup\{\hat{1}\})$, which is spanned by all possible linear combinations of $\{\hat{1}\}$ and the members of $\{A\}$ together with the associative binary commutation operation $[]$ defined in (\ref{T4-eq4},\ref{T4-eq5},\ref{T4-eq20}). The ordered pair $(\{{S}\},[])$ is now a semigroup.

Having therefore these ten commutators (\ref{T4-eq20}) known, we may proceed now to composing the second-order approximation to the nonlinear Langevin equations (\ref{T4-eq16}), from which a much more accurate solution could be obtained. Here, the corresponding Langevin equations may be constructed at each step by setting both $\hat{z}$ and $\hat{x}$ in (\ref{T4-eq10}) equal to either of the 15 operators, while the noise input terms for cross operators is a simple product of related individual noise terms. The linear damping rates of higher-order operators is furthermore simply the sum of individual damping rates of corresponding single operators, which completes the needed parameter set of Langevin equations.

\subsection{Optomechanical Interaction \& Drive Terms}

The method described in the above can be simultaneously used if other terms such as the standard optomechanical interaction $\mathbb{H}_{\textrm{OM}}$ is non-zero, or there exists a coherent pumping drive term which can be expressed as  $\mathbb{H}_{\textrm{d}}=\sum_k F_k \hat{b}^\dagger+F_k^*\hat{b}$, where $F_k$ are time-dependent drive amplitudes. While $\mathbb{H}_{\textrm{d}}$ does not appear directly in the Langevin equations, treatment of $\mathbb{H}_{\textrm{OM}}$ requires inclusion of additional Langevin equations for $\hat{a}$ and $\hat{b}$ where appropriate, as well as few extra terms in the rest. This can be done in a pretty standard way, and is not repeated here for the sake of brevity \cite{S2-1,S2-2,S2-Noise1,S2-Noise0,S2-Noise2,S2-Noise3}. 

\subsection{Multi-mode Fields}

The analysis is also essentially unaltered if there are more than one mechanical mode to be considered \cite{S4-17,S4-39,S3-28}, and the method is still easily applicable with no fundamental change. Suppose that there are a total of $M$ mechanical modes with the corresponding bosonic operators $\hat{b}_k$ and $\hat{b}_k^\dagger$ where $k\in[1,M]$. Then, these modes are mutually independent in the sense that $[\hat{b}_j,\hat{b}_k]=0$ and $[\hat{b}_j,\hat{b}_k^\dagger]=\delta_{jk}$. The set of commutators (\ref{T4-eq5}) will be usable for all $M$ modes individually and as a result (\ref{T4-eq19}) and therefore (\ref{T4-eq20}) may be still used. The first and second order perturbations will respectively result in $3+3M=3(M+1)$ and $3+3M+9M=3(4M+1)$ equations. The redefined set of operators will be respectively now $\{A\}^{\rm T}=\{\hat{c},\hat{c}^\dagger,\hat{n},\hat{d}_k,\hat{d}_k^\dagger,\hat{m}_k;k\in[1,M]\}$ and $\{A\}^{\rm T}=\{\hat{c},\hat{c}^\dagger,\hat{n},\hat{d}_k,\hat{d}_k^\dagger,\hat{m}_k,\hat{c}\hat{d}_k,\hat{c}\hat{d}_k^\dagger,\hat{c}\hat{m}_k,\hat{c}^\dagger\hat{d}_k,\hat{c}^\dagger\hat{d}_k^\dagger,\hat{c}^\dagger\hat{m}_k,\hat{n}\hat{d}_k,\hat{n}\hat{d}_k^\dagger,\hat{n}\hat{m}_k;k\in[1,M]\}$.

Similarly, in case of $N$ optical modes satisfying $[\hat{a}_j,\hat{a}_k]=0$ and $[\hat{a}_j,\hat{a}_k^\dagger]=\delta_{jk}$, the set of commutators (\ref{T4-eq4}) can be used and the operator set should be now expanded as $\{A\}^{\rm T}=\{\hat{c}_j,\hat{c}_j^\dagger,\hat{n}_j,\hat{d}_k,\hat{d}_k^\dagger,\hat{m}_k;j\in[1,N];k\in[1,M]\}$ and $\{A\}^{\rm T}=\{\hat{c}_j,\hat{c}_j^\dagger,\hat{n}_j,\hat{d}_k,\hat{d}_k^\dagger,\hat{m}_k,\hat{c}_j\hat{d}_k,\hat{c}_j\hat{d}_k^\dagger,\hat{c}_j\hat{m}_k,\hat{c}_j^\dagger\hat{d}_k,\hat{c}_j^\dagger\hat{d}_k^\dagger,\hat{c}_j^\dagger\hat{m}_k,\hat{n}_j\hat{d}_k,\hat{n}_j\hat{d}_k^\dagger,\hat{n}_j\hat{m}_k;j\in[1,N];k\in[1,M]\}$ respectively for first and second order perturbations. Hence, the corresponding dimensions will be now respectively either $3(N+M)$ or $3(N+M+3NM)$. Higher-order commutators (\ref{T4-eq19}) and (\ref{T4-eq20}) can be still used again by only addition of appropriate photonic $j$ and phononic $k$ mode indices to the respective operators contained in the expanded operator basis set $\{A\}$.

\subsection{The Husimi-Kano Q-functions}

It is mostly appropriate that moments of operators are known, which are scalar functions and much easier to work with. The particular choice of $Q-$functions \cite{S3-71} is preferred when dealing with ladder operators, and are obtained by taking the expectation value of density operator with respect to a complex coherent state $\ket{\alpha}$ and dividing by $\pi$. This definition leads to a non-negative real valued function $Q(\alpha)=Q(\Re[\alpha],\Im[\alpha])$ of $\ket{\alpha}$. Then, obtaining $Q-$function moments of any expression containing the ladder operators would be straightforward \cite{S3-71}. However, it must be antinormally ordered, with creators be moved to the right. In $\{A\}^{\rm T}$ above all operators are actually in the normal form, except $\hat{n}^2$. It is possible to put the nontrivial members of $\{A\}$ in the antinormal order
\begin{eqnarray}
\label{T4-eq35a}
\hat{n}&=&\hat{a}\hat{a}^\dagger-1,\\ \nonumber
\hat{n}^2&=&\hat{a}\hat{a}\hat{a}^\dagger\hat{a}^\dagger-2\hat{a}\hat{a}^\dagger,\\ \nonumber
\hat{n}\hat{c}&=&\frac{1}{2}\hat{a}\hat{a}\hat{a}\hat{a}^\dagger-\frac{3}{2}\hat{a}\hat{a},\\ \nonumber
\hat{c}^\dagger\hat{n}&=&\frac{1}{2}\hat{a}\hat{a}^\dagger\hat{a}^\dagger\hat{a}^\dagger-\frac{3}{2}\hat{a}^\dagger\hat{a}^\dagger.
\end{eqnarray}
While evaluating $Q-$function moments, $\hat{a}$ and $\hat{a}^\dagger$ are replaced with $\alpha$ and $\alpha^\ast$ respectively as
\begin{eqnarray}
\label{T4-eq35b}
\braket{\hat{n}}&=&|\alpha|^2-1,\\ \nonumber
\braket{\hat{n}^2}&=&|\alpha|^4-2|\alpha|^2,\\ \nonumber
\braket{\hat{n}\hat{c}}&=&\frac{1}{2}\alpha^2|\alpha|^2-\frac{3}{2}\alpha^2,\\ \nonumber
\braket{\hat{c}^\dagger\hat{n}}&=&\frac{1}{2}\alpha^{\ast 2}|\alpha|^2-\frac{3}{2}\alpha^{\ast 2}.
\end{eqnarray}
\noindent
All remains now is to redefine the array of $Q-$functions bases, using common terms from which the original $Q-$functions could be readily restored. These are $\{\braket{A}\}^{\rm T}=\{\alpha^2,\alpha^{\ast 2},|\alpha|^2,|\alpha|^4,\alpha^4,\alpha^{\ast 4},\alpha^2|\alpha|^2,\alpha^{\ast 2}|\alpha|^2\}$ . This translates into a set of scalar differential equations which conveniently could be solved. Fluctuations of noise terms also vanish while taking the expectation values, and only their average values survive. To illustrate this, suppose that the system is driven by a coherent field $\hat{a}_{\rm in}$ with the normalized electric field amplitude $\beta=\alpha/\sqrt{2}$ and at the frequency $\omega$. Then, the $Q-$function moments of the input fields after defining the loss rates $\Gamma_3=2\Gamma_2=4\Gamma_1$ become $\braket{\hat{a}_{\rm in}}=\sqrt{2\Gamma_1}\beta$, $\braket{\hat{c}_{\rm in}}=\sqrt{\Gamma_2}\beta$, $\braket{\hat{n}_{\rm in}}=\sqrt{\Gamma_2}(2|\beta|^2+1)$, $\braket{\hat{c}^2_{\rm in}}=\sqrt{\Gamma_3}\beta^2$, and $\braket{\hat{n}_{\rm in}\hat{c}_{\rm in}}=\sqrt{\Gamma_3}\beta^2(2|\beta|^2+3)$.

\subsection{Time-dependence}

Under external drive, periodicity, or dynamical control $[\textbf{M}(t)]$ in (\ref{T4-eq6}) is time-dependent \cite{S4-36f,S4-51}. For instance, the ultimate optomechanical cooling limit is a function of system dynamics \cite{S4-56}. Then, integration should be done numerically, since exact analytical solutions without infinite perturbations exist only for very restricted cases. However, there exists the approximate yet accurate expression
\begin{equation}
\label{T4-eq40}
\{A(t)\}\approx\exp\left(\int_{0}^{t}[\textbf{M}(\tau)]d\tau\right)\{A(0)\}-\int_{0}^{t}\exp\left(\int_{\tau}^{t}[\textbf{M}(s)]ds\right)\sqrt{[\Gamma]}\{A_{\rm in}(\tau)\}d\tau,
\end{equation}
\noindent
where $\exp(\cdot)$ is matrix exponentiation. In general (\ref{T4-eq40}) is exact if a time-ordering operator $\mathbb{T}$ is applied to $\exp(\cdot)$, however, analytical evaluation of (\ref{T4-eq40}) will be no longer possible. Nevertheless, (\ref{T4-eq40}) under certain sufficient conditions \cite{S4-59,S4-60} could still be exact, since the trace and product of eigenvalues of $\exp(\cdot)$ remain unaltered if $\mathbb{T}$ is dropped \cite{S4-59,S4-60}. This hints the fact that $\mathbb{T}$ might actually cause an orthogonal transformation which preserves the eigenvalues. Anyhow, (\ref{T4-eq40}) remains sufficiently accurate for all practical purposes.

In the end, it has to be mentioned that under external drive, periodicity, or dynamical control $[\textbf{M}(t)]$ in (\ref{T4-eq6}) is time-dependent \cite{S4-36f,S4-51}. For instance, the ultimate optomechanical cooling limit is a function of system dynamics \cite{S4-56}. Then, integration should be done numerically, since exact analytical solutions without infinite perturbations exist only for very restricted cases. This is, however, beyond the scope of the current study.

In this section, a new method was described to solve quadratic quantum interactions using perturbative truncation schemes, by including higher-order operators in the solution space. Transformation to scalar forms using $Q-$functions was also presented. Spectral densities of square operators, calculation of the second-order correlation function, as well as quantum limited amplifiers, nondemolition measurements, and quantum anharmonic oscillator are to be demonstrated in the next three sections \S\ref{Section-5} through \S\ref{Section-7}, followed by in depth discussions of quantum optomechanics, quadratic optomechanics, and cross-Kerr interactions in \S\ref{Section-8} through \S\ref{Section-11}.

\section{Nonlinear Noise Spectra}\label{Section-5}
This section deals with the calculation of nonlinear noise spectra for squared noise operators such as $\hat{a}_{\rm in}^2(t)$ and so on. While the results are of relevance in part to the higher-order operator analysis, mostly we shall come across multiplicative noise such as $\hat{a}\hat{a}_{\rm in}$ and make use of linearized approximations such as $\bar{a}\hat{a}_{\rm in}$. This is quite tempting since it makes the method of higher-order operators a lot easier to apply. However, still the knowledge of noise spectra for squared operators is much useful and of relevance, at least when nonlinear oscillations are to be studied around their equilibrium.  We show that not only it is possible to estimate the spectral noise density of squared noise operators analytically but also further application of results can yield previously unknown quantities such as photon cavity occupation number at lasing threshold. These are all to be discussed below in sufficient details.  

The required noise spectra \cite{S4-39b} of cross operators is clearly a product of each of the individual terms, since the nature of particles are different. However, the noise spectra of quadratic operators themselves need to be appropriately expressed. For instance, $\hat{d}_{\textrm{in}}$ actually corresponds to the spectral input noise of the square operator $\hat{d}=\hat{b}\hat{b}/2\sqrt{\Gamma}$ from (\ref{T4-eq3}), which clearly satisfies $\hat{d}_{\textrm{in}}(t)=\frac{1}{2}\hat{b}_{\textrm{in}}(t)\hat{b}_{\textrm{in}}(t)/\sqrt{\Gamma}$, or $\hat{d}_{\textrm{in}}(w)=\frac{1}{2}\hat{b}_{\textrm{in}}(w)\ast \hat{b}_{\textrm{in}}(w)/\sqrt{\Gamma}$ in the frequency domain, where $\ast$ merely represents the convolution operation. Therefore, once $\hat{a}_{\textrm{in}}(w)$ and $\hat{b}_{\textrm{in}}(w)$ are known, all relevant remaining input noise spectra could be obtained accordingly using simple convolutions or products in frequency domain.

As a result, the corresponding spectral density of the noise input terms to the cross operators can be determined from the relevant vacuum noise fluctuations and performing a Fourier transform. For instance, we have $S_{CDCD}[w]=S_{CC}[w]S_{DD}[w]$ where $S_{CC}[w]=\frac{1}{4}S_{A^2A^2}[w]$ and $S_{DD}[w]=\frac{1}{4}S_{B^2B^2}[w]$. Then Isserlis-Wick theorem \cite{S3-64,S4-40} could be exploited to yield the desired expressions. If we assume 
\begin{eqnarray}
\label{S4-eq21}
\left<\hat{f}(t)\hat{f}(\tau)\right>&=&\zeta(t-\tau), \\ \nonumber
\left<\hat{f}(t)\hat{f}^\dagger(\tau)\right>&=&\psi(t-\tau), \\ \nonumber
[\hat{f}(t),\hat{f}^\dagger(\tau)]&=&\hat{\upsilon}(t-\tau),
\end{eqnarray}  where the dimensionless correlation integrator runs on phase, instead of time, as
\begin{equation} 
\label{S4-eq22}
\left<\hat{f}(t)\hat{g}(\tau)\right>=\int \hat{f}(t+\tau)\hat{g}(\tau) d(\omega \tau),
\end{equation}
\noindent
then the functions $\zeta(\cdot),$ $\psi(\cdot),$ and the operator $\hat{v}(\cdot)$ should be all having the dimension of $\hat{f}^2(\cdot)$ as well. That means if $\hat{f}$ is dimensionless, which is the case for the choice of ladder operators, then $\zeta(\cdot),$ $\psi(\cdot),$ and $\hat{v}(\cdot)$ become dimensionless, too. The functions $\zeta(\cdot)$ and $\psi(\cdot)$ together can cause squeezing or thermal states if appropriately defined \cite{S2-Noise1,S2-Noise3}. By Isserlis-Wick theorem applied to scalars we have $\left<x_1 x_2 x_3 x_4\right>=\left<x_1 x_2\right>\left<x_3 x_4\right>+\left<x_1 x_3\right>\left<x_2 x_4\right>+\left<x_1 x_4\right>\left<x_3 x_4\right>$. This gives for the operators
\begin{eqnarray}
\label{S4-eq23}
S_{F^2F^2}[w]&=&\frac{1}{2\pi}\int_{-\infty}^{+\infty}\left<\hat{f}^2(t)\hat{f}^{2\dagger}(0)\right>e^{iwt} dt =\frac{1}{2\pi}\int_{-\infty}^{+\infty}\left<\hat{f}(t)\hat{f}(t)\hat{f}^{\dagger}(0)\hat{f}^{\dagger}(0)\right>e^{iwt} dt \\ \nonumber
&=&\frac{1}{2\pi}\int_{-\infty}^{\infty}\left\{\left<\hat{f}^2(t)\right>\left<\hat{f}^{2\dagger}(0)\right>+2\left<\hat{f}(t)\hat{f}^\dagger(0)\right>^2+2\left<\hat{f}(t)\left[\hat{f}(t),\hat{f}^\dagger(0)\right]\hat{f}^\dagger(0)\right>\right\} e^{iwt} dt \\ \nonumber
&=&\frac{1}{2\pi}\int_{-\infty}^{\infty}\left[\zeta(0)\zeta^\ast(0)+2\psi^2(t)+2\left<\hat{f}(t)\hat{\upsilon}(t)\hat{f}^\dagger(0)\right>\right] e^{iwt} dt\\ \nonumber 
&=&|\zeta(0)|^2\delta(w)+\frac{1}{\pi}\int_{-\infty}^{\infty}\left[\psi^2(t)+\left<\hat{f}(t)\hat{\upsilon}(t)\hat{f}^\dagger(0)\right>\right] e^{iwt} dt. 
\end{eqnarray} 
\noindent
Hence, for a given stochastic process where $\left<\hat{f}(t)\hat{f}(\tau)\right>=0$, $\left<\hat{f}(t)\hat{f}^\dagger(\tau)\right>=\Psi(t-\tau)$, and having the scalar commutator $[\hat{f}(t),\hat{f}^\dagger(\tau)]=\Upsilon (t-\tau)$, we simply get
\begin{equation}
\label{S4-eq24}
S_{F^2F^2}[w]=\frac{1}{\pi}\int_{-\infty}^{\infty}\Psi^2(t) e^{iwt} dt+\frac{1}{\pi}\int_{-\infty}^{\infty}\Upsilon (t)\Psi(t) e^{iwt} dt. 
\end{equation} 

Now, suppose that we have a coherent field of photons at the angular frequency $\omega$ with an initial Gaussian distribution, in which $\Psi(t)=\exp(-\chi^2\omega^2 t^2/2)\exp(-i\omega t)$ and $\Upsilon (t)=\Psi(t)$, while having the linewidth $\Delta f=\frac{1}{2\pi}\chi\omega$. Clearly, $\chi$ is a dimensionless and positive real number. In the limit of $\chi\rightarrow 0^{+}$, the expected relationship $\Psi(t)=\sqrt{2\pi}\delta(\omega t)/\chi$ is easily recovered. 

This particular definition of the correlating function $\Psi(t)$ ensures that the corresponding spectral density is appropriately normalized, that is
\begin{eqnarray}
\label{S4-eq25}
\int_{-\infty}^{+\infty} S_{FF}[w]dw&=&\int_{-\infty}^{\infty}\left[\frac{1}{2\pi}\int_{-\infty}^{+\infty}\left<\hat{f}(t)\hat{f}^\dagger(0)\right>e^{iwt} dt\right]dw\\ \nonumber
&=&1.
\end{eqnarray}
\noindent
Hence, one may obtain the following spectral density
\begin{equation}
\label{S4-eq26}
S_{F^2F^2}[w]=\frac{\chi}{\pi\sqrt{\pi}\omega}\exp\left[-\frac{(w-2\omega)^2}{4\chi^2\omega^2}\right],
\end{equation}
\noindent
which is centered at the doubled frequency $2\omega$, has a linewidth of $\sqrt{2}\Delta f$, and satisfies the property 
\begin{equation}
\label{S4-eq27}
\int_{-\infty}^{+\infty} S_{F^2F^2}[w]dw=\frac{2}{\pi}\chi^2.
\end{equation}

Once the spectral densities of input noise terms are found, spectral densities of all output fields immediately follows (\ref{T4-eq8},\ref{T4-eq9}) as $\{A[w]\}_{\rm out}=[\textbf{S}^\dagger(w)\textbf{S}(w)]\{A[w]\}_{\rm in}$, in which $[\textbf{S}^\dagger(w)\textbf{S}(w)]=[|S_{ij}(w)|^2]$,  $\{A[w]\}_{\rm in}$ is an array containing the spectral densities of inputs, and similarly $\{A[w]\}_{\rm out}$ is the array of spectral densities at each of the output fields.

\subsection{Estimation of $g^{(2)}(0)$}

Many of the important features of an interacting quantum system is given by its second-order correlation function $g^{(2)}(0)$ at zero time-delay \cite{S4-36a,S4-36b,S4-36c} defined as
\begin{equation}
\label{S4-eq28}
g^{(2)}(0)=\frac{\langle\hat{a}^\dagger(0)\hat{a}^\dagger(0)\hat{a}(0)\hat{a}(0)\rangle}{\langle \hat{a}^\dagger(0)\hat{a}(0)\rangle^2}.
\end{equation}
\noindent
It is fairly easy to estimate this function once the spectral densities of all higher order operators of the nonlinear system are calculated. For this purpose, we may first employ the definition (\ref{T4-eq3}) to rewrite
\begin{equation}
\label{S4-eq29}
g^{(2)}(0)=4\frac{\langle\hat{c}^\dagger(0)\hat{c}(0)\rangle}{\langle \hat{n}(0)\rangle^2}=\frac{4}{\bar{n}^2}\langle\hat{c}^\dagger(0)\hat{c}(0)\rangle=\frac{4}{\bar{n}^2}\left[\langle\hat{c}(0)\hat{c}^\dagger(0)\rangle-\bar{n}-\frac{1}{2}\right].
\end{equation}
\noindent
Estimation of the average within brackets can be done by having $S_{CC}[w]=\frac{1}{4}S_{A^2A^2}[w]$ corresponding to the higher-order operator $\hat{c}$. This can be assumed to has been already found from knowledge of the scattering matrix $[\textbf{S}(w)]$, spectral densities of input fields $\{A[w]\}_{\rm in}$, and subsequent derivation of spectral density array of output fields $\{A[w]\}_{\rm out}$. Then, $S_{CC}[w]$ will be simply an element of the vector $\{A[w]\}_{\rm out}$. Using (\ref{S4-eq24}), this results in a fairly brief representation 
\begin{equation}
\label{S4-eq30}
g^{(2)}(0)=\frac{4}{\bar{n}^2}\left(\int_{-\infty}^{+\infty}S_{CC}[w]dw\right)-\frac{4\bar{n}+2}{\bar{n}^2}=\frac{2}{\bar{n}^2}\Psi(0)\left[\Psi(0)+\Upsilon(0)\right]-\frac{4\bar{n}+2}{\bar{n}^2}.
\end{equation}
\noindent
With the assumptions above for an ideal initial Gaussian distribution, we have $\Psi(0)=\Upsilon(0)=1$ and thus $g^{(2)}(0)=4(\frac{1}{2}-\bar{n})/\bar{n}^2$. One should have in mind that this relationship cannot be readily used for a coherent radiation, since for a practical laser the true statistics is Poissonian and not Gaussian. This analysis thus reveals that the cavity occupation number of such an ideal laser with the threshold defined as $g^{(2)}(0)=1$ is exactly $\bar{n}=\sqrt{6}-2\approx 0.450$. This is in contrast to the widely used assumption of quantum threshold condition $\bar{n}=1$ \cite{S4-40a,S4-40b,S4-40b1,S4-40b2,S4-40b3,S4-40b4}. Interestingly, a new study \cite{S4-40c} of photon statistics in weakly nonlinear optical cavities based on extensive density matrix calculations \cite{S4-40d,S4-40e} yields the value $\bar{n}=0.4172$, which is in reasonable agreement to our estimate. An earlier investigation on quantum-dot photonic crystal cavity lasers \cite{S4-40f,S4-40g} also gives the value $\bar{n}=0.485$.

\section{Quantum Read-out Circuits} \label{Section-6}
This brief section presents the method of higher-order operators as applied to two of the very common quantum circuits used in read-out process: quantum limited amplifiers, and quantum non-demolition measurements. A more extensive and focused study of this subject through cross-Kerr interaction combined with parametric amplification though follows later in \S\ref{Section-11}, which is shown to admit a mathematically exact solution through the proper utilization of the higher-order operators. 

\subsection{Quantum Limited Amplifiers} \label{Amplifier}
The method of higher-order operators can be extended to the quantum limited amplifiers, which in the general form coincides with the expression (\ref{S4-eq31}), but is usually solved using a zeroth-order perturbation \cite{S4-36e}. For the single-mode degenerate quantum limited amplifier \cite{S2-5,S4-36e,S4-36e1}, the corresponding Hamiltonian is slightly different given by 
\begin{equation}
\label{S6-1}
\mathbb{H}=\hbar\omega\hat{n}+\hbar(g\hat{c}+g^\ast\hat{c}^\dagger), 
\end{equation}
with the 3-dimensional basis $\{A\}^{\rm T}=\{\hat{n},\hat{c},\hat{c}^\dagger\}$ which satisfies closedness. Then, the second-order accurate Langevin equations with inclusion of the self-energy $\mathbb{H}_{\rm self}=\hbar\omega\hat{n}$ can be shown to be unconditionally stable with $\Re\{{\rm eig}[{\bf M}]\}<0$, given by
\begin{eqnarray}
\dot{\hat{n}}&=&-i2(g\hat{c}-g^\ast\hat{c}^\dagger),\\ \nonumber
\dot{\hat{c}}&=&(-2i\omega-\frac{2\bar{n}+1}{4}\Gamma_2)\hat{c}-ig^\ast\hat{n}-i\frac{1}{2}g^\ast-(\bar{n}+\frac{1}{2})\sqrt{\Gamma_2}\hat{c}_{\rm in},\\ \nonumber
\dot{\hat{c}}^\dagger&=&(2i\omega-\frac{2\bar{n}+1}{4}\Gamma_2)\hat{c}^\dagger+ig\hat{n}+i\frac{1}{2}g-(\bar{n}+\frac{1}{2})\sqrt{\Gamma_2}\hat{c}^\dagger_{\rm in}.
\end{eqnarray}
In presence of Kerr nonlinearity \cite{S4-36f} as $\mathbb{H}=\hbar\omega\hat{n}+\hbar(g\hat{c}+g^\ast\hat{c}^\dagger)+\hbar\gamma\hat{c}^\dagger\hat{c}$, one may use $4\hat{c}^\dagger\hat{c}=\hat{n}^2-\hat{n}$, $[\hat{n}^2,\hat{c}]\approx-\frac{1}{2}(6\bar{n}+7)\hat{c}$, and the basis $\{A\}^{\rm T}=\{\hat{n},\hat{n}^2,\hat{c},\hat{c}^\dagger\}$ to construct a set of $4\times 4$ integrable Langevin equations. The rest of necessary commutators are already found in (\ref{T4-eq4}), and later below in (\ref{S7-eq34}) and (\ref{S4-eq35}).

\subsection{Quantum Nondemolition Measurements} \label{QND}
Quantum nondemolition measurements of states require a cross-Kerr nonlinear interaction of the type
\begin{eqnarray}
\label{S6-2}
\mathbb{H}&=&\hbar\omega\hat{a}^\dagger\hat{a}+\hbar\Omega\hat{b}^\dagger\hat{b}+\hbar\chi\hat{a}^\dagger\hat{a}\hat{b}^\dagger\hat{b}\\ \nonumber
&=&\hbar\omega\hat{n}+\hbar\Omega\hat{m}+\hbar\chi\hat{n}\hat{m},
\end{eqnarray} 
in which $\hat{a}$ and $\hat{b}$ fields respectively correspond to the probe and signal \cite{S4-36g,S4-36h}. This system can be conveniently analyzed by the preferred choice \cite{S4-36g} of the higher-order operators $\{A\}^{\rm T}=\{\hat{n},\hat{m},\hat{C},\hat{S}\}$, where
\begin{eqnarray}
\hat{C}&=&\frac{1}{2}\left[(\hat{n}+1)^{-\frac{1}{2}}\hat{a}+\hat{a}^\dagger(\hat{n}+1)^{-\frac{1}{2}}\right], \\ \nonumber
\hat{S}&=&\frac{1}{2i}\left[(\hat{n}+1)^{-\frac{1}{2}}\hat{a}-\hat{a}^\dagger(\hat{n}+1)^{-\frac{1}{2}}\right],
\end{eqnarray}
\noindent
are quadratures of the readout observable. It is straightforward to show by induction that $[f(\hat{a}^\dagger),\hat{a}]=-f'(\hat{a}^\dagger)$ and $[\hat{a}^\dagger,f(\hat{a})]=-f'(\hat{a})$ with $f(\cdot):\mathcal{R}\mapsto\mathcal{R}$ being a real function of its argument. Now, the non-zero commutators of the basis $\{A\}^{\rm T}$ can be found after some algebra as $[\hat{n},\hat{C}]=-i\hat{S}$, $[\hat{n},\hat{S}]=i\hat{C}$, and $[\hat{C},\hat{S}]=\frac{1}{2}i(\hat{n}+2)^{-1}$. All remains to construct the Langevin equations now, is to linearize the last commutators as $[\hat{C},\hat{S}]\approx \frac{1}{2}i(\bar{n}+2)^{-1}$, by which the basis $\{A\}^{\rm T}=\{\hat{n},\hat{m},\hat{C},\hat{S}\}$ would satisfy closedness. Input noise terms to the operators $\hat{C}$ and $\hat{S}$ should be constructed by linear combinations of $\hat{a}_{\rm in}$ and $\hat{a}_{\rm in}^\dagger$ while replacing the multiplier term $1/\sqrt{\hat{n}+1}$ with the linearized form $1/\sqrt{\bar{n}+1}$.

In practice, both of the quantum limited amplification and cross-Kerr interaction are needed at once to perform a quantum non-demolition read-out of system state. Hence, a practical system makes use of a combined Hamiltonian of (\ref{S6-1}) and (\ref{S6-2}). Solution of such system becomes a lot more complicated, however, it still admits a mathematically exact solution through nonlinear analysis of higher-order operators as shown later in \S\ref{Section-11}.

\section{Quantum Anharmonic Oscillator} \label{Section-7}
The quantum anharmonic oscillator appears in many nonlinear systems including quadratic optomechanics \cite{S4-40h,S4-40i}, where our method here is applicable, and this section discusses the application of higher-order operator algebra to tackle this system. 

The anharmonic Kerr Hamiltonian is \cite{S4-41,S4-42}
\begin{equation}
\label{S4-eq31}
\mathbb{H}=\hbar\omega\hat{a}^\dagger\hat{a}+\frac{1}{2}\hbar\zeta\hat{a}^{\dagger 2}\hat{a}^2=\hbar\omega\hat{a}^\dagger\hat{a}+2\hbar\zeta\hat{c}^{\dagger}\hat{c}=\hbar\left(\omega-\frac{1}{2}\zeta\right)\hat{n}+\frac{1}{2}\hbar\zeta\hat{n}^2,
\end{equation}
\noindent
in which $\zeta$ is a constant. It is well known that in case of $\zeta>2\omega$ this system exhibits an effective bistable potential, and is otherwise monostable. However, we are here much interested in a slightly different but more complicated form given by \cite{S4-2DM}
\begin{equation}
\label{S4-eq32}
\mathbb{H}=\hbar\omega\hat{a}^\dagger\hat{a}-\frac{1}{2}\hbar\zeta\left(\hat{a}^\dagger+\hat{a}\right)^4,
\end{equation}
\noindent
which is monostable or bistable if both $\omega$ and $\zeta$ are respectively positive or negative. This type of nonlinearity is of particular importance in fourth-order analysis of qubits \cite{S4-43,S3-70,S4-45,S4-46,S4-47,S4-49,S4-50}. While the Hamiltonian (\ref{S4-eq32}) is for a single-mode field, the case of multi-mode electromagnetic field could be easily devised following the existing interaction Hamiltonians \cite{S4-2DM} and the presented method. Nevertheless, the above expression after some algebraic manipulations can be put into the form 
\begin{equation}
\label{S4-eq33}
\mathbb{H}=\hbar(\omega-3\zeta)\hat{n}-3\hbar\zeta\hat{n}^2-2\hbar\zeta\left[\hat{c}^2+\hat{c}^{\dagger 2}+3 \left(\hat{c}+\hat{c}^\dagger\right)\right]-4\hbar\zeta\left(\hat{n}\hat{c}+\hat{c}^\dagger\hat{n}\right),
\end{equation}
\noindent
where a trivial constant term $\hbar\zeta$ is dropped. Here, we may proceed with the 8-dimensional basis operator set $\{A\}^{\rm T}=\{\hat{c},\hat{c}^\dagger,\hat{n},\hat{n}^2,\hat{c}^2,\hat{c}^{\dagger 2},\hat{n}\hat{c},\hat{c}^\dagger\hat{n}\}$, resulting in a second order perturbation accuracy.

Treating this problem using the Langevin equation (\ref{T4-eq10}), regardless of the values of $\zeta$ and $\omega$, is possible, only if the following non-trivial exact commutators
\begin{eqnarray}
\label{S7-eq34}
[\hat{n},\hat{c}^2]&=&-4\hat{c}^2, \\ \nonumber
[\hat{n}^2,\hat{c}]&=&-3\hat{n}\hat{c}-\frac{7}{2}\hat{c}, \\ \nonumber
[\hat{n}^2,\hat{c}^2]&=&4(\hat{n}-2)\hat{n}\hat{c}^2, \\ \nonumber
[\hat{c}^2,\hat{c}^\dagger]&=&2\hat{n}\hat{c}+3\hat{c}, \\ \nonumber
[\hat{c}^2,\hat{c}^{\dagger 2}]&=&\hat{n}^3+\frac{3}{2}\left(\hat{n}^2+1\right)+\frac{1}{4}\hat{n}, \\ \nonumber
[\hat{c}^2,\hat{c}^\dagger\hat{n}]&=& 3\left(\hat{n}+2\right)\hat{n}\hat{c}+6\hat{c},\\ \nonumber
[\hat{c},\hat{c}^\dagger\hat{n}]&=&\frac{3}{2}\hat{n}^2, \\ \nonumber
[\hat{n}\hat{c},\hat{c}^\dagger\hat{n}]&=&\frac{1}{2}\left(4\hat{n}^2-3\hat{n}+2\right)\hat{n},
\end{eqnarray}
\noindent
are known, which may be found after significant algebra. The rest of required commutators which are not conjugates of those in the above, can either directly or after factorization of a common term be easily found from (\ref{T4-eq4}). Again, the set of commutators (\ref{S7-eq34}) does not yet satisfy the closedness property within $\{S\}={\rm span}(\{A\}\cup\{\hat{1}\})$, unless the approximate linearization 

\begin{eqnarray}
\label{S4-eq35}
[\hat{n},\hat{c}^2]&=&-4\hat{c}^2, \\ \nonumber
[\hat{n}^2,\hat{c}]&=&-3\hat{n}\hat{c}-\frac{7}{2}\hat{c}, \\ \nonumber
[\hat{n}^2,\hat{c}^2]&=&4(\bar{n}-2)\bar{n}\hat{c}^2, \\ \nonumber
[\hat{c}^2,\hat{c}^\dagger]&=&2\hat{n}\hat{c}+3\hat{c}, \\ \nonumber
[\hat{c}^2,\hat{c}^{\dagger 2}]&=&\frac{1}{2}\left(2\bar{n}+3\right)\hat{n}^2+\frac{1}{4}\hat{n}+\frac{3}{2}, \\ \nonumber
[\hat{c}^2,\hat{c}^\dagger\hat{n}]&=& 3\left(\bar{n}+2\right)\hat{n}\hat{c}+6\hat{c},\\ \nonumber
[\hat{c},\hat{c}^\dagger\hat{n}]&=&\frac{3}{2}\hat{n}^2, \\ \nonumber
[\hat{n}\hat{c},\hat{c}^\dagger\hat{n}]&=&\frac{1}{2}\left(4\bar{n}-3\right)\hat{n}^2+\hat{n},
\end{eqnarray}
\noindent
is employed. The rest of the process is identical to the one described under (\ref{T4-eq20}). Construction of the respective noise terms is also possible by iterated use of the results in \S\ref{Section-5} and so on.
\section{Quantum Optomechanics}  \label{Section-8}

This section concerns the application of the method of higher-order operators to conventional optomechanics, which is a third-order nonlinear interaction. The nonlinearity arises from radiation pressure acting upon the mirror, which also in turn causes backaction unto mechanical oscillator. The combination of nonlinearity and backaction is what is required to properly explain the behavior of optomechanical systems and time-domain evolution of operators through Langevin equations. The tricky part is to recover the measureable spectral density by means of higher-order algebra, and this is shown to be well doable. Once the appropriate formulation is set, no only the linearized results can be recovered and the higher-order algebra is consistent with linearized optomechanics, but also, additional correction terms and explicit expressions could be found as a result of nonlinearity. The coherent phonon population $\bar{m}=\braket{\hat{b}^\dagger\hat{b}}$, which turns out to be significantly different from mean-field approximated value of $|\bar{b}|^2=|\braket{\hat{b}}|^2$, and side-band inequivalence in \S\ref{Section-9}, which is a counter-intuitive symmetry breaking in frequency shifts of Stokes/anti-Stokes side-bands, are clear manifestions of higher-order effects in nonlinear optomechanics.

Similarly, a higher-order resonance shift exists appearing as changes in both of the optical and mechanical resonances. We provide the first known method to explicitly estimate the population of coherent phonons. We also calculate corrections to spring effect due to higher-order interactions and coherent phonons, and show that these corrections can be quite significant in measurement of single-photon optomechanical interaction rate. It is shown that there exists non-unique and various choices for the higher-order operators to solve the optomechanical interaction with different multiplicative noise terms, among which a minimal basis offers exactly linear Langevin equations, while decoupling one Langevin equation and thus leaving the whole standard optomechanical problem exactly solvable by explicit expressions. We finally present a detailed treatment of multiplicative noise as well as nonlinear dynamic stability phases by the method of higher-order operators. Similar approach can be used outside the domain of standard optomechanics to quadratic and all other types of nonlinear interactions in quantum physics.

Nonlinear quantum interactions with stochastic noise input stand among the most difficult analytical challenges to solve in the context of stochastic differential equations. While linearized interactions remain accurate for description of many experiments, a certain class of quadratic and higher-order physical phenomena cannot be normally understood under linearized approximations. While in classical problems the resulting Langevin equations are scalar functions, in quantum problems one has to deal with nonlinear operator differential equations. If expanded unto base kets, bosonic operators can assume infinite-dimensional matrix forms, rendering the solution entirely intractable.

Such classes of nonlinear operator problems can be addressed by construction of Fokker-Planck or nonlinear Schr\"{o}dinger equations, among which there exists a one-to-one correspondence. The Fokker-Planck equation \cite{S4-33.20,S4-34.15,S4-34.16,S4-34.16a,S8-Fokker5} is actually equivalent to the nonlinear Schr\"{o}dinger equation with bosonic operator algebra, and its moments \cite{S4-34.17} translate into nonlinear Langevin equations. The method of master equations \cite{S4-33a,S4-33b} also can be used in combination with the quasi-probablity Wigner functions \cite{S4-33c,S4-33d} to deal with nonlinear quantum interactions. The master equation approach is reasonably accurate as long as Born and Markov approximations are not employed \cite{S8-Master5}. But none of these methods is probably as convenient as the method of Langevin equations \cite{S2-Noise1,S2-Noise0,S2-Noise2,S2-Noise3}, which has found popularity in the context of quantum optoemchanics \cite{S2-5,S8-Macri,S2-1,S2-3,S2-6,S8-Opto1,S8-Opto2,S4-38,S8-Polariton,S8-Multi1,S8-Multi2,S8-Multi3,S3-16,S3-47}. 

Being an inherently nonlinear interaction among photonic and phononic baths \cite{S3-6,S3-7,S4-36d,S3-17,S8-Kerr5,S3-53,S3-19,S3-20,S8-Girvin}, the standard quantum optomechanics is normally described by linearized Langevin equations \cite{S2-Noise1,S2-Noise0,S2-Noise2,S2-Noise3}. This will suffice to address a majority of complex experimental situations such as optomechanical-induced transparency \cite{S8-Lemonde2,S8-OMIT1,S8-OMIT2} and polaron anti-crossing \cite{S8-Polaron}, but effects such as non-classical states of light \cite{S4-33c,S4-33d,S4-36a,S8-Lemonde1}, optomechanical emission of real photons from vacuum \cite{S8-Lemonde3}, photon blockade \cite{S4-36a}, nonlinear self-oscillations \cite{S8-Self1,S8-Self2,S8-Self3,S8-Self4,S8-Self5}, and chaos \cite{S8-Chaos1,S8-Chaos2} are all among manifestations of nonlinear regimes in standard optomechanics, which need description using nonlinear algebra. Also, biquadratic interactions (mostly referred to as quadratic interactions) among bosonic baths remain a hurdle. In quadratic optomechanics \cite{S3-57,S3-56,S3-57,S3-58,S2-9,S3-59,S2-7,S4-33.3,S4-33.6.0,S2-8,S3-61,S3-60,S2-Bruschi}, which is a topic of growing interest in the recent year, having an analytical tool capable of addressing such kinds of nonlinearity is advantageous. A perturbation technique based on the expansion of time-evolution operators \cite{S2-Bruschi} is employed to investigate quadratic interactions and it has been shown that for mechanical frequencies exceeding optical frequencies a new unexplored regime appears in which the roles of optical and mechanical partitions are interchanged.

Recently, the author has reconsidered the theoretical description of optomechanics \cite{S2-Paper1} and shown that quadratic interactions are subject to two corrections resulting from momentum conservation and relativistic effects. Such types of quadratic corrections become significant when the mechanical frequency is within the same order of or exceeds electromagnetic frequency. Furthermore, an analytical approach is proposed to tackle nonlinear quantum interactions \cite{S2-Paper2} and a method of expansion unto higher-order operators is proposed and investigated in details.

In this section, the higher-order operator approach recently proposed by the author \cite{S2-Paper2,S2-Paper4,S2-Paper6,S2-Paper5} is employed to address the standard optomechanics, and it is shown that there exists a minimal choice of higher-order operator basis which leads to exactly linear and fully separable Langevin equations with multiplicative input noise terms \cite{S2-Noise4}. We also present a full mathematical treatment of multiplicative noise terms, which turn out to play a crucial rule in higher-order quantum optomechanics. This allows one to provide an exact and explicit solution using an operator-based method to solve the optomechanical interactions in the nonlinear regime. There exists higher-order effects appearing at high optical pump rates, and can be predicted using the method discussed here. These include inequivalent red and blue detunings, higher-order resonance shift and spring effects, and also zero-point-field induced optomechanical shift of mechanical frequency. The inequivalency of red and blue detuned side-bands, which appears as a counter-intuitive difference in their respective frequency shifts, is different from the well-known anomalous Stokes-Anti-Stokes symmetry breaking \cite{S8-Stokes1,S8-Stokes2,S8-Stokes3} which is connected to different scattering amplitudes. The same method of higher-order operator algebra has been recently used independently as well \cite{S2-Quad}. 

We also show for the first time that the introduced method of higher-order operators can be used to estimate the coherent population of phonons in the optomechanical cavity, here referred to as the coherent phonon number. This quantity can not only be calculated explicitly in terms of optomechanical parameters, but also, can be found by fitting the expressions of corrected spring effect to the experimental observations. Also, dynamic linear and nonlinear stability phases in red and blue-detuned drives can be well computed and estimated using the method of higher-order operators.

\subsection{Choice of Basis}

The standard optomechanical Hamiltonian after transformation into rotating frame of cavity resonance, and inclusion of non-interacting self-energies $\mathbb{H}_s$ and interacting term $\mathbb{H}_{\rm OM}$ reads \cite{S2-5,S8-Macri,S2-1,S2-3,S2-6}
\begin{eqnarray}
\label{T8-eq1}
\mathbb{H}&=&\mathbb{H}_s+\mathbb{H}_{\rm OM}\\ \nonumber
&=&\hbar \Omega\hat{m}-\hbar \Delta\hat{n}-\hbar g_0\hat{n}\left(\hat{b}+{\hat{b}}^{\dagger }\right),
\end{eqnarray}
\noindent
where $\hat{n}=\hat{a}^\dagger\hat{a}$ and $\hat{m}=\hat{b}^\dagger\hat{b}$ are photon and phonon number operators with $\hat{a}$ and $\hat{b}$ respectively being the photon and phonon annihilators, $\Omega$ is the mechanical frequency, $\Delta$ is optical detuning from cavity resonance, and $g_0$ is the single-photon optomechanical interaction rate. The interaction $\mathbb{H}_{\text{OM}}$ is not quadratic, but is still cubic nonlinear. It is normally solved by a straightforward linearization \cite{S2-1,S2-3,S2-6}, but can be also solved at the second-order accuracy using the higher-order operators described in \S\ref{Section-4}. 

In order to form a closed basis of operators, we may choose either the higher-order operators
\begin{equation}
\label{T8-eq1a}
\{A\}^\text{T}=\left\{\hat{a},\hat{a}\hat{b},\hat{a}{\hat{b}}^{\dagger }\right\},
\end{equation}
of the second-degree, which forms a $3\times 3$ system of Langevin equations, or
\begin{equation}
\label{T8-eq3}
\{A\}^\text{T}=\left\{\hat{a},\hat{b},\hat{a}\hat{b},\hat{a}{\hat{b}}^{\dagger },\hat{n},\hat{c}\right\},
\end{equation}
\noindent 
which forms a $6\times 6$ system of Langevin equations. Here, we adopt the definition $\hat{c}=\frac{1}{2}\hat{a}^2$ \cite{S2-Paper1,S2-Paper2}. 

It is easy to verify that this system is exactly closed, by calculation of all possible commutation pairs between the elements. Out of the $6!$ commutators, the non-zero ones are $[\hat{a},\hat{n}]=-[\hat{a}{\hat{b}}^{\dagger },\hat{b}]=\hat{a}$, $[\hat{a}\hat{b},\hat{n}]=\hat{a}\hat{b}$, $[\hat{a}{\hat{b}}^{\dagger },\hat{n}]=\hat{a}{\hat{b}}^{\dagger }$, and $[\hat{a}\hat{b},\hat{a}{\hat{b}}^{\dagger }]=[\hat{c},\hat{n}]=2\hat{c}$, which is obviously a closed basis. Now, one may proceed with composition of the Langevin equations. 

The applicability of the basis (\ref{T8-eq1a}) becomes readily clear by calculating the braket $[\hat{a},\mathbb{H}]$ as appears in the corresponding Langevin equation. The terms involving the second-degree operators $\hat{a}\hat{b}$ and $\hat{a}\hat{b}^\dagger$ immediately show up. The key in the method of higher-order operators is to keep these operator pairs, triplets and so on together, as each combination has a clear corresponding physical process. While $\hat{a}$ and $\hat{b}$ refer to individual ladder operators, $\hat{a}\hat{b}$ and $\hat{a}\hat{b}^\dagger$ respectively construct the blue and red 1-photon/1-phonon processes. For this reason, it is probably more appropriate to call these higher-degree operator combinations as processes. 

The Langevin equations for the blue $\hat{a}\hat{b}$ and red $\hat{a}\hat{b}^\dagger$ processes do not close on themselves, because of the appearance of third-order blue- and red-like processes $\hat{a}\hat{b}^2$ and $\hat{a}\hat{b}^{\dagger 2}$, describing 1-photon/2-phonon processes. Similarly, every $j$-th order blue- or red-like process such as $\hat{a}\hat{b}^j$ and $\hat{a}\hat{b}^{\dagger j}$ will lead to the $j+1$-order process. Hence, the infinite-dimensional basis $\{\hat{a}\}\cup\{\forall\hat{a}\hat{b}^j,\hat{a}\hat{b}^{\dagger j};j\in\mathscr{N}\}$ can provide an exact solution to the optomechanics. Furthermore, the convergence of solutions basis on such expansions would be questionable when $g_0<<\Omega$ is violated. In general, the $j$-th order processes correspond to the 1-photon/$j$-phonon interactions and contribute to the $j+1$-order sidebands. Here, it has been shown that under practical conditions, it is unnecessary to take account of the processes $j\geq 2$ and the $3\times 3$ basis (\ref{T8-eq1a}) is rather sufficient for most practical purposes. Nonetheless, $j=2$ processes contribute significantly to nonlinear stability and second-order mechanical sidebands. While the use of an infinite-dimensional basis is surprisingly unnecessary in still a higher-order formulation, using the compact minimal basis to be discussed in the following can lead to the mathematically exact solution.

The choice of basis is not unique, and every non-degenerate linear combination of bases leads to another equivalent form. One may for instance arbitrate the three-dimensional linear basis $\{A\}^{\rm T}=\{\hat{a},\hat{b},\hat{b}^\dagger\}$ or the four-dimensional linear basis $\{A\}^{\rm T}=\{\hat{a},\hat{b},\hat{a}^\dagger,\hat{b}^\dagger\}$ as is taken in the context of linearized standard optomechanics \cite{S2-1,S2-3,S2-6}, the five-dimensional all-Hermitian basis $\{A\}^{\rm T}=\{\hat{n},\hat{m},\hat{n}^2,\hat{n}(\hat{b}+\hat{b}^\dagger),i\hat{n}(\hat{b}-\hat{b}^\dagger)\}$ \cite{S2-Bruschi}, and ultimately the minimal three-dimensional basis 
\begin{equation}
\label{T8-eqMinimal}
\{A\}^{\rm T}=\{\hat{n}^2,\hat{n}\hat{b},\hat{n}\hat{b}^\dagger\}=\{\hat{N},\hat{B},\hat{B}^\dagger\},
\end{equation}
\noindent
assumed here, which is of the fourth-degree. We shall later observe that while (\ref{T8-eq3}) is necessary to construct the closed Langevin equations, a second-order linearization will be needed to decouple three operators, leaving only the basis (\ref{T8-eq1a}) in effect. Quite remarkably, however, and in a similar manner, the use of minimal basis (\ref{T8-eqMinimal}) turns out to be fairly convenient to construct the optomechanical Langevin equations. This is not only since the Langevin equations take on exactly linear forms, but also eventually the equation for $\hat{N}$ and $\hat{B}^\dagger$ will decouple. This leaves the whole standard optomechanical interaction exactly solvable through integration of only one linear differential equation in terms of $\hat{B}$. The main difference between using various choices of higher-order operator bases \cite{S2-Paper2} is the noise terms. It turns out that the definition and higher-order operators lead to multiplicative noise inputs, which once known, the problem will be conveniently solvable. Full mathematical treatment of multiplicative noise terms is necessary for description of some various phenomena and this will be discussed in \S\ref{S2-Noise4}. 

\subsection{Outline of Results}

Since the comprehensive treatment of quantum optomechanical effects cannot be described all at once, we prefer to first present a quick overview of main deductions and later we expand the derivations and show the calculation details.

\subsubsection{Coherent Phonon Population}

The overall phonon population inside an optomechanical cavity can be parted into two groups: the non-coherent phonons with population $m_{\rm th}$ which are mostly due to the random thermal fluctuations, and coherent phonons with population $\bar{m}$ which are driven by radiation pressure due to optomechanical interaction. It may be correctly anticipated for a single-mode 1:1 optomechanical cavity that while $\bar{m}$ should depend on the intracavity photon population $\bar{n}$, $m_{\rm th}$ is determined by temperature and possible presence a cooling mechanism. Mostly, $m_{\rm th}=1/[\exp(\hbar\Omega/k_{\rm B}T)-1]$ at thermal equilibrium can be kept well under unity by bringing temperature low enough and also a cooling tone can effectively deplete the cavity from noisy thermal phonons.

Now, it is possible to show that $\bar{m}(\Delta)=\braket{\hat{b}^\dagger\hat{b}}$ is the coherent phonon population given by
\begin{equation}
\label{T8-m8Roadster} 
\bar{m}(\Delta)\approx\frac{32 g_0^2 \Omega ^2 \left(\gamma ^2+\gamma  \Gamma +4 \Delta ^2\right)}{\left(\gamma ^2+4 \Delta ^2\right) \left(\Gamma ^2+4 \Omega ^2\right)^2}\bar{n}^2(\Delta )=g_0^2\zeta(\Delta)\bar{n}^2(\Delta),
\end{equation}
where $\Gamma$ is the mechanical decay rate, and $\gamma=\kappa+\Gamma$ is the total optomechanical decay rate with $\kappa$ being the optical decay rate, as proved in details in \S\ref{S8-BMWm8} using the method of higher-order operators. Also, $\bar{n}(\Delta)$ can be found from numerical solution of a third-order algebraic equation  (\ref{T8-eq12}). The relationship $\bar{m}\propto\bar{n}^2$ signifies the fact that mechanical oscillations are nonlinearly driven by optical radiation pressure. While the expression (\ref{T8-m8Roadster}) is significantly different from $|\bar{b}|^2=\braket{\hat{b}}\braket{\hat{b}^\dagger}=|\braket{\hat{b}}|^2$, numerical tests show that $|\bar{b}|^2$ is within $50\%$ error range of the actual value (\ref{T8-m8Roadster}).

\subsubsection{Weakly Nonlinear Approximation of Side-band Inequivalence}
Defining $\Delta_b$ and $\Delta_r$ respectively as the blue and red frequency shifts of sidebands, it is possible to show that these two quantities do not necessarily agree in magnitude, such that $\Delta_b+\Delta_r\neq 0$. As shown in \S\ref{S4-Suppl}, an explicit relation for the side-band inequivalence $\delta=\Delta_\text{r}+\Delta_\text{b}-2\Delta$, also alternatively defined as $2\delta\Delta=\frac{1}{2}(\Delta_\text{r}+\Delta_\text{b})-\Delta$, can be found through series expansion of the eigenvalues of the coefficient matrix from (\ref{T8-eq18}). With some algebra, it is possible to show that for $g_0\sqrt{\bar{n}}<<\Omega$ correct to the fourth-order in the weakly nonlinear limit, we get
\begin{equation}
\label{T8-Sideband}
\frac{\delta\Omega}{\Omega}=\frac{1}{2}\frac{\delta}{\Omega}\approx \left(\frac{g_0}{\Omega}\right)^2\left(\bar{n}+\frac{1}{2}\right)-2\left(\frac{g_0}{\Omega}\right)^4\left(\bar{n}+\frac{1}{2}\right)\left(\bar{m}+\frac{1}{2}\right).
\end{equation}
\noindent
The typical behavior of this equation in the weakly nonlinear regime is that the side-band inequivalence $\delta$ should increase linearly with $\bar{n}$ before $\bar{m}$ gets large enough to bring it down. However, the actual accuracy of (\ref{T8-m8Roadster}) is limited to when the normalized side-band inequivalence $\bar{\delta}=\delta/\Omega$ does not exceed a few percent. A more accurate analysis of side-band inequivalence shall be thus needed, to be presented in \S\ref{Section-9}, showing that side-band inequivalence exhibits a near-resonant behavior with respect to pump.

There is a related polaritonic splitting effect \cite{S8-Lemonde2}, as a result of anti-crossing between the optomechanically interacting optical and mechanical resonances generated across either of the mechanical side-bands, amount of which happens to be exactly $2\delta\Delta$. This has nothing to do with the side-band asymmetry, which happens to occur on the two opposite sides of the main cavity resonance. It should be mentioned that observation of this phenomenon in superconducting electromechanics \cite{S8-Mika} as well as parametrically actuated nano-string resonators \cite{S8-Weig1,S8-Weig2} can potentially yield the most clear results due to various experimental conditions. In fact, intracavity photon numbers as large as $10^6$ and $10^{8}$ and more are attainable respectively in superconducting electromechanics and optically-trapped nano-particle optomechanics. 

A close inspection of a very high-resolution measurement on a side-band resolved microtoroidal disk\cite{S8-Kip3} yields a side-band inequivalence of $\delta\Delta\approx 2\pi\times(142\pm 36) \text{Hz}$, which perfectly complies to (\ref{T8-Sideband}) if $\bar{n}=(5.1\pm 1.3)\times 10^3$. Unfortunately, further such a high resolution measurements on deeply side-band resolved optomechanical cavities are not reported elsewhere to the best knowledge of authors. Nevertheless, clear signatures of side-band inequivalence can be easily verified in few other experiments \cite{S8-Mika,S8-Weig1,S8-Teufel}. Remarkably, recent measurements on Stokes-Anti-Stokes scattering from multi-layered $\text{MoTe}_2$ exhibits a difference in frequency shift as large as 7\% for the five-layered sample \cite{S8-Stokes2}, which corresponds to $0.88 \pm 0.11\text{cm}^{-1}$. Also, a recent landmark experiment on room-temperature quantum optomechanical correlations \cite{S8-Vivishek} has reported measurements which coincidentally exhibit a sideband inequivalence up to $4\text{kHz}$ and roughly agree to the approximation $\delta\approx 2g_0^2\bar{n}/\Omega$. Much further experimental evidence is to be presented in \S\ref{Section-9}.

In any experimental optomechanical attempt to measure this phenomenon, a side-band resolved cavity could be driven on resonance and noise spectra of the two mechanical side bands be measured with extreme precision in a heterodyne setup. Even in case of well-known thermo-optical effects and two-photon dispersion or absorption which cause drifts in the optical resonance and other optomechanical parameters \cite{S8-Painter}, this effect should be still observable in principle. The reason is that the amount of inequivalence is actually independent of the exact pump frequency as long as intracavity photon population does not change significantly. So, it should be sufficient only if the cavity is driven on or close to the optical resonance for the side-bands to be sufficiently different in their frequency shifts.

Quantum optomechanical experiments are typically difficult and require stringent fabrication constraints. To this end, equivalent experimental setups which actually share identical third-order nonlinear interaction Hamiltonian could be used. These include Brillouin/Raman scattering, ion/Paul traps, and electrooptic/acoustooptic modulation.

\subsubsection{Higher-order Resonance Shift}

The contribution of the off-diagonal terms to the mechanical frequency $\Omega$ in the coefficients matrix of optomechanical Langevin equations (\ref{T8-eq19}), can be ultimately held responsible for the so-called optomechanical spring effect \cite{S2-1,S2-3,S2-6,S2-8,S8-Spring1,S8-Spring2,S8-Spring3,S8-Spring4,S8-Spring5}. As the result of optomechanical interaction, both of the optical and mechanical resonance frequencies and damping rates undergo shifts. Even at the limit of zero input optical power $\alpha=0$ and therefore zero cavity photon number $\bar{n}=0$, it is possible to show that there is a temperature-dependent shift in the mechanical resonance frequency, markedly different from the lattice-expansion dependent effect. This effect is solely due to the optomechanical interaction with virtual cavity photons, which completely vanishes when $g_0=0$. In close relationship to the shift of resonances, we can also study the optomechanical spring effect with the corrections from higher-order interactions included.

The analysis of spring effect is normally done by consideration of the effective optomechanical force acting upon the damped mechanical oscillator, thus obtaining a shift in squared mechanical frequency $\delta(\Omega^2)$, whose real and imaginary parts give expressions for $\delta\Omega$ and $\delta\Gamma$. Corrections to these two terms due to higher-order interactions are discussed in \S\ref{S8-AppA}. Here, we demonstrate that the analysis using higher-order operator algebra can recover some important lost information regarding the optical and mechanical resonances when the analysis is done on the linearized basis $\{A\}^\text{T}=\{\hat{a},\hat{a}^\dagger,\hat{b},\hat{b}^\dagger\}$. 

To proceed, we consider finding eigenvalues of the matrix $\textbf{M}$ as defined in (\ref{T8-eq18}). Ignoring all higher-order nonlinear effects beyond the basis $\{A\}^\text{T}=\{\hat{a},\hat{a}\hat{b},\hat{a}\hat{b}^\dagger\}$, we set $s=0$. This enables us to search for the eigenvalues of the coefficients matrix $\textbf{M}$ as
\begin{eqnarray}
\label{T8-eqSpring1Copy}
\text{eig}[\textbf{M}]&=&\text{eig}\left[ 
\begin{array}{ccc}
i\Delta-\frac{\kappa }{2} & ig_0 & ig_0 \\ 
i(G+f^+) &  -i(\Omega-\Delta)-\frac{\gamma }{2} & 0 \\ 
-i(G-f^-) & 0 & i(\Omega+\Delta)-\frac{\gamma }{2}  
\end{array}
\right]\\ \nonumber &=&i\left\{
\begin{array}{c}
\Delta+\lambda_1+i\gamma_1 \\
\Delta+\lambda_2+i\gamma_2 \\
\Delta+\lambda_3+i\gamma_3 
\end{array}
\right\}\\ \nonumber &=&i\left\{
\begin{array}{c}
\Delta+\eta_1(\Delta,T) \\
\Delta+\eta_2(\Delta,T) \\
\Delta+\eta_3(\Delta,T) 
\end{array}
\right\},
\end{eqnarray}
\noindent
in which $G=g_0\bar{n}$, $f^\pm=g_0(\bar{m}+\frac{1}{2})\pm \frac{1}{2}g_0$, $\lambda_j=\Re[\eta_j]$ and $\gamma_j=\Im[\eta_j]$ with $j=1,2,3$ are real valued functions of $\Delta$ and bath temperature $T$. The temperature $T$ determines $\bar{m}$ while $\bar{n}$ is a function of $\Delta$ as well as input photon rate $\alpha$. 	In general, the three eigenvalues $\eta_j=\lambda_j(\Delta,T)+i\gamma_j(\Delta,T), j=1,2,3$ are expected to be deviate from the three free-running values $\psi_1=i\frac{1}{2}\kappa$, $\psi_2=-\Omega+i\frac{1}{2}\gamma$, and $\psi_3=\Omega+i\frac{1}{2}\gamma$, as $\eta_j\approx \psi_j-\Delta$ because of non-zero $g_0$. Solving the three equations therefore gives the values of shifted optical and mechanical frequencies and their damping rates compared to the bare values in absence of optomechanical interactions with $g_0=0$, given by $\delta\Omega=-\frac{1}{2}\Re[\eta_2-\eta_3]-\Omega$, $\delta\omega=-\frac{1}{2}\Re[\eta_2+\eta_3]$, $\delta\Gamma=\Im[-2\eta_1+\eta_2+\eta_3]-\Gamma$, and $\delta\kappa=2\Im[\eta_1]-\kappa$. This  method to calculate the alteration of resonances, does not regard the strength of the optomechanical interaction or any of the damping rates. In contrast, the known methods to analyze this phenomenon normally require $g<<\kappa$ and $\Gamma+\delta\Gamma<<\kappa$ \cite{S2-3}.

\subsubsection{Corrections to Optical Spring Effect}\label{S8-AppA}

As shown in \S\ref{S8-Spring}, the full expression for corrected spring effect is given as 
\begin{eqnarray}
\label{T8-eqA10Copy}
\delta\Omega(w,\Delta)&=& \frac{g_0^2\bar{n}\Omega}{w}\left[\frac{\Delta+w}{(\Delta+w)^2+\frac{1}{4}\kappa^2}+\frac{\Delta-w}{(\Delta-w)^2+\frac{1}{4}\kappa^2}\right]\\ \nonumber
&+&\frac{g_0^2\Re[\mu(w)]\Omega}{w}\left[\frac{\Delta+w}{(\Delta+w)^2+\frac{1}{4}\kappa^2}+\frac{\Delta-w}{(\Delta-w)^2+\frac{1}{4}\kappa^2}\right]\\ \nonumber
&+&\frac{g_0^2\Im[\mu(w)]\Omega}{w}\left[\frac{\kappa}{(\Delta+w)^2+\frac{1}{4}\kappa^2}-\frac{\kappa}{(\Delta-w)^2+\frac{1}{4}\kappa^2}\right],\\ 
\delta\Gamma(w,\Delta)&=&\frac{g_0^2\bar{n}\Omega}{w}\left[\frac{\kappa}{(\Delta+w)^2+\frac{1}{4}\kappa^2}-\frac{\kappa}{(\Delta-w)^2+\frac{1}{4}\kappa^2}\right]\\ \nonumber
&+&\frac{g_0^2\Re[\mu(w)]\Omega}{w}\left[\frac{\kappa}{(\Delta+w)^2+\frac{1}{4}\kappa^2}-\frac{\kappa}{(\Delta-w)^2+\frac{1}{4}\kappa^2}\right]\\ \nonumber
&-&\frac{g_0^2\Im[\mu(w)]\Omega}{w}\left[\frac{\Delta+w}{(\Delta+w)^2+\frac{1}{4}\kappa^2}+\frac{\Delta-w}{(\Delta-w)^2+\frac{1}{4}\kappa^2}\right].
\end{eqnarray}
Here, the second and third terms on the rights hand sides of both equations are corrections to the spring effect due to the higher-order interactions, resulting from the temperature-dependent expressions 
\begin{eqnarray}
\label{T8-eqA11Copy}
\Re[\mu(w)]&=&\frac{w}{\Omega}\left(\bar{m}+\frac{1}{2}\right)+\frac{1}{2}, \\ \nonumber
\Im[\mu(w)]&=&\frac{\Gamma}{2\Omega}\left(\bar{m}+\frac{1}{2}\right).
\end{eqnarray}
The temperature-dependence of (\ref{T8-eqA11Copy}) causes dependence of the spring effect on temperature as well. The influence of additional terms in (\ref{T8-eqA10Copy}) due to higher-order interactions can strongly influence any measurement of $g_0$ through spring effect, as most easily can be observable in the weak coupling limit for Doppler cavities.

In the weakly coupled operation mode and far Doppler regime where $g_0<<\Omega$ and $\kappa>>\Omega>>\Gamma$ hold \cite{S8-Transduction,S3-64}, using  (\ref{T8-eq12}) with $\bar{n}\approx(\Delta^2+\frac{1}{4}\kappa^2)^{-1}|\alpha|^2$, the spring equations are obtained from (\ref{T8-eqA10Copy}) by setting $w=\Omega$ as
\begin{equation}
\label{T8-eqA14}
\delta\Omega(\Omega,\Delta)\approx 2\Delta g_0^2\frac{\bar{n}(\Delta)+\bar{m}(\Delta)+1}{\Delta^2+\frac{1}{4}\kappa^2}
\approx g_0^2\left[\frac{2\Delta|\alpha|^2}{\left(\Delta^2+\frac{1}{4}\kappa^2\right)^2}\right]+ g_0^4\left[\frac{2\Delta\zeta(\Delta)|\alpha|^4}{\left(\Delta^2+\frac{1}{4}\kappa^2\right)^3}\right].
\end{equation}
\noindent
Here, $|\alpha|$ is photon input rate to the cavity with $\alpha$ being complex drive amplitude, and $\Re[\mu(\Omega)]=\bar{m}+1$ and $\Im[\mu(\Omega)]\approx 0$ from (\ref{T8-eqA11Copy}). The importance of this equation is that the optical spring effect is actually proportional to $\delta\Omega\propto g_0^2(\bar{n}+\bar{m})\propto g_0^2\bar{n}(1+g_0^2\zeta\bar{n})$ where $\zeta(\Delta)$ is already defined in (\ref{T8-m8}). This shows that if $g_0$ is to be determined from experimental measurement of the optical spring effect, then the experiment should be done at the lowest optical power possible, otherwise the term $\bar{m}\propto g_0^2\bar{n}^2$ becomes large and would result in an apparent change in $g_0$. This fact also can explain why the measured $g_0$ through optical spring effect using uncorrected standard expressions (\ref{T8-eqA12}) is always different from the design value, which could be attributed to the absence of the second term proportional to $g_0^4$ in the corrected optical spring effect using the higher-order algebra.

The above equation together with the fact that on the far red detuning $\Delta\rightarrow+\infty$ we have $\bar{n}(\Delta)\rightarrow 0$, $\bar{m}(\Delta)\rightarrow 0$, and $\delta\Omega(\Omega,\Delta)\rightarrow 0$, provides an alternate approximation for the resonant coherent phonon number $\bar{m}(0)$ at zero-detuning as
\begin{eqnarray}
\label{T8-eqA15}
\bar{m}(0)&\approx&\frac{\kappa^2}{8g_0^2}\left[\frac{\partial\delta\Omega(\Omega,\Delta)}{\partial\Delta}\right]_{\Delta=0}-4\frac{|\alpha|^2}{\kappa^2}-1 \approx \frac{32 g_0^2 Q^2_\text{m}}{\Gamma^2}\bar{n}^2(0) \approx \frac{512 g_0^2 Q^2_\text{m}}{\Gamma^2\kappa^4}|\alpha|^4,
\end{eqnarray}
\noindent
where $Q_\text{m}=\Omega/\Gamma$ is the mechanical quality factor, and the expression within the brackets can be measured experimentally, and represents the slope of frequency displacement due to the spring effect  versus detuning. The second expression proportional to $\bar{n}^2(0)$ follows (\ref{T8-m8}) from \S\ref{S8-BMWm8} where an explicit and accurate formula for $\bar{m}(\Delta)$ is found. 

Noting $|\alpha|^2\propto P_\text{op}$ reveals that while the intracavity photon number is propotional to the optical power as $\bar{n}(0)\propto P_\text{op}$, the coherent phonon population is proportional to the square of the optical power as $\bar{m}(0)\propto P_\text{op}^2$. This implies that the effects of coherent mechanical field gets important only at sufficiently high optical powers, and also marks the fact that in the low optical power limit where linear optomechanics is expected to work well, effects of coherent phonons do not appear. This also explains why this quantity has not been so far noticed in the context of quantum optomechanics. Because it does not show up anywhere in the corresponding fully linearized Langevin equations. 

\subsubsection{Corrections to Spectral Noise Density}

As shown in \S\ref{S2-Noise4}, a fairly convenient but approximate solution to the symmetrized spectral density of output optical field due to multiplicative noise is given as
\begin{eqnarray}
\label{T8-Noise18Copy}
S(\omega)&=&|Y_{11}(\omega)|^2 S_{AA}(\omega)\\ \nonumber
&+&\frac{1}{\gamma^2}\left|\left[Y_{12}(\omega)+Y_{13}(\omega)\right]\ast\bar{a}(\omega)\right|^2 S_{BB}(\omega) +\frac{1}{\theta^2}\left|\left[Y_{14}(\omega)\ast\overline{ab}(\omega)+Y_{15}(\omega)\ast\overline{ab^\ast}(\omega)\right]\right|^2 S_{BB}(\omega),
\end{eqnarray}
where spectral power densities $S_{AA}$ and $S_{BB}$ are already introduced in (\ref{T8-eq13}) and convolutions $\ast$ take place over the entire frequency axis. In practice it is far easier to use numerical integration, however, this can cause numerical instabilities when $|\omega-\Delta|>\frac{1}{2}\Omega$. Owing to the fractional polynomial expressions for the elements of scattering matrix elements as well as the multiplicative terms, it is possible to evaluate the integrals exactly using complex residue techniques. Here, we proceed using numerical integration of the convolution integrals. The third term involving the functions $\overline{ab}(\omega)$ and $\overline{ab^\ast}(\omega)$ are unnecessary for the $3\times 3$ second-order formalism, and arise only in the $5\times 5$ third-order formalism.

In the above equation, every term adds up the contribution from linear, second-order, and third-order optomechanics. These respectively are due to the processes of photon creation-annihilation $\{\hat{a},\hat{a}^\dagger\}$, the 1-photon/1-phonon blue $\{\hat{a}\hat{b},\hat{a}^\dagger\hat{b}^\dagger\}$ and red $\{\hat{a}\hat{b}^\dagger,\hat{a}^\dagger\hat{b}\}$ processes, and the 1-photon/2-phonon second-order blue-like $\{\hat{a}\hat{b}^2,\hat{a}^\dagger\hat{b}^{\dagger 2}\}$ and red-like $\{\hat{a}\hat{b}^{\dagger 2},\hat{a}^\dagger\hat{b}^2\}$ sideband processes. Apparently the Hermitian conjugate operators do not exist in the original $5\times 5$ higher-order formalism (\ref{T8-Noise14}), since they are completely uncoupled from their Hermitian counterparts. However, calculation of the noise spectral densities necessitates their presence, so that a real-valued and positive definite spectral density has actually already taken care of these conjugate processes. Obviously, the first term contributes to the $w=\Delta$ resonance, while the second term contributes to the first-order mechanical side-bands at $w=\Delta=\pm\Omega$. Similarly, the third term constitutes the second-order mechanical sidebands at $w=\Delta=\pm2\Omega$.

It has to be mentioned that the spectral density (\ref{T8-Noise18Copy}) is not mathematically exact, since the multiplicative operators appearing behind Weiner noise terms, are approximated by their time-averaged frequency-dependent terms (\ref{T8-Noise12}).

Simulations of the noise spectrum across the red mechanical sideband and optical resonance generated in an optomechanical experiment on the whispering galley mode of an optical micro-toroid, reported in a very remarkable experiment \cite{S8-Polaron} has been done, and results using linearized and higher-order optomechanics are calculated. The simulations using $3\times 3$ linear optomechanics with the basis $\{\hat{a},\hat{b},\hat{b}^\dagger\} $, and $4\times 4$ linearized optomechanics using the basis $\{\hat{a},\hat{a}^\dagger,\hat{b},\hat{b}^\dagger\}$, and $3\times 3$ higher-order optomechanics with the basis $\{\hat{a},\hat{a}\hat{b},\hat{a}\hat{b}^\dagger\}$ and $5\times 5$ higher-order optomechanics with the basis $\{\hat{a},\hat{a}\hat{b},\hat{a}\hat{b}^\dagger,\hat{a}\hat{b}^2,\hat{a}\hat{b}^{\dagger 2}\}$ are all observed. Agreement between the two higher-order formalisms using second-order $3\times 3$ and third-order $5\times 5$ formalisms is noticed to be remarkably good. 

\subsubsection{Dynamic Stability Maps}

It is possible to employ the method of higher-order operators to investigate the dynamic stability of optomechanical systems in the side-band resolved operation limit. A stable optomechanical system can be still perturbed by thermal effects and they appear to be dominant in driving the cavity into instability for Doppler samples. However, for side-band resolved samples, thermal effects are much less pronounced and the major contribution to the instability comes from inherent nonlinear dynamics of the optomechanical interactions. That implies that optomechanical interactions are linearly stable, but they can become nonlinearly unstable at a certain interaction order to be discussed below.

The dynamic stability can be done by inspecting eigenvalues of the coefficients matrix $[\textbf{M}]$. If the real part of at least one of the eigenvalues is positive, then the system is unstable and its response to any perturbation grows indefinitely in time. The linear formalisms of optomechanics fails to describe this phenomenon, since they always yield constant eigenvalues. Even the second-order higher-operator method with $3\times 3$ formalism, which describes the nonlinear 1-photon/1-phonon processes, fails to reproduce the correct expected stability phases. This only can be understood by employing at least the third-order $5\times 5$ operator method, which includes the nonlinear 1-photon/2-phonon processes. Hence, surprisingly enough, it is the 1-photon/2-phonon process and beyond, which contributes to the instability of an optomechanical system.

Detailed verification of the linear stability from $4\times 4$ full linear formalism, reveals that not surprisingly, the linear and nonlinear stability diagrams remarkably are different. The linear instability starts at moderate resonant pump powers, while it starts rapidly growing exactly over the blue detuning at a slightly higher power. 

Firstly, across almost the entire domains of linear stability, the system is also nonlinearly stable. Secondly, by careful observation of the nonlinear stability and linear stability, it can be seen that in most of the domain of linear stability, the system is already nonlinearly stable. Hence, any attempt to drive the system within the region of linearly unstable but nonlinearly stable, ultimately results in significant growth of mechanical amplitude and therefore side-bands. Any further increase in the amplitude of side-bands become limited due to nonlinear stability. Hence, four possible stability scenarios could be expected:
\begin{itemize}
	\item  Linearly and Nonlinearly Stable: The intersect of the linear and nonlinear stability domains, marks a shared domain of unconditionally stable optomechanical interaction. Unless the system is influenced by thermal or other nonideal effects, the stability is always guaranteed.
	\item  Linearly Unstable, but Nonlinearly Stable: By inspection, an optomechanical system can be linearly unstable while nonlinearly stable. This corresponds to the domains where any attempt to drive the system in these regions causes immediate but limited growth in the amplitude of mechanical oscillations.
	\item Linearly and Nonlinearly Unstable: Under this scenario, the optomechanical cavity is always unstable regardless of the other nonideal effects. This happens only at remarkably high drive powers.
	\item Linear Stable, but Nonlinearly Unstable: The unlikely and surprising case of linear stability and nonlinear instability is also possible according to the stability maps at some portions of non-resonant high drive powers. This strange behavior corresponds to the case when the system remains stable only at infinitesimal optical powers. Any fluctuation beyond tiny amplitudes shall drive the system into unstable growing and large amplitudes.
\end{itemize}

There is a threshold power $P_\text{th}$ at which instabilities start to appear. As it expected and in agreement to experimental observations, the unstable domain mostly covers the blue domain with $\Delta<0$. However, at higher optical powers than $P_\text{th}$, instability phase can diffuse well into the red detunings as well $\Delta>0$. Therefore, the general impression that instability always occurs on the blue side at every detuning above a certain threshold power is not correct. Interestingly, the boundary separating the dynamically stable and dynamically unstable phases for this side-band resolved sample with $\Omega>>\kappa>>\Gamma$ can be well estimated using
\begin{equation}
\label{T8-Boundary}
\bar{n}(\Delta)>n_\text{cr},
\end{equation}
\noindent
in which $n_\text{cr}$ is a critical cavity photon number. Extensive numerical tests for cavities within deep side-band resolved $\Omega>>\kappa$, deep Doppler $\Omega>>\kappa$, and intermediate $\Omega\sim\kappa$ regimes reveal the existence of such a critical intracavity photon number limit $n_\text{cr}$, beyond which dynamical instability takes over. However, for a cavity in deep side-band resolved regime, it can be estimated in a phenomenological way, and is roughly given by
\begin{equation}
\label{T8-CriticalDensity}
n_\text{cr}\sim \frac{4}{\mathcal{C}_0}\left(\frac{\Omega}{\kappa}\right)^2. 
\end{equation}
\noindent
Here, $\mathcal{C}_0$ is the single-photon cooperativity given by
\begin{equation}
\label{T8-C0}
\mathcal{C}_0=g_0^2/\kappa\Gamma.
\end{equation}
If the cavity is not side-band resolved, (\ref{T8-CriticalDensity}) cannot be used, but numerical computations can still yield the limiting number $n_\text{cr}$. 

For Doppler cavities, no such dynamic instability can be observed, and therefore thermal effects should dominate over dynamical effects in driving a Doppler cavity toward instability. Meanwhile, it is the nonlinear optomechanical dynamics which seems to be dominant in driving a side-band resolved cavity into instability. As a result, the existence of such a critical maximum intracavity photon number is not related to thermal effects, but rather to the nonlinear stability. 

Now, we are all set to dive deep into the details of mathematics of higher-order analysis of nonlinear optomechanics.

\subsection{Optomechanical Hamiltonian}\label{S1-Suppl}
The Langevin equations for the Hamiltonian $\mathbb{H}=\hbar \Omega\hat{m}-\hbar \Delta\hat{n}-\hbar g_0\hat{n}(\hat{b}+{\hat{b}}^{\dagger }) $ with the basis $\{A\}^\text{T}=\left\{\hat{a},\hat{b},\hat{a}\hat{b},\hat{a}{\hat{b}}^{\dagger },\hat{n},\hat{c}\right\}$ are given exactly by 
\begin{eqnarray}
\nonumber
&&\left[ \begin{array}{c}
\begin{array}{cccccc}
i\Delta-\frac{\kappa }{2} & 0 & ig_0 & ig_0 & 0 & 0 \\ 
0 & -(i\Omega+\frac{\Gamma}{2}) & 0 & 0 & ig_0 & 0 \\ 
ig_0\left(\hat{m}+\hat{n}+1\right) & 0 & i\left(\Delta_B+g_0 \hat{b}\right)-\frac{\gamma }{2} & 0 & 0 & 0 \\ 
ig_0\left(\hat{m}-\hat{n}\right) & 0 & 0 & i\left(\Delta_R+g_0\hat{b}^\dagger\right)-\frac{\gamma }{2} & 0 & 0 \\ 
0 & 0 & 0 & 0 & -\kappa  & 0 \\ 
0 & 0 & ig_0\hat{a} & ig_0\hat{a} & 0 & 2i\left(\Delta+g_0\hat{B}\right)-\kappa  \end{array}
\end{array}
\right]\\ 
\label{T8-eq4}
&&\times
\left\{ \begin{array}{c}
\hat{a} \\ 
\hat{b} \\ 
\hat{a}\hat{b} \\ 
\hat{a}{\hat{b}}^{\dagger } \\ 
\hat{n} \\ 
\hat{c}
\end{array}
\right\}
-\left\{ \begin{array}{c}
\sqrt{\kappa }{\hat{a}}_{\text{in}} \\ 
\sqrt{\Gamma}{\hat{b}}_{\text{in}} \\ 
\sqrt{\gamma }{\left(\hat{a}\hat{b}\right)}_{\text{in}} \\ 
\sqrt{\gamma }{\left(\hat{a}{\hat{b}}^{\dagger }\right)}_{\text{in}} \\ 
\sqrt{2\kappa }{\hat{n}}_{\text{in}} \\ 
\sqrt{2\kappa }{\hat{c}}_{\text{in}} 
\end{array}
\right\}=\frac{d}{dt}\left\{ \begin{array}{c}
\hat{a} \\ 
\hat{b} \\ 
\hat{a}\hat{b} \\ 
\hat{a}{\hat{b}}^{\dagger } \\ 
\hat{n} \\ 
\hat{c} 
\end{array}
\right\},
\end{eqnarray}
\noindent 
\noindent 
in which $\gamma =\kappa +\Gamma$ is the total optomechanical decay rate, $\hat{B}=(\hat{b}+\hat{b}^\dagger)$, $\Delta_R=\Delta+\Omega$, and $\Delta_B=\Delta-\Omega$. We have set $\hat{x}=\hat{a}$ in all equations except the second and third where both of the bath operators $\hat{x}=\hat{a}$ and $\hat{x}=\hat{b}$ are taken separately to construct the noise terms, $\sqrt{2}{\hat{n}}_\text{in}={\hat{a}}^{\dagger }{\hat{a}}_\text{in}+{{\hat{a}}^{\dagger }}_\text{in}\hat{a}$, $(\hat{a}\hat{b}^\dagger)_\text{in}={\hat{a}}_\text{in}\hat{b}^\dagger+\hat{a}\hat{b}_\text{in}^\dagger$, $(\hat{a}\hat{b})_\text{in}={\hat{a}}_\text{in}\hat{b}+\hat{a}{\hat{b}}_\text{in}$ and $\sqrt{2}{\hat{c}}_\text{in}=\hat{a}{\hat{a}}_\text{in}$. 

The system (\ref{T8-eq4}) is still nonlinear and non-integrable because of the dependence of the coefficients matrix on the operators. But it can be simplified by first noting that from the fifth equation we could expect any disturbance in $\hat{n}$ would decay as $\delta\hat{n}(t)\sim\exp(-\kappa t)$ on time scales smaller than $\kappa^{-1}$. This can be further approximated as $\hat{n}\sim\bar{n}$ at steady input. Similar argument goes with $\delta\hat{m}\sim\exp(-\Gamma t)$ in response to a disturbance on time scales smaller than $\Gamma^{-1}$, which enables us to make the approximate replacement  $\hat{m}\sim\bar{m}$ at equilibrium. 

For the phononic mechanical operators $\hat{b}$ and $\hat{b}^\dagger$ appearing within the brackets, approximate decays $\delta\hat{b}(t)\sim\exp[-(i\Omega+\frac{1}{2}\Gamma) t]$ and $\delta\hat{b}^\dagger(t)\sim\exp[(i\Omega-\frac{1}{2}\Gamma) t]$ in response to disturbances hold, making the coefficients matrix time-dependent. But these can be nevertheless dropped in whole if we notice that $g_0\bar{b}<<\Omega$ which is the normal experimental condition of weakly-coupling in optomechanics. Otherwise, they can replaced by constant amplitudes $\bar{b}$ and $\bar{b}^\ast$ given below in (\ref{T8-eq9}) on sufficiently longer time scales than $\Gamma^{-1}$ for strongly-coupled systems. 

Such types of approximations are in fact quite highly in use within the context of continuous wave standard optomechanics. Therefore, once the steady state solution to (\ref{T8-eq4}) around the equilibrium values due to optical drive $\langle\hat{a}_\text{in}\rangle=\alpha$ is sought, the coefficients matrix can be kept time-independent, keeping only the fluctuations of input terms as the only source. The case of time dependent drive $\alpha=\alpha(t)$ for pulsed experiments shall be discussed later below.

Having said that, all the operators $\hat{n}$, $\hat{m}$, $\hat{b}$, and $\hat{b}^\dagger$ in the coefficients matrix can be replaced by their respective average values to proceed with the second-order accurate optomechanical system of equations as
\begin{eqnarray}
\nonumber
\left[ \begin{array}{c}
\begin{array}{cccccc}
i\Delta-\frac{\kappa }{2} & 0 & ig_0 & ig_0 & 0 & 0 \\ 
0 & -(i\Omega+\frac{\Gamma}{2}) & 0 & 0 & ig_0 & 0 \\ 
iL^+ & 0 & i\left(\Delta_B+s\right)-\frac{\gamma }{2} & 0 & 0 & 0 \\ 
iL^- & 0 & 0 & i\left(\Delta_R+s^\ast\right)-\frac{\gamma }{2} & 0 & 0 \\ 
0 & 0 & 0 & 0 & -\kappa  & 0 \\ 
0 & 0 & ig & ig & 0 & 2i(\Delta+2\Re[s])-\kappa  \end{array}
\end{array}
\right]\\ 
\label{T8-eq5}
\times
\left\{ \begin{array}{c}
\hat{a} \\ 
\hat{b} \\ 
\hat{a}\hat{b} \\ 
\hat{a}{\hat{b}}^{\dagger } \\ 
\hat{n} \\ 
\hat{c}
\end{array}
\right\}
-\left\{ \begin{array}{c}
\sqrt{\kappa }{\hat{a}}_{\text{in}} \\ 
\sqrt{\Gamma}{\hat{b}}_{\text{in}} \\ 
\sqrt{\gamma }{\left(\hat{a}\hat{b}\right)}_{\text{in}} \\ 
\sqrt{\gamma }{\left(\hat{a}{\hat{b}}^{\dagger }\right)}_{\text{in}} \\ 
\sqrt{2\kappa }{\hat{n}}_{\text{in}} \\ 
\sqrt{2\kappa }{\hat{c}}_{\text{in}} 
\end{array}
\right\}=\frac{d}{dt}\left\{ \begin{array}{c}
\hat{a} \\ 
\hat{b} \\ 
\hat{a}\hat{b} \\ 
\hat{a}{\hat{b}}^{\dagger } \\ 
\hat{n} \\ 
\hat{c} 
\end{array}
\right\},
\end{eqnarray}
\noindent 
in which $g=g_0\sqrt{\bar{n}}$, $s=g_0\bar{b}$ with $\Re[\bar{b}]=\bar{x}/2x_{\rm zp}$ and $x_{\rm zp}$ being the zero-point displacement,  $L^+=g_0(\bar{m}+\bar{n}+1)$, and $L^-=g_0(\bar{m}-\bar{n})$. These can be further approximated by $L^\pm \approx \pm g_0\bar{n}=\pm F$ under normal experimental conditions of an ultracold cavity with sufficiently high pumping. The average mirror displacement $\bar{x}$ is due to the average radiation pressure. The fact that $L^+\neq-L^-$ provides the quantum mechanical asymmetry between blue and red sidebands. 

It is easy to verify that this way of linearization decouples the state operators and reduces the space into a 3-dimensional one spanned by $\{A\}^{\rm T}=\{\hat{a},\hat{a}\hat{b},\hat{a}\hat{b}^\dagger\}$. This will be discussed in further details later.

In the absence of red-side-band optical cooling tone as well as any other interaction, the average population value is $\bar{m}={1}/{\left[\exp\left({\hbar \Omega}/{k_\text{B}T}\right)-1\right]}$, while $\bar{n}$ can be obtained from the steady state solution of the first row by replacements of input noise term $\sqrt{\kappa}{\hat{a}}_\text{in}\to \alpha+\sqrt{\kappa}{\hat{a}}_\text{in}$. Here, $\alpha $ is the input photon flux originally due to an undisplayed resonant drive term $\mathbb{H}_\text{d}=\hbar(\alpha\hat{a}+\alpha^\ast\hat{a}^\dagger)$ added to the Hamiltonian $\mathbb{H}_\text{OM}$. Furthermore, $\alpha$ has some non-zero phase taken from the cavity population $\bar{n}$ away. Now that the drive term $\mathbb{H}_\text{d}$ has been dropped from $\mathbb{H}_\text{OM}$, and $\hat{a}_\text{in}$ now only contains the fluctuations with zero-average $\langle\hat{a}_\text{in}\rangle=0$. 

As it will be shown later, the quantity $\bar{m}$ here being referred to as the coherent phonon population, can enter the optomechanical interaction processes due to higher-order effects, where its value normally needs to be fitted for cavities in the Doppler regime. Hence, the coherent phonon population $\bar{m}$ is independent of the simple thermal equilibrium value $m$, and actually represents those number of phonons who take part in the optomechanical interaction. We will observe that the negative detunings with the blue process can actually lead to a relatively constant phonon population, whereas on the red detunings it starts to decrease with the detuning.

Defining  $K=\bar{n}\kappa $ we may use the substitutions for the noise and input terms as
\begin{eqnarray}
\label{T8-eq6}
\sqrt{\gamma }{\left(\hat{a}\hat{b}\right)}_\text{in}&\to& \sqrt{\Gamma\bar{n}}{\hat{b}}_\text{in}+\sqrt{\kappa }\bar{b}{\hat{a}}_\text{in}+\bar{b}\alpha,\\ \nonumber
{\sqrt{\gamma }\left(\hat{a}{\hat{b}}^{\dagger }\right)}_\text{in}&\to& \sqrt{\Gamma\bar{n}}{\hat{b}}_\text{in}^\dagger+\sqrt{\kappa }\bar{b}^\ast{\hat{a}}_\text{in}+\bar{b}^\ast\alpha,\\ \nonumber
\sqrt{\kappa}{\hat{n}}_\text{in}&\to& \sqrt{K}{\hat{a}}_\text{in}+\sqrt{K}{\hat{a}}_\text{in}^\dagger+2\sqrt{\bar{n}}\Re[\alpha],\\ \nonumber
\sqrt{\kappa}{\hat{c}}_{\rm in}&\to&\sqrt{K}{\hat{a}}_\text{in}+\sqrt{\bar{n}}\alpha.
\end{eqnarray}
\noindent
These substitutions follow the fact that terms such as $\hat{a}\hat{a}_{\rm in}$ which contain the interaction of a time-dependent operator $\hat{a}(t)$ and a purely white Weiner noise process with zero average $\langle\hat{a}_{\rm in}\rangle=0$, can be fairly well approximated by noting first that $\hat{a}(t)\sim\bar{a}\exp(i\Delta t)$ around the equilibrium, and then noting that shifting the noise process $\hat{a}_{\rm in}$ in frequency to the amount of $\Delta$ has essentially no effect by definition. Hence, the sinusoidal time dependence $\exp(i\Delta t)$ is irrelevant and can be dropped. Similar arguments hold for the phononic operator $\hat{b}(t)\sim\bar{b}\exp(-i\Omega t)$ and their Hermitian adjoints interacting with a white noise term with uniform spectrum.

This allows us to ultimately rewrite the Langevin equations (\ref{T8-eq5}) as 
\begin{eqnarray}
\nonumber
&&\left[ \begin{array}{c}
\begin{array}{cccccc}
i\Delta-\frac{\kappa }{2} & 0 & ig_0 & ig_0 & 0 & 0 \\ 
0 & -(i\Omega+\frac{\Gamma}{2}) & 0 & 0 & ig_0 & 0 \\ 
iL^+ & 0 & i\left(\Delta_B+s\right)-\frac{\gamma }{2} & 0 & 0 & 0 \\ 
iL^- & 0 & 0 & i\left(\Delta_R+s^\ast\right)-\frac{\gamma }{2} & 0 & 0 \\ 
0 & 0 & 0 & 0 & -\kappa  & 0 \\ 
0 & 0 & ig & ig & 0 & 2i(\Delta+2\Re[s])-\kappa  \end{array}
\end{array}
\right]
\left\{ \begin{array}{c}
\hat{a} \\ 
\hat{b} \\ 
\hat{a}\hat{b} \\ 
\hat{a}{\hat{b}}^{\dagger } \\ 
\hat{n} \\ 
\hat{c}
\end{array}
\right\}\\ \label{T8-eq7}
&&-\left[ 
\begin{array}{cccc}
\sqrt{\kappa} & 0 & 0 & 0\\
0 & 0 & \sqrt{\Gamma} & 0\\
\sqrt{\kappa}\bar{b} & 0 & \sqrt{\Gamma\bar{n}} & 0\\
\sqrt{\kappa}\bar{b}^\ast & 0 & 0 & \sqrt{\Gamma\bar{n}}\\
\sqrt{K} & \sqrt{K} & 0 & 0\\
\sqrt{K} & 0 & 0 & 0\\
\end{array}
\right]
\left\{ \begin{array}{c}
\hat{a}_\text{in} \\ 
\hat{a}_\text{in}^\dagger \\ 
\hat{b}_\text{in} \\ 
\hat{b}_\text{in}^\dagger 
\end{array}
\right\}-
\left[ 
\begin{array}{cc}
1 & 0\\
0 & 0\\
\bar{b} & 0\\
\bar{b}^\ast & 0\\
\sqrt{\bar{n}} & \sqrt{\bar{n}}\\
\sqrt{\bar{n}} & 0\\
\end{array}
\right]
\left\{ 
\begin{array}{c}
\alpha \\ 
\alpha^\ast
\end{array}
\right\}
=\frac{d}{dt}\left\{ \begin{array}{c}
\hat{a} \\ 
\hat{b} \\ 
\hat{a}\hat{b} \\ 
\hat{a}{\hat{b}}^{\dagger } \\ 
\hat{n} \\ 
\hat{c} 
\end{array}
\right\}.
\end{eqnarray}
\noindent
The second term on the right is the noise fluctuations due to the optical and mechanical fields with zero average $\langle\hat{a}_\text{in}\rangle=\langle\hat{b}_\text{in}\rangle=0$, and the last term in the above is proportional to the input photon flux $|\alpha|=\sqrt{\epsilon\kappa P/\hbar\omega}$ where $P$ is the incident radiation power and $\epsilon$ is the coupling efficiency. As it will be mentioned briefly later, the average values $\bar{n}$ and $\bar{x}$ have to be solved by setting $d/dt=0$ on the left and taking average values, which eliminates the noise fluctuations, causing the replacements $\hat{a}\to\sqrt{\bar{n}}$, $\hat{b}\to\bar{b}$, $\hat{a}\hat{b}\to\bar{b}\sqrt{\bar{n}}$, $\hat{a}\hat{b}^\dagger\to\bar{b}^\ast\sqrt{\bar{n}}$, $\hat{n}\to\bar{n}$, and $\hat{c}\to\bar{n}/2$.

Hence, the average values $\bar{a}=\sqrt{\bar{n}}$ and $\bar{b}$ get nonlinearly coupled to the input flux $\alpha$ through the system of algebraic relations as
\begin{eqnarray}
\nonumber
\left[ \begin{array}{c}
\begin{array}{cccccc}
i\Delta-\frac{\kappa }{2} & 0 & ig_0 & ig_0 & 0 & 0 \\ 
0 & -(i\Omega+\frac{\Gamma}{2}) & 0 & 0 & ig_0 & 0 \\ 
iL^+ & 0 & i\left(\Delta_B+g_0\bar{b}\right)-\frac{\gamma }{2} & 0 & 0 & 0 \\ 
iL^- & 0 & 0 & i\left(\Delta_R+g_0\bar{b}^\ast\right)-\frac{\gamma }{2} & 0 & 0 \\ 
0 & 0 & 0 & 0 & -\kappa  & 0 \\ 
0 & 0 & ig & ig & 0 & i\left(\Delta+g_0\bar{B}\right)-\frac{\kappa}{2}  \end{array}
\end{array}
\right]\\ \label{T8-eq8} 
\times
\left\{ \begin{array}{c}
\bar{a} \\ 
\bar{b} \\ 
\bar{a}\bar{b} \\ 
\bar{a}\bar{b}^\ast \\ 
\bar{a}^2 \\ 
\bar{a}^2
\end{array}
\right\}
=\left\{ 
\begin{array}{cc}
1 & 0\\
0 & 0\\
\bar{b} & 0\\
\bar{b}^\ast & 0\\
\bar{a} & \bar{a}\\
\bar{a} & 0\\
\end{array}
\right\}
\left\{ 
\begin{array}{c}
\alpha \\ 
\alpha^\ast
\end{array}
\right\},
\end{eqnarray}
with $\bar{B}=\bar{b}+\bar{b}^\ast$. With a given input photon flux $|\alpha|$, this system can be now solved to obtain the phase $\angle \alpha$ in such a way that $\angle\bar{a}=0$. Then $\bar{a}$ and $\bar{b}$ can be obtained in an algebraic manner. This sets up a system of equations in terms of the total of four unknowns $\angle\alpha$, $\bar{a}=\sqrt{\bar{n}}$, $\bar{b}$, and $\bar{b}^\ast$. 

In the above system, the second equation is independent of $\alpha$, while together the fifth they yield 
\begin{eqnarray}
\label{T8-eq9}
\bar{b}&=&\frac{ig_0}{i\Omega+\frac{1}{2}\Gamma}\bar{a}^2,\\ \nonumber
\bar{a}&=&-\frac{1}{\kappa}(\alpha+\alpha^\ast).
\end{eqnarray}
\noindent
This also already solves $\bar{b}^\ast$ in terms of $\bar{a}$. Plugging in the results into the first equation leads to the third-order algebraic equation which can be now solved. Doing this and some algebraic manipulation gives the equation
\begin{equation}
\label{T8-eq10}
ig_0^2\frac{2\Omega}{\Omega^2+\frac{1}{4}\Gamma^2}\bar{a}^3+\left(i\Delta-\frac{\kappa}{2}\right)\bar{a}=\alpha.
\end{equation}
\noindent
This equation in general is expected to yield only real-valued $\bar{a}$. Separating the real and imaginary parts gives
\begin{eqnarray}
\label{T8-eq11}
\Re[\alpha]&=&-\kappa\frac{\bar{a}}{2}\\
\nonumber
\Im[\alpha]&=&g_0^2\frac{2\Omega}{\Omega^2+\frac{1}{4}\Gamma^2}\bar{a}^3+\Delta\bar{a}.
\end{eqnarray}
The first of these is the same as the second of (\ref{T8-eq9}). The above two equations can be now iteratively solved to yield $\angle\alpha$ and $\bar{a}$ for a given $|\alpha|$. One may also discard $\angle\alpha$ by combining the above two, resulting in 
\begin{equation}
\label{T8-eq12}
|\alpha|^2=\left[\frac{\kappa^2}{4}+\left(\frac{2g_0^2\Omega}{\Omega^2+\frac{1}{4}\Gamma^2}\bar{n}+\Delta\right)^2\right]\bar{n}.
\end{equation}
\noindent
Only real and positive-valued roots of (\ref{T8-eq12}) for $\bar{n}$ are acceptable. Sufficiently large blue-detuning with $\Delta<\Delta_\text{b}<0$ causes the well-known bistability. It is easy to find the negative blue detuning $\Delta_\text{b}<0$ at which bistability starts to appear, by looking for the only negative real root of the cubic equation
\begin{equation}
\label{T8-eqDetuning}
-\Delta_\text{b}\left(\Delta_\text{b}^2+\frac{9}{4}\kappa^2\right)=\frac{27g_0^2\Omega}{\Omega^2+\frac{1}{4}\Gamma^2}|\alpha|^2.
\end{equation}

These two noise terms we assume have the flat shot-noise uncorrelated spectral power densities 
\begin{eqnarray}
\label{T8-eq13}
S_{AA}(\omega)=\frac{1}{2}, \\ \nonumber
S_{BB}(\omega)=m+\frac{1}{2},
\end{eqnarray}
\noindent
which are identical on both positive and negative frequencies. The ultimate difference of noise power spectral densities will be later maintained by the asymmetry caused by $L^+ -L^-=2F+g_0 > 0$. Here, $m=1/\left[\exp(\hbar\Omega/k_\text{B}T)-1\right]$ is the population of incoherent phonons under thermal equilibrium, which contribute to the random fluctuations of thermal noise. This quantity is not to be mistaken with $\bar{m}$ which here denotes the population of coherent phonons, contributing coherently to the optomechanical interaction, and are driven by the optical radiation pressure. This shall be discussed later in \S\ref{S8-BMWm8}.

\subsubsection{Perturbative Solution}

At this moment, the system of equations (\ref{T8-eq7}) can be perturbed around equilibrium values found above. This procedure and taking a Fourier transform gives out the solution. Let us define first
\begin{equation}
\label{T8-eq14}
\textbf{M}=\left[ \begin{array}{c}
\begin{array}{cccccc}
i\Delta-\frac{\kappa }{2} & 0 & ig_0 & ig_0 & 0 & 0 \\ 
0 & -(i\Omega+\frac{\Gamma}{2}) & 0 & 0 & ig_0 & 0 \\ 
iL^+ & 0 & i\left(\Delta_B+s\right)-\frac{\gamma }{2} & 0 & 0 & 0 \\ 
iL^- & 0 & 0 & i\left(\Delta_R+s^\ast\right)-\frac{\gamma }{2} & 0 & 0 \\ 
0 & 0 & 0 & 0 & -\kappa  & 0 \\ 
0 & 0 & ig & ig & 0 & 2i(\Delta+2\Re[s])-\kappa  \end{array}
\end{array}
\right],
\end{equation}
\noindent
as well as
\begin{eqnarray}
\label{T8-eq15}
\left\{\delta A(\omega)\right\}^{\text{T}}&=&\left\{ 
\delta\hat{a}(\omega), 
\delta\hat{b}(\omega),  
\delta(\hat{a}\hat{b})(\omega),
\delta(\hat{a}{\hat{b}}^{\dagger })(\omega), 
\delta\hat{n}(\omega), 
\delta\hat{c}(\omega) 
\right\},\\ \nonumber
\left\{A_\text{in}(\omega)\right\}^\text{T}&=&\left\{ 
\hat{a}_\text{in}(\omega) ,
\hat{a}_\text{in}^\dagger(\omega) , 
\hat{b}_\text{in}(\omega) ,
\hat{b}_\text{in}^\dagger(\omega) 
\right\},\\ \nonumber
\left[\sqrt{\Gamma}\right]&=&\left[ 
\begin{array}{cccc}
\sqrt{\kappa} & 0 & 0 & 0\\
0 & 0 & \sqrt{\Gamma} & 0\\
\sqrt{\kappa}\bar{b} & 0 & \sqrt{\Gamma\bar{n}} & 0\\
\sqrt{\kappa}\bar{b}^\ast & 0 & 0 & \sqrt{\Gamma\bar{n}}\\
\sqrt{K} & \sqrt{K} & 0 & 0\\
\sqrt{K} & 0 & 0 & 0\\
\end{array}
\right].
\end{eqnarray}

Then, taking $\textbf{I}_j$ as the $j\times j$ identity matrix, we get
\begin{eqnarray}
\label{T8-eq16}
\left\{A_\text{out}(\omega)\right\}&=&\left\{A_\text{in}(\omega)\right\}-\left[\sqrt{\Gamma}\right]^\text{T}\left\{\delta A(\omega)\right\}=\textbf{Y}(\omega)\left\{A_\text{in}(\omega)\right\}, \\ \nonumber
\left\{\delta A(\omega)\right\}&=&\textbf{Z}(\omega)\left\{A_\text{in}(\omega)\right\}, \\ \nonumber
\textbf{Z}(\omega)&=&\left[\textbf{M}-i\omega\textbf{I}_6\right]^{-1}\left[\sqrt{\Gamma}\right], \\ \nonumber
\textbf{Y}(\omega)&=&\textbf{I}_4-\left[\sqrt{\Gamma}\right]^\text{T}\textbf{Z}(\omega).
\end{eqnarray}
\noindent
Here, $\textbf{Y}(\omega)$ is the scattering matrix connecting the input and output ports. Now, the spectral density of reflected light from the cavity can be found using (\ref{T8-eq13}) by the expression
\begin{equation}
\label{T8-eq17}
S(\omega)=\left[|Y_{11}(\omega)|^2+|Y_{12}(\omega)|^2\right]S_{AA}(\omega)+\left[|Y_{13}(\omega)|^2+|Y_{14}(\omega)|^2\right]S_{BB}(\omega),
\end{equation}
\noindent
as long as the noise processes of $\hat{a}_\text{in}$ and $\hat{b}_\text{in}$ have zero cross-correlation \cite{S2-6}.

\subsection{Linearized Optomechanics}\label{S2-Suppl}

It is fairly easy to see that the system of equations (\ref{T8-eq18}) when simplified and rewritten for the basis $\{\hat{a},\hat{b},\hat{b}^\dagger\}$ reproduces the widely used linearized optomechanical equations \cite{S2-3}. To demonstrate this, we ignore the perturbation matrix $\delta\textbf{N}$, as well as $\bar{b}/\sqrt{\bar{n}}$ in the noise terms, and then employ the substitutions 
\begin{eqnarray}
\label{T8-eq19a}
\hat{a}\hat{b}&\to&\exp\left[(i\Delta-\frac{1}{2}\kappa)t\right]\bar{a}\hat{b}=\exp\left[(i\Delta-\frac{1}{2}\kappa)t\right]\sqrt{\bar{n}}\hat{b},\\ \nonumber
\hat{a}\hat{b}^\dagger&\to&\exp\left[(i\Delta-\frac{1}{2}\kappa)t\right]\bar{a}\hat{b}^\dagger=\exp\left[(i\Delta-\frac{1}{2}\kappa)t\right]\sqrt{\bar{n}}\hat{b}^\dagger.
\end{eqnarray}
\noindent
This will immediately result in rewriting (\ref{T8-eq18}) as
\begin{equation}
\label{T8-eq19b}
\frac{d}{dt}
\left\{ 
\begin{array}{c}
\delta\hat{a} \\ 
\delta\hat{b} \\ 
\delta\hat{b}^{\dagger }
\end{array}
\right\}=\left[ 
\begin{array}{ccc}
i\Delta-\frac{\kappa }{2} & ig_0 & ig_0 \\ 
0 &  -i\Omega-\frac{\Gamma }{2} & 0 \\ 
0 & 0 & i\Omega-\frac{\Gamma }{2}  
\end{array}
\right]
\left\{ 
\begin{array}{c}
\delta\hat{a} \\ 
\delta\hat{b} \\ 
\delta\hat{b}^{\dagger }
\end{array}
\right\}+\left[ 
\begin{array}{ccc}
\sqrt{\kappa} & 0 & 0\\
0 & \sqrt{\Gamma} & 0\\
0 & 0 & \sqrt{\Gamma}
\end{array}
\right]\left\{ \begin{array}{c}
\hat{a}_\text{in} \\ 
\hat{b}_\text{in} \\ 
\hat{b}_\text{in}^\dagger 
\end{array}
\right\},
\end{equation}
\noindent
which is nothing but exactly the linearized state equations of optomechanics. Hence, the method of higher-order operators \cite{S2-Paper2} is mathematically able to reproduce the less approximate linearized optomechanics.

\subsection{Pulsed Drive}\label{S3-Suppl}

Under the situation of pulsed drive, one may assume the input photon rate $\alpha(t)$ to be a function of time. If the input drive varies on a time-scale or longer than the mechanical period with $|d\alpha(t)/dt|<\Omega\alpha$, then one may assume $\bar{n}(t)$ is solved through (\ref{T8-eq12}) at each moment with updated momentary mechanical frequency $\Omega(t)$ and linewidth $\Gamma(t)$ to yield an effective time dependent coefficients matrix $\textbf{M}(t)$. This offers the solution 
\begin{eqnarray}
\label{T8-Pulsed}
\{A(t)\}&=&\exp\left[\int_{0}^{t}\textbf{M}(\tau)d\tau\right]\{A(0)\} +\int_{0}^{t}\exp\left[\int_{0}^{t-\tau}\textbf{M}(\nu)d\nu\right][\beta(\tau)]\left\{\alpha(\tau)\right\}d\tau \\ \nonumber
&+&\int_{0}^{t}\exp\left[\int_{0}^{t-\tau}\textbf{M}(\nu)d\nu\right][\Gamma(\tau)]\{A_\text{in}(\tau)\}d\tau\\ \nonumber
[\beta(t)]^{\text{T}}&=&\left[ 
\begin{array}{cccccc}
1 & 0 & \bar{b}(t) & \bar{b}^\ast(t) & \sqrt{\bar{n}(t)} & \sqrt{\bar{n}(t)}\\
0 & 0 & 0 & \sqrt{\bar{n}(t)} & 0
\end{array}
\right], \\ \nonumber
\{\alpha(t)\}^{\text{T}}&=&\left\{
\alpha(t), 
\alpha^\ast(t)
\right\}.
\end{eqnarray}

\subsection{Side-band Inequivalence in Weakly Nonlinear Limit}\label{S4-Suppl}

Let us go back to the set of equations (\ref{T8-eq14}) and only retain the first, third, and fourth equations.  This reduction gives a $3\times 3$ system of equations, identical to (\ref{T8-eq16}) with the redefinitions 
\begin{eqnarray}
\label{T8-eq18}
\textbf{M}&=&\textbf{N}+\delta\textbf{N}\\ \nonumber
\textbf{N}&=&\left[ 
\begin{array}{ccc}
i\Delta-\frac{\kappa }{2} & ig_0 & ig_0 \\ 
iF &  -i(\Omega-\Delta)-\frac{\gamma }{2} & 0 \\ 
-iF & 0 & i(\Omega+\Delta)-\frac{\gamma }{2}  
\end{array}
\right],\\ \nonumber
\left\{\delta A(\omega)\right\}^{\text{T}}&=&\left\{ \delta\hat{a}(\omega), \delta(\hat{a}\hat{b})(\omega), \delta(\hat{a}{\hat{b}}^{\dagger })(\omega) \right\},\\ \nonumber
\left\{A_\text{in}(\omega)\right\}^{\text{T}}&=&\left\{ \hat{a}_\text{in}(\omega),\hat{b}_\text{in}(\omega), 
\hat{b}_\text{in}^\dagger(\omega) \right\},\\ \nonumber
\left[\sqrt{\Gamma}\right]&=&\left[ 
\begin{array}{ccc}
\sqrt{\kappa} & 0 & 0\\
\sqrt{\kappa}\bar{b} & \sqrt{\Gamma\bar{n}} & 0\\
\sqrt{\kappa}\bar{b}^\ast & 0 & \sqrt{\Gamma\bar{n}}
\end{array}
\right].
\end{eqnarray}
\noindent
Here, the perturbation matrix $\delta\textbf{N}$ is defined through the relation
\begin{equation}
\label{T8-eq19}
\delta\textbf{N}=\left[ 
\begin{array}{ccc}
0 & 0 & 0 \\ 
if^+ & is & 0 \\ 
if^- & 0 & is^\ast  
\end{array}
\right],
\end{equation}
in which $f^+=g_0(\bar{m}+1)$ and $f^-=g_0\bar{m}$. It is quite apparent that the second and third rows of $\textbf{N}$ in (\ref{T8-eq18}) are complex conjugates. 

By setting $\Delta=0$ in (\ref{T8-eq18}) one would expect identically displaced sidebands at $\pm\Omega$. However, this is contingent on the fact that the eigenvalues of $\textbf{N}$ be either complex conjugates as $\Im[\eta]=\pm\Omega$ corresponding to the frequencies of the sidebands, or $\Im[\eta]=0$ corresponding to the resonant pump. However, the presence of perturbation matrix $\delta\textbf{N}$ breaks this symmetry between the sidebands. This causes a very tiny displacement of sidebands so that $\Delta_\text{r}+\Delta_\text{b}\neq0$. First figure  illustrates the side-band asymmetry for various intracavity photon numbers $\bar{n}=(g/g_0)^2$ and coherent phonon numbers $\bar{m}$, when $g_0/\Omega=10^{-3}$. This effect is actually due to the higher-order optomechanical spring effect analyzed in the following. 

It has to be noticed that the horizontal axes are nonlinear functions of the incident light intensity and therefore $\alpha$. Typically, an inequivalence would be observable in a heterodyne side-band resolved experiment if the effect is large enough to allow clear and measurable motion of side-bands. This condition requires $|\Delta_\text{r}+\Delta_\text{b}|>\Gamma=\Omega/Q_\text{m}$, in which $Q_\text{m}$ is the mechanical quality factor. If $Q_\text{m}>10^5$, then an intracavity occupation number of $\bar{n}>10^4$ should be sufficient to detect any such inequivalence.

The side-band inequivalence should not be mistaken with the fundamental energy conservation and time reversal symmetry. Firstly according to these, the spectral density on the negative frequencies of the spectrum should be mirror symmetric with respect to the positive frequencies. Normally, the actual optical frequency $\omega$ is much larger than the mechanical frequency $\Omega$, so that the observed red- and blue-detuned sidebands within the range $\Delta\in(-\Omega,+\Omega)$ actually entirely correspond to the positive absolute frequencies. So the speculation that $\delta\Delta$ could be non-zero has nothing to do with the time-reversal symmetry. Secondly, side-band inequivalence is a purely nonlinear effect and is therefore strictly forbidden in any linearized approximation of the optomechanical Hamiltonian. 

\subsection{Resonance Shift}\label{S5-Suppl}
The contribution of the terms $\pm iF+if^{\pm}$ to the mechanical frequency $\Omega$ in the second and third equations of (\ref{T8-eq19}), can be held responsible for the so-called optomechanical spring effect \cite{S2-1,S2-3,S2-6,S2-8,S8-Spring1,S8-Spring2,S8-Spring3,S8-Spring4,S8-Spring5}. As the result of optomechanical interaction, both of the optical and mechanical resonance frequencies and damping rates undergo shifts. Even at the limit of zero input optical power $\alpha=0$ and therefore zero cavity photon number $\bar{n}=0$, it is possible to show that there is a temperature-dependent shift in the mechanical resonance frequency, markedly different from the lattice-expansion dependent effect. This effect is solely due to the optomechanical interaction with virtual cavity photons, which completely vanishes when $g_0=0$. In close relationship to the shift of resonances, we can also study the optomechanical spring effect with the corrections from higher-order interactions included.

The analysis of spring effect is normally done by consideration of the effective optomechanical force acting upon the damped mechanical oscillator, thus obtaining a shift in squared mechanical frequency $\delta(\Omega^2)$, whose real and imaginary parts give expressions for $\delta\Omega$ and $\delta\Gamma$. Corrections to these two terms due to higher-order interactions are discussed in the above. Here, we demonstrate that the analysis using higher-order operator algebra can recover some important lost information regarding the optical and mechanical resonances when the analysis is done on the linearized basis $\{A\}^\text{T}=\{\hat{a},\hat{a}^\dagger,\hat{b},\hat{b}^\dagger\}$. 

To proceed, we consider finding eigenvalues of the matrix $\textbf{M}$ as defined in (\ref{T8-eq18}). Ignoring all higher-order nonlinear effects beyond the basis $\{A\}^\text{T}=\{\hat{a},\hat{a}\hat{b},\hat{a}\hat{b}^\dagger\}$, we set $s=0$. This enables us to search for the eigenvalues of the coefficients matrix $\textbf{M}$ as
\begin{eqnarray}
\label{T8-eqSpring1}
\text{eig}\{\textbf{M}\}&=&\text{eig}\left[ 
\begin{array}{ccc}
i\Delta-\frac{\kappa }{2} & ig_0 & ig_0 \\ 
i(G+f^+) &  -i(\Omega-\Delta)-\frac{\gamma }{2} & 0 \\ 
-i(G-f^-) & 0 & i(\Omega+\Delta)-\frac{\gamma }{2}  
\end{array}
\right]\\ \nonumber
&=&i\left\{
\begin{array}{c}
\Delta+\lambda_1(\Delta,T)+i\gamma_1(\Delta,T) \\
\Delta+\lambda_2(\Delta,T)+i\gamma_2(\Delta,T) \\
\Delta+\lambda_3(\Delta,T)+i\gamma_3(\Delta,T) 
\end{array}
\right\}\\ \nonumber
&=&i\left\{
\begin{array}{c}
\Delta+\eta_1(\Delta,T) \\
\Delta+\eta_2(\Delta,T) \\
\Delta+\eta_3(\Delta,T) 
\end{array}
\right\},
\end{eqnarray}
\noindent
in which $\lambda_j=\Re[\eta_j]$ and $\gamma_j=\Im[\eta_j]$ with $j=1,2,3$ are real valued functions of $\Delta$ and bath temperature $T$. The temperature $T$ determines $\bar{m}$ while $\bar{n}$ is a function of $\Delta$ as well as input photon rate $\alpha$. 

In general, the three eigenvalues $\eta_j=\lambda_j(\Delta,T)+i\gamma_j(\Delta,T), j=1,2,3$ are expected to be deviate from the three free-running values $\psi_1=i\frac{1}{2}\kappa$, $\psi_2=-\Omega+i\frac{1}{2}\gamma$, and $\psi_3=\Omega+i\frac{1}{2}\gamma$, as $\eta_j\approx \psi_j-\Delta$ because of non-zero $g_0$. Solving the three equations therefore gives the values of shifted optical and mechanical frequencies and their damping rates compared to the bare values in absence of optomechanical interactions with $g_0=0$ 
\begin{eqnarray}
\label{T8-eqSpring2}
\delta\Omega&=&-\frac{1}{2}\Re[\eta_2-\eta_3]-\Omega, \\ \nonumber
\delta\omega&=&-\frac{1}{2}\Re[\eta_2+\eta_3], \\ \nonumber
\delta\Gamma&=&\Im[-2\eta_1+\eta_2+\eta_3]-\Gamma, \\ \nonumber
\delta\kappa&=&2\Im[\eta_1]-\kappa. 
\end{eqnarray}
\noindent
This  method to calculate the alteration of resonances, does not regard the strength of the optomechanical interaction or any of the damping rates. In contrast, the known methods to analyze this phenomenon normally require $g<<\kappa$ and $\Gamma+\delta\Gamma<<\kappa$ \cite{S2-3}.

A very simple way to estimate the shift in eigenvalues is by separating the real and imaginary parts of the optomechanical interaction, as $\Omega\to\Omega-\Re[s]$ and $\gamma\to\gamma+2\Im[s]$. These shifts in mechanical frequency and damping rates can be approximated using (\ref{T8-eq9}) as
\begin{eqnarray}
\label{T8-eq20}
\delta\Omega+\delta\omega&\approx&-g_0\Re[\bar{b}]=-\frac{g^2\Omega}{\Omega^2+\frac{1}{4}\Gamma^2},\\ \nonumber
\delta\Gamma+\delta\kappa&\approx&g_0\Im[\bar{b}]=\frac{g^2\Gamma}{2(\Omega^2+\frac{1}{4}\Gamma^2)},
\end{eqnarray}
\noindent
where $g$ has been defined under (\ref{T8-eq5}). This approximation requires the optomechanical processes $\{\hat{a}\hat{b},\hat{a}\hat{b}^\dagger\}$ being independent of the other state variables. Since this decoupling is not exact, relations (\ref{T8-eq20}) also will remain approximate. However, the accuracy of these are still quite remarkable as is demonstrated here. 

First of all, it is noticed through extensive numerical computations that the shifts in optical and mechanical frequencies take place primarily in the optical part.  That implies the resonance shift is typically much stronger in the optical partition of the system instead of the mechanical partition, leading to marked change in reflection spectra of optomechanical cavities. Extensive numerical calculations for various configurations establish the fact that it is actually the optical resonance frequency which receives the optomechanical interaction effect. The asymmetry of this shift in cavity optical frequency across the zero-detuning $\Delta=0$, is well exhibited for relatively large intracavity photon numbers. This clear signature underlines the fact that the well-known asymmetry of cavity optical response at high intensities could be in part actually a result of this higher-order spring effect, rather than thermally induced instabilities.

Summarizing, any such higher-order resonance shift will cause a change in mechanical frequency $\delta\Omega$ and decay rate $\delta\Gamma$, as well as optical detuning $\delta\omega=-\delta\Delta$ and decay rate $\delta\kappa$. While all these four components are non-zero, it is $\delta\Delta$ which is ultimately dominant over the three others in the bistability relation (\ref{T8-eq12}). This will make the cavity response to follow the bistability and therefore appear to be asymmetric at high illumination drive intensities.

As a result of higher-order spring effect and $\delta\Delta$, a shift in intracavity photon number follows $\delta\bar{n}$, which immediately shifts the higher-order spring effect and therefore $\delta\Delta$. The infinite cycle of shifts in intracavity photon number and optical resonance frequency establishes a deterministic chaotic behavior, which is also a well known experimental observation in the community.

Furthermore, the zero-point optical field can change the mechanical frequency, which is being in close relationship with the Dynamical Casimir effect \cite{S8-Macri}. This value could be in principle measured if temperature-induced expansion and the resulting change of mechanical frequency is much smaller. The thermal expansion coefficient of Silicon is roughly $2.6\times10^{-6}\text{K}^{-1}$, roughly equivalent to $2.6\text{kHz/K}$. The contribution of zero-point field can be therefore larger or at least within the same order of magnitude. Here, we have assumed that the thermal expansion coefficient of Silicon is independent of temperature and also $\Omega$ shifts linearly with temperature. 

The same calculation gives out typical values much less than the temperature expansion drift for the same structure. This phenomenon has been also noticed and referred to as the Nonlinear Transduction \cite{S8-Transduction} where the photon-phonon coupling can induce a temperature-dependent change in the resonance frequency of the cavity, even on the order of cavity linewidth. 

\subsection{Coherent Phonon Population}\label{S8-BMWm8}
It is here shown that the method of higher-order operators allows one to find an explicit expression for $\bar{m}$. In order to do this, we need to write down the $3\times 3$ reduced set of higher-order optomechanical equations with the fluctuations terms dropped, which reads
\begin{equation}
\label{T8-m1}
\frac{d}{dt}\left\{\begin{array}{c}
\hat{a}\\
\hat{a}\hat{b}\\
\hat{a}\hat{b}^\dagger
\end{array}
\right\}=\left[\begin{array}{ccc}
i\Delta-\frac{1}{2}\kappa & ig_0 & ig_0 \\
ig_0(\bar{m}+\bar{n}+1) & i\Delta_B-\frac{1}{2}\gamma & 0\\
ig_0(\bar{m}-\bar{n}) & 0 & i\Delta_R-\frac{1}{2}\gamma
\end{array}\right]\left\{\begin{array}{c}
\hat{a}\\
\hat{a}\hat{b}\\
\hat{a}\hat{b}^\dagger
\end{array}
\right\}-\left\{\begin{array}{c}
\alpha \\ 
\bar{b}\alpha \\
\bar{b}^\ast\alpha^\ast
\end{array}
\right\}.
\end{equation}
Here, $\alpha$ should be taken as a complex number from (\ref{T8-eq11}), $\bar{b}$ is substituted from (\ref{T8-eq9}) in terms of $\bar{a}$, where $\bar{a}=\sqrt{\bar{n}}$ is taken as a real number and $\bar{n}$ can already by found from the solution of the third-order algebraic equation (\ref{T8-eq12}). 

We are here interested in the steady state solutions, so that the time derivative $d/dt=0$ can be set to zero. Then we arrive at the system of equations
\begin{equation}
\label{T8-m2}
\left[\begin{array}{ccc}
i\Delta-\frac{1}{2}\kappa & ig_0 & ig_0 \\
ig_0(\bar{m}+\bar{n}+1) & i\Delta_B-\frac{1}{2}\gamma & 0\\
ig_0(\bar{m}-\bar{n}) & 0 & i\Delta_R-\frac{1}{2}\gamma
\end{array}\right]\left\{\begin{array}{c}
\bar{a}\\
\overline{ab}\\
\overline{ab^\ast}
\end{array}
\right\}=\left\{\begin{array}{c}
\alpha \\ 
\bar{b}\alpha \\
\bar{b}^\ast\alpha^\ast
\end{array}
\right\}.
\end{equation}
In the above system of equations, $\overline{ab}$ corresponds to the time-average of the operators $\braket{\hat{a}\hat{b}}$, while $\overline{ab^\ast}$ corresponds to the time-average of the operators $\braket{\hat{a}\hat{b}^\dagger}$. Quite obviously, $\overline{ab}\approx\bar{a}\bar{b}$ and  $\overline{ab^\ast}\approx\bar{a}\bar{b}^\ast$ can approximately hold based on the mean-field approximation. We shall here furthermore observe that this approximation does not any longer hold for the coherent phonons as $\bar{m}=\braket{\hat{b}^\dagger\hat{b}}\neq\bar{b}^\ast\bar{b}$ for the reasons discussed below.

We now can rearrange (\ref{T8-m2}) in terms of the unknown quantities $\bar{m}$, $\overline{ab}$, and $\overline{ab^\ast}$ as 
\begin{equation}
\label{T8-m3}
\left[\begin{array}{ccc}
0 & ig_0 & ig_0 \\
ig_0\bar{a} & i\Delta_B-\frac{1}{2}\gamma & 0\\
ig_0\bar{a} & 0 & i\Delta_R-\frac{1}{2}\gamma
\end{array}\right]\left\{\begin{array}{c}
\bar{m}\\
\overline{ab}\\
\overline{ab^\ast}
\end{array}
\right\}=\left\{\begin{array}{c}
\alpha-\left(i\Delta-\frac{1}{2}\kappa\right)\bar{a} \\ 
\bar{b}\alpha-ig_0\left(\bar{n}+1\right)\bar{a} \\
\bar{b}^\ast\alpha^\ast+ig_0 \bar{n}\bar{a}
\end{array}
\right\}.
\end{equation}
This linear system of equations after appropriate substitutions from (\ref{T8-eq9}) and (\ref{T8-eq11}) now can be solved to find
\begin{equation}
\label{T8-m4}
\bar{m}(\Delta)=\frac{64 g_0^2 \Omega ^2 \bar{n}^2(\Delta ) \left(\gamma ^2+\gamma  \Gamma +4 \Delta ^2\right)-\left[(\Gamma ^2+4 \Omega ^2\right)^2 \left(\gamma ^2+4 \Delta  (\Delta +\Omega )\right]}{2 \left(\gamma ^2+4 \Delta ^2\right) \left(\Gamma ^2+4 \Omega ^2\right)^2}.
\end{equation}
Here, a small imaginary part remains which has to be dropped and results from the inexactness of (\ref{T8-eq9}) and (\ref{T8-eq11}) coming from linearized optomechanics, and not being in complete consistency with the higher-order formalism. 

In a similar manner, one may find
\begin{eqnarray}
\label{T8-m5}
\overline{ab}&=&
\frac{i \sqrt{\bar{n}} \left[g_0^2 \left(8 \bar{n}+4\right)+(2 \Delta +i \kappa ) (2 (\Delta +\Omega )+i \gamma )\right]-2 i \alpha  \left[\frac{8 \Gamma  g_0^2 \bar{n}}{\Gamma ^2+4 \Omega ^2}+\gamma -2 i (\Delta +\Omega )\right]}{4 g_0 (\gamma -2 i \Delta )}, \\ \nonumber
\overline{ab^\ast}&=&\frac{2 \alpha  \left[\frac{8 i \Gamma  g_0^2 \bar{n}}{\Gamma ^2+4 \Omega ^2}-i \gamma -2 \Delta +2 \Omega \right]-\sqrt{\bar{n}} \left[4 i g_0^2 \left(2 \bar{n}+1\right)+(\kappa -2 i \Delta ) (i \gamma +2 \Delta -2 \Omega )\right]}{4 g_0 (\gamma -2 i \Delta )}.
\end{eqnarray}

The expression (\ref{T8-m4}) for $\bar{m}$ is accurate within half a quanta $\pm\frac{1}{2}$, so that in order to satisfy the zero limits at infinite detuning 
\begin{equation}
\label{T8-m6}
\lim_{\Delta\rightarrow\infty}\bar{m}(\Delta)=0, 
\end{equation}
a half-quanta must be added to (\ref{T8-m4}). Then it will read
\begin{eqnarray}
\label{T8-m8} 
\bar{m}(\Delta)&=&\frac{32 g_0^2 \Omega ^2 \left(\gamma ^2+\gamma  \Gamma +4 \Delta ^2\right)}{\left(\gamma ^2+4 \Delta ^2\right) \left(\Gamma ^2+4 \Omega ^2\right)^2}\bar{n}^2(\Delta ) -\frac{2 \Delta  \Omega  }{\gamma ^2+4 \Delta ^2 }\pm\frac{1}{2}\\ \nonumber
&\approx&\frac{32 g_0^2 \Omega ^2 \left(\gamma ^2+\gamma  \Gamma +4 \Delta ^2\right)}{\left(\gamma ^2+4 \Delta ^2\right) \left(\Gamma ^2+4 \Omega ^2\right)^2}\bar{n}^2(\Delta ) \\ \nonumber
&=&g_0^2\zeta(\Delta)\bar{n}^2(\Delta).
\end{eqnarray}
The approximation holds well if $\bar{n}$ is well above unity. Hence, we can infer from (\ref{T8-m8}) that $\bar{m}\propto\bar{n}^2$. In the lossless limit, where $\gamma\approx 0$ and $\Gamma\approx 0$, one may even further simplify (\ref{T8-m8}) to obtain the simple expression $\bar{m}\approx 2g_0^2\bar{n}^2/\Omega^2$ which is typically accurate within 10\% of the actual value or better. It is not difficult to check the resonant coherent phonon number $\bar{m}(0)$. In the practical limit of $\kappa>>\Gamma$, it is easy to verify that (\ref{T8-m8}) actually simplifies to $\bar{m}(0)\approx 32 [g_0 Q_\text{m}\bar{n}(0)/\Gamma]^2$ with $Q_\text{m}=\Omega/\Gamma$ being the mechanical quality factor. In the next section, we point out a straightforward method to measure this quantity through experiment on the well-known optical spring effect.

In practice, the expression (\ref{T8-m8}) is sensitive to the choice of optomechanical parameters and in particular $g_0$. A slight variation in the basic optomechanical parameters $\left\{g_0,\omega,\kappa,\Omega,\Gamma\right\}$ with $\gamma=\kappa+\Gamma$ as small as few percent can make a pronounced effect in expected behavior in $\bar{m}$.

For the side-band resolved systems in the lossless limit, it is within 10\% of the relationship $\bar{m}\propto 2|\bar{b}|^2$, meanwhile for Doppler cavities, the agreement is roughly within 3\% or better. This result perfectly agrees to the large-amplitude oscillation limit of $\bar{b}(t)\approx\delta\hat{b}(t)+[\bar{b}+\bar{b}\exp(-i\Omega t)]$ where $\delta\hat{b}$ represents the random fluctuations in the mechanical field with $\braket{\delta\hat{b}}=0$. This also tells that the coherent oscillations of the mechanical field are not differential in amplitude, and can vary in the range $(0,2|\bar{b}|)$. So, the amplitude of coherent mechanical oscillations is just as big as their average. This large-amplitude coherent mechanical wave is driven and waked by the optical coherent field inside the cavity, through optomechanical interactions. The mean field approximations $\overline{ab}\approx\bar{a}\bar{b}$ and $\overline{ab^\ast}\approx\bar{a}\bar{b}^\ast$ seem however to always hold better than 0.1\% for Doppler cavities. This accuracy breaks down for side-band resolved cavities.

If there are more than one mechanical fields available $j=1,2,\cdots$, the coherent phonon population of each mode $\bar{m}_j$ shall be determined with the corresponding sets of optomechanical parameters $\left\{g_{0,j},\omega_j,\kappa_j,\Omega_j,\Gamma_j\right\}$, with the expected approximate result $\bar{m}_j\approx 2g_{0,j}^2\bar{n}_j^2/\Omega_j^2$ as long as the mechanical modes are almost uncorrelated. The case of coherent phonon numbers of two or more correlated mechanical modes needs a separate study.

It is here again stressed out that the oscillations of the mechanical field can be decomposed into the incoherent and coherent parts. The incoherent part results from random thermal fluctuations with the thermal occupancy $m$, as well as half a quanta contributing from the quantum noise of the coherent part, while the coherent oscillations correspond to the coherent phonon number $\bar{m}$. The same also should be true for the optical field, however, the random fluctuations of a coherent light is only half a quanta, and the thermal optical occupancy $n$ of optomechanical cavity is normally negligible under practical considerations and working temperatures. 

\subsection{Higher-order Spring Effect}\label{S8-Spring}

It is possible to calculate the optomechanical spring effect due to the standard linearized and higher-order interactions. In order to do this, we start from the matrix $[\textbf{M}]$ given in (\ref{T8-eq18}), and after dropping the noise and drive input terms we notice the expansion of first Langevin equation for the operator $\hat{a}$. That reads
\begin{equation}
\label{T8-eqA1}
\frac{d}{dt}\hat{a}=(i\Delta-\frac{1}{2}\kappa)\hat{a}+ig_0\hat{a}(\hat{b}+\hat{b}^\dagger).
\end{equation}
From the second and third equations we get
\begin{eqnarray}
\label{T8-eqA2}
\hat{a}\frac{d}{dt}\hat{b}+\left[(i\Delta-\frac{1}{2}\kappa)\hat{a}+ig_0\hat{a}(\hat{b}+\hat{b}^\dagger)\right]\hat{b}&=&i(f^+ +F)\hat{a}+\left[i(\Delta_B+g_0\hat{b})-\frac{1}{2}\gamma\right]\hat{a}\hat{b}, \\ \nonumber
\hat{a}\frac{d}{dt}\hat{b}^\dagger+\left[(i\Delta-\frac{1}{2}\kappa)\hat{a}+ig_0\hat{a}(\hat{b}+\hat{b}^\dagger)\right]\hat{b}^\dagger&=&i(f^- -F)\hat{a}+\left[i(\Delta_R+g_0\hat{b})-\frac{1}{2}\gamma\right]\hat{a}\hat{b}^\dagger.
\end{eqnarray}
This is equivalent to 
\begin{eqnarray}
\label{T8-eqA3}
\frac{d}{dt}\hat{b}&=&i(f^+ +F+\hat{b}^\dagger\hat{b})-(i\Omega+\frac{1}{2}\Gamma)\hat{b}, \\ \nonumber
\frac{d}{dt}\hat{b}^\dagger&=&i(f^- -F+\hat{b}\hat{b}^\dagger)+(i\Omega-\frac{1}{2}\Gamma)\hat{b}^\dagger.
\end{eqnarray}
These two equations can be now combined after dropping the nonlinear terms by addition and subtraction, and then taking the Fourier transform to yield the system
\begin{equation}
\label{T8-eqA4}
\left[
\begin{array}{cc}
-iw+\frac{1}{2}\Gamma & i\Omega \\ 
i\Omega & -iw+\frac{1}{2}\Gamma
\end{array}
\right]\left\{
\begin{array}{c}
\hat{b}+\hat{b}^\dagger \\
\hat{b}-\hat{b}^\dagger
\end{array}
\right\}=2ig_0\left\{
\begin{array}{c}
\bar{m}+\frac{1}{2} \\
\bar{n}+\frac{1}{2}
\end{array}
\right\}.
\end{equation}
This can be solved now to yield the expression for $\delta\hat{x}=x_\text{zp}(\hat{b}+\hat{b}^\dagger)$ as 
\begin{equation}
\label{T8-eqA5}
\delta\hat{x}(w)=2ix_\text{zp}g_0\frac{(-iw+\frac{1}{2}\Gamma)(\bar{m}+\frac{1}{2})-i\Omega(\bar{n}+\frac{1}{2})}{(-iw+\frac{1}{2}\Gamma)^2-(-i\Omega)^2}.
\end{equation}
A rearrangement of this expression yields
\begin{equation}
\label{T8-eqA6}
\delta\hat{x}(w)=\frac{2x_\text{zp}g_0\Omega}{-(w+i\frac{1}{2}\Gamma)^2+\Omega^2}\left\{\bar{n}+\left[\left(\frac{w}{\Omega}+i\frac{\Gamma}{2\Omega}\right)\left(\bar{m}+\frac{1}{2}\right)+\frac{1}{2}\right]\right\}.
\end{equation}
It is straightforward now to see that the term within brackets contributes to the necessary corrections to the spring effect. This will change the mechanical response function $\Sigma(w)$ \cite{S2-1,S2-3,S2-6,S8-SKip2} as
\begin{equation}
\label{T8-eqA7}
\Sigma(w,\Delta)=2\Omega g_0^2 \left[\frac{1}{(\Delta+w)+\frac{i}{2}\kappa}+\frac{1}{(\Delta-w)-\frac{i}{2}\kappa}\right][\bar{n}+\mu(w)],
\end{equation}
where a term with the dimension of mass in the numerator, which in the following calculation ultimately cancels out, and is equal to the effective motion mass $m_\text{eff}$, is not shown for simplicity. We have also
\begin{equation}
\label{T8-eqA8}
\mu(w)=\frac{1}{\Omega}\left(w+\frac{i}{2}\Gamma\right)\left(\bar{m}+\frac{1}{2}\right)+\frac{1}{2},
\end{equation}
represents corrections to yield the effective cavity photon number $\bar{n}_\text{eff}=\bar{n}+\mu(w)$ because of higher-order interactions. This corrections is easy to see that are important if the pump level is not too high. Typically, for $\bar{n}<10^2$ higher-order spring effects are quite significant, and when $\bar{n}>10^3$ the higher-order effects are suppressed by the standard spring effect.

The spring effect modifies the measured mechanical frequency $\Omega$ and linewidth $\Gamma$ as
\begin{eqnarray}
\label{T8-eqA9}
\delta\Omega(w,\Delta)&=&\frac{1}{2w}\Re[\Sigma(w,\Delta)], \\ \nonumber
\delta\Gamma(w,\Delta)&=&-\frac{1}{w}\Im[\Sigma(w,\Delta)].
\end{eqnarray}
Put together combined, we get
\begin{eqnarray}
\label{T8-eqA10}
\delta\Omega(w,\Delta)&=& \frac{g_0^2\bar{n}\Omega}{w}\left[\frac{\Delta+w}{(\Delta+w)^2+\frac{1}{4}\kappa^2}+\frac{\Delta-w}{(\Delta-w)^2+\frac{1}{4}\kappa^2}\right]\\ \nonumber
&+&\frac{g_0^2\Re[\mu(w)]\Omega}{w}\left[\frac{\Delta+w}{(\Delta+w)^2+\frac{1}{4}\kappa^2}+\frac{\Delta-w}{(\Delta-w)^2+\frac{1}{4}\kappa^2}\right] \\ \nonumber
&+&\frac{g_0^2\Im[\mu(w)]\Omega}{w}\left[\frac{\kappa}{(\Delta+w)^2+\frac{1}{4}\kappa^2}-\frac{\kappa}{(\Delta-w)^2+\frac{1}{4}\kappa^2}\right],\\ \nonumber
\delta\Gamma(w,\Delta)&=&\frac{g_0^2\bar{n}\Omega}{w}\left[\frac{\kappa}{(\Delta+w)^2+\frac{1}{4}\kappa^2}-\frac{\kappa}{(\Delta-w)^2+\frac{1}{4}\kappa^2}\right]\\ \nonumber
&+&\frac{g_0^2\Re[\mu(w)]\Omega}{w}\left[\frac{\kappa}{(\Delta+w)^2+\frac{1}{4}\kappa^2}-\frac{\kappa}{(\Delta-w)^2+\frac{1}{4}\kappa^2}\right]\\ \nonumber
&-&\frac{g_0^2\Im[\mu(w)]\Omega}{w}\left[\frac{\Delta+w}{(\Delta+w)^2+\frac{1}{4}\kappa^2}+\frac{\Delta-w}{(\Delta-w)^2+\frac{1}{4}\kappa^2}\right].
\end{eqnarray}
Here, the second and third terms on the rights hand sides of both equations are corrections to the spring effect due to the higher-order interactions, resulting from the temperature-dependent expressions 
\begin{eqnarray}
\label{T8-eqA11}
\Re[\mu(w)]&=&\frac{w}{\Omega}\left(\bar{m}+\frac{1}{2}\right)+\frac{1}{2}, \\ \nonumber
\Im[\mu(w)]&=&\frac{\Gamma}{2\Omega}\left(\bar{m}+\frac{1}{2}\right).
\end{eqnarray}
The temperature-dependence of (\ref{T8-eqA11}) causes dependence of the spring effect on temperature as well.

The uncorrected standard expressions read \cite{S2-3,S8-SKip2}
\begin{eqnarray}
\label{T8-eqA12}
\delta\Omega(w,\Delta)&=& \frac{g_0^2\bar{n}\Omega}{w}\left[\frac{\Delta+w}{(\Delta+w)^2+\frac{1}{4}\kappa^2}+\frac{\Delta-w}{(\Delta-w)^2+\frac{1}{4}\kappa^2}\right],\\ \nonumber
\delta\Gamma(w,\Delta)&=&\frac{g_0^2\bar{n}\Omega}{w}\left[\frac{\kappa}{(\Delta+w)^2+\frac{1}{4}\kappa^2}-\frac{\kappa}{(\Delta-w)^2+\frac{1}{4}\kappa^2}\right],
\end{eqnarray}
from which we may observe
\begin{eqnarray}
\label{T8-eqA13}
\delta\Omega(w,\Delta)&=&-\delta\Omega(w,-\Delta),\\ \nonumber
\delta\Omega(-\Delta,\Delta)&=&\frac{g_0^2\bar{n}\Omega}{2}\frac{\Delta}{\Delta^2+\frac{1}{16}\kappa^2},
\end{eqnarray}
which do hold for the standard spring effect at sufficiently high optical powers.

\subsection{Minimal Basis}\label{S8-Minimal}

Complete solution of optomechanical interaction $\mathbb{H}_\text{OM}$ can be attained analytically using the minimal basis $\{A\}^{\rm T}=\{\hat{n}^2,\hat{n}\hat{b},\hat{n}\hat{b}^\dagger\}=\{\hat{N},\hat{B},\hat{B}^\dagger\}$. Construction of Langevin equations leads to the system
\begin{equation}
\label{T8-eq21}
\frac{d}{dt}\left\{\begin{array}{c}
\hat{N}\\
\hat{B}\\
\hat{B}^\dagger
\end{array}\right\}=
\left[
\begin{array}{ccc}
-2\kappa & 0 & 0\\
ig_0 & -i\Omega-\frac{\gamma}{2} & 0\\
ig_0 & 0 & i\Omega-\frac{\gamma}{2}
\end{array}
\right]
\left\{\begin{array}{c}
\hat{N}\\
\hat{B}\\
\hat{B}^\dagger
\end{array}\right\}-\left\{\begin{array}{c}
\sqrt{4\kappa}\hat{N}_{\rm in}\\
\sqrt{\gamma}\hat{B}_{\rm in}\\
\sqrt{\gamma}\hat{B}_{\rm in}^\dagger
\end{array}\right\}.
\end{equation}
\noindent
Here, the multiplicative noise terms are defined as 
\begin{eqnarray}
\label{T8-eq22}
\sqrt{4\kappa}\hat{N}_\text{in}&=&2\sqrt{\kappa}\left(\hat{n}\hat{a}^\dagger\hat{a}_\text{in}+\hat{a}_\text{in}^\dagger\hat{a}\hat{n}\right),\\ \nonumber
\sqrt{\gamma}\hat{B}_\text{in}&=&\sqrt{2\kappa}\hat{b}\hat{n}_\text{in}+\sqrt{\Gamma}\hat{n}\hat{b}_\text{in},
\end{eqnarray} 
\noindent
where $\hat{n}_\text{in}$ is already defined under (\ref{T8-eq4}), and the spectral density of which can be estimated using the method described elsewhere \cite{S8-SLoo}. A very effective method to deal with multiplicative noise is to be discussed in \S\ref{S2-Noise4}. This can be immediately noticed to be reducible as 
\begin{equation}
\label{T8-eq23}
\frac{d}{dt}\left\{\begin{array}{c}
\hat{N}\\
\hat{B}
\end{array}\right\}=
\left[
\begin{array}{cc}
-2\kappa & 0 \\
ig_0 & -i\Omega-\frac{\gamma}{2}
\end{array}
\right]
\left\{\begin{array}{c}
\hat{N}\\
\hat{B}
\end{array}\right\}-\left\{\begin{array}{c}
\sqrt{4\kappa}\hat{N}_{\rm in}\\
\sqrt{\gamma}\hat{B}_{\rm in}
\end{array}\right\}.
\end{equation}
\noindent
These will make the evaluation of spectral densities $S_{NN}(\omega)$ and $S_{BB}(\omega)$ possible. Interestingly, (\ref{T8-eq23}) is actually decoupled, since the equation for $\hat{N}$ is already independent of $\hat{B}$, which admits the solution
\begin{equation}
\label{T8-eq24}
\hat{N}(t)=\hat{N}(0)e^{-2\kappa t}-2\sqrt{\kappa}e^{-2\kappa t}\int_{0}^{t}\hat{N}_\text{in}(\tau)e^{2\kappa \tau}d\tau.
\end{equation}
\noindent
We can be now plug (\ref{T8-eq24}) in the second equation of (\ref{T8-eq23}) to solve exactly for $\hat{B}$. We define $\vartheta=i\Omega+\frac{\gamma}{2}$ and may write down
\begin{equation}
\label{T8-eq25}
\hat{B}(t)=\hat{B}(0)e^{-\vartheta t}-e^{-\vartheta t}\int_{0}^{t}e^{\vartheta \tau}\left[ig_0\hat{N}(\tau)+\sqrt{\gamma}\hat{B}_\text{in}(\tau)\right]d\tau.
\end{equation}

The treatment of multiplicative noise terms (\ref{T8-eq22}) can be quite difficult in the most general form, especially that they demand prior knowledge of photonic and phononic ladder operators. However, assuming that the extra ladder operators can be replaced by their mean values, we can do the zeroth order approximations 
\begin{eqnarray}
\label{T8-eq26}
\sqrt{4\kappa}\hat{N}_\text{in}&\approx&\sqrt{\kappa\bar{n}}\bar{n}\left(\hat{a}_\text{in}+\hat{a}_\text{in}^\dagger\right)\to 2\sqrt{\kappa\bar{n}}\bar{n}\check{a}_\text{in},\\ \nonumber
\sqrt{\gamma}\hat{B}_\text{in}&\approx&\sqrt{\kappa\bar{n}}\bar{b}\left(\hat{a}_\text{in}+\hat{a}_\text{in}^\dagger\right)+\sqrt{\Gamma}\bar{n}\hat{b}_\text{in}\to 2\sqrt{\kappa\bar{n}}\bar{b}\check{a}_\text{in}+\sqrt{\Gamma}\bar{n}\hat{b}_\text{in}.
\end{eqnarray} 
\noindent
Here, the real-valued Weiner process $\check{a}_\text{in}(t)$ with the \textit{symmetrized} classical spectral density $\check{a}_\text{in}(\omega)$ is obtained as
\begin{eqnarray}
\label{T8-eq27}
\check{a}_\text{in}(t)&=&\frac{\hat{a}_\text{in}(t)+\hat{a}_\text{in}^\dagger(t)}{2},\\ \nonumber
\check{a}_\text{in}(\omega)&=&\Re[\hat{a}_\text{in}(\omega)].
\end{eqnarray}

While this type of approximations in multiplicative noise could be useful for many cases, there are some phenomena which cannot be reproduced without correct treatment of multiplicative noise. This shall be discussed in details in \S\ref{S2-Noise4}. Nevertheless, it is also instructive the take the expectation values of (\ref{T8-eq21}) to obtain the classical system
\begin{equation}
\label{T8-eq28}
\frac{d}{dt}\left\{\begin{array}{c}
N(t)\\
B(t)
\end{array}\right\}=
\left[
\begin{array}{cc}
-2\kappa & 0 \\
ig_0 & -i\Omega-\frac{\gamma}{2}
\end{array}
\right]
\left\{\begin{array}{c}
N(t)\\
B(t)\\
\end{array}\right\}+2\sqrt{\bar{n}}\left\{\begin{array}{c}
\bar{n}\\
\bar{b}
\end{array}\right\}\Re[\alpha].
\end{equation}
\noindent
Together with (\ref{T8-eq9},\ref{T8-eq12}), and setting the time-derivative on the left of the above to zero, makes the evaluation of steady-state values $N(\infty)=\bar{n}^2$ and $B(\infty)=\overline{nb}\approx\bar{n}\bar{b}$ readily possible. Doing this and solving for $\bar{n}$ and $\bar{b}$ precisely gives back (\ref{T8-eq9}). This not only is in agreement with the equilibrium equation (\ref{T8-eq12}), but also confirms the general finding that the equilibrium intracavity photon population $\bar{n}(\Delta)$ is independent of the coherent phonon population $\bar{m}(\Delta)$. However, the opposite is not correct, and as it was shown in the previous sections, $\bar{m}(\Delta)$ can actually be either determined from $\bar{n}(\Delta)$ and fitting to the experimental data, or directly estimated using the expression (\ref{T8-m8}) in \S\ref{S8-BMWm8}. 

Existence of such an exact transformation which puts the optomechanical interaction into exactly linear form should be connected to the polaron transformation \cite{S2-3} which leaves behind a Kerr nonlinear term as $\hat{n}^2$ in the transformed optomechanical Hamiltonian. It furthermore highlights the fact that usage of higher-order operators ultimately can reach a fully linear system at which convergence of this method to the exact solution is evident.

\subsection{Higher-Order Sidebands}\label{S9-Suppl}

When the optical frequency is much larger than the mechanical frequency, apart from the mechanical sidebands which are roughly placed at $\Delta^{(1)}\approx\pm\Omega$, there exist higher-order sidebands such as $\Delta^{(2)}\approx\pm2\Omega$ and so on. The occurrence of these higher-order sidebands, which are observable for sideband-resolved experiments, is obviously stringent on the existence of two- and multi-phonon processes. Normally, one would expect that these could be studied by constructing the Langevin equations for the operators $\hat{b}^2$, $\hat{b}^{\dagger 2}$ and so on. But this guess turns out to be incorrect, since the corresponding Langevin equations would be totally independent of the one for $\hat{a}$, implying that the second- and higher-order sidebands could not be reconstructed via the fully linearized Langevin equations. This has already been shown to be a nonlinear process which does not naturally appear in the Hamiltonian of the fully linearized optomechanics \cite{S8-Girvin}. But this difficulty can be appropriately addressed by the method of Higher-order Operators, too.

In order to investigate this phenomenon, let us restrict the case only to the second-order sidebands roughly located at $\Delta^{(2)}\approx\pm2\Omega$. In order to study these, it is sufficient to extend the basis $\{A\}^\text{T}=\left\{\hat{a},\hat{a}\hat{b},\hat{a}{\hat{b}}^{\dagger }\right\}$ to 
\begin{equation}
\label{T8-eq54}
\{A\}^{\rm T}=\{\hat{a},\hat{a}\hat{b},\hat{a}\hat{b}^\dagger,\hat{a}\hat{b}^2,\hat{a}\hat{b}^{\dagger 2} \},
\end{equation}
where the third-rank higher-order operators $\hat{a}\hat{b}^2$ and $\hat{a}\hat{b}^{\dagger 2}$ take care of the one-photon two-phonon processes, ultimately leading to formation of second-order sidebands at $\Delta^{(2)}\approx\pm2\Omega$. The Langevin equations for this basis within the zeroth-order approximation of multiplicative noise reads
\begin{eqnarray}
\nonumber
\left[ 
\begin{array}{ccc|cc}
i\Delta-\frac{\kappa }{2} & ig_0 & ig_0 & 0 & 0 \\ 
ig_0(\bar{m}+\bar{n}+1) & i\Delta_B-\frac{\gamma }{2} & 0 & ig_0 & 0 \\ 
ig_0(\bar{m}-\bar{n}) & 0 & i\Delta_R-\frac{\gamma }{2} & 0 & ig_0 \\ \hline
0 & ig_0(\bar{m}+2\bar{n}+2) & 0 & i\Delta_{BB}-\frac{\theta}{2} & 0 \\ 
0 & 0 & ig_0(\bar{m}-2\bar{n}-1) & 0 & i\Delta_{RR}-\frac{\theta}{2}  
\end{array}
\right]\left\{ \begin{array}{c}
\hat{a} \\ 
\hat{a}\hat{b} \\ 
\hat{a}{\hat{b}}^{\dagger } \\ \hline
\hat{a}\hat{b}^2 \\
\hat{a}\hat{b}^{\dagger 2}
\end{array}
\right\}\\ \label{T8-eq55}
-\left[ 
\begin{array}{ccc}
\sqrt{\kappa} & 0 & 0\\
\sqrt{\frac{1}{2}\kappa\bar{m}} &  \sqrt{\Gamma\bar{n}} & 0\\
\sqrt{\frac{1}{2}\kappa\bar{m}} &  0 & \sqrt{\Gamma\bar{n}}\\ \hline
\frac{1}{2}\sqrt{\kappa}\bar{m} & \sqrt{\Gamma\bar{n}\bar{m}} & 0\\
\frac{1}{2}\sqrt{\kappa}\bar{m} & 0 & \sqrt{\Gamma\bar{n}\bar{m}}
\end{array}
\right]
\left\{ \begin{array}{c}
\hat{a}_\text{in} \\ 
\hat{b}_\text{in} \\ 
\hat{b}_\text{in}^\dagger 
\end{array}
\right\}+
\left[ 
\begin{array}{cc}
1 & 0\\
\bar{b} & 0\\
\bar{b}^\ast & 0\\ \hline
\bar{b}^2 & 0\\
\bar{b}^{\ast 2} & 0\\
\end{array}
\right]
\left\{ 
\begin{array}{c}
\alpha \\ 
\alpha^\ast
\end{array}
\right\} 
=\frac{d}{dt}\left\{ \begin{array}{c}
\hat{a} \\ 
\hat{a}\hat{b} \\ 
\hat{a}{\hat{b}}^{\dagger } \\  \hline
\hat{a}\hat{b}^{2} \\ 
\hat{a}\hat{b}^{\dagger 2} 
\end{array}
\right\}.
\end{eqnarray}
\noindent
Here, $\theta=\kappa+2\Gamma$ is the decay rate associated with the third-rank one-photon two-phonon processes $\hat{a}\hat{b}^2$ and $\hat{a}\hat{b}^{\dagger 2}$, $\Delta_{BB}=\Delta-2\Omega$ and $\Delta_{RR}=\Delta+2\Omega$. The approximation $2|\bar{b}|^2\approx\bar{m}$ is used following the discussions in \S\ref{S8-BMWm8}. We do observe that this treatment of multiplicative noise causes non-negligible error in some cases, and is due to be discussed later in \S\ref{S2-Noise4}.

Vertical and horizontal partitions separate single-phonon $\{\hat{a}\hat{b},\hat{a}\hat{b}^\dagger\}$ and two-phonon processes $\{\hat{a}\hat{b}^2,\hat{a}\hat{b}^{\dagger^2}\}$. So by retaining only the first $3\times 3$ blocks and first $3$ rows what remains is nothing but the equations of first-order optomechanics in terms of the second-rank single-phonon operator basis $\{\hat{a},\hat{a}\hat{b},\hat{a}\hat{b}^\dagger\}$. 

The first-order $\delta\Delta^{(1)}$ and second-order $\delta\Delta^{(2)}$ sideband inequivalences take on similar expansions as
\begin{equation}
\label{T8-eq56}
\frac{\delta\Delta^{(1)}}{\Omega}\approx-\frac{\delta\Delta^{(2)}}{2\Omega}\approx \left(\frac{g_0}{\Omega}\right)^2\left(\bar{n}+\frac{1}{2}\right)-2\left(\frac{g_0}{\Omega}\right)^4\left(\bar{n}+\frac{1}{2}\right)\left(\bar{m}+\frac{1}{2}\right) \approx\mathcal{G}^2-4\mathcal{G}^4\mathcal{G}_0^2.
\end{equation}
Here, $\mathcal{G}_0=g_0/\Omega$ and $\mathcal{G}=g/\Omega$ are normalized interaction rates with respect to the mechanical frequency, where $g=g_0\sqrt{\bar{n}}$ is the enhanced optomechanical interaction rate. Furthermore, $\bar{m}$ is approximated from (\ref{T8-m8}) in the above.

Results of noise spectrum calculations using the fully linearized and higher-order formulations of optomechanics is already verified at input powers being high-enough to cause the cavity to exhibit asymmetric reflectivity, a very clear hallmark of bistability seen easily in experiments. This may be calculated for various linear and higher-order formulations resulting from simulating a scanning pump experiment. In practice, the fully-linearized optomechanics cannot reasonably reproduce the highly asymmetric and non-Lorentzian lineshape of cavity under strong pump. As a simple measure of reflectivity, one may use the Langevin equation for photons $\hat{a}$ with the semi-classical substitutions  $\hat{a}\rightarrow\bar{a}$ and $\hat{b}\rightarrow\bar{b}$, where $\bar{b}$ is correspondingly given from (\ref{T8-eq9}), and $\bar{n}$ can be nonlinearly solved from (\ref{T8-eq12}). For a side-coupled cavity where the reflectivity is not identity, and external coupling rate $\kappa_\text{ex}$ is known, we have $\eta=\kappa_\text{ex}/\kappa$, leading to the approximation
\begin{equation}
\mathcal{R}(\omega,\Delta)=1-\frac{i\kappa_\text{ex}}{\omega+\Delta+2g_0^2\bar{n}\frac{\Omega}{\Omega^2+\frac{1}{4}\Gamma^2}+i\frac{1}{2}\kappa},
\end{equation} 
where even more accurate solutions can be found by taking the scattering matrix element $\mathcal{R}=Y_{11}$.

The existence of tiny second-order sidebands around $\pm 2\Omega$ is successfully reproduced in reflectivity near the corresponding resonances using higher-order algebra, too, although the depths of these resonances are rather small. 

\subsection{Multiplicative Noise}\label{S2-Noise4}

Rewriting (\ref{T8-eq55}) without the zeroth-order approximation for multplicative noise, gives the exact higher-order set of Langevin equations 
\begin{eqnarray}
\nonumber
\frac{d}{dt}\left\{ \begin{array}{c}
\hat{a} \\ 
\hat{a}\hat{b} \\ 
\hat{a}{\hat{b}}^{\dagger } \\ 
\hat{a}\hat{b}^{2} \\ 
\hat{a}\hat{b}^{\dagger 2} 
\end{array}
\right\}&=&\left[ 
\begin{array}{ccccc}
i\Delta-\frac{\kappa }{2} & ig_0 & ig_0 & 0 & 0 \\ 
ig_0(\bar{m}+\bar{n}+1) & i\Delta_B-\frac{\gamma }{2} & 0 & ig_0 & 0 \\ 
ig_0(\bar{m}-\bar{n}) & 0 & i\Delta_R-\frac{\gamma }{2} & 0 & ig_0 \\ 
0 & ig_0(\bar{m}+2\bar{n}+2) & 0 & i\Delta_{BB}-\frac{\theta}{2} & 0 \\ 
0 & 0 & ig_0(\bar{m}-2\bar{n}-1) & 0 & i\Delta_{RR}-\frac{\theta}{2}  
\end{array}
\right]
\\ \label{T8-Noise1}
&\times&
\left\{ \begin{array}{c}
\hat{a} \\ 
\hat{a}\hat{b} \\ 
\hat{a}{\hat{b}}^{\dagger } \\ 
\hat{a}\hat{b}^2 \\
\hat{a}\hat{b}^{\dagger 2}
\end{array}
\right\}
-\left[ 
\begin{array}{ccc}
\sqrt{\kappa} & 0 & 0\\
\sqrt{\kappa}\hat{b} &  \sqrt{\Gamma}\hat{a} & 0\\
\sqrt{\kappa}\hat{b}^\dagger &  0 & \sqrt{\Gamma}\hat{a}\\ 
\sqrt{\kappa}\hat{b}^2 & \sqrt{\Gamma}\hat{a}\hat{b} & 0\\
\sqrt{\kappa}\hat{b}^{\dagger 2} & 0 & \sqrt{\Gamma}\hat{a}\hat{b}^\dagger
\end{array}
\right]
\left\{ \begin{array}{c}
\hat{a}_\text{in} \\ 
\hat{b}_\text{in} \\ 
\hat{b}_\text{in}^\dagger 
\end{array}
\right\}+
\left[ 
\begin{array}{cc}
1 & 0\\
\bar{b} & 0\\
\bar{b}^\ast & 0\\ 
\bar{b}^2 & 0\\
\bar{b}^{\ast 2} & 0\\
\end{array}
\right]
\left\{ 
\begin{array}{c}
\alpha \\ 
\alpha^\ast
\end{array}
\right\}.
\end{eqnarray}

To illustrate how the multiplicative noise terms on the second line are to be treated, let us assume that a simple equation is given as 
\begin{equation}
\label{T8-Noise2}
\frac{d}{dt}\mathscr{A}(t)=(i\Delta-\frac{1}{2}\kappa)\mathscr{A}(t)-\sqrt{\kappa}\hat{x}(t)\hat{a}_\text{in}(t),
\end{equation}
\noindent
where $\hat{x}$ is some dimensionless and time-dependent operator and $\hat{a}_\text{in}$ corresponds to a white noise random process. The spectral density of the zero-average operator $\mathscr{A}$ by definition is
\begin{equation}
\label{T8-Noise3}
S_{\mathscr{A}\mathscr{A}}(w)=\int_{-\infty}^{\infty}d\tau e^{iw\tau}\braket{\mathscr{A}^\dagger(t)\mathscr{A}(t+\tau)}.
\end{equation}
\noindent
Therefore, the symmetrized spectral density via symmetrization operator $\mathcal{S}$ which is the actual quantity measured in experiments is
\begin{eqnarray}
\mathcal{S}S_{\mathscr{A}\mathscr{A}}(w)&=&\int_{-\infty}^{\infty}d\tau e^{iw\tau}\braket{\mathcal{S}\{\mathscr{A}^\dagger(t)\mathscr{A}(t+\tau)\}}\\ \nonumber
&=&\int_{-\infty}^{\infty}d\tau e^{iw\tau}\braket{\mathscr{A}^\dagger(t)\mathscr{A}(t+\tau)}_\text{S}.
\end{eqnarray}
The equation (\ref{T8-Noise2}) admits a formal solution
\begin{equation}
\label{T8-Noise4}
\mathscr{A}(t)=-\sqrt{\kappa}\mathbb{L}(t)\hat{x}(t)\hat{a}_\text{in}(t),
\end{equation}
\noindent
where $\mathbb{L}$ is given as
\begin{equation}
\label{T8-Noise5}
\mathbb{L}(t)=\left(\frac{d}{dt}-i\Delta+\frac{1}{2}\kappa\right)^{-1},
\end{equation}
\noindent
and is an operator which can be understood as an inverse Fourier transform such as
\begin{equation}
\label{T8-Noise6}
\mathbb{L}(t)=\mathcal{F}^{-1}\left(\frac{1}{iw-i\Delta+\frac{1}{2}\kappa}\right)(t)=\mathcal{F}^{-1}\{L(w)\}(t).
\end{equation}
Here, we are not interested in an explicit form of $\mathbb{L}$ although it is easy to be evaluated or looked up from table of Fourier transforms.

The formal solution (\ref{T8-Noise4}) gives rise to the symmetrized spectral density
\begin{equation}
\label{T8-Noise7}
\mathcal{S}S_{\mathscr{A}\mathscr{A}}(w)=\kappa\int_{-\infty}^{\infty}d\tau e^{iw\tau}\braket{\hat{a}^\dagger_\text{in}(t)\hat{x}^\dagger(t)\mathbb{L}^\dagger(t)\mathbb{L}(t+\tau)\hat{x}(t+\tau)\hat{a}_\text{in}(t+\tau)}_\text{S}.
\end{equation}
\noindent
Now, we can employ the Isserlis-Wick theorem to decompose the expectation value as \cite{S2-Paper2,S8-SWick1,S8-SWick2,S4-40}
\begin{eqnarray}
\label{T8-Noise8}
\braket{\hat{a}^\dagger_\text{in}(t)\hat{y}^\dagger(t)\hat{y}(t+\tau)\hat{a}_\text{in}(t+\tau)}_\text{S}&=&\braket{\hat{a}^\dagger_\text{in}(t)\hat{a}_\text{in}(t+\tau)}_\text{S}\braket{\hat{y}^\dagger(t)\hat{y}(t+\tau)}_\text{S}\\ \nonumber
&+&\braket{\hat{a}^\dagger_\text{in}(t)\hat{y}^\dagger(t)}_\text{S}\braket{\hat{y}(t+\tau)\hat{a}_\text{in}(t+\tau)}_\text{S}\\ \nonumber
&+&\braket{\hat{a}^\dagger_\text{in}(t)\hat{y}(t+\tau)}_\text{S}\braket{\hat{y}^\dagger(t)\hat{a}_\text{in}(t+\tau)}_\text{S}.
\end{eqnarray}
\noindent
where $\hat{y}(t)=\mathbb{L}(t)\hat{x}(t)$ is adopted for shorthand notation. Since $\hat{a}_\text{in}$ is a white noise Wiener random process, we may expect that to a very good approximation the second and third terms vanish and thus
\begin{equation}
\label{T8-Noise9}
\mathcal{S}S_{\mathscr{A}\mathscr{A}}(w)=\kappa\int_{-\infty}^{\infty}d\tau e^{iw\tau}\braket{\hat{a}^\dagger_\text{in}(t)\hat{a}_\text{in}(t+\tau)}_\text{S}\braket{\hat{y}^\dagger(t)\hat{y}(t+\tau)}_\text{S}.
\end{equation}
The random nature of a Weiner process requires that \cite{S2-6} 
\begin{eqnarray}
\label{T8-Noise10}
\mathcal{S}S_{\mathscr{A}\mathscr{A}}(w)&=&\left|\kappa\int_{-\infty}^{\infty}d\tau e^{iw\tau}\braket{\hat{y}^\dagger(t)\hat{y}(t+\tau)}_\text{S}\right|^2\mathcal{S}S_{AA}(w)\\ \nonumber
&=&\left|L(w)\ast\hat{x}(w)\right|^2\mathcal{S}S_{AA}(w).
\end{eqnarray}
\noindent
Here, $\ast$ represents convolution in frequency and $L(w)$ is defined in (\ref{T8-Noise6}) and actually represents the equivalent to the scattering matrix element. The expression (\ref{T8-Noise10}) presents a mathematically exact solution to the spectral density problem of multiplicative noise (\ref{T8-Noise2}).

In the context of higher-order quantum optomechanics and referring to (\ref{T8-Noise1}) the operator $\hat{x}$ may represent either of the operators within the set $\{\hat{a},\hat{b},\hat{b}^\dagger,\hat{a}\hat{b},\hat{a}\hat{b}^\dagger,\hat{b}^2,\hat{b}^{\dagger 2}\}$. However, not only these are not yet known, but also, they are influenced by random processes from the correspondingly lower-order interactions with the optical field and mechanical bath. The only approximation needed here is to replace these with corresponding non-operator functions which can be already obtained from the solution to lower-order equations. Doing this results in a set of equations for $\{\bar{a},\bar{b},\bar{b}^\ast,\overline{ab},\overline{ab^\ast},\bar{b}^2,\bar{b}^{\ast 2}\}$, where solutions for $\{\bar{b},\bar{b}^\ast,\bar{b}^2,\bar{b}^{\ast 2}\}$ can be obtained by having $\bar{b}$. This is here calculated from the $3\times 3$ linearized optomechanics, giving rise to the expression 
\begin{equation}
\label{T8-Noise11}
\bar{b}(w)=\frac{\alpha }{i (w+\Omega)+\frac{1}{2}\Gamma}.
\end{equation}
In a similar manner to \S\ref{S8-BMWm8}, the next required expressions can be explicitly obtained  by ${\sf Mathematica}$ as
\begin{eqnarray}
\nonumber
\bar{a}(w)&=&\frac{-\alpha ^2 g_0 (w -\Delta -\Omega-i \frac{1}{2}\gamma)}{(w+\Omega-\frac{1}{2}i \Gamma) \left\{2 g_0^2\left[ (\Delta-w+\frac{1}{2} i \gamma)  \tilde{m}+\Omega\tilde{n}\right]+(w-\Delta -\frac{1}{2}i \kappa) \left[\left(w-\Delta -\frac{1}{2}i \gamma \right)^2-\Omega ^2\right]\right\}}, \\ \nonumber
\overline{ab}(w)&=&\frac{\alpha ^2 \left[g_0^2 (\bar{m}-\bar{n})-(w-\Delta -\frac{1}{2}i \kappa) (w -\Delta -\Omega -\frac{1}{2}i \gamma)\right]}{(w+\Omega-\frac{1}{2}i \Gamma) \left\{2 g_0^2 \left[(\Delta-w+\frac{1}{2}i \gamma)\tilde{m}+\Omega\tilde{n}\right]+(w-\Delta -\frac{1}{2}i \kappa) \left[\left(w-\Delta -\frac{1}{2}i \gamma \right)^2-\Omega ^2\right]\right\}}, \\ \nonumber
\overline{ab^\ast}(w)&=&\frac{-\left| \alpha \right| ^2 \left[g_0^2 (\bar{m}+\bar{n}+1)-(w-\Delta -\frac{1}{2}i \kappa) (w-\Delta+\Omega-\frac{1}{2}i \gamma)\right]}{(w+\Omega+\frac{1}{2}i \Gamma) \left\{2 g_0^2 \left[(\Delta-w+\frac{1}{2}i \gamma)\tilde{m}+\Omega\tilde{n}\right]+(w-\Delta -\frac{1}{2}i \kappa) \left[\left(w-\Delta -\frac{1}{2}i \gamma \right)^2-\Omega ^2\right]\right\}},
\\
\label{T8-Noise12}
\end{eqnarray}
\noindent
where $\tilde{n}=\bar{n}+\frac{1}{2}$ and $\tilde{m}=\bar{m}+\frac{1}{2}$.
In the above equations, it has to be noticed that $\alpha$ is a complex quantity which satisfies $|\alpha|=\sqrt{\kappa_\text{ex}P_\text{op}/\hbar\omega}$, and also by (\ref{T8-eq11}) we have
\begin{equation}
\label{T8-Noise13}
\alpha=\sqrt{\bar{n}}\left[-\frac{\kappa}{2}+i\left(\Delta+\frac{2g_0^2\Omega}{\Omega^2+\frac{1}{4}\Gamma^2}\bar{n}\right)\right].
\end{equation}

Now, we can rewrite the Langevin equations (\ref{T8-Noise1}) as
\begin{eqnarray}
\label{T8-Noise14}
\frac{d}{dt}\{\delta A\}&=&[\textbf{M}]\{\delta A\}-[\hat{\textbf{G}}]\{ A_\text{in}\},\\ \nonumber
[\textbf{M}]&=&\left[ 
\begin{array}{ccccc}
i\Delta-\frac{\kappa }{2} & ig_0 & ig_0 & 0 & 0 \\ 
ig_0(\bar{m}+\bar{n}+1) & i\Delta_B-\frac{\gamma }{2} & 0 & ig_0 & 0 \\ 
ig_0(\bar{m}-\bar{n}) & 0 & i\Delta_R-\frac{\gamma }{2} & 0 & ig_0 \\ 
0 & ig_0(\bar{m}+2\bar{n}+2) & 0 & i\Delta_{BB}-\frac{\theta}{2} & 0 \\ 
0 & 0 & ig_0(\bar{m}-2\bar{n}-1) & 0 & i\Delta_{RR}-\frac{\theta}{2}  
\end{array}
\right], \\ \nonumber
\{\delta A\}^\text{T}&=&\left\{\delta\hat{a},\delta(\hat{a}\hat{b}),\delta(\hat{a}{\hat{b}}^{\dagger }),\delta(\hat{a}\hat{b}^{2}),\delta(\hat{a}\hat{b}^{\dagger 2})\right\},\\ \nonumber
\{ A_\text{in}\}^\text{T}&=&\left\{\hat{a}_\text{in},\hat{b}_\text{in},\hat{b}_\text{in}^\dagger\right\}, \\ \nonumber
[\hat{\textbf{G}}]&=&\left[ 
\begin{array}{ccc}
\sqrt{\kappa} & 0 & 0\\
\sqrt{\kappa}\hat{b} &  \sqrt{\Gamma}\hat{a} & 0\\
\sqrt{\kappa}\hat{b}^\dagger &  0 & \sqrt{\Gamma}\hat{a}\\ 
\sqrt{\kappa}\hat{b}^2 & \sqrt{\Gamma}\hat{a}\hat{b} & 0\\
\sqrt{\kappa}\hat{b}^{\dagger 2} & 0 & \sqrt{\Gamma}\hat{a}\hat{b}^\dagger
\end{array}
\right].
\end{eqnarray}
After defining the decay matrix
\begin{equation}
\label{T8-Noise15}
[\sqrt{\Gamma}]=\left[\begin{array}{ccccc}
\sqrt{\kappa} & 0 & 0 & 0 & 0\\
0 & \sqrt{\gamma} & 0 & 0 & 0\\
0 & 0 & \sqrt{\gamma} & 0 & 0\\
0 & 0 & 0 & \sqrt{\theta} & 0\\
0 & 0 & 0 & 0 & \sqrt{\theta}
\end{array}\right],
\end{equation}
\noindent
taking the Fourier transform, and using the input-output relation 
\begin{equation}
\label{T8-Noise16}
\{A_\text{out}(\omega)\}=\{A_\text{in}(\omega)\}+[\sqrt{\Gamma}]^\text{T}\{\delta A(\omega)\},
\end{equation}
\noindent
we arrive at the definition of the scattering matrix
\begin{equation}
\label{T8-Noise17}
[\textbf{Y}(\omega)]=\textbf{I}-[\sqrt{\Gamma}]^\text{T}\left([\textbf{M}]-i\omega[\textbf{I}]\right)^{-1}[\sqrt{\Gamma}],
\end{equation}
by which and (\ref{T8-Noise10}) we can evaluate the desired symmetrized spectral density of output optical field as
\begin{eqnarray}
\label{T8-Noise18}
S(\omega)&=&|Y_{11}(\omega)|^2 S_{AA}(\omega)\\ \nonumber
&+&\frac{1}{\gamma^2}\left|\left[Y_{12}(\omega)+Y_{13}(\omega)\right]\ast\bar{a}(\omega)\right|^2 S_{BB}(\omega) \\ \nonumber
&+&\frac{1}{\theta^2}\left|\left[Y_{14}(\omega)\ast\overline{ab}(\omega)+Y_{15}(\omega)\ast\overline{ab^\ast}(\omega)\right]\right|^2 S_{BB}(\omega),
\end{eqnarray}
where spectral power densities $S_{AA}$ and $S_{BB}$ are already introduced in (\ref{T8-eq13}) and convolutions $\ast$ take place over the entire frequency axis. 

\subsection{Elements of Higher-order Scattering Matrices}\label{S11-Suppl}
This section reports the explicit elements of the first row of scattering matrix $[\textbf{Y}]$ in (\ref{T8-Noise17}), as needed for calculation of the spectral density according to (\ref{T8-Noise18}). These might be useful only when the method of residues are to be used for exact evaluation of complex convolution integrals, otherwise full numerical simulation of (\ref{T8-Noise18}) is much preferable.  

The elements of the scattering matrix are explicitly found using ${\sf Mathematica}$, and after some simplification they take the form
\begin{eqnarray}
Y_{11}(\omega)&=&1-\frac{2i\kappa_\text{ex}\left[\left(\omega-\Delta -\frac{1}{2}i \gamma \right)^2-\Omega ^2\right]}{2 g_0^2\left[ (\Delta-\omega+\frac{1}{2} i \gamma)  \tilde{m}+\Omega\tilde{n}\right]+(\omega-\Delta -\frac{1}{2}i \kappa) \left[\left(\omega-\Delta -\frac{1}{2}i \gamma \right)^2-\Omega ^2\right]}, \\ \nonumber
Y_{12}(\omega)&=&\frac{-ig_0\sqrt{\gamma\kappa_\text{ex}}\left(\omega-\Delta-\Omega -\frac{1}{2}i \gamma \right)}{2 g_0^2\left[ (\Delta-\omega+\frac{1}{2} i \gamma)  \tilde{m}+\Omega\tilde{n}\right]+(\omega-\Delta -\frac{1}{2}i \kappa) \left[\left(\omega-\Delta -\frac{1}{2}i \gamma \right)^2-\Omega ^2\right]}, \\ \nonumber
Y_{13}(\omega)&=&\frac{-ig_0\sqrt{\gamma\kappa_\text{ex}}\left(\omega-\Delta+\Omega -\frac{1}{2}i \gamma \right)}{2 g_0^2\left[ (\Delta-\omega+\frac{1}{2} i \gamma)  \tilde{m}+\Omega\tilde{n}\right]+(\omega-\Delta -\frac{1}{2}i \kappa) \left[\left(\omega-\Delta -\frac{1}{2}i \gamma \right)^2-\Omega ^2\right]}.
\end{eqnarray}
These expressions are useful in speed up of the code, as well as wherever the method of residues is to be used.

The convergence of $3\times 3$ is sufficiently good for most practical purposes, and also the explicit expressions for $5\times 5$ matrices decompose into products of fourth-order polynomials in terms of $\omega$ in their denominators, which severely limits the usefulness of applicability of the method of residues. For this reason, their explicit expressions are not included here. 

Summarizing this section, a new analytical method was shown to solve the standard optomechanical interaction with cubic nonlinearity interaction, based on the higher-order operators. It was demonstrated that not only the higher-order operator method can reproduce the linear optomechanics, but also it can predict and provide estimates to unnoticed effects such as a new type of symmetry breaking in frequency, here referred to as side-band inequivalence, and yield new explicit expressions for quantities such as the coherent phonon population and higher-order spring effect. Corrections to the standard spring effect due to higher-order interactions have been found, and it has been shown that such corrections arise mainly because of the coherent phonons and can significantly influence measurement of single-photon optomechanical interaction rate through spring effect. A minimal basis has been defined which allows exact and explicit solution to standard nonlinear optomechanics, using the method of higher-order operators. This method can be finally used to investigate the dynamic nonlinear stability of optomechanical systems, and it has been demonstrated that at least the third-order nonlinear processes are prerequisite for occurrence of dynamic instability. We have shown that there is a reasonable correspondence between the onset of nonlinear dynamic instability and a critical intracavity photon number limit, which remains independent of thermal effects.



\section{Side-band Inequivalence} \label{Section-9}
The goal of this section is to revisit the side-band inequivalence obtained in the previous section within the weakly coupled limit, and extend it into a more general form. Frequency shifts of red- and blue-scattered (Stokes/anti-Stokes) side-bands in quantum optomechanics are shown to be counter-intuitively inequal, resulting in an unexpected symmetry breaking. This difference is referred to as side-band inequivalence, which normally leans towards red, and being a nonlinear effect it depends on optical power or intracavity photon number. Also there exists a maximum attainable side-band inequivalence at an optimal operation point. The mathematical method employed here is a combination of operator algebra equipped with harmonic balance, which allows a clear understanding of the associated nonlinear process. This reveals the existence of three distinct operation regimes in terms of pump power, two of which have immeasurably small side-band inequivalence. Compelling evidence from various experiments sharing similar interaction Hamiltonians, including quantum optomechanics, ion/Paul traps, electrooptic modulation, Brillouin scattering, and Raman scattering unambiguously confirm existence of a previously unnoticed side-band inequivalence.

In the weak coupling approximation, side-band inequivalence exhibits linear increase with intracavity photon number, and this behavior has to change at sufficiently high pump levels or sufficiently strong nonlinearity. In order to accomplish this, one needs to do a theoretical analysis from a different perspective.

So far, we have analyzed the phenomenon of side-band inequivalence through a different theoretical means, and together with the analysis to be presented in this section, we manage to obtain side-band inequivalence through three essentially different theoretical tools. These are
\begin{enumerate}
	\item Comparison of asymmetric shifts in dynamic eigenvalues corresponding to the generation of side-bands, which was done in \S\ref{S4-Suppl}. This analysis breaks down in the strong nonlinear regime.
	\item Harmonic Higher-order Operator analysis of quantum optomechanical Langevin equations. The method presented here in \S\ref{S4-SI} though mathematically not being exact, it is still able to recover the weakly nonlinear behavior and also reasonably explain the strongly nonlinear behavior. 
	\item Study of Akhmediev breather analytical solution to classical optomechanical nonlinear equations. This results in an approximate soliton-like solutions to the system discussed in \S\ref{Akhmediev}, which of course do encompass side-band inequivalence.
\end{enumerate} 
In all these very different three types of analyses, two things are noticeably common. Firstly, side-band inequivalence is always towards the red or Stokes side-band. Secondly, in the weakly nonlinear approximation it increases almost linearly with enhanced optomechanical interaction rate $g$, which is the product of the square-root of intracavity photon population $\sqrt{\bar{n}}$ and single-photon optomechanical interaction rate $g_0$.

It is fairly straightforward to verify the existence of side-band inequivalence at least in the strongly nonlinear regime through numerical analysis of optomechanical equations, too. This can be done either through study of time-domain behavior of classical optomechanical equations for electromagnetic and mechanical amplitudes, or solving the quantum master equation which is much more reliable from a fundamental point of view, but demands excessive computational hardware. Surprisingly or not, side-band inequivalence shows up in either way.  

The nonlinearity of optomechanical interaction \cite{S2-3,S2-1,S2-2,S2-4,S2-6,S9-0f,S9-0g,S9-0h,S9-0i} causes scattering of incident photons with the annihilator $\hat{a}$ from the cavity unto either red- or blue-shifted photons through annihilation or generation of a cavity phonon with the annihilator $\hat{b}$, giving rise to the first-order mechanical side-bands. Taking the optical frequency $\omega$ to be at a detuning $\Delta=\omega_{\rm c}-\omega$ from cavity resonance $\omega_{\rm c}$, the $\nu-$th order sidebands are naturally expected to occur at the detunings $\Delta_{\pm\nu}=\Delta\mp \nu\Omega$, where $\Omega$ represents the mechanical frequency. As a results, the first-order mechanical side-bands of scattered red $\Delta_{+1}$ and blue $\Delta_{-1}$ processes must average out back to the original pump detuning $\Delta$. 

Defining the side-band inequivalence as the deviation of this average from $\Delta$, as $\delta=(\Delta_{+1}+\Delta_{-1})-2\Delta$, then one may conclude $\delta=0$. A non-zero $\delta$ would have otherwise implied the so-called side-band inequivalence. This type of asymmetry appears to have a classical nonlinear nature.

There is, however, another well-known type of side-band asymmetry between the red and blue side-bands in the context of optomechanics, which has a quantum nature and may be used for instance to accurately determine the absolute temperature through a reference-free optomechanical measurement \cite{S9-Purdy1,S9-Purdy2}. This is based on the ratio of Stokes to anti-Stokes Raman transition rates, which is equal to $\exp(\hbar\Omega/k_{\rm B}T)$ where $k_{\rm B}$ is the Boltzmann's constant and $T$ is the absolute temperature \cite{S9-asym1,S9-asym2}. Clearly, side-band inequivalence is quite different from this type of side-band asymmetry.

While both time-reversal symmetry and energy conservation are fundamentally preserved in this scattering process, a nonlinear analysis of quantum optomechanics using the recently developed method of higher-order operators \cite{S2-Paper3,S2-Paper1,S2-Paper2,S2-Paper4,S2-Paper5} necessitates a slight difference among detunings of blue and red-scattered photons, the amount of which was initially found to increase roughly proportional to the intracavity photon number $\bar{n}$. Here, $\bar{n}$ is defined as the steady-state mean-value of the number operator $\hat{n}=\hat{a}^\dagger\hat{a}$. 

Surprisingly enough, this disagreement satisfying $\delta\neq 0$ does not violate the energy conservation law, actually allowed by the finite cavity linewidth as well as the single-photon/single-phonon nature of the process involved. Moreover, the time-reversal symmetry is also preserved.

Among the pool of available experimental data, only a handful of side-band resolved cavities reveal this disagreement \cite{S2-Paper3}. Some initial trial experiments recently done at extremely high intracavity photon numbers $\bar{n}$, and/or extremely large single-photon optomechanical interaction rates $g_0$, though, failed to demonstrate its existence. This may raise the speculation that whether side-band inequivalence would have been merely a mathematical artifact, or something has been missing due to not doing the operator analysis to the highest-order.

A careful analysis of this phenomenon, however, confirms the latter, thus classifying the quantum optomechanical interaction into three distinct regimes with different behaviors:
\begin{itemize}
	\setlength\itemsep{-0.4em}
	\item Fully Linear: This regime can be investigated using the lowest-order analysis and first-order operators, which is conventionally done by linearizing the Hamiltonian around equilibrium points. This will require the four-dimensional basis of first-order ladder operators $\{\hat{a},\hat{a}^\dagger,\hat{b},\hat{b}^\dagger\}$ and is indeed quite sufficient to understand many of the complex quantum optomechanical phenomena \cite{S2-Paper6}.
	\item Weakly Nonlinear: This regime requires higher-order operator analysis of at least second-order. This can be done using the three-dimensional reduced basis \cite{S2-Paper3,S2-Paper6} given as $\{\hat{a},\hat{a}\hat{b},\hat{a}\hat{b}^\dagger\}$.
	\item Strongly Nonlinear: Full understanding of this interaction regime requires the highest-order analysis using third-order operators. Referred as to the minimal basis \cite{S2-Paper3,S2-Paper6}, the convenient reduced choice is the two-dimensional basis $\{\hat{n}^2,\hat{n}\hat{b}\}$. While most of the quantum optomechanical experiments happen to fall in this regime, the striking behavior of governing equations is in such a way that a fully-linearized analysis of fluctuations mostly happens to work.
\end{itemize}

side-band inequivalence is essentially forbidden in the fully linear regime, and it also quickly fades away in the strongly nonlinear regime. But it may only happen in the weakly nonlinear regime. This is now also confirmed both by the higher-order operator method and extensive calculations. It typically does not exceed one part in million to one part in ten thousand in optomechanics, and therefore, it is a very delicate phenomenon and elusive to observe. As it will be shown, Raman scattering experiments may instead exhibit much stronger inequivalences.

This letter provides a direct route towards clear understanding of this complex nonlinear phenomenon. Using a combination of operator algebra and harmonic balance (used in analysis of laser diodes) \cite{S9-Cabon}, we obtain a closed form expression for side-band inequivalence $\delta$ as a function of intracavity photon number $\bar{n}$, which is expected to be valid through all three above operation regimes, and for any arbitrarily chosen set of optomechanical parameters. Not only the findings of this work reproduce the approximate linear expression found earlier through second-order operators \cite{S2-Paper3} for the weakly nonlinear regime, but also we can show that there is an optimal point at which the side-band inequivalence attains a maximum. Moving away from the optimal point, both at the much smaller and much larger pump rates, $\delta$ attains much smaller values, tending to zero in the limit of very large $\bar{n}$. 

This will be greatly helpful to designate the investigation range of experimental parameters given any available optomechanical cavity. Furthermore, it marks a clear and definable border among the three above-mentioned operation regimes. 

As further unexpected results, we show a closely associated symmetry breaking in particle pair generation as the time-reversed optomechanics. Even more unexpectedly, we can demonstrate that the Stokes/anti-Stokes ratio as a nonlinear quantity also is a function of the nonlinearity strength. Hence, it can change in a nonlinear way and does not necessarily require the quantum mechanical and Bose-Einstein statistics to exceed unity. This phenomenon can very well mask the quantum mechanical effect in the strong nonlinear regime. Therefore, the accuracy of optomechanical or Raman thermometry should be limited to the weakly nonlinear optomechanical systems.

\subsection{Harmonic Higher-order Operator Method}\label{S4-SI}

The analysis of side-band inequivalence proceeds with considering the behavior of optomechanical cavity under steady-state conditions. We will focus only on the first-order side-bands and discard all other contributions coming from or to the second- and higher-order side-bands. We consider a single-frequency pump with ideally zero linewidth at a given detuning $\Delta$, which normally gives rise to two stable blue- and red- side-bands. Hence, the time-dependence of the photon annihilator will look like
\begin{equation}
\label{T9-eq1}
\hat{a}(t)=\hat{a}_0 e^{i\Delta t}+\hat{a}_b e^{i(\Delta-\Omega+\frac{1}{2}\delta)t}+\hat{a}_r e^{i(\Delta+\Omega+\frac{1}{2}\delta)t}+\cdots,
\end{equation}
\noindent
where $\hat{a}_0$, $\hat{a}_b$, and $\hat{a}_r$ respectively correspond to the central excitation resonance at pump frequency, and blue- and red-detuned side-bands. The steady-state time-average of central excitation satisfies $\braket{\hat{a}_0}=\sqrt{\bar{n}}$, where $\bar{n}$ can be determined by solution of a third-order algebraic equation once the optical power pump rate $P_{\rm op}$, detuning $\Delta$, external coupling $\eta$ and all other optomechanical parameters are known. The standard set of basic optomechanical parameters needed here are mechanical frequency $\Omega$, optical decay rate $\kappa$, and mechanical decay rate $\Gamma$. Therefore, the photon number operator up to the first side-bands will behave as
\begin{eqnarray}
\label{T9-eq2}
\hat{n}(t)&=&\hat{a}_0^\dagger\hat{a}_0+\hat{a}_b^\dagger\hat{a}_b+\hat{a}_r^\dagger\hat{a}_r\\ \nonumber
&+&\hat{a}_b^\dagger\hat{a}_0 e^{-i(-\Omega+\frac{1}{2}\delta)t}+\hat{a}_0^\dagger\hat{a}_b e^{i(-\Omega+\frac{1}{2}\delta)t}+\hat{a}_r^\dagger\hat{a}_0 e^{-i(\Omega+\frac{1}{2}\delta)t}+\hat{a}_0^\dagger\hat{a}_r e^{i(\Omega+\frac{1}{2}\delta)t}+\cdots,
\end{eqnarray}
\noindent
while the mechanical annihilator will exhibit a closely spaced doublet around the mechanical frequency spaced within $\delta$ as
\begin{equation}
\label{T9-eq3}
\hat{b}(t)=\hat{b}_0+\hat{b}_b e^{-i(\Omega-\frac{1}{2}\delta)t}+\hat{b}_r e^{-i(\Omega+\frac{1}{2}\delta)t}+\cdots .
\end{equation}
Here, the average mechanical displacement satisfies
\begin{equation}
\label{T9-eq4}
b_0=\braket{\hat{b}_0}=\frac{i g_0 \bar{n}}{i\Omega+\frac{1}{2}\Gamma}.
\end{equation} 
Now, let us get back to the Langevin equation for mechanical motions, which simply is
\begin{equation}
\frac{d}{dt}\hat{b}(t)=(-i\Omega-\frac{1}{2}\Gamma)\hat{b}(t)+i g_0\hat{n}(t)+\sqrt{\Gamma}\hat{b}_{\rm in}(t),
\end{equation}
\noindent
where $\hat{b}_{\rm in}(t)$ is the operator for mechanical fluctuations. For the purpose of our analysis here, all fluctuations can be discarded since they are irrelevant to the formation of side-band frequencies and average out to zero. Using (\ref{T9-eq2}) and (\ref{T9-eq3}) we get
\begin{eqnarray}
\label{T9-eq6}
&&-i(\Omega+\frac{\delta}{2})\hat{b}_r e^{-i(\Omega+\frac{\delta}{2})t}-i(\Omega-\frac{\delta}{2})\hat{b}_b e^{-i(\Omega-\frac{\delta}{2})t} \\ \nonumber
&&\approx-(i\Omega+\frac{\Gamma}{2})\hat{b}_r e^{-i(\Omega+\frac{\delta}{2})t}-(i\Omega+\frac{\Gamma}{2})\hat{b}_b e^{-i(\Omega-\frac{\delta}{2})t}+i g_0 \hat{a}_0^\dagger\hat{a}_b e^{-i(\Omega-\frac{\delta}{2})t}+i g_0 \hat{a}_r^\dagger\hat{a}_0 e^{-i(\Omega+\frac{\delta}{2})t}+\cdots .
\end{eqnarray}
\noindent
From the above, we obtain two key operator equations
\begin{eqnarray}
\label{T9-eq7}
\hat{b}_r&=&\frac{i 2 g_0}{-i\delta+\Gamma}\hat{a}_r^\dagger\hat{a}_0, \\ \nonumber
\hat{b}_b&=&\frac{i 2 g_0}{i\delta+\Gamma}\hat{a}_0^\dagger\hat{a}_b.
\end{eqnarray}

In a similar manner, the Langevin equation for the photon annihilator is
\begin{equation}
\label{T9-eq8}
\frac{d}{dt}\hat{a}(t)=(i\Delta-\frac{1}{2}\kappa)\hat{a}(t)+ig_0\hat{a}(t)[\hat{b}(t)+\hat{b}^\dagger(t)]+\sqrt{\kappa}\hat{a}_{\rm in}.
\end{equation}
\noindent
Using (\ref{T9-eq1}) and (\ref{T9-eq3}) we obtain
\begin{eqnarray}
\label{T9-eq9}
&&i\Delta\hat{a}_0 e^{i\Delta t}+i(\Delta-\Omega+\frac{\delta}{2})\hat{a}_b e^{i(\Delta-\Omega+\frac{\delta}{2})t}+i(\Delta+\Omega+\frac{\delta}{2})\hat{a}_r e^{i(\Delta+\Omega+\frac{\delta}{2})t}\approx \\ \nonumber
&&(i\Delta-\frac{\kappa}{2})\left[\hat{a}_0 e^{i\Delta t}+\hat{a}_b e^{i(\Delta-\Omega+\frac{\delta}{2})t}+\hat{a}_r e^{i(\Delta+\Omega+\frac{\delta}{2})t}\right] \\ \nonumber
&&+i g_0\left[\hat{a}_0 e^{i\Delta t}+\hat{a}_b e^{i(\Delta-\Omega+\frac{\delta}{2})t}+\hat{a}_r e^{i(\Delta+\Omega+\frac{\delta}{2})t}\right]\times \\ \nonumber
&&\hspace{8mm} \left[\hat{x}_0+\hat{b}_r e^{-i(\Omega+\frac{\delta}{2})t}+\hat{b}_b e^{-i(\Omega-\frac{\delta}{2})t}+\hat{b}_r^\dagger e^{i(\Omega+\frac{\delta}{2})t}+\hat{b}_b^\dagger e^{i(\Omega-\frac{\delta}{2})t}\right].
\end{eqnarray}
\noindent
where $\hat{x}_0=\hat{b}_0+\hat{b}_0^\dagger$. This will yield the further operator equations as
\begin{eqnarray}
\label{T9-eq10}
\frac{\kappa}{2i g_0}\hat{a}_0&=& \hat{a}_0\hat{x}_0+\hat{a}_b\hat{b}_b^\dagger+\hat{a}_r\hat{b}_r, \\ \nonumber
\left[\frac{i(-\Omega+\frac{\delta}{2})+\frac{\kappa}{2}}{ig_0}\right]\hat{a}_b&=&\hat{a}_b\hat{x}_0+\hat{a}_0\hat{b}_b, \\ \nonumber
\left[\frac{i(\Omega+\frac{\delta}{2})+\frac{\kappa}{2}}{ig_0}\right]\hat{a}_r&=&\hat{a}_r\hat{x}_0+\hat{a}_0\hat{b}_r^\dagger.
\end{eqnarray}

Now, substituting whatever we have in hand in the second equation of (\ref{T9-eq10}), and taking expectation values at the end, we obtain a key algebraic equation in terms of $\delta$ as
\begin{equation}
\label{T9-eq11}
i\left(-\Omega+\frac{1}{2}\delta\right)+\frac{1}{2}\kappa=ig_0 x_0+ig_0\sqrt{\bar{n}}\frac{2ig_0\sqrt{\bar{n}}}{i\delta+\Gamma}.
\end{equation}
\noindent
with $x_0=b_0+b_0^\ast$. Rearrangement of the above gives rise to the equation
\begin{equation}
\label{T9-eq12}
\delta^2-\left[2\Omega+i\gamma+2g_0 x_0\right]\delta+\left[(2i\Omega-\kappa)\Gamma-4g_0^2\bar{n}+2i\Gamma g_0 x_0\right]=0,
\end{equation}
in which $x_0=b_0+b_0^\ast$ and $\gamma=\kappa+\Gamma$ is the total optomechanical decay rate \cite{S2-Paper3,S2-Paper2}.

This approximate nature of this equation will yield complex values for $\delta$ the imaginary value of which has to be discarded. Furthermore, it leaves room to ignore the square terms $\delta^2$, to admit the solution
\begin{equation}
\label{T9-eq13}
\delta=\Re\left[\frac{(2i\Omega-\kappa)\Gamma-4g_0^2\bar{n}+2i\Gamma g_0 x_0}{2\Omega+i\gamma+2g_0 x_0}\right].
\end{equation}  

This solution can be put into the more convenient form using (\ref{T9-eq4}) and further simplification as
\begin{eqnarray}
\label{T9-eq14}
\delta(\bar{n})&=&\Re\left[\frac{A+B\bar{n}}{C-i D\bar{n}}\right] \\ \nonumber
&=&\frac{\Re[A C^\ast]+(B\Re[C]-\Im[A]D)\bar{n}}{|C|^2 -2\Im[C]D\bar{n}+D^2 \bar{n}^2}\\ \nonumber
&=&\frac{[2\Gamma^2\Omega]+2\Omega(B-\Gamma D)\bar{n}}{|C|^2 -4\Omega D\bar{n}+D^2 \bar{n}^2},
\end{eqnarray}
where
\begin{eqnarray}
\label{T9-eq15}
A&=&\Gamma(2i\Omega-\kappa), \\ \nonumber
B&=&\frac{4g_0^2(\Omega-\frac{1}{2}\Gamma)^2}{\Omega^2+\frac{1}{4}\Gamma^2}=B^\ast, \\ \nonumber
C&=&i\gamma+2\Omega, \\ \nonumber
D&=&\frac{4g_0^2\Omega}{\Omega^2+\frac{1}{4}\Gamma^2}=D^\ast.
\end{eqnarray}

One should keep in mind that the expansions (\ref{T9-eq1}) and (\ref{T9-eq2}) are not mathematically exact. It just has been tried here to keep the physically relevant terms and ignore the rest. It is believed that there should be infinitely and increasingly less significant doublets at multiples of $\delta$ around the mechanical frequency in (\ref{T9-eq3}), since they can be easily seen in numerical simulations for unphysically large $g_0$.

So, there are no chances for existence of an exact and explicit mathematical expression for side-band inequivalence. However, it can be noticed that both of the second and third equations of (\ref{T9-eq10}) should be satisfied and able to sit on equal footings only if we take optical spring effect into account as well. In order to do this, one needs to modify (\ref{T9-eq3}) with doublets placed asymmetrically as $\Omega+\delta_1$ and $\Omega-\delta_2$. Then the average of $\delta_1$ and $\delta_2$ will yield the side-band inequivalence, while their difference will yield the optical spring shift.

The reason this has been ignored here is three fold:
\begin{enumerate}
	\item  This will lead to third-order algebraic equations for $\delta$ instead of the quadratic (\ref{T9-eq12}), which is a lot more difficult to solve with useless results. Since both the quadratic and algebraic equations will be nevertheless approximate and eventual expressions for roots will be large physically uninteresting to the researchers, which remain still inexact.
	\item  The optical spring effect is well studied and is known to vanish at resonant pump, while side-band inequiavlence does not. Hence, one may assume a resonant pump and simply ignore the spring effect while it will leave undoubtedly some error in the ultimate result. The same arguments go for the imaginary part of $\delta$ which has been ignored, too.
	\item The assumption for existence of a non-zero side-band inequivalence term $\delta$ in (\ref{T9-eq1}), (\ref{T9-eq2}), and (\ref{T9-eq3}) is supported by the fact that a non-zero $\delta$ value is actually admissible by the governing equations although it seems not possible to obtain an exact explicit solution.
\end{enumerate}

The expression (\ref{T9-eq14}) obtained for the side-band inequivalence has interesting properties at the limits of zero and infinite intracavity photon number. We may obtain here after some simplification easily the asymptotic expressions
\begin{eqnarray}
\label{T9-eq16}
\lim\limits_{\bar{n}\rightarrow\infty} \delta(\bar{n})&\sim&\frac{\Omega}{\beta\bar{n}}, \\ \nonumber
\lim\limits_{\bar{n}\rightarrow 0} \delta(\bar{n})&\sim&\frac{2\Gamma^2\Omega}{4\Omega^2+\gamma^2}\approx 0,
\end{eqnarray}
where $\beta=2g_0^2/\Omega^2$, while noting that $\Gamma<<\Omega$ and also for a side-band resolved cavity $\kappa<<\Omega$, together which we have $\kappa<\gamma<<\Omega$. 

One should take into account the fact that for Doppler cavities, side-bands normally resolve well enough for a decisive measurement \cite{S2-Paper6}, and the concept of side-band inequivalence is only practically meaningful for side-band resolved cavities. Therefore, the following approximations are valid
\begin{eqnarray}
\label{T9-eq17}
A&\approx&2i\Gamma\Omega, \\ \nonumber
B&\approx&4g_0^2, \\ \nonumber
C&\approx&2\Omega, \\ \nonumber
D&\approx&\frac{4g_0^2}{\Omega},\\ \nonumber
\delta(\bar{n})&\approx&\frac{2\Gamma^2\Omega+8g_0^2\Omega\bar{n}}{\gamma^2+4\Omega^2\left[1-2\left(g_0/\Omega\right)^2\bar{n}\right]^2}. 
\end{eqnarray}

Any final result in any sense if given in an explicit expression in this context will be approximate. However, the obtained expression (\ref{T9-eq17}) above is able to explain two operation regimes: When it is considered a function of intracavity photon number $\bar{n}$, and when it is considered as a function of $\Omega$. The limit of large $\bar{n}$ may not be experimentally feasible to study, but surprisingly enough, the limit of large $\Omega$ is easily accessible. For this to happen, one may notice that  $\Omega$ represents the modulation frequency which is although not tunable in optomechanical cavities as the frequency of mechanical mode, it actually is a controllable quantity in electrooptic and acoustooptic modulators.

As long as satisfies $\bar{n}<<2|\Im[C]|/D=\Omega^2/g_0^2$, then second order term $\bar{n}^2$ in the denominator of (\ref{T9-eq15}) is negligible and can be ignored. Under this regime, the side-band inequivalence varies almost linearly with $\bar{n}$ as
\begin{eqnarray}
\label{T9-eq18}
\delta(\bar{n})&\approx&\frac{2\Gamma^2\Omega}{4\Omega^2+\gamma^2}+\frac{8g_0^2\Omega(4\Omega^2+\gamma^2+\Gamma^2)}{(\gamma^2+4\Omega^2)^2}\bar{n} \nonumber \\ 
&\approx&\frac{2g_0^2}{\Omega}\bar{n}.
\end{eqnarray}
This result is also well in complete agreement with the expression obtained earlier for the side-band inequivalence \cite{S2-Paper3} in the limit of $g_0<<\Omega$ given as $g_0^2/\Omega$, considering that a factor of $\frac{1}{2}$ must be added as a result of different definition of $\delta$. 

The first immediate conclusion which can be obtained from (\ref{T9-eq18}) is that the side-band inequivalence $\delta$ is always positive, meaning that the detuning frequency of red-sideband should be always a bit larger in magnitude than the blue-sideband. This also agrees with the previous findings of higher-order operator algebra \cite{S2-Paper3}.

The unique mathematical form of (\ref{T9-eq14}) which is composed of a first- and second-order polynomials in terms of $\bar{n}$ respectively, offers a clear maximum at a certain optimum intracavity photon number $\bar{n}_{\rm max}$. To do this, let us first define the dimensionless constants $\alpha=4g_0^2/\Gamma^2$ and $\beta$ already defined under (\ref{T9-eq16}), $\vartheta=\gamma/2\Omega$, and $\psi=\Gamma^2/2\Omega^2$. Then, the side-band inequivalence (\ref{T9-eq17}) can be rewritten as 
\begin{equation}
\label{T9-eq19}
\delta(\bar{n})=\Omega\psi\frac{1+\alpha\bar{n}}{\vartheta^2+(1-\beta\bar{n})^2}.
\end{equation}
This offers the optimum intracavity photon number and thus the maximum attainable side-band inequivalence as
\begin{eqnarray}
\label{T9-eq21}
\bar{n}_{\rm max}&=&\frac{\sqrt{(\alpha+\beta)^2+\alpha^2\vartheta^2}}{\alpha\beta}-\frac{1}{\alpha}\approx\frac{1}{\beta}
=\frac{\Omega^2}{2g_0^2}, \\ \nonumber
\delta_{\rm max}&=&\delta(\bar{n}_{\rm max})\approx\frac{4\Omega^3}{\gamma^2}.
\end{eqnarray}
We should take note of the fact that the maximum practically measureable side-band inequivalence, which occurs at the optimum intracavity photon number $\bar{n}_{\rm max}=\Omega^2/2g_0^2$, is actually at the onset of bistability, and under practical conditions, heating due to optical losses in dielectric. 

Variation of side-band inequivalence versus intracavity photon number and in terms of different settings for input parameters $\{\alpha,\beta,\vartheta\}$ can be easily therefore deduced.

Another very important result which can be drawn from the above discussions, is marking the boundaries of linear, weakly nonlinear, and strongly nonlinear interaction regimes in quantum optomechanics. This follows by normalizing $\delta$ with respect to the mechanical frequency $\Omega$ first, as $\bar{\delta}=\delta/\Omega$.  

\begin{itemize}
	\setlength\itemsep{-0.4em}
	
	\item Fully Linear: This regime is easily given by $\bar{n}<<\bar{n}_{\rm max}$, where intracavity photon number is essentially too low to cause any appreciable side-band inequivalence. Here, the behavior of normalized side-band inequivalence is proportional to $\bar{n}$.
	
	\item Weakly Nonlinear: This regime is next given by $\bar{n}\sim\bar{n}_{\rm max}$ around the optimum operation point, where the side-band inequivalence rises to attain a maximum. The behavior of normalized side-band inequivalence is nearly Lorentzian centered at $\bar{n}=\bar{n}_{\rm max}$, with an intracavity photon number linewidth of $\Delta\bar{n}=\vartheta\bar{n}_{\rm max}$. 
	
	\item Stongly Nonlinear: This regime at larger intracavity photon numbers satisfying $\bar{n}>>\bar{n}_{\rm max}$ will push the system into strongly nonlinear regime where the side-band inequivalence quickly start to fade away. Here, the behavior of normalized side-band inequivalence is inversely proportional to $\bar{n}$.
\end{itemize}
These three behaviors in above operation regimes can be respectively displayed as
\begin{equation}
\label{T9-eq21a}
\bar{\delta}(\bar{n})\sim\left\{
\begin{matrix}
\left(\frac{\bar{n}}{\bar{n}_{\rm max}}\right)\bar{\delta}_{\rm max}, & \bar{n}\ll\bar{n}_{\rm max},\\ 
\left[1+\vartheta^{-2}\left(\frac{\bar{n}}{\bar{n}_{\rm max}}-1\right)^2\right]^{-1}\bar{\delta}_{\rm max}, & \bar{n}\sim\bar{n}_{\rm max},\\ 
\left(\frac{\bar{n}_{\rm max}}{\bar{n}}\right)\bar{\delta}_{\rm max}, & \bar{n}\gg\bar{n}_{\rm max}.
\end{matrix}
\right.
\end{equation}

It is easy to verify that the side-band inequivalence does not violate the two fundamental symmetries of the nature. Here, both the time-reversal symmetry as well as the conservation of energy are preserved. The energy of scattered red- and blue- photons $\hbar\omega\mp\hbar\Omega$ is normally expected to be within the energy of one phonon $\hbar\Omega$ where $\omega$ is the angular frequency of incident electromagnetic radiation. Per every annihilated photon, exactly one phonon is either annihilated, giving rise to a blue-shifted photon, or one phonon is created, giving rise to a red-shifted photon.

However, not all phonons are having exactly the same energies. This is permissible by the non-vanishing mechanical linewidth $\Gamma>0$ of the cavity. One should expect that once this quantity vanishes, the side-band inequivalence is gone, since it is by (\ref{T9-eq19}) proportional to $\Gamma^2$. Hence, basically it should be not contradictory to have a possible non-zero side-band inequivalence.

With regard to the time-reversal symmetry, one must take notice of the fact that all optical frequencies are physically positive, since we first must move back out of the rotating reference frame. For instance, the blue- and red-scattered photons have frequencies given by $\omega_b=\omega_c+\Omega-\frac{1}{2}\delta$ and $\omega_r=\omega_c-\Omega-\frac{1}{2}\delta$. Therefore, blue and red processes are not time-reversed processes of each other, as they both stay on the positive frequency axis. Negative frequency images corresponding to both processes do however exist and exactly satisfy the time-reversal.

\subsection{Amplitude Asymmetry}

It is a well-known fact that there is an amplitude asymmetry between red and blue side-bands due to their different frequencies, and therefore, different thermal populations following the standard Bose-Einstein statistics. Being a quantum mechanical effect, this fact can be used to estimate the temperature of a reservoir through a reference-free mechanism. However, this quantum asymmetry is gradually replaced by a classical asymmetry biased towards red once the nonlinearity is sufficiently strong, henceforth, the accuracy of quantum thermometry should be limited to weakly nonlinear optomechanical reservoirs. Now, let us examine carefully what happens to the amplitude asymmetry for strongly nonlinear optomechanics.

We can plug-in (\ref{T9-eq4}) and (\ref{T9-eq7}) into the first of (\ref{T9-eq10}). Some simplifications, while ignoring the inequilibrium quantum thermal effects on the population of side-bands, gives the equation
\begin{eqnarray}
\label{T9-eq22}
\bar{n}_r-\bar{n}_b&\approx&\left(\frac{\Omega\bar{n}}{\Omega^2+\frac{1}{4}\Gamma^2}\right)\delta \\ \nonumber
&\approx&\frac{\bar{n}}{\Omega}\delta.
\end{eqnarray}
\noindent
Here, $\bar{n}_r=|\braket{\bar{a}_r}|^2$ and $\bar{n}_b=|\braket{\bar{a}_b}|^2$ respectively refer to the number of scattered photons unto red and blue side-bands. Then, from (\ref{T9-eq19}), and assuming that $\bar{N}=(\bar{n}_r-\bar{n}_b)/\bar{n}$ denotes the normalized asymmetry of side-bands, we get
\begin{equation}
\label{T9-eq23}
\bar{N}(\bar{n})=\frac{\psi(1+\alpha\bar{n})}{\vartheta^2+(1-\beta\bar{n})^2}.
\end{equation}
Accordingly the asymmetry is increases up to a positive maximum, before decreasing back to zero at sufficiently high powers. This is another striking result and unexpected aspect of higher-order operator analysis. More surprisingly, this asymmetry also is biased towards red, and can in part explain why Stokes/anti-Stokes ratio is also a strong power of incident power, and that is not all about thermal quantum effects, but rather the inherent nonlinearity of optomechanical interaction.

\subsection{Pair Generation}

Finally, it is easy to see that the same nonlinear symmetry breaking can lead to asymmetry in the particle pair production or parametric down conversion, which can be considered as the dual of optomechanical process \cite{S9-Pair1,S9-Pair2}. In order to observe this fact, consider an optomechanical system with a mechanical frequency roughly double the optical frequency $\Omega\approx2\omega$. If the mechanics is driven strong enough at the frequnecy $\Omega$, then the effective interaction Hamiltonian will be simply $\mathbb{H}_{\rm eff}=i\hbar g(\hat{a}^\dagger\hat{a}^\dagger\hat{b}-\hat{a}\hat{a}\hat{b}^\dagger)$, where a phonon with energy $\hbar\Omega$ dissociates into two photons with energies $\hbar(\omega\pm\delta)$ with $\delta$ representing the corresponding symmetry breaking in pair frequencies caused by side-band inequivalence. Parametric down conversion for phonons has recently been observed and reported, too \cite{S9-Pair3}.

Also, based on the duality of effective interaction Hamiltonian in linear electro-optic modulation (within the validity of rotating wave approximation), with the optomechanical Hamiltonian \cite{S9-EOV1,S9-EOV2,S9-EOV3}, one could predict that the same nonlinear inequivalence to appear in relevant experiments, too. Optical modulation of Hydrogen \cite{S9-MOM} may already be shown to exhibit side-band inequivalence. Hence, further implications could be expected in communications technology and filtering, where precise positioning of side-bands are of importance. Similar arguments should be valid for enhanced Raman scattering of single molecules by localized plasmonic resonances as well \cite{S3-78}.

\subsection{Breathing Solutions}\label{Akhmediev}

In the fully classical approximation, the nonlinear optomechanical equations read
\begin{eqnarray}
\label{T9-eq24-SI}
\frac{d}{dt}\Upsilon&=&\left(i\Delta-\frac{1}{2}\kappa\right)\Upsilon+ig_0\Upsilon(\Phi+\Phi^\ast),\\ \nonumber
\frac{d}{dt}\Phi&=&\left(-i\Omega-\frac{1}{2}\Gamma\right)\Phi+ig_0\Upsilon^\ast\Upsilon.
\end{eqnarray}
Under the assumption of slowly varying amplitude $|d\Upsilon/dt|<<\Omega|\Upsilon|$ and large optical quality factor $\kappa<<|\Delta|$, the set of equations (\ref{T9-eq24-SI}) admits an analytical solution for the optical field of the form
\begin{equation}
\label{T9-eq25-SI}
\Upsilon(t)=\frac{\theta e^{i\theta^2 \tau}}{\sqrt{2D/|\Delta|}}\left[\frac{2\iota^2 \cosh(\vartheta\tau)-i2\iota\vartheta\sinh(\vartheta\tau)}{\cosh(\vartheta\tau)-\sqrt{1-\iota^2}}-1\right],
\end{equation}
referred to Akhmediev breathers \cite{S9-Breather1,S9-Breather2}. Here, $\tau=|\Delta| t$ is normalized time, $\theta$ is a real constant to be determined, $D$ is defined in (\ref{T9-eq15}), and $\vartheta=-2\theta^2\iota\sqrt{1-\iota^2}$. For $\iota<1$ solutions are breathing soliton-like, while for $\iota>1$ solutions become nonlinear oscillatory waves.

When $\iota<1$, then the expression within the brackets is composed of an even real part and an odd imaginary part. Hence, the Fourier transform of this hyperbolic expression within the brackets is real-valued, which we here denoted by $\mathcal{G}(w)$. Obviously, the magnitude of Fourier transform, or the spectrum $|\mathcal{G}(w)|$ no longer needs to be even. Hence, the Fourier transform of $\Upsilon(t)$, which is given by $\mathcal{J}(w)=\mathcal{F}\{\Upsilon(t)\}(w)$ is simply the shifted transform $\mathcal{G}(w-\theta^2|\Delta|)$, and the amount of frequency shift is nothing but the side-band inequivalence. This fact is a direct result of the multiplying term $\exp(i\theta^2\tau)$ in the analytical solution (\ref{T9-eq25-SI}), and thus the side-band inequivalence. 

While, the algebraic hyperbolic form of (\ref{T9-eq25-SI}) with $\iota<1$ disallows analytical evaluation of the spectrum $\mathcal{J}(w)$, however, calculation of $\mathcal{J}(w)$ from (\ref{T9-eq25-SI}) with $\iota>1$, which leads to periodic nonlinear oscillatory solutions, becomes explicitly possible as shown in the following. To do so, we may proceed with the substitution $\iota=|(\Omega+i\Gamma/2)/g|\approx\Omega/g$ where $g=g_0\sqrt{n}$ is the enhanced optomechanical interaction rate, which in the weakly-coupled limit obviously satisfies $\iota>>1$. Hence, we have $i\vartheta=2\theta^2\iota\sqrt{\iota^2-1}\approx2\theta^2\iota^2$. Now, we can rearrange (\ref{T9-eq25-SI}) as
\begin{equation}
\label{T9-eq26-SI}
\Upsilon(t)=\frac{2\iota\theta e^{i\theta^2 \tau}}{\sqrt{2D|\Delta|}}\left[\frac{\iota|\Delta|\cos(\varpi t)+i\varpi\sin(\varpi t)}{\cos(\varpi t)-i\sqrt{\iota^2-1}}-\frac{|\Delta|}{2\iota}\right],
\end{equation}
\noindent
where the replacement $i\vartheta\tau=\varpi t$ has been made and $\varpi=\Omega+\delta\Omega$ is the shifted mechanical frequency while taking the optical spring effect $\delta\Omega$ into account. The expression within brackets in (\ref{T9-eq26-SI}) can be simplified using trigonometric identities as 
\begin{eqnarray}
\label{T9-eq27-SI}
\Upsilon(t)&=&\Xi e^{i\frac{1}{2}\delta t}\left[\frac{(\iota|\Delta|+\varpi)e^{i\varpi t}+(\iota|\Delta|-\varpi)e^{-i\varpi t}}{e^{i\varpi t}+e^{-i\varpi t}-2i\sqrt{\iota^2-1}}-\frac{|\Delta|}{2\iota}\right]\\ \nonumber 
&=&\Xi e^{i\frac{1}{2}\delta t}\left[\frac{2i\sqrt{\iota^2-1}(\iota|\Delta|+\varpi)-2\varpi e^{-i\varpi t}}{e^{i\varpi t}+e^{-i\varpi t}-2i\sqrt{\iota^2-1}}+ \frac{2\iota(\iota|\Delta|+\varpi)-|\Delta|}{2\iota}\right]
\\ \nonumber 
&=&2\Xi e^{i\frac{1}{2}\delta t}\left[\frac{i\varsigma-\varpi e^{-i\varpi t}}{e^{i\varpi t}+e^{-i\varpi t}-2i\sqrt{\iota^2-1}}\right]+\Xi\left[ \frac{2\iota(\iota|\Delta|+\varpi)-|\Delta|}{2\iota}\right]e^{i\frac{1}{2}\delta t}
\\ \nonumber
&=&i\Psi e^{\tfrac{1}{2}\delta t}f(t)+\Lambda e^{i\tfrac{1}{2}\delta t},
\end{eqnarray}
in which $\Xi=\sqrt{2}\iota\theta/\sqrt{D}$, $\Lambda=\Xi(2|\Delta|\iota^2-2\varpi\iota-|\Delta|)/2\iota$, $\varsigma=\sqrt{\iota^2-1}(\iota|\Delta|+\varpi)$, and $\Psi=2\Xi\varsigma$. We here may notice that $\delta=\theta^2 |\Delta|>0$ is nothing but the side-band inequivalence. The function $f(t)$ can be rewritten as
\begin{eqnarray}
\label{T9-eq28-SI}
f(t)&=&\frac{1+i\frac{\varpi}{\varsigma} e^{-i\varpi t}}{e^{i\varpi t}+e^{-i\varpi t}-2i\sqrt{\iota^2-1}}=\frac{e^{i\varpi t}+i\varkappa}{(e^{i\varpi t}-i\sqrt{\iota^2-1})^2+\iota^2}\\ \nonumber
&=&\frac{e^{i\varpi t}+i\varkappa}{(e^{i\varpi t}-i\zeta^+)(e^{i\varpi t}-i\zeta^-)}=\frac{1}{2\iota}\frac{\zeta^++\varkappa}{e^{i\varpi t}-i\zeta^+}-\frac{1}{2\iota}\frac{\zeta^-+\varkappa}{e^{i\varpi t}-i\zeta^-}\\ \nonumber
&=&h(e^{i\varpi t}),
\end{eqnarray}
where the replacement $\zeta^\pm=\pm\iota+\sqrt{\iota^2-1}$ has been made and $\varkappa=\varpi/\varsigma$ with $\zeta^+\zeta^-=-1$, $\zeta^+>0$ and $\zeta^-<0$ respectively lying in the upper and lower half-complex planes. Obviously, the pole at $i\zeta^+$ falls outside the unit circle, and only the pole at $i\zeta^-$ remains inside the unit circle. Hence, $\varkappa<<-\zeta^-<<1<<\zeta^+$. These will be needed later to derive the spectrum of (\ref{T9-eq26-SI}).

All remains now is to solve $i\vartheta \tau=\varpi t$, which gives rise to the approximate solution
\begin{equation}
\label{T9-eq29-SI}
\delta\approx\frac{2\varpi}{\iota\sqrt{\iota^2-1}}=2g^2\frac{\Omega+\delta\Omega}{\sqrt{\Omega^2+\frac{1}{4}\Gamma^2}\sqrt{\Omega^2+\frac{1}{4}\Gamma^2-g^2}}.
\end{equation} 
This can be further written as
\begin{equation}
\label{T9-eq30-SI}
\frac{\delta}{2}\approx \frac{\Omega+\delta\Omega}{\iota^2}=g^2\frac{\Omega+\delta\Omega}{\Omega^2+\frac{1}{4}\Gamma^2}\approx\frac{g^2}{\Omega}+\frac{g^2}{\Omega^2}\delta\Omega\approx\frac{g^2}{\Omega}-\frac{2g^4\Delta}{\Omega^2}\frac{(\Omega^2-\frac{1}{4}\kappa^2)}{(\Omega^2+\frac{1}{4}\kappa^2)^2},
\end{equation}
since $\delta\Omega\approx -2g^2\Delta(\Omega^2-\frac{1}{4}\kappa^2)/(\Omega^2+\frac{1}{4}\kappa^2)^2$. It is here seen that the correction arising from spring effect is $O(g^4)$ while the side-band inequivalence is $O(g^2)$. Hence, the normalized side-band inequivalence $\bar{\delta}=\delta/\Omega$ can be finally approximated for the side-band resolved regime $\Omega>>\kappa$ as
\begin{equation}
\label{T9-eq31-SI}
\bar{\delta}\approx\frac{2g^2}{\Omega^2}\left[1-\frac{2g^2\Delta(\Omega^2-\frac{1}{4}\kappa^2)}{\Omega(\Omega^2+\frac{1}{4}\kappa^2)^2}\right]\approx\frac{2g^2}{\Omega^2}\left[1-\frac{2g^2}{\Omega^3}\Delta\right],
\end{equation}
showing that the correction of optomechanical spring effect to the ansatz (\ref{T9-eq1}), ultimately yielding the expression for side-band inequivalence (14) had been safely ignored indeed. There is also a higher-order correction to the optomechanical spring effect $\delta\Omega$ \cite{S2-3} as a result of non-zero coherent phonon population $\bar{m}$, which results in an extra correction to (\ref{T9-eq30-SI}) by replacing $g=g_0\sqrt{\bar{n}}$ within the brackets of (\ref{T9-eq31-SI}) with $g\approx g_0\sqrt{\bar{n}+\bar{m}+1}$. But this leaves our derivations and conclusions regarding the side-band inequivalence unchanged. 

This solution (\ref{T9-eq27-SI}) is a bi-periodic and complex-valued product of two periodic functions $f(t)$ and $\exp(i\delta t)$. Fourier transform of the periodic function $f(t)=f(t+2\pi/\varpi)$ defined as $F(w)=\tfrac{1}{2\pi}\int_{-\infty}^{\infty}f(t)e^{-iwt}dt$ is starightforward to obtain. In fact, one should have $F(w)=\sum_\nu f_\nu\updelta(w-\nu\varpi)$, where $\updelta(\cdot)$ are Dirac's delta functions, and $f_0 =f(1)$ while $f_\nu =\exp(i2\pi/\nu), \nu\neq 0$. Hence, it is straightforward to see that how its spectrum looks like. Defining the spectrum of $f(t)$ as $|F(w)|$ in the Fourier domain $w$ simply is $F(w)=\sum_\nu |f_\nu|\updelta(w-\nu\varpi)$, which consists of Dirac deltas at $w=\pm \nu\varpi\approx\pm\nu\Omega, \nu\in\mathbb{N}$ corresponding to the side-bands. A practical system obviously does not exactly follow the breather solution (\ref{T9-eq25-SI}) and hence side-bands all have finite non-zero linewidths. Ultimately, we have $I(w)=|\mathcal{J}(w)|$.

Therefore the ultimate spectrum of the cavity within the approximation of breather solutions is $I(w)\approx|F(w-\tfrac{1}{2}\delta)|+R(w)$, where $R(w)$ is the reflection from cavity at central resonance $w=0$. Since the reflected central resonance $R(w)$ normally masks out the zeroth harmonic, therefore the side-bands appear to be positioned asymmetrically in frequency equal to the side-band inequivalence $\delta$. Hence, $I(w)$ may be written conveniently as
\begin{eqnarray}
\label{T9-eq32-SI}
I(w)&=&|\tfrac{1}{2\pi}\Lambda+i\Psi f_0|\updelta(w-\tfrac{1}{2}\delta)+\Psi\sum_{\nu=1}^{\infty}\left[|f_\nu|\updelta(w-\tfrac{1}{2}\delta-\nu\varpi)+|f_{-\nu}| \updelta(w-\tfrac{1}{2}\delta+\nu\varpi)\right], \\ \nonumber
f_\nu&=&\frac{1}{2i\varpi\pi}\oint\frac{h(z)}{z^{\nu+1}}dz, \\ \nonumber
&=&-\left[\frac{\zeta^- +\varkappa}{2\varpi\iota (i\zeta^-)^{\nu+1}}\right] +\frac{1}{\varpi\nu!}\left[\frac{d^\nu}{dz^{\nu}}h(z)\right]_{z=0}u(\nu)\\ \nonumber
&=&\left[\frac{i(\zeta^- +\varkappa)}{2\varpi\iota i^\nu(\zeta^-)^{\nu+1}}\right] -\frac{\zeta^-+\varkappa}{2\iota\varpi\nu!}\left[\frac{(-1)^\nu \nu!}{i^{\nu+1}(\zeta^-)^{\nu+1}}\right]_{z=0}u(\nu)\\ \nonumber
&=&\left[\frac{i(\zeta^- +\varkappa)}{2\varpi\iota i^\nu(\zeta^-)^{\nu+1}}\right]\left[1+(-1)^\nu u(\nu)\right].
\end{eqnarray}
\noindent
where the change of variables $z=\exp(i\varpi t)$ has taken place, and the integration is taken counter-clockwise on the unit circle in the complex $u-$plane. The only contributing pole of $h(z)$ is at $z=i\zeta^-\approx-i/2\iota^2$. Furthermore, $u(\cdot)$ is the unit-step function, which allows the second term to contribute only if $\nu\geq 0$. The function $h(z)=(z-\varkappa)/(z-i\zeta^+)(z-i\zeta^-)$ was also defined in (\ref{T9-eq28-SI}). In (\ref{T9-eq32-SI}), the positive odd harmonics identically vanish, and $f_\nu$ and $f_{-\nu}$ respectively correspond to Stokes and anti-Stokes amplitudes. There are no odd-ordered Stokes components in the breather nonlinear oscillatory wave (\ref{T9-eq26-SI}), and also anti-Stokes components diminish in strength with their order $-\nu$ increasing as $(\zeta^-)^{-\nu}$ according to (\ref{T9-eq32-SI}).

The total power $P=\int_{-\infty}^\infty I(w)dw$ is now simply $P\approx\sum_\nu |f_\nu|$, and total harmonic distortion shall be given by the simple expression ${\rm THD}\approx\sum_{|\nu|\geq 2} |f_\nu|/\sum_{|\nu|\geq 1} |f_\nu|$. The first-order mechanical side-bands correspond to $f_{\pm1}$ with sharp peaks located at $\pm\varpi-\tfrac{1}{2}\delta\approx\pm\Omega-\tfrac{1}{2}\delta$, confirming the initial ansatz (\ref{T9-eq1}) and speculation regarding the existence of side-band inequivalence. In summary, the breathing analytical solution (\ref{T9-eq25-SI}) actually highlights the existence of a non-zero side-band inequivalence, simply because of the multiplying term $\exp(i\theta^2\tau)$ and its bi-periodic form, and furthermore side-band inequivalence has to be always towards red (Stokes) simply because $\theta^2>0$ is always positive.

The ratios of coefficients $f_\nu$ in (\ref{T9-eq32-SI}) can be estimated using binomial expansion of denominator in (\ref{T9-eq27-SI}) and the original form, resulting in approximate expressions for the ratios of side-band powers. For instance, the ratio of optical amplitude in the first-order side-bands with respect to the central resonance is roughly
\begin{eqnarray}
\label{T9-eq33-SI}
\frac{I_1}{I_0}&=&\frac{|f_1|+|f_{-1}|}{2\left|\tfrac{1}{2\pi}\Lambda+i\Psi f_0\right|}=\frac{\pi}{\varpi\iota}\frac{\zeta^-+\varkappa}{\left[\Lambda-(\zeta^-+\varkappa)\pi\Psi/(\varpi\iota\zeta^-)\right]}\\ \nonumber
&\approx&\frac{\pi\zeta^-}{\varpi\iota\Lambda-\pi\Psi}.
\end{eqnarray}
The accuracy of breathing solutions for second- and higher-order harmonics is insufficient to obtain a meaningful ratio such as (\ref{T9-eq33-SI}), nevertheless, it exhibits a positive and unmistakable side-band inequivalence towards red.

In summary, in this section we presented a complete analysis of side-band inequivalence in quantum optomechanics, and showed it undergoes a maximum and obtained closed-form expressions for optimum intracavity photon number as well as maximum attainable side-band inequivalence. We classified the operation into the linear, weakly nonlinear, and strongly nonlinear regimes, in which the behavior of system is markedly different. The results of this investigation can provide the accuracy constraints as well as necessary experimental set up to resolve the elusive side-band inequivalence. Analysis of high resolution measurements of Brillouin scattering and Raman scattering for different materials as well as ion traps, Paul traps, electrooptic modulators, and acoustooptic modulators confirms the existence of side-band inequivalence. One could speculate that precise measurement of the variation of side-band inequialence in terms of various system parameters could provide further insight into unexplored nonlinear properties of different materials. 

We successfully and consistently demostrated the existence of side-band inequivalence through three essentially very different theoretical methods, which agree at least in the weakly nonlinear regime. As byproducts of this analysis presented here, we notice the existence of a similar symmetry breaking in particle pair generation, which can be considered as the time-reversed process of optomechanics. More surprisingly, there is a symmetry breaking in Stokes/anti-Stokes ratio which is nonlinear and does not depend on the quantum mechanical and temperature-dependent Bose-Einstein statistics. This phenomenon can actually mask the quantum mechanical effect in the strong nonlinear regime.


\section{Quadratic Optomechanics} \label{Section-10}

This section presents a full operator analytical method for studying the quadratic nonlinear interactions in quantum optomechanics. The method is based on the application of higher-order operators, using a six-dimensional basis of second order operators which constitute an exactly closed commutators. We consider both types of standard position-field and the recently predicted non-standard momentum-field quadratic interactions, which is significant when the ratio of mechanical frequency to optical frequency is not negligible. This unexplored regime of large mechanical frequency can be investigated in few platforms including the superconducting electromechanics and simulating quantum cavity electrodynamic circuits. It has been shown that the existence of non-standard quadratic interaction could be observable under appropriate conditions.

The field of quantum optomechanics \cite{S2-1,S2-3,S2-6} is flourishing as one of the modern applications of quantum physics, where interactions of optical field and mechanical motion inside a confined cavity is being studied. Nonlinear interactions in optomechanics play a critical role in a growing number of studied physical phenomena, which include emergence of second-order mechanical side-bands \cite{S4-40e,S10-q12}, nonlinear optomechanical induced transparency \cite{S8-Lemonde2,S10-q3,S4-40h,S10-q8}, phonon-laser \cite{S10-q5,S10-q13}, nonlinear reciprocity \cite{S10-q6}, and parameteric phonon-phonon coupling for cascaded optical transparency \cite{S10-q7}. Recently, existence of optomechanical chaos has been confirmed \cite{S3-53}, which is contingent on and can be explained only using the nonlinearity of interactions. Meterology with optomechanical symmetry breaking \cite{S10-q10} as well as Kerr-type nonlinear interactions \cite{S10-q8,S4-40h} are other examples of optomechanical phenomena which call for nonlinear analysis. Interestingly, it has been also shown using an extensive analysis, that the gravitational constant can be measured by increased precision using nonlinear optomechanics \cite{S10-q14}, reaching an ideal fundamental sensitivity of $10^{-15}\text{ms}^{-2}$ for state-of-the-art parameters.

Many of the studies in this field utilize linearization of photonic $\hat{a}$ and phononic  $\hat{b}$ ladder operators around their mean values as $\hat{a}\to\bar{a}+\hat{a}$ and $\hat{b}\to\bar{b}+\hat{b}$, where the substituted ladder operators now represent field fluctuations around their respective mean values. This way of linearization is however insufficient for quadratic \cite{S2-7,S3-57,S10-Quad4,S2-9,S3-59,S4-33.3,S2-8,S2-Bruschi} and higher-order interactions where the resulting Langevin equations \cite{S2-Noise1,S2-Noise0,S2-Noise2,S2-Noise3} are expected to be strongly nonlinear. Full linearization essentially transforms back every such nonlinear interaction picture into the simple linearized form of $(\hat{a}+\hat{a}^\dagger)(\hat{b}+\hat{b}^\dagger)$. As a result, the behavior of systems under study due to quadratic and higher-order interactions becomes indifferent to that of ordinary linearized optomechanics, implying that in this way of linearization some of the important underlying nonlinear physics may be lost.

Recently, theory of optomechanics has been revisited by the author \cite{S2-Paper1} as well as others for nonlinear \cite{S10-Sala} and quadratic \cite{S10-Thesis} interactions. It has been shown that a non-standard quadratic term could exist due to momentum-field interaction and relativistic effects, the strength of which is proportional to $(\Omega/\omega)^2$ with $\Omega$ and $\omega$ respectively being the mechanical and optical frequencies \cite{S2-Paper1}. Normally, such momentum-field interactions are not expected to survive under the regular operating conditions of large optical frequencies $\omega>>\Omega$. However, it would be a matter of question that whether they could be observable in spectral response of a cavity, given that the $\omega$ and $\Omega$ could be put within the same order of magnitude? The answer is Yes.

Superconducting electromechanics provides a convenient means to observe quadratic effects at such conditions. Not only high photon cavity numbers could be attained, but also mechanical and radio or microwave frequencies could be set within the same order of magnitude with relative ease. This conditions have been actually met at least in one reported experiment \cite{S3-79}, where the mechanical and superconducting circuit frequencies are designed to be equal at $\omega=\Omega=2\pi\times 720\text{kHz}$. Tuning to the lowest-order odd-profiled mechanical mode, or using the membrane-in-the-middle setup \cite{S4-40h,S3-79} in optomechanical experiments could altogether eliminate the standard optomechanical interaction, leaving only the quadratic terms and higher. 

Apart from experimental considerations for observation of quadratic effects, there remains a major obstacle in theoretical analysis of combined standard and non-standard quadratic interactions. To the best knownledge of the author, this regime has been investigated theoretically for the case of only standard quadratic interaction using the time-evolution operators \cite{S2-Bruschi}. 

Growing out of the context of quantum optomechanics, the method of higher-order operators developed by the author \cite{S2-Paper2} can address problems with any general combination of nonlinearity, stochastic input, operator quantities, and spectral estimation. Higher-order operators have been already used in analysis of nonlinear standard optomechanics \cite{S2-Paper3}, where its application has uncovered effects known as sideband inequivalence, quantities such as coherent phonon population, as well as corrections to the optomechanical spring effect, zero-point field optomechanical interactions, and a minimal basis with the highest-order which allows exact integration of optomechanical Hamiltonian subject to multiplicative noise input. Also, it has been independently used \cite{S2-Quad} for investigation of quadratic effects. But this method has not been verified yet for non-standard quadratic optomechanics, which is the central topic of this study.

We employ a six-dimensional basis of higher-order operators, all being second order, which satisfy an exact closed commutation relations. This basis can be used to analyze the quadratic interactions of both standard and non-standard types, which has been so far not done. It has been shown that the momentum-field interaction, referred to as the non-standard quadratic term, does have observable effects on the spectral response of the optomechanical cavity, if the design criteria could violate $\Omega<<\omega$. Hence, the effect of this interaction should not be overlooked when the ratio $\Omega/\omega$ is non-negligible. The non-standard term appears to survive even under weak quadratic coupling. This study paves the way for probing a previously unexplored domain of quantum optomechanics.

\subsection{Quadratic Interaction Hamiltonian}
In this section, we discuss the model Hamiltonian for the quadratic interaction in quantum optomechanics. As it will be shown, it is composed of two contributing terms. The first term $\mathbb{H}_{1}$ is the well-known standard quadratic term, resulting from the product of photon number $\hat{n}=\hat{a}^\dagger\hat{a}$ and squared displacement $\mathbb{X}^2=(\hat{b}+\hat{b}^\dagger)^2$. The second term $\mathbb{H}_2$, which is also quadratic in order, represents the non-standard term and describes the momentum-field interaction among the squared momentum of the mirror $\mathbb{P}^2=(\hat{b}-\hat{b}^\dagger)^2$ and squared second quadrature of the electromagnetic field $\mathcal{P}^2=(\hat{a}-\hat{a}^\dagger)^2$ \cite{S2-Paper1}, thus admitting the form $\mathcal{P}^2\mathbb{P}^2$. It should be mentioned that the phase of $\hat{a}$ can be arbitrarily shifted by $\pi/2$, allowing one to rewrite the latter interaction as $\mathcal{X}^2\mathbb{P}^2$, where $\mathcal{X}^2=(\hat{a}+\hat{a}^\dagger)^2$. In that sense, the referral to the field quadratures as either $\mathcal{X}$ or $\mathcal{P}$ is quite arbitrary.

For the purpose of this section, we consider that the standard optomechanical interaction $\mathbb{H}_\text{OM}$, also denoted by $\mathbb{H}_0$ in the paper, vanishes due to appropriate design with $g_0=0$. This not only simplifies the description of the problem and reduces the dimension of basis significantly, but is also favorable from an experimental point of view, since the lowest order surviving interaction now would be quadratic. As mentioned in the above, this criterion can be easily met in superconducting electromechanics by tuning the electroagnetic frequency to the first odd-profiled mechanical mode, or in optomechanics by using arrangements such as the membrane-in-the-middle setup \cite{S4-40h,S3-79}. In general, the condition $g_0=0$ might not be exactly achieved and it may cause some difficulty in observation of quadratic effects, since standard optomechanical interactions could still mask out the much weaker quadratic interactions. However, for the purpose of current study, we may neglect this effect in the same way is being done by other researchers \cite{S2-7,S3-57,S10-Quad4,S2-9,S3-59,S4-33.3,S2-8,S2-Bruschi}. The total Hamiltonian in absence of $\mathbb{H}_0$ is thus given by $\mathbb{H}=\mathbb{H}_{s}+\mathbb{H}_1+\mathbb{H}_2+\mathbb{H}_\text{d}$ where
\begin{eqnarray}
\label{T10-eq1}
\mathbb{H}_{s}&=&\hbar\tilde{\omega}\hat{a}^\dagger\hat{a}+\hbar\tilde{\Omega}\hat{b}^\dagger\hat{b}=\hbar\tilde{\omega}\hat{n}+\hbar\tilde{\Omega}\hat{m},\\ \nonumber
\mathbb{H}_{1}&=&\hbar \tfrac{1}{2}\varepsilon\hat{a}^\dagger\hat{a}(\hat{b}+\hat{b}^\dagger)^2=\hbar \tfrac{1}{2}\varepsilon\hat{n}\mathbb{X}^2, \\ \nonumber
\mathbb{H}_{2}&=&-\hbar \tfrac{1}{2}\beta (\hat{a}-\hat{a}^\dagger)^2(\hat{b}-\hat{b}^\dagger)^2=-\hbar \tfrac{1}{2}\beta\mathcal{P}^2\mathbb{P}^2, \\ \nonumber
\mathbb{H}_\text{d}&=&\hbar\gamma(\alpha e^{i\omega_\text{d} t}\hat{a}+\alpha^\ast e^{-i\omega_\text{d} t}\hat{a}^\dagger).
\end{eqnarray}
\noindent
Here, $\tilde{\omega}$ and $\tilde{\Omega}$ are respectively the bare unperturbed optical and mechanical frequencies. We notice that the relative frequency notation of optical detuning, which is useful in standard optomechanical and quadratic interactions \cite{S2-Paper3} is not to be used here. In (\ref{T10-eq1}), furthermore, we assume the existence of only one drive term with the complex amplitude $\alpha$ and frequency $\omega_\text{d}$. In general, it is possible to assume the existence of multiple drive terms at different frequencies, but this does not alter the mathematical approach under consideration. We furthermore notice that the non-standard quadratic term can be also written as $\mathbb{H}_2=-\hbar\tfrac{1}{2}\beta\mathcal{X}^2\mathbb{P}^2$ with basically no physically significant difference, as mentioned in the above section, too. Hence, the non-standard term can take on either of the sign conventions $--$ as $\mathcal{P}^2\mathbb{P}^2$, or $+-$ as $\mathcal{X}^2\mathbb{P}^2$. We shall proceed with the latter. Also, $\varepsilon$ is the strength of the standard quadratic interaction and $\beta$ is the strength of the non-standard quadratic interaction. These are shown to be related as \cite{S2-Paper1}
\begin{equation}
\label{T10-eq2}
\beta=\frac{1}{4}\left(\frac{\pi^2}{3}+\frac{1}{4}\right)\left(\frac{\tilde{\Omega}}{\tilde{\omega}}\right)^2\varepsilon\equiv \tfrac{1}{2}\rho\varepsilon.
\end{equation}
It is easily seen that in the regime of large optical frequency $\tilde{\omega}>>\tilde{\Omega}$, we get $\beta\approx 0$ and the non-standard term vanishes. This is what has actually been probed in nearly all experiments on quadratic optomechanical interactions so far \cite{S2-7,S3-57,S10-Quad4,S2-9,S3-59,S4-33.3,S2-8,S2-Quad}. The large mechanical frequency regime of standard quadratic interactions has been however recently probed \cite{S2-Bruschi} and it has been suggested that the roles of optical and mechanical parts are expected to interchange without consideration of the non-standard quadratic effect. Hence, the condition defining the critical mechanical frequency as $\tilde{\Omega}=\frac{1}{2}\sqrt{\frac{1}{3}\pi^2+\frac{1}{4}}\tilde{\omega}\approx 0.941\tilde{\omega}$ marks a critical value for the transition border, across which the regimes of large and small mechanical frequency with respect to the given electromagnetic frequency are distinguished. This happens to be remarkably close to the identical frequencies as $\tilde{\Omega}=\tilde{\omega}$, too. It is not difficult to see that the above Hamiltonian with the sign convention taken as $+-$ can be rewritten as
\begin{equation}
\label{T10-eq3}
\mathbb{H}=\hbar\omega\hat{n}+\hbar{\Omega}\hat{m}+\hbar\epsilon\left[\hat{n}(\hat{m}+\hat{d}+\hat{d}^\dagger)-\rho(\hat{n}+\hat{c}+\hat{c}^\dagger)(\hat{d}+\hat{d}^\dagger-\hat{m})\right]+\mathbb{H}_\text{d},
\end{equation}
\noindent
in which $\hat{c}=\frac{1}{2}\hat{a}^2$ and $\hat{d}=\frac{1}{2}\hat{b}^2$ are defined and discussed extensively in the preceding sections and articles \cite{S2-Paper1,S2-Paper2,S2-Paper3}, and time-independent non-interacting terms are dropped which are irrelevant to the behavior of system dynamics. Furthermore, the altered effective optical and mechanical frequencies due to the quadratic optomechanical interaction assume different forms, and now read
\begin{eqnarray}
\label{T10-eq4}
\omega&=&\tilde{\omega}+\tfrac{1}{2}\varepsilon+\beta, \\ \nonumber \Omega&=&\tilde{\Omega}+\beta.
\end{eqnarray}

The importance of the non-standard quadratic term $\mathbb{H}_2$ \cite{S2-Paper1} is that it describes a non-vanishing correction to the field-mirror interaction, beyond simple nonlinear quantum back-action of mirror on the reflected light. In standard picture of quantum optomechanics, the light gets reflected off a displaced mirror which already has shifted the resonance frequency of cavity. Higher-order corrections to combination of these effects give rise to quadratic interactions. But in quadratic quantum optomechanics, either momenta of field and mirror do not apparently come into consideration, or their contributions are somehow lost because of the approximations used in the expansions. It is expected that normally such an interaction should take care of momentum exchange. The exchange and conservation of momenta under standard quadratic interaction is uncertain and actually not quite obvious, since momentum operators do not show up in the interaction. 

\subsection{Six-dimensional Basis}

Analysis of the Hamiltonian (\ref{T10-eq3}) here is going to be based on the six-dimensional space spanned by the basis operators $\{A\}^\text{T}=\{\hat{c},\hat{c}^\dagger,\hat{n},\hat{d},\hat{d}^\dagger,\hat{m}\}$ \cite{S2-Paper2}. This basis can be easily seen to be the smallest possible set, with closed commutators, capable of describing the system modeled by (\ref{T10-eq1}). The commutation properties of this basis \cite{S2-Paper2} is here given for the sake of convenience
\begin{eqnarray}
\label{T10-eq5}
[\hat{c},\hat{c}^\dagger]&=&\hat{n}+\tfrac{1}{2},\\ \nonumber
[\hat{c},\hat{n}]&=&2\hat{c}, \\ \nonumber
[\hat{n},\hat{c}^\dagger]&=&2\hat{c}^\dagger, \\ \nonumber
[\hat{d},\hat{d}^\dagger]&=&\hat{m}+\tfrac{1}{2}, \\ \nonumber
[\hat{d},\hat{m}]&=&2\hat{d}, \\ \nonumber
[\hat{m},\hat{d}^\dagger]&=&2\hat{d}^\dagger.
\end{eqnarray}
\noindent
All commutators among photonic $\{\hat{c},\hat{c}^\dagger,\hat{n}\}$ and phononic operators $\{\hat{d},\hat{d}^\dagger,\hat{m}\}$ are clearly zero. This fact together with the set of commutators (\ref{T10-eq5}) establishes the closedness property of our basis. The basis $\{A\}$ under consideration is called to be second-order, since its operators are all products of two single ladder operators. Other bases of the third- and higher-orders are discussed elsewhere, respectively for standard optomechanics \cite{S2-Paper3} and quadratic interactions \cite{S2-Paper2}.

Furthermore, we will need $[\hat{c},\hat{a}]=[\hat{c}^\dagger,\hat{a}^\dagger]=0$ along with the pair of relationships
\begin{eqnarray}
\label{T10-eq6}
[\hat{c},\hat{a}^\dagger]&=&[\hat{a},\hat{n}]=\hat{a}, \\ \nonumber 
[\hat{a},\hat{c}^\dagger]&=&[n,\hat{a}^\dagger]=\hat{a}^\dagger, 
\end{eqnarray}
to evaluate the effect of drive and input noise terms later. Assumption of a resonant drive term here requires $\omega_\text{d}=\omega$, which highlights a constant shift in cavity optical frequency because of the presence of quadratic optomechanical interactions. This will greatly simplify the mathematics involved in construction of Langevin equations. These latter relations (\ref{T10-eq5},\ref{T10-eq6}) show that the eight-dimensional basis $\{\hat{a},\hat{a}^\dagger\}\bigcup\{A\}$, which is of the mixed first and second-order, also constitutes a closed commutation relationships. However, for the purpose of our study here, the original six-dimensional basis $\{A\}$ is sufficient.

The task of construction of Langevin equations proceeds with the original equation \cite{S2-1,S2-3,S2-6,S2-Noise1,S2-Noise0,S2-Noise2,S2-Noise3}
\begin{equation}
\label{T10-eq7}
\dot{\hat{z}}=-\frac{i}{\hbar}[\hat{z},\mathbb{H}]-[\hat{z},\hat{x}^\dagger]\left(\tfrac{1}{2}\gamma\hat{x}+\sqrt{\gamma}\hat{x}_\text{in}\right)+\left(\tfrac{1}{2}\gamma\hat{x}^\dagger+\sqrt{\gamma}\hat{x}_\text{in}^\dagger\right)[\hat{z},\hat{x}],
\end{equation}
\noindent
where $\hat{z}$ is an arbitrary operator belonging to the basis, $\hat{x}$ is any system annihilator associated with the decay rate $\gamma$, and $\hat{x}_\text{in}$ is the corresponding input field including contributions from both of the deterministic drive and stochastic noise terms. Also, we may set either $\hat{x}=\hat{a}$ with $\gamma=\kappa$ and $\hat{x}_\text{in}=\hat{a}_\text{in}$, or $\hat{x}=\hat{b}$ with $\gamma=\Gamma$ and $\hat{x}_\text{in}=\hat{b}_\text{in}$ to allow simple and straightforward construction of noise terms. This particular choice avoids appearance of squared noises, which are otherwise required \cite{S2-Paper2}.

\subsection{Six-dimensional Langevin Equations}

There are a total of six Langevin equations for the system (\ref{T10-eq1}), which can be constructed one by one, corresponding to the six members of $\{A\}$. These equations can be obtained using the commutators (\ref{T10-eq5},\ref{T10-eq6}) after some straightforward algebra as
\begin{eqnarray}
\label{T10-eq8}
\dot{\hat{c}}&=&-i2\omega\hat{c}-\kappa \hat{c}-i\varepsilon\left[2(\hat{m}+\hat{d}+\hat{d}^\dagger)\hat{c}-\rho(\hat{d}+\hat{d}^\dagger-\hat{m})(2\hat{c}+\hat{n}+\tfrac{1}{2})\right]+i\alpha^\ast e^{-i\omega t}\hat{a}-\sqrt{\kappa}\hat{a}\hat{a}_\text{in} ,\\ \nonumber
\dot{\hat{c}}^\dagger&=&i2\omega\hat{c}^\dagger-\kappa \hat{c}^\dagger+i\varepsilon\left[2(\hat{m}+\hat{d}+\hat{d}^\dagger)\hat{c}^\dagger-\rho(\hat{d}+\hat{d}^\dagger-\hat{m})(2\hat{c}^\dagger+\hat{n}+\tfrac{1}{2})\right]-i\alpha e^{i\omega t}\hat{a}^\dagger-\sqrt{\kappa}\hat{a}_\text{in}^\dagger\hat{a}^\dagger ,\\ \nonumber
\dot{\hat{n}}&=&-i2\beta(\hat{c}-\hat{c}^\dagger)(\hat{d}+\hat{d}^\dagger-\hat{m})-\kappa\hat{n}+i(\alpha e^{i\omega t}\hat{a}-\alpha^\ast e^{-i\omega t}\hat{a}^\dagger)-\sqrt{\kappa}(\hat{a}^\dagger\hat{a}_\text{in}+\hat{a}_\text{in}^\dagger\hat{a}), \\ \nonumber
\dot{\hat{d}}&=&-i2\Omega\hat{d}-\Gamma \hat{d}-i\varepsilon\left[\hat{n}(2\hat{d}+\hat{m}+\tfrac{1}{2})-\rho(\hat{c}+\hat{c}^\dagger+\hat{n})(\hat{m}+\tfrac{1}{2}-2\hat{d})\right]-\sqrt{\Gamma}\hat{b}\hat{b}_\text{in} ,\\ \nonumber
\dot{\hat{d}}^\dagger&=&i2\Omega\hat{d}^\dagger-\Gamma \hat{d}^\dagger+i\varepsilon\left[\hat{n}(2\hat{d}^\dagger+\hat{m}+\tfrac{1}{2})-\rho(\hat{c}+\hat{c}^\dagger+\hat{n})(\hat{m}+\tfrac{1}{2}-2\hat{d}^\dagger)\right]-\sqrt{\Gamma}\hat{b}_\text{in}^\dagger\hat{b}^\dagger ,\\ \nonumber
\dot{\hat{m}}&=&-i2\varepsilon(\hat{d}-\hat{d}^\dagger)\left[(\rho-1)\hat{n}+\rho(\hat{c}+\hat{c}^\dagger)\right]-\sqrt{\Gamma}(\hat{b}^\dagger\hat{b}_\text{in}+\hat{b}_\text{in}^\dagger\hat{b}).
\end{eqnarray}

\subsection{Steady-state Equilibrium}

The special algebraic form of non-standard interaction together with the presence of drive term makes the analysis requiring a bit of care. Ambiguities could be avoided by using explicit decompositions and replacements
\begin{eqnarray}
\label{T10-eq9}
\hat{a}(t)&\to&\bar{a}e^{-i\omega t}+\hat{a}(t), \\ \nonumber
\hat{a}^\dagger(t)&\to&\bar{a}^\ast e^{i\omega t}+\hat{a}^\dagger(t), \\ \nonumber
\hat{n}(t)&\to&\bar{n}+\hat{n}(t), \\ \nonumber
\hat{m}(t)&\to&\bar{m}+\hat{m}(t),
\end{eqnarray}
\noindent
where $\bar{n}=|\bar{a}|^2$ is the average cavity photon number and $\bar{m}$ is the average cavity phonon number. From now on, $\hat{n}$ and $\hat{m}$ will represent only deviations from the average steady-state populations. In a similar manner, we can employ the decompositions and replacements
\begin{eqnarray}
\label{T10-eq10}
\hat{c}(t)&\to&\bar{c}e^{-2i\omega t}+\hat{c}(t), \\ \nonumber
\hat{c}^\dagger(t)&\to&\bar{c}^\ast e^{2i\omega t}+\hat{c}^\dagger(t), \\ \nonumber
\hat{d}(t)&\to&\bar{d}e^{-2i\Omega t}+\hat{d}(t), \\ \nonumber
\hat{d}^\dagger(t)&\to&\bar{d}^\ast e^{2i\Omega t}+\hat{d}^\dagger(t).
\end{eqnarray}
\noindent
It has to be mentioned that the replacements (\ref{T10-eq9},\ref{T10-eq10}) are not linearization, but rather an algebraic convention which further allows distinction of steady-state cavity photon and phonon populations under resonant drive. The first separated scalar terms represent the major non-oscillating parts $\bar{n}$ and $\bar{m}$, and oscillating parts for the rest.

One may now proceed to construct the equations for steady-state populations $\bar{n}$ and $\bar{m}$. This requires employing the replacements (\ref{T10-eq9},\ref{T10-eq10}) in Langevin equations (\ref{T10-eq8}), taking the derivatives on the left, multiplying both sides of the first and fourth equations of (\ref{T10-eq8}) by respectively $e^{i2\omega t}$ and $e^{i2\Omega t}$, and discarding all remaining time-dependent terms. This is equivalent to a Rotating-Wave Approximation (RWA) analysis, but carried out at double frequencies. Obviously, all subsequent derivations and calculations in the steady-state will retain their validity within the constraint of RWA condition.

Under resonance conditions where optical and mechanical frequencies are the same $\omega=\pm\Omega$, there remain a few extra terms. After discarding stochastic noise input, application of these steps to the first, third, fourth, and sixth equations of (\ref{T10-eq8}) and simplifying, gives the following four nonlinear algebraic equations for steady state terms
\begin{eqnarray}
\label{T10-eq11}
i\alpha^\ast\sqrt{\bar{n}}&=&\tfrac{1}{2}\kappa\bar{n}+i\varepsilon\left[\tfrac{1}{2}(1+\rho)\bar{m}\bar{n}-\rho(\bar{n}+\tfrac{1}{2})(\delta_{\omega,\Omega}\bar{d}+\delta_{\omega,-\Omega}\bar{d}^\ast)\right],\\ \nonumber
\bar{m}^2-\bar{m}&=&4|\bar{d}|^2,\\ \nonumber
-\Gamma\bar{d}&=&i\varepsilon\left[2(1+\rho)\bar{n}\bar{d}-\rho(\bar{m}+\tfrac{1}{2})(\delta_{\omega,\Omega}+\delta_{\omega,-\Omega})\bar{n}\right], \\ \nonumber
\kappa\bar{n}&=&i(\alpha-\alpha^\ast)\sqrt{\bar{n}}-i2\beta(\bar{d}^\ast-\bar{d})\bar{n}(\delta_{\omega,\Omega}-\delta_{\omega,-\Omega}).
\end{eqnarray}
\noindent
Here, $\delta_{\theta,\varphi}$ represents the Kronecker's delta, and equals $1$ if $\theta=\varphi$ and $0$ otherwise. Also, $\alpha$ is the complex drive amplitude as already defined in the above under (\ref{T10-eq1}). These are four equations in terms of $\bar{n}$, $\bar{m}$, $\bar{d}$, and $\angle\alpha$, and we notice that $|\alpha|=\eta P_\text{op}/\hbar\omega$ is the photon flux incident on the cavity due to the optical power $P_\text{op}$, where $\eta$ is the input coupling efficiency. When there is no resonance between optics and mechanics with $\delta_{\pm\omega,\pm\Omega}=0$, the system (\ref{T10-eq11}) reduces to the fairly simple pair 
\begin{eqnarray}
\label{T10-eq12}
|\alpha|^2&=&\varepsilon^2(1+\rho)^2\bar{m}^2\bar{n}, \\ \nonumber
\bar{n}|\alpha|^2&=&\varepsilon^2(\bar{m}^2-\bar{m}).
\end{eqnarray}
These two nonlinear algebraic equations can be numerically solved for non-negative real roots of $\bar{n}$ and $\bar{m}$.

\subsection{Integrable System \& Stability}

The Langevin equations (\ref{T10-eq8}) are still nonlinear and thus non-integrable, but they are instead expressed in terms of second-order operators. Hence, even after taking out the constant oscillating parts using the replacements (\ref{T10-eq9},\ref{T10-eq10}), and ignoring the remaining fourth- and higher-order nonlinear terms, the basic nonlinear quadratic interaction still survives. This advantage in using higher-order operators has been noticed by the author \cite{S2-Paper2} as well as others \cite{S2-Quad}. 

Following the one-dimensional analysis provided in the preceding study \cite{S2-Paper1}, the strength of subsequent nonlinear nth-order terms decrease typically as $(x_\text{zp}/l)^n$ where $x_\text{zp}$ is the mechanical zero-point fluctuations of vacuum and $l$ is the typical length scale of the cavity. Having that said, even and odd powers grow almost proportionally. So, once the membrane-in-the-middle or any equivalent setup which allows vanishing $g_0$ is employed, there would be no non-zero odd-order optomechanical interaction at all. Similarly, the lowest non-vanishing nonlinear term after the quadratic term would be the sextic (6th) order interaction which is expected to be weaker by a factor of $(x_\text{zp}/l)^n$ at least. This ratio shall be normally too small to be of any physical significance. Doing this leaves the linearized integrable form
\begin{eqnarray}
\label{T10-eq13}
\dot{\hat{c}}&=&[i2\omega-\kappa-i2(\varepsilon+\beta)\bar{m}]\hat{c}+i\beta(\bar{n}+\tfrac{1}{2})(\hat{d}+\hat{d}^\dagger-\hat{m})-i\beta\bar{m}\hat{n}-\sqrt{\tfrac{1}{2}\bar{n}\kappa}\hat{a}_\text{in}, \\ \nonumber
\dot{\hat{c}}^\dagger&=&[-i2\omega-\kappa+i2(\varepsilon+\beta)\bar{m}]\hat{c}^\dagger-i\beta(\bar{n}+\tfrac{1}{2})(\hat{d}+\hat{d}^\dagger-\hat{m})+i\beta\bar{m}\hat{n}-\sqrt{\tfrac{1}{2}\bar{n}\kappa}\hat{a}_\text{in}^\dagger, \\ \nonumber
\dot{\hat{n}}&=&-\kappa\hat{n}+i2\beta\bar{m}(\hat{c}-\hat{c}^\dagger)-\sqrt{\kappa\bar{n}}(\hat{a}_\text{in}+\hat{a}_\text{in}^\dagger),\\ \nonumber
\dot{\hat{d}}&=&[-i2\Omega-\Gamma-i2(\varepsilon+\beta)\bar{n}]\hat{d}+i\beta(\bar{m}+\tfrac{1}{2})(\hat{c}+\hat{c}^\dagger)-i(\varepsilon-\beta)[\bar{n}\hat{m}+(\bar{m}+\tfrac{1}{2})\hat{n}]-\sqrt{2\Gamma|\bar{d}|}\hat{b}_\text{in},\\ \nonumber
\dot{\hat{d}}^\dagger&=&[i2\Omega-\Gamma+i2(\varepsilon+\beta)\bar{n}]\hat{d}^\dagger-i\beta(\bar{m}+\tfrac{1}{2})(\hat{c}+\hat{c}^\dagger)+i(\varepsilon-\beta)[\bar{n}\hat{m}+(\bar{m}+\tfrac{1}{2})\hat{n}]-\sqrt{2\Gamma|\bar{d}|}\hat{b}_\text{in}^\dagger,\\ \nonumber
\dot{\hat{m}}&=&-\Gamma\hat{m}+i2(\varepsilon-\beta)\bar{n}(\hat{d}-\hat{d}^\dagger)-\sqrt{\Gamma|\bar{d}|}(\hat{b}_\text{in}+\hat{b}_\text{in}^\dagger).
\end{eqnarray} 
\noindent
Here, we have furthermore employed the white noise approximation to the Weiner processes $\hat{a}_\text{in}$ and $\hat{b}_\text{in}$. This Markovian approximation makes these noise processes insensitive to any frequency shift, or multiplication by purely oscillating terms such as $\exp(\pm i\omega t)$ and $\exp(\pm i\Omega t)$. This fact facilitates the construction of noise processes, avoiding the burden of higher-order noise terms. Otherwise, terms such as $\hat{a}_\text{in}^2$ and $\hat{b}_\text{in}^2$ enter the Langevin equations \cite{S2-Paper2} which need a careful and very special treatment to evaluate their corresponding spectral densities. Now, the system of Langevin equations (\ref{T10-eq13}) is fully linearized and can be integrated to obtain the spectral densities of each variable. In order to do this, we first define the input vector
\begin{equation}
\label{T10-eq14}
\{A_\text{in}\}^\text{T}=\left\{
\hat{a}_\text{in},
\hat{a}_\text{in}^\dagger, 
\tfrac{1}{2}(\hat{a}_\text{in}+\hat{a}_\text{in}^\dagger), 
\hat{b}_\text{in},
\hat{b}_\text{in}^\dagger,
\tfrac{1}{2}(\hat{b}_\text{in}+\hat{b}_\text{in}^\dagger) 
\right\},
\end{equation}
\noindent
and the diagonal matrix
\begin{equation}
\label{T10-eq15}
[\gamma]=\text{Diag}\left[\bar{n}\kappa,\bar{n}\kappa,4\bar{n}\kappa,2|\bar{d}|\Gamma,2|\bar{d}|\Gamma,4|\bar{d}|\Gamma\right].
\end{equation}
\noindent
The Langevin equations now read
\begin{equation}
\label{T10-eq16}
\frac{d}{dt}\{A\}=[\textbf{M}]\{A\}-\sqrt{[\gamma]}\{A_\text{in}\},
\end{equation}
\noindent
where the coefficients matrix $[\textbf{M}]$ is given as
\begin{equation}
\label{T10-eq17}
[\textbf{M}]= 
\left[
\begin{array}{cccccc}
i2(\omega-\zeta\bar{m})-\kappa & 0 & -i\beta\bar{m} & i\beta\tilde{n} & i\beta\tilde{n} & -i\beta\tilde{n} \\ 
0 & -i2(\omega-\zeta\bar{m})-\kappa & i\beta\bar{m} & -i\beta\tilde{n} & -i\beta\tilde{n} & i\beta\tilde{n} \\ 
i2\beta\bar{m} & -i2\beta\bar{m} & -\kappa & 0 & 0 & 0 \\
i\beta\tilde{m} & i\beta\tilde{m} & -i\chi\tilde{m} & -i2(\Omega+\zeta\bar{n})-\Gamma & 0 & -i\chi\bar{n} \\
-i\beta\tilde{m} & -i\beta\tilde{m} & i\chi\tilde{m} & 0 & i2(\Omega+\zeta\bar{n})-\Gamma & i\chi\bar{n} \\
0 & 0 & 0 & i2\chi\bar{n} & -i2\chi\bar{n} & -\Gamma
\end{array}
\right],
\end{equation}
\noindent
Here, $\zeta=\varepsilon+\beta$, $\chi=\varepsilon-\beta$, $\tilde{n}=\bar{n}+\frac{1}{2}$, and $\tilde{m}=\bar{m}+\frac{1}{2}$ similar to the definitions under (\ref{T8-Noise12}). This system is fully integrable and linearized, describing a nonlinear Hamiltonian, and can be Fourier transformed to find the spectral densities. Dynamical stability can be easily determined by inspection of the loci of eigenvalues of (\ref{T10-eq17}). The optomechanical system is unconditionally stable if all six eigenvalues have negative real values. 

It is remarkable that any linearization on the operators $\hat{a}$ and $\hat{b}$ done on the Hamiltonian (\ref{T10-eq1}) shall put the interaction in the well recognized forms of either $\mathcal{X}\mathbb{X}$ or $\mathcal{P}\mathbb{P}$, which reverts back to the basic conventional physics of optomechanics \cite{S2-1,S2-3,S2-6}. Hence, any linearization process on the first-order operators will eventually discard the physics of higher-order interactions. The method of higher-order operators discussed here and previous researches \cite{S2-Paper2,S2-Paper3,S2-Quad} is targeted to alleviate this problem.

\subsection{Spectral Density}

The input-output relations \cite{S2-Noise1,S2-Noise0,S2-Noise2} connects the incident and reflection waves of a weak probe beam at frequency $w$ (not to be confused with optical frequency $\omega$). This relationship in Fourier domain together with (\ref{T10-eq16}) yield
\begin{eqnarray}
\label{T10-eq18}
\{A_\text{out}(w)\}&=&\{A_\text{in}(w)\}+\sqrt{[\gamma]}\{A(w)\},\\ \nonumber
\{A_\text{out}(w)\}&=&[\textbf{S}(w)]\{A_\text{in}(w)\},\\ \nonumber
[\textbf{S}(w)]&=&[\textbf{I}]-\sqrt{[\gamma]}\left([\textbf{M}]-iw [\textbf{I}]\right)\sqrt{[\gamma]},
\end{eqnarray}
\noindent
where $[\textbf{I}]$ is the identity matrix. This can be now summarized to yield the spectral density of $\hat{c}$ for positive frequencies $w>0$ as
\begin{equation}
\label{T10-eq19}
S_{CC}(w)=|S_{11}(w)|^2+|S_{13}(w)|^2\tfrac{1}{2}+|S_{14}(w)|^2(\bar{m}+1)+|S_{15}(w)|^2\bar{m}+|S_{14}(w)|^2(\bar{m}+\tfrac{1}{2}),
\end{equation}
\noindent
where $S_{ij}(w)$ are frequency dependent elements of the $6\times 6$ scattering matrix $[\textbf{S}(w)]$ defined in (\ref{T10-eq18}). In the above, we have made use of the fact that spectral density of the input vector $\{\hat{A}_\text{in}(w)\}$ (\ref{T10-eq14}) is 
\begin{eqnarray}
\label{T10-eq20}
\{ S_\text{in}(w>0)\}^{\rm T}&=&\{1,0,\tfrac{1}{2},\bar{m}+1,\bar{m},\bar{m}+\tfrac{1}{2}\}, \\ \nonumber
\{ S_\text{in}(w<0)\}^{\rm T}&=&\{0,1,\tfrac{1}{2},\bar{m},\bar{m}+1,\bar{m}+\tfrac{1}{2}\}.
\end{eqnarray}
\noindent
There are quite a few assumptions needed to obtain (\ref{T10-eq19}), including validity of Gaussian white noise processes for both photons and phonons, complete independence of stochastic processes for phonons and photons, and ignoring higher-order noise terms arising from multiplicative noise terms in (\ref{T10-eq8}) leading to the mean-field approximation for nonlinear multiplicative noise. 

All remains now is to recover the spectral density $S_{AA}(w)$ of the first-order operator $\hat{a}$ from the calculated spectral density $S_{CC}(w)$ of the second-order operator. Assuming that $S_{CC}(w)$ is composed of well isolated Gaussian peaks as $S_{CC}(w)=\sum s_j \exp[-(w-\omega_j)^2/\Delta\omega_j^2]$ with $s_j$ and $\Delta\omega_j$ respectively being the amplitude and spread of the $j$-th Lorentzian, $S_{AA}(w)$ can be related \cite{S2-Paper2} to $S_{CC}(w)$ roughly as
\begin{equation}
\label{T10-eq21}
S_{AA}(w)\approx 2\pi\sqrt{\pi}\sum_j \frac{s_j\omega_j^2}{\Delta\omega_j} \exp\left[-\frac{\left(w-\frac{1}{2}\omega_j\right)^2}{2\Delta\omega_j^2}\right].
\end{equation}
\noindent
Therefore, the spectral density of the requested system operator can be approximately recovered from the calculated spectral density of the higher-order system operator. A factor 4 has been already been absorbed in (\ref{T10-eq21}) because of the definition $\hat{c}=\frac{1}{2}\hat{a}^2$.

Results of solving the system (\ref{T10-eq16}) using the scattering matrix formalism (\ref{T10-eq18}) and computation of spectral densities is possible for resonant $\omega=\Omega$, near-resonant $\omega\approx\Omega$, and off-resonant cases $\omega>\Omega$ and $\omega<\Omega$, in weak, strong, and ultrastrong coupling regimes. The cases of strong and ultrastrong coupling are here somehow idealized, and there might be difficulty in attaining those conditions under practical experimental situations. Nevertheless, these can be obtained using appropriately designed circuit cavity-electrodynamics quantum simulation of quadratic optomechanics \cite{S3-61}. For an initial study, the parameters may be taken from a recent study on superconducting electromechanics \cite{S3-79} where radio-and mechanical frequencies are accurately tuned and set to equal values.  

Since no quadratic interaction is investigated in the article under consideration, we can proceed to assume that the interaction with the second mechanical mode is being studied so that $g_0$ may be effectively set to zero. Furthermore, $\varepsilon$ is taken to be one to two orders of magnitude below the typical measured $g_0$ in the system, where $g_0$ is the single-photon optomechanical interaction rate. Typical parameters can be for instance chosen as similar as possible to the experiment \cite{S3-79} where available. This is the only known experimental configuration to the author, where mechanical frequency $\Omega$ and electromagnetic circuit frequency $\omega$ are designed to be within the same order of magnitude, here identical as $\omega=\Omega$, using a carefully designed membrane-in-the-middle setup. Having that said, not only it is actually possible to probe the regime of large mechanical frequency and even $\omega=\Omega$, as already has been shown, but also a membrane-in-the-middle configuration could provide direct access to quadratic interactions. So, not only the effect of momentum-field interaction term as was derived earlier in (\ref{S3-41}) for the non-standard quadratic optomechanics is in principle observable, but also, it can cause a bit of squeezing, also allowed by the calculations in \S\ref{S3-Squeeze}.

The results of this research could be relevant to sensing applications where nonlinear optomechanics could in principle produce pronounced sensitivities, such as those reported recently for quantum gravimetry \cite{S10-q14,S10-Thesis}.  Also possible generation of continuous squeezed electromagnetic radiation seems to be feasible by carefully optimized design of a membrane-in-the-middle superconducting electromechanics setup. Further applications could be still plausible but will need more in depth study. Experiments can unambiguously determine the existence of non-standard quadratic optomechanics.

In summary, we presented a detailed theoretical and numerical analysis of the quadratic optomechanical interactions, using the method of higher-order operators. We have studied both types of standard quadratic and the predicted non-standard quadratic interactions, and established clear signatures for existence of non-standard interactions, while stability of the system under study has also been established. Such types of quadratic interactions can be probed in a carefully designed experiment where the optical and mechanical frequencies fall within the same order of magnitude. 
\section{Cross-Kerr Interaction} \label{Section-11}
This section discusses a full operator approach to treatment of the cross-Kerr interaction combined with parametric amplification. It is shown that this problem can be exactly integrated using the method of higher-order operators. While the initial basis is infinite-dimensional, an orthogonal transformation can reduce the problem exactly into a six-dimensional basis which can be integrated conveniently. In some sense, the topic of this section covers both sub-categories of the quantum read-out circuits discussed previously in \S\ref{Section-6}, which includes quantum limited amplifiers \S\ref{Amplifier} and quantum nondemolition measurement through cross-Kerr interaction \S\ref{QND}. It turns out that presence of combined cross-Kerr interaction and parametric amplification is a lot more complex to solve, yet it admits exact solution to be discussed here.

The cross-Kerr Hamiltonian \cite{S4-38a,S4-36g,S4-36h} is among one of the mostly used nonlinear quantum interactions between two bosonic fields, which describes a wide range of phenomena. In the case of superconducting circuits, this interaction is of primary importance in modeling nonlinearity of quantum circuits, such as quantum bits and parametric amplifiers. Usually, one field represents a strong or pump field while the other refers to the weak or probe field. In the context of quantum optomechanics \cite{S4-38a} the physical nature of these two interacting fields could be quite different, referring to the photons and phonons. When combined with a parametric amplification term, then the total interaction Hamiltonian could be a lot more difficult to solve. 

Here, we demonstrate that the cross-Kerr interaction with parametric amplification could be exactly solvable using the method of higher-order operators \cite{S2-Paper2,S2-Paper3}, which has evolved out of the rich domain of quadratic optomechanics \cite{S2-Bruschi,S2-Quad,S2-9,S3-59,S4-33.6.1,S2-7,S4-33.6.5}. This method employs a different basis than the simple bath ladder operators, and quite recently has been independently also reported elsewhere \cite{S2-Quad}. In the context of superconducting quantum circuits, the interaction of two pump-probe microwave fields with the transmon qubits is effectively a cross-Kerr nonlinear interaction \cite{S11-Hoi}, and for all practical reasons it has to be followed immediately by a quantum-limited parametric amplifier stage. This combination leads to a cross-Kerr term with parametric amplification, the solution of which is the purpose of the present study. 

The importance of this contribution is two-fold. On the one hand, one may obtain the time evolution of the number of quanta in time. This enables accurate modeling of quantum non-demolition measurements, for instance, where the number of quanta is measured indirectly through an interaction of cross-Kerr type. Secondly, when optomechanical systems are being considered and the nature of the two interacting bosonic baths are different, the noise spectral density is the actual measurable quantity, the estimation of which is discussed here.

Further contributions of this section are connected to the treatment of multiplicative noise terms, which normally arise in the method of higher-order operators. It has been demonstrated that for the purpose of calculation of the noise spectral density, these can be exactly simplified to a great extent, where the multiplicative operators can be conveniently replaced by their silent or noiseless non-operator parts. 

Consider the cross-Kerr interaction \cite{S4-38a,S4-36g,S4-36h} with parametric amplification \cite{S4-36e,S4-36e1,S4-36f}, defined as 
\begin{align}
\mathbb{H}=\hbar\omega\hat{a}^\dagger\hat{a}+\hbar\Omega\hat{b}^\dagger\hat{b}+\hbar g\hat{a}^\dagger\hat{a}\hat{b}^\dagger\hat{b}+\hbar f\left(\hat{b}^2+\hat{b}^{\dagger 2}\right).
\end{align}
This Hamiltonian is usually analyzed using the basis $\{A\}^\text{T}=\{\hat{n},\hat{m},\hat{C},\hat{S} \}$ where $\hat{m}=\hat{b}^\dagger\hat{b}$, $\hat{n}=\hat{a}^\dagger\hat{a}$, and 
\begin{align}
\hat{C}&=\frac{1}{2}\left[(\hat{n}+1)^{-\frac{1}{2}}\hat{a}+\hat{a}^\dagger (\hat{n}+1)^{-\frac{1}{2}}\right], \\ \nonumber
\hat{S}&=\frac{1}{2i}\left[(\hat{n}+1)^{-\frac{1}{2}}\hat{a}-\hat{a}^\dagger (\hat{n}+1)^{-\frac{1}{2}}\right],
\end{align}
are the quadrature operators, satisfying the commutators $[\hat{n},\hat{C}]=-i\hat{S}$, $[\hat{n},\hat{S}]=i\hat{C}$, and $[\hat{C},\hat{S}]=\frac{1}{2}i(\hat{n}+2)^{-1}$. Usage of these quadrature operators might be advantageous in studying some cases, but construction of Langevin equations would require further approximation since these do not form a closed Lie algebra. As a result, their usage normally needs further linearization procedures which as a result deviates from the mathematically exact solution. In our analysis, however, we use a different basis with closed Lie algebra, which not only admits exact solution, but also allows inclusion of a parametric amplification to either of the $\hat{a}$ or $\hat{b}$ fields.

In the present formulation, we exclude the drive term from the Hamiltonian, and instead feed it through the input noise terms to the system. In particular, when the input terms also fluctuate around a non-zero input or drive term, this approach is accurate. Besides simplicity and the rather convenience involved, the other reason is that the drive term normally contains the simple ladder operator such as $\hat{a}$, whose presence changes the operator basis significantly. Any method to circumvent this difficulty could be much helpful in mathematical description of the problem. Furthermore, this picture where noise and drive terms are fed through the same channel to the system is physically consistent and correct.

We try to analyze this type of interaction in an open-system using Langevin equations \cite{S2-3,S2-6,S2-4,S2-Noise1,S2-Noise0}
\begin{align}
\frac{d}{dt}\hat{x}&=-\frac{i}{\hbar}[\hat{x},\mathbb{H}]-\sum_j[\hat{x},\hat{a}_j^\dagger]\left(\frac{1}{2}\kappa_j\hat{a}_j+\sqrt{\kappa_j}\hat{a}_{j,\text{in}}\right)+\sum_j\left(\frac{1}{2}\kappa_j\hat{a}_j^\dagger+\sqrt{\kappa_j}\hat{a}_{j,\text{in}}^\dagger\right)[\hat{x},\hat{a}_j],
\end{align}
where $j$ denotes the bosonic bath, $\hat{a}_j$ is the corresponding annihilator, and $\kappa_j$ is the associated coupling/loss rate. Hence, choosing $j=a,b$ implies $\hat{a}_a=\hat{a}$ and $\hat{a}_b=\hat{b}$, and also $\kappa_a=\kappa$ and $\kappa_b=\Gamma$, respectively corresponding to pump and probe, strong and weak fields, or photons and phonons, depending on the nature of the system under study. Furthermore, $\hat{a}_{j,\text{in}}$ is the input quantum noise from the bosonic bath $j$, and $\hat{x}$ is any operator in the system.

Choosing the infinite dimensional closed Lie algebra of higher-order operators
\begin{align}
\{A\}^\text{T}=\{\hat{m},\hat{d},\hat{d}^\dagger,\hat{n}\hat{m},\hat{n}\hat{d},\hat{n}\hat{d}^\dagger,\dots,\hat{n}^j\hat{m},\hat{n}^j\hat{d},\hat{n}^j\hat{d}^\dagger,\dots \},
\end{align}
where $\hat{d}=\frac{1}{2}\hat{b}^2$, with $[\hat{d},\hat{m}]=2\hat{d}$, $[\hat{m},\hat{d}^\dagger]=2\hat{d}^\dagger$, $[\hat{d},\hat{b}^\dagger]=\hat{b}$ and $[\hat{d},\hat{d}^\dagger]=\hat{m}+\frac{1}{2}$ \cite{S2-Paper2} allows construction of linear infinite-dimensional Langevin equations, given as
\begin{align}
\frac{1}{2\Omega}\frac{d}{dt}\left\{A\right\}=\left(i\left[\textbf{M}\right]-\left[\Gamma\right]\right)\left\{A\right\}-i\frac{\alpha}{2}\left\{A_\text{c}\right\}-\left[\sqrt{\Gamma}\right]\left\{A_\text{in}\right\},
\end{align}
in which $\alpha=f/\Omega$ and 
\begin{align}
\left\{A_\text{c}\right\}&=\left\{A_{\text{c},j};j\in\mathscr{N}\right\}=\{0,1,-1,0,\hat{n},-\hat{n},\dots,0,\hat{n}^l,-\hat{n}^l,\dots \},
\end{align}
and
\begin{align}
\left[\Gamma\right]&=\left[\Gamma_j\delta_{ij};i,j\in\mathscr{N}\right] =\text{diag}\{\gamma_1,\gamma_1,\gamma_1,\dots,\gamma_l,\gamma_l,\gamma_l,\dots\},
\end{align}
is a diagonal matrix of normalized loss rates with $\gamma_l=\left[\Gamma+(l-1)\kappa\right]/2\Omega$. Furthermore, the noise input vector is
\begin{align}
&\{A_\text{in}\}^\text{T}=\{\hat{A}_{\text{in},j};j\in\mathscr{N}\}^\text{T}=\frac{1}{\sqrt{2\Omega}}\{\hat{m}_\text{in},\hat{d}_\text{in},\hat{d}^\dagger_\text{in},\dots,(\hat{n}^j\hat{m})_\text{in},(\hat{n}^j\hat{d})_\text{in},(\hat{n}^j\hat{d}^\dagger)_\text{in},\dots \}, 
\end{align}
in which the combined noise terms are constructed following \S\ref{S11-Noise} according to 
\begin{align}
\sqrt{\gamma_{j+1}}(\hat{n}^j\hat{d})_\text{in}&=\sqrt{j\frac{\kappa}{2\Omega}}\hat{n}^j_\text{in}\hat{d}+\sqrt{\frac{\Gamma}{2\Omega}}\hat{n}^j\hat{d}_\text{in}; j\in\mathscr{Z}^+, \\ \nonumber
\sqrt{\gamma_{j+1}}(\hat{n}^j\hat{d}^\dagger)_\text{in}&=\sqrt{j\frac{\kappa}{2\Omega}}\hat{n}^j_\text{in}\hat{d}^\dagger+\sqrt{\frac{\Gamma}{2\Omega}}\hat{n}^j\hat{d}^\dagger_\text{in}; j\in\mathscr{Z}^+, \\ \nonumber
\sqrt{\gamma_{j+1}}(\hat{n}^j\hat{m})_\text{in}&=\sqrt{j\frac{\kappa}{2\Omega}}\hat{n}^j_\text{in}\hat{m}+\sqrt{\frac{\Gamma}{2\Omega}}\hat{n}^j\hat{m}_\text{in}; j\in\mathscr{Z}^+. 
\end{align}
The single terms are given as
\begin{align}
\hat{n}^{j+1}_\text{in}&=\sqrt{j+1}\left(\hat{n}^j\hat{a}^\dagger\hat{a}_\text{in}+\hat{a}_\text{in}^\dagger\hat{a}\hat{n}^j\right); j\in\mathscr{Z}^+, \\ \nonumber
\hat{m}_\text{in}&=\hat{b}^\dagger\hat{b}_\text{in}+\hat{b}_\text{in}^\dagger\hat{b}, \\ \nonumber
\hat{d}_\text{in}&=\frac{1}{2}\hat{b}\hat{b}_\text{in}+\frac{1}{2}\hat{b}_\text{in}\hat{b}.
\end{align}
At this point, there are three very important facts to take notice of: 
\begin{enumerate}
	\item 
	Firstly, the contributing part of the multiplicative operators which operate on the white Gaussian noise processes $\hat{a}_\text{in}$ and $\hat{b}_\text{in}$ as shown in \S\ref{S11-Silence} are actually the silent or noiseless parts of these operators, which can be found by solving the corresponding Langevin equations with all zero-mean stochastic processes dropped and only keeping the drive terms. The calculation of silent terms will thus be no longer an operator problem, and can be addressed by any appropriate analytical or numerical approach.
	\item
	Secondly, the order of multiplicative terms, as whether they appear on the left or right of the noise terms is found to be immaterial within the accuracy of Langevin equations. This latter and rather important conclusion can be drawn from the last equation which silent operators actually commute with any Gaussian White noise process, following the construction procedure discussed in \S\ref{S11-Noise}, and is furthermore compatible with the commutation of multiplicative terms with noise operators. 
	\item
	The third issue is connected to the Hermitian conjugates of noise processes, such as $\hat{a}^\dagger_\text{in}(t)$ as opposed to $\hat{a}_\text{in}(t)$. In the frequency domain these are time-reversed conjugates of each other, which happen to be identical by the general laws of the expectation values of Gaussian noise, given by \cite{S2-Noise1,S2-Noise0} $\braket{\hat{a}^\dagger_\text{in}(w)\hat{a}_\text{in}(W)}=\delta(w+W)$ and $\braket{\hat{a}_\text{in}(w)\hat{a}_\text{in}(W)}=0$. Therefore, while the spectral densities of $\hat{a}_\text{in}(w)$ and $\hat{a}_\text{in}^\dagger(w)$ are evidently equal, they share the same Fourier transform, too. As a result, the Hermitian conjugate can be arbitrarily dropped from or added to the Gaussian White noise processes as long as the noise spectral density is going to be the quantity to be calculated.
\end{enumerate}

Hence, for the purpose of calculation of noise spectral density at non-zero frequencies, the replacements $\hat{d}_\text{in}=b\hat{b}_\text{in}$, and similarly $\hat{n}^{j+1}_\text{in}=(a^\ast +a)n^j\hat{a}_\text{in}$ and $\hat{m}_\text{in}=(b^\ast+b)\hat{b}_\text{in}$ are admissible, where all multiplicative operators can effectively be replaced with their silent contributions. Knowledge of these expressions is extremely helpful in any computation of noise spectral density, especially in the context of the method higher-order operators, where occurrence of multiplicative noise terms is inevitable. 

Also, the dimensionless coefficients matrix $\left[\textbf{M}\right]$ may be decomposed into real-valued $3\times 3$ partitions as
\begin{align}
\left[\textbf{M}\right]=\left[\begin{array}{c|c|c|c|c|c}
\textbf{A} & \textbf{B} & \textbf{0} & \textbf{0} & \textbf{0} & \dots \\ \hline
\textbf{0} & \textbf{A} & \textbf{B} & \textbf{0} & \textbf{0} & \dots \\ \hline
\textbf{0} & \textbf{0} & \textbf{A} & \textbf{B} & \textbf{0} & \dots \\ \hline
\textbf{0} & \textbf{0} & \textbf{0} & \textbf{A} & \textbf{B} & \dots \\ \hline
\vdots & \vdots & \vdots & \vdots & \vdots & \ddots 
\end{array}\right],
\end{align}
in which the partitions are given by
\begin{align}
\left[\textbf{A}\right]=\left[\begin{array}{ccc}
0 & 2\alpha & -2\alpha \\
-\alpha & -1 & 0 \\
\alpha & 0 & 1
\end{array}
\right],
\end{align}
and 
\begin{align}
\left[\textbf{B}\right]=\left[\begin{array}{ccc}
0 & 0 & 0 \\
0 & -\beta & 0 \\
0 & 0 & \beta
\end{array}
\right],
\end{align}
with $\beta=g/\Omega$. Similarly, the normalized decay matrix $[\Gamma]$ can be written as
\begin{align}
\left[\Gamma\right]=\left[\begin{array}{c|c|c|c|c}
\textbf{G}_1 & \textbf{0} & \textbf{0} & \textbf{0} & \dots \\ \hline
\textbf{0} & \textbf{G}_2 & \textbf{0} & \textbf{0}  & \dots \\ \hline
\textbf{0} & \textbf{0} & \textbf{G}_3 & \textbf{0} & \dots \\ \hline
\vdots & \vdots & \vdots & \vdots & \ddots 
\end{array}\right],
\end{align}
in which the partitions are given by $\textbf{G}_j=\text{diag}\{\gamma_j,\gamma_j,\gamma_j\}$.

Here, we can show that there exist $3\times 3$ matrices $[\textbf{U}]$ and $[\textbf{V}]$ in such a way that if the $9\times 9$ unimodular transformation matrix $[\textbf{P}]$ with $|[\textbf{P}]|=1$ is constructed as
\begin{align}
[\textbf{P}]=\left[\begin{array}{c|c|c}
\textbf{I} & \textbf{0} & \textbf{U} \\ \hline
\textbf{0} & \textbf{I} & \textbf{V} \\ \hline
\textbf{0} & \textbf{0} & \textbf{I} 
\end{array}
\right],\\ \nonumber
[\textbf{P}]^{-1}=\left[\begin{array}{c|c|c}
\textbf{I} & \textbf{0} & -\textbf{U} \\ \hline
\textbf{0} & \textbf{I} & -\textbf{V} \\ \hline
\textbf{0} & \textbf{0} & \textbf{I} 
\end{array}
\right],
\end{align}
where single lines separate $3\times 3$ blocks, and 
\begin{align}
[\textbf{Q}]=\left[\begin{array}{c||c|c|c|c|c}
\textbf{P} & \textbf{0} & \textbf{0} & \textbf{0} & \textbf{0} & \dots \\ \hline\hline
\textbf{0} & \textbf{I} & \textbf{0} & \textbf{0} & \textbf{0} & \dots \\ \hline
\textbf{0} & \textbf{0} & \textbf{I} & \textbf{0} & \textbf{0} & \dots \\ \hline
\textbf{0} & \textbf{0} & \textbf{0} & \textbf{I} & \textbf{0} & \dots \\ \hline
\vdots & \vdots & \vdots & \vdots & \vdots & \ddots 
\end{array}\right],
\end{align}
where double lines separate $9\times 9$ blocks, and then
\begin{align}
[\textbf{Q}]^{-1}&\left(i[\textbf{M}]-[\Gamma]\right)[\textbf{Q}]=\left[\begin{array}{c|c|c||c|c|c}
i\textbf{A}-\textbf{G}_1 & i\textbf{B} & \textbf{0} & \textbf{0} & \textbf{0} & \dots \\ \hline
\textbf{0} & i\textbf{A}-\textbf{G}_2 & \textbf{0} & \textbf{0} & \textbf{0} & \dots \\ \hline
\vdots & \vdots & \vdots & \vdots & \vdots & \ddots 
\end{array}\right].
\end{align}
This orthogonal transformation reduces the coefficients matrix $i[\textbf{M}]-[\Gamma]$ in such a way that the Langevin equations for the first six elements of $\{A\}$ are isolated. That therefore will reduce the infinite dimensional problem exactly into a six-dimensional problem in the basis
\begin{align}
\{A_6\}^\text{T}=\{\hat{m},\hat{d},\hat{d}^\dagger,\hat{n}\hat{m},\hat{n}\hat{d},\hat{n}\hat{d}^\dagger\}.
\end{align}
To show the existence of such a transformation, we can evaluate the transformed matrix first, given as
\begin{equation}
[\textbf{R}]=[\textbf{Q}]^\textbf{T}\left(i[\textbf{M}]-[\Gamma]\right)[\textbf{Q}],
\end{equation}
and then set the first two rows of the third column of the $3\times 3$ partition blocks to zero. This gives to the set of algebraic equations
\begin{align}
i\left(\textbf{A}\textbf{U}-\textbf{U}\textbf{A}+\textbf{B}\textbf{V}\right)-\textbf{G}_1\textbf{U}+\textbf{U}\textbf{G}_3=\textbf{0},\\ \nonumber
i\left(\textbf{A}\textbf{V}-\textbf{V}\textbf{A}+\textbf{B}\right)-\textbf{G}_2\textbf{V}+\textbf{V}\textbf{G}_3=\textbf{0}.
\end{align}
When expanded, these give rise to a total of $18=2\times 9=2\times 3\times 3$ linear algebraic equations in terms of the elements of $\textbf{U}$ and $\textbf{V}$, which conveniently offers a unique solution for nonzero decay matrix $[\Gamma]$. Explicit expressions are not useful and numerical solution can help if needed. But it is not difficult to calculate $\textbf{V}$ from the second equation. Doing this gives
\begin{align}
\textbf{V}&=\frac{1}{\lambda(4\alpha^2-\lambda^2-1)} \left[
\begin{array}{ccc}
0 & 2\alpha\beta(i-\lambda) & -2\alpha\beta(i+\lambda)\\
-\alpha\beta(i+\lambda) & -i\beta(1+\lambda^2) & 0 \\ 
\alpha\beta(i-\lambda) & 0 & i\beta(1+\lambda^2)
\end{array}
\right],
\end{align}
with $\lambda=\kappa/2\Omega$. However, once it is known that $\textbf{U}$ and $\textbf{V}$ do exist, then it is actually unnecessary to calculate them any longer, since the top left $6\times 6$ block of $\textbf{P}$ is nothing but the identity matrix. That means, very surprisingly, that the truncated system of Langevin equations in terms of the operator basis $\{A_6\}$ as in (18) is already exact. Hence, the $6\times 6$ truncated Langevin equations are actually already exact and integrable for the case of cross-Kerr interaction with parametric amplification.

When the pump field $\hat{a}$ is so strong that its quantum nature could be neglected, a more compact representation of the cross-Kerr interaction can be obtained. The same procedure can be exactly applied to the first $3\times 3$ block by solving the equation $i(\textbf{A}\textbf{V}-\textbf{V}\textbf{A}+\textbf{B})+\textbf{G}_1\textbf{V}-\textbf{V}\textbf{G}_2=\textbf{0}$ in terms of the elements of $\textbf{V}$. That will make the truncated $3\times 3$ Langevin equations in terms of the operators $\{A_3\}^\text{T}=\{\hat{m},\hat{d},\hat{d}^\dagger\}$ exact and integrable again. This will lead to the relatively simple expression for the $6\times 6$ unimodular matrix $[\textbf{P}]$ as
\begin{align}
[\textbf{P}]=\left[\begin{array}{c|c}
\textbf{I} & \textbf{V} \\ \hline
\textbf{0} & \textbf{I}  
\end{array}
\right],\\ \nonumber
[\textbf{P}]^{-1}=\left[\begin{array}{c|c}
\textbf{I} &  -\textbf{V} \\ \hline
\textbf{0} & \textbf{I}
\end{array}
\right],
\end{align}
while $\textbf{V}$ is again already known from (19). But this will not pull out any information regarding the second other field expressed by the bosonic population operator $\hat{n}$. In the end, it is appropriate therefore and makes sense to assign $\hat{n}$ to the strong field and $\hat{m}$ to the weak field. Under the circumstances where the strong field could be treated classically, then this $3\times 3$ choice of basis is convenient.

Once the system is made integrable, calculation of Noise Spectral Density and time-evolution of operators becomes straightforward, as discussed in \S\ref{S11-SpectralDensity} and \S\ref{S11-TimeDomain}, respectively.

ppose that $\hat{a}$ represents the strong pump field. Then, $\sqrt{\kappa\eta}\braket{\hat{a}_\text{in}}$ is the photon input rate to the cavity, which after normalization corresponds to the input optical power as 
\begin{align}
\xi=\frac{1}{2\Omega}\sqrt{\frac{\kappa\eta }{\hbar \omega}P_\text{op}}.
\end{align}
Here, $\eta$ and $P_\text{op}$ respectively are the coupling efficiency and input optical power. Under steady-state where $d/dt=0$, the operators relax to their mean values. Then one may construct a system of equations in terms of the mean field values $\{\bar{m},\bar{d},\bar{d}^\ast,\overline{nm},\overline{nd},\overline{nd^\ast} \}$. Using the further approximation $\bar{a}=\sqrt{\bar{n}}$, $\overline{nm}\approx\bar{n}\bar{m}$ and $\overline{nd}\approx\bar{n}\bar{d}$, as well as $\braket{\hat{b}_\text{in}}=0$, and after significant but straightforward algebra, one may construct the nonlinearly coupled steady state algebraic equations, which can be then solved to yield
\begin{align}
\bar{m}&=\frac{2\alpha^2}{\left(1+\beta\bar{n}\right)^2+\gamma^2-4\alpha^2}, \\ \nonumber
\bar{d}&=-\frac{i\alpha}{i(1+\beta\bar{n})+\gamma}\left(\bar{m}+\frac{1}{2}\right).
\end{align}
Here, $\gamma=\gamma_1$. The mean value of $\bar{n}$ can be obtained by numerical solution of the implicit equation
\begin{align}
\lambda^2\bar{n}\left(\frac{\bar{m}}{2|\bar{d}|}\right)^2=\xi^2.
\end{align}
The above quintic equation in terms of $\bar{n}$ is nonlinearly linked to the normalized pump $\xi$. Here, $\bar{m}$ and $\bar{d}$ are taken from the previous equations (23). The expression within the parentheses is numerically of the order of $4$ for typical choice of cavity parameters, and the quintic equation conveniently offers only one single positive real root for $\bar{n}$ for most range of the input power. This is while in standard optomechanics, this ratio has been shown to be roughly or extremely close to $2$ for respectively side-band resolved or Doppler cavities.

Now that the steady-state equations are known, all operators are replaced by their respective variations around their mean values, and non-zero mean drive and constant terms can be dropped. Doing this, simplifies the problem as the $3\times 3$ set of normalized dimensionless Langevin equations, given by
\begin{align}
\frac{d}{d\tau}\left\{
\begin{array}{c}
\delta\hat{d} \\
\delta\hat{d}^\dagger\\
\delta\hat{m} 
\end{array}
\right\}
&=\left[
\begin{array}{ccc}
-i(1+\beta\bar{n})-\gamma & 0 & -i\alpha \\
0 & i(1+\beta\bar{n})-\gamma & i\alpha \\
2i\alpha & -2i\alpha & -\gamma
\end{array}
\right] \left\{
\begin{array}{c}
\delta\hat{d} \\
\delta\hat{d}^\dagger\\
\delta\hat{m} 
\end{array}
\right\}
-\sqrt{\gamma}\left\{
\begin{array}{c}
\bar{b}\hat{y}_\text{in}\\
\bar{b}^\ast\hat{y}_\text{in}^\dagger\\
\bar{b}\hat{y}_\text{in}^\dagger+\bar{b}^\ast\hat{y}_\text{in}
\end{array}
\right\}.
\end{align}
Here, $\tau=2\Omega t$ is the normalized time, $\hat{y}_\text{in}=\hat{b}_\text{in}/\sqrt{2\Omega}$ is the normalized noise input with the normalized symmetrized spectral density $S_{YY}(w)=\frac{1}{2}$, and $\bar{b}=\sqrt{2\bar{d}}$ is known from solution of (24) and then (23).

We now adopt the definitions
\begin{align}
\{\delta\hat{A}\}^\text{T}&=\{\delta\hat{d},\delta\hat{d}^\dagger,\delta\hat{m} \}, \\ \nonumber
[\textbf{N}]&=\left[
\begin{array}{ccc}
-i(1+\beta\bar{n})-\gamma & 0 & -i\alpha \\
0 & i(1+\beta\bar{n})-\gamma & i\alpha \\
2i\alpha & -2i\alpha & -\gamma
\end{array}
\right], \\ \nonumber
\{\hat{A}_\text{in}\}^\text{T}&=\left\{
\bar{b}\hat{y}_\text{in},
\bar{b}^\ast\hat{y}_\text{in}^\dagger,
\bar{b}\hat{y}_\text{in}^\dagger+\bar{b}^\ast\hat{y}_\text{in}
\right\},
\end{align} 
which allows us to rewrite (25) in the compact form
\begin{align}
\frac{d}{d\tau}\{\delta\hat{A}(\tau)\}=[\textbf{N}]\{\delta\hat{A}(\tau)\}-\sqrt{\gamma}\{\hat{A}_\text{in}(\tau)\}.
\end{align}
These equations can be numerically integrated to study the evolution of number of quanta $\hat{m}(\tau)$, where $\hat{y}_\text{in}$ is the stochastic noise input to the system.

Taking the Fourier transform in normalized frequency units of $w=\omega/2\Omega$ gives
\begin{align}
\{\delta\hat{A}(w)\}=\sqrt{\gamma}\left([\textbf{N}]-iw[\textbf{I}]\right)^{-1}\{\hat{A}_\text{in}(w)\}.
\end{align}
Using the input-output relation \cite{S2-Noise1,S2-Noise0} we have
\begin{align}
\{\hat{A}_\text{out}(w)\}&=\left\{[\textbf{I}]-\gamma\left([\textbf{N}]-iw[\textbf{I}]\right)^{-1}\right\}\{\hat{A}_\text{in}(w)\}=[\textbf{S}(w)]\{\hat{A}_\text{in}(w)\}.
\end{align}
Here, we refer $[\textbf{S}(w)]$ as the scattering matrix. Once $[\textbf{S}(w)]$ is known, we can obtain $S_{DD}(w)$ from \cite{S2-6}
\begin{align}
&S_{DD}(w)=|\bar{b}[S_{11}(w)+S_{13}(w)]+\bar{b}^\ast [S_{12}(w)+S_{13}(w)]|^2 S_{YY}(w).
\end{align}
It is ultimately possible to recover $\bar{S}_{BB}(w)$ from $S_{DD}(w)$, which is the desired measurable spectrum, as shown in \S\ref{S11-SpectralDensity} through the transformation
\begin{align}
S_{BB}(w)&=\frac{1}{2}+\mathscr{F}\left\{\sqrt{\frac{1}{2}\mathscr{F}^{-1}\left\{S_{DD}(w)-\frac{1}{2}\right\}(t)}\right\}(w),
\end{align}
in which $\mathscr{F}$ denotes the Fourier transformation, and based on which we may now define $\bar{S}_{BB}(w)=\frac{1}{2}[S_{BB}(w)+S_{BB}(-w)]$ as the symmetrized noise spectrum.

We first notice that the Langevin equation for $\delta\hat{d}$ is independent of $\delta\hat{d}^\dagger$ and vice versa, which greatly simplifies the analysis. However, the same is not true for the scattering matrix $[\textbf{S}(w)]$, whose top-left $2\times 2$ block must be diagonalized first to correctly separate contributions from $\hat{d}$ and $\hat{d}^\dagger$.

Let us assume that $[\Sigma(w)]$ is the reflection scattering matrix defined as
\begin{align}
[\Sigma(w)]=[\textbf{I}]+\gamma\left([\textbf{N}]-iw[\textbf{I}]\right)^{-1}.
\end{align}
This scattering matrix is different from $[\textbf{S}(w)]$ defined in (29), since input shines from the outside whereas for the purpose of noise spectral density calculations, noise is generated from within the cavity. Therefore, defining $R(w)=b_\text{out}(w)/|b_\text{in}(w)|$ with $\phi=\angle R(w)$, we have
\begin{align}
\frac{1}{2}R^2(w)&=\frac{1}{2}\left[\Sigma_{11}(w)e^{2i\phi}+\Sigma_{12}(w)e^{-2i\phi}\right]+\Sigma_{13}(w),\\ \nonumber
\frac{1}{2}R^{\ast 2}(w)&=\frac{1}{2}\left[\Sigma_{21}(w)e^{2i\phi}+\Sigma_{22}(w)e^{-2i\phi}\right]+\Sigma_{23}(w).
\end{align}
These can be solved to find the phase $\phi$ as
\begin{align}
\sin(2\phi)=i\frac{\Sigma_{13}(w)-\Sigma_{23}^\ast(w)}{\Sigma_{11}(w)-\Sigma_{21}^\ast(w)-\Sigma_{12}(w)+\Sigma_{22}^\ast(w)}.
\end{align}
This offers the solution
\begin{align}
e^{\pm 2i\phi}&=\frac{\Sigma_{23}^\ast(w)-\Sigma_{13}(w)}{\Sigma_{11}(w)-\Sigma_{21}^\ast(w)-\Sigma_{12}(w)+\Sigma_{22}^\ast(w)}\mp\sqrt{1+\left[\frac{\Sigma_{23}^\ast(w)-\Sigma_{13}(w)}{\Sigma_{11}(w)-\Sigma_{21}^\ast(w)-\Sigma_{12}(w)+\Sigma_{22}^\ast(w)}\right]^2}.
\end{align}
The reflectivity $\mathcal{R}(w)$ and transmissivity $\mathcal{T}(w)$ now can be easily found from the relationship
\begin{align}
\mathcal{R}(w)&=|R^2(w)|,\\ \nonumber
\mathcal{T}(w)&=1-|R^2(w)|.
\end{align}

Setting up the fully linearized Langevin equations for (1) in terms of both operators $\hat{a}$ and $\hat{b}$ gives an identical set of equations to that of fully linearized optomechanics. In fact, all nonlinear interaction Hamiltonian between two bosonic operators, such as standard optomechanics, standard and non-standard quadratic optomechanics, and cross-Kerr interaction, take identical set of fully linearized equations. This is a well-known fact in nonlinear quantum mechanics.

Here, we proceed only by linearization of the probe beam $\hat{b}$ and leave the pump $\hat{a}$ out of basis. This will give the set of equations (33), which after some further linearization becomes
\begin{align}
\frac{d}{d\tau}\left\{\begin{array}{c}
\delta\hat{b} \\
\delta\hat{b}^\dagger
\end{array}\right\}&=\frac{1}{2}
\left[\begin{array}{cc}
-i(2\beta\bar{n}+1)-\frac{1}{2}\gamma & -i4\alpha \\
i4\alpha & i(2\beta\bar{n}+1)-\frac{1}{2}\gamma
\end{array}\right] 
\left\{\begin{array}{c}
\delta\hat{b} \\
\delta\hat{b}^\dagger
\end{array}\right\}-\sqrt{\gamma}\left\{\begin{array}{c}
\hat{y}_\text{in}\\
\hat{y}_\text{in}^\dagger
\end{array}
\right\}.
\end{align} 
Quite clearly, there is no way to determine the operator mean field values $\bar{n}$ and $\bar{b}$ from this analysis, since the pump field $\hat{a}_\text{in}$ is absent. Let us for the moment assume that $\bar{n}$ is determined from the same equation as (24) found in the above for the extended higher-order basis.

We assume $\omega=2\Omega=2\pi\times 2\text{GHz}$, and the quality factors for both modes are set to $100$. We furthermore set the coupling efficiency as $\eta=0.4$, while the cross-Kerr interaction rate is $g=2\pi\times 100\text{kHz}$ and the parametric amplification rate is $f=2\pi\times 50\text{MHz}$. The ratio $f/g$ is swept across various input pump optical powers $P_\text{op}$ from close to zero up to $4\text{fW}$. At microwave frequencies, the input optical power of $P_\text{op}=1\text{fW}$ corresponds to a normalized photon input rate of $\xi=0.0155$.

\subsection{Construction of Noise Terms}\label{S11-Noise}

It is straightforward to see how the noise terms of higher-order operators should be constructed. In the case that the operators are treated fully nonlinearly regardless of their mean values, then the corresponding noise spectral densities could be non-trivial to calculate. This issue, for the case of the squared operator $\hat{d}$ has been discussed in details elsewhere \cite{S2-Paper2}. But this is appropriate only when the noise spectral density is under consideration. When other quantities are to be measured, for which the non-squared ladder operators might be needed, an iterative approach like $\hat{b}_{j+1}=\hat{d}+\hat{b}_{j}-\frac{1}{2}\hat{b}_{j}^2$ with $\hat{b}_{0}=\hat{1}$ provides uniform convergence to the ladder operator $\hat{b}$.

In order to do this, one may use, for instance, the Langevin equation for $\hat{b}$ and multiply both sides by $\hat{b}$ from left and right. Summing up together, leads to a Langevin equation for $\hat{b}^2$ with a noise input term such as those displayed in (10). The Langevin equation for $\hat{d}$ can be also directly constructed as shown in the above, and that ends up in a noise term as $\hat{d}_\text{in}$ with a decay rate of $\Gamma$. Within the accuracy of Langevin equations, these two noise terms coming from the two approaches should be identical, and this is how one can obtain all the noise terms in (10) in such an iterative manner alike.

One should keep in mind that the Langevin equations are neither exact nor rigorous by nature, as their construction necessitates at least two approximations of non-dispersive coupling and Gaussian white noise. The discussion around this topic is outside of the scope of the present study.

\subsection{Silence of Multiplicative Operators}\label{S11-Silence}

The first issue to notice in treatment of the multiplicative noise is the dependence of the multiplying operators to the noise terms. These operators also are determined from lower order Langevin equations in which similar noise terms are fed in. Iteratively going back to the lowest order determines that these multiplying operators appear as an infinite series such as 
\begin{equation}
\label{T11-B1}
\hat{a}(t)=\hat{a}_0(t)+a(t)+a_1(t)\hat{a}_\text{in}(t)+a_2(t)\hat{a}_\text{in}^2(t)+\dots,
\end{equation}
where $\hat{a}_0(t)=\exp\left[( i[\textbf{M}]-[\Gamma] ) t\right]\hat{a}_0(0)$ is the decaying operator term of the homogeneous solution to the system of Langevin equations, which decays to zero, thus taking no part in the steady state solution, and is excluded from contributing to the noise spectral density. Furthermore, $a(t)=a(t)\hat{1}$ is the noiseless or silent part of the operator $\hat{a}(t)$ driven by the external classical pump field, which can in principle be determined from solving the Langevin equations with the stochastic terms dropped, while only keeping the drive terms as input. 

The above term ultimately gets multiplied to another Gaussian white noise term, such as $\hat{a}_\text{in}(t)$ again. Such a multiplicative noise term as $\hat{a}(t)\hat{a}_\text{in}(t)$ will have an expansion given by 
\begin{equation}
\label{T11-B2}
\hat{a}(t)\hat{a}_\text{in}(t)=a(t)\hat{a}_\text{in}(t)+a_1(t)\hat{a}_\text{in}^2(t)+a_2(t)\hat{a}_\text{in}^3(t)+\dots,
\end{equation}
and so on. It is not difficult to see that as long as a lower-order Gaussian noise term is present, the higher-order terms will have negligible contribution to non-zero absolute (and not detuned) frequencies in the ultimate noise spectral density. This is discussed in \S\ref{S11-HigherOrderNoise}.

In order to establish this, we may define the second-order noise corresponding to the squared process $\hat{c}(t)=\frac{1}{2}\hat{a}^2(t)$, which clearly has a decay rate of $2\kappa$. The corresponding stochastic noise process is
\begin{equation}
\label{T11-B3}
\hat{c}_\text{in}(t)=\frac{1}{\sqrt{2\kappa}}\hat{a}_\text{in}^2(t).
\end{equation}
The stochastic process $\hat{c}_\text{in}(t)$ is no longer Gaussian white although $\hat{a}_\text{in}(t)$ is a Gaussian white stochastic process by assumption with a symmetrized auto-correlation $\braket{\hat{a}^\dagger(t)\hat{a}(\tau)}_\text{S}=\frac{1}{2}\delta(t-\tau)$. The symmetrized autocorrelation of this higher-order stochastic process in light of the Isserlis-Wick theorem \cite{S2-Paper2} is thus given by
\begin{align}
\label{T11-B4}
\braket{\hat{c}_\text{in}^\dagger(t)\hat{c}_\text{in}(\tau)}_\text{S}&=\frac{1}{2\kappa}\braket{\hat{a}_\text{in}^\dagger(t)\hat{a}_\text{in}^\dagger(\tau)}_\text{S}\braket{\hat{a}_\text{in}(t)\hat{a}_\text{in}(\tau)}_\text{S} +2\times\frac{1}{2\kappa}\braket{\hat{a}_\text{in}^\dagger(t)\hat{a}_\text{in}(\tau)}_\text{S}\braket{\hat{a}_\text{in}^\dagger(t)\hat{a}_\text{in}(\tau)}_\text{S}\\ \nonumber
&=\frac{1}{\kappa}\braket{\hat{a}_\text{in}^\dagger(t)\hat{a}_\text{in}(\tau)}_\text{S}^2=\frac{1}{4\kappa}\delta^2(t-\tau).
\end{align}
The corresponding spectral density of this noise process, being its Fourier transform, simply causes a Dirac delta at zero frequency \cite{S2-Paper2}. Similarly, all higher-power noise processes will have no contribution to the non-zero frequency of the noise spectral density. As a result, the multiplicative noise (\ref{T11-B2}) can be  effectively truncated as 
\begin{equation}
\label{T11-A5}
\hat{a}(t)\hat{a}_\text{in}(t)=a(t)\hat{a}_\text{in}(t),
\end{equation}
without causing any error in the non-zero frequencies of the resulting noise spectral density. A more general treatment of the second-order noise processes with Gaussian resonances is discussed elsewhere \cite{S2-Paper2}. 

\subsection{Noise Spectral Density}\label{S11-SpectralDensity}

Following the general approach to construction of the scattering matrix based on the input-output formalism \cite{S2-Noise1,S2-Noise0,S2-Paper2}, one may easily show that
\begin{equation}
\label{T11-C1}
\{\hat{A}_\text{out}(w)\}=\textbf{S}(w)\{\hat{A}_\text{in}(w)\},
\end{equation}
where $\textbf{S}(w)$ is the $6\times 6$ scattering matrix given by
\begin{equation}
\label{T11-C2}
\textbf{S}(w)=\textbf{I}-[\sqrt{\Gamma}]\left(i[\textbf{M}]-[\Gamma]-iw\textbf{I}\right)^{-1}[\sqrt{\Gamma}].
\end{equation}
Expansion of the output operator array gives
\begin{equation}
\label{T11-C3}
\hat{A}_{\text{out},j}(w)=\sum_{l=1}^6 S_{jl}(w)\hat{A}_{\text{in},l}(w),
\end{equation}
where $\hat{A}_{\text{in},l}(w)$ are multiplicative noise terms such as $a_l(w)\hat{a}_{\text{in},l}(w)$, where $\hat{a}_{\text{in},l}$ stand for white Guassian White stochastic processes $\hat{a}_{\text{in}}$, $\hat{b}_{\text{in}}$, and their conjugates $\hat{b}_{\text{in}}^\dagger$, $\hat{b}_{\text{in}}^\dagger$, and also $a_l(w)$ are the corresponding Fourier-transformed silent multiplicative terms. This can be correspondingly shown to lead to the noise spectral densities
\begin{equation}
\label{T11-C4}
S_{AA,j}(w)=\sum_{l=1}^6 \left|\frac{1}{\gamma_l}S_{jl}(w)\ast a_l(w)\right|^2 S_{A_l A_l}(w),
\end{equation}
with the understanding that the terms corresponding to conjugate noise operators are grouped together under the absolute value. Here, $S_{A_l A_l}(w)$ are the symmetrized noise spectral densities of the Gaussian White processes $\hat{a}_{\text{in},l}$. The spectral densities of these processes are typically constants as $\bar{n}_l+\frac{1}{2}$ with $\bar{n}_l$ being thermal occupation number of bosons. For an optical bosonic bath, one may conveniently set $\bar{n}_l=0$, while for phonons $\bar{n}_l$ can be estimated from Bose-Einstein distribution \cite{S2-3}. Furthermore, the symbol $\ast$ represents convolution in the frequency domain.

The approach provided here, leads to the noise spectral densities of higher-order operators $\hat{d}=\frac{1}{2}\hat{b}^2$, $\hat{d}^\dagger=\frac{1}{2}\hat{b}^{\dagger 2}$ and $\hat{m}$. Neither of these is the directly measurable spectrum, but it is rather the noise spectral density of ladder operator $\hat{b}$ for photons, which can be measured. These necessitates a way to recover the information through what is calculable by the method of higher-order operators.

The symmetrized noise spectral density of $\hat{d}=\frac{1}{2}\hat{b}^2$ is by definition given in terms of the Fourier transform of the corresponding symmetrized auto-correlation function, which is
\begin{align}
\label{T11-C5}
S_{DD}(w)&=\frac{1}{2\pi}\int_{-\infty}^{+\infty}e^{i w t}\braket{d^\dagger(\tau)d^\dagger(t+\tau)}_\text{S}dw=\frac{1}{8\pi}\int_{-\infty}^{+\infty}e^{i w t}\braket{b^\dagger(\tau)b^\dagger(\tau)b^\dagger(t+\tau)b^\dagger(t+\tau)}_\text{S}dw\\ \nonumber
&=\frac{1}{4\pi}\int_{-\infty}^{+\infty}e^{i w t}\braket{b^\dagger(\tau)b^\dagger(t+\tau)}_\text{S}^2 dw.
\end{align}
where the last expression is found by application of the Isserlis-Wick theorem and $\braket{b^\dagger(\tau)b^\dagger(\tau)}_\text{S}=0$. By noting the definition of Fourier and inverse Fourier transforms, we get
\begin{align}
\label{T11-C6}
S_{BB}(w)&=\frac{1}{2}+\mathscr{F}\left\{\sqrt{\frac{1}{2}\mathscr{F}^{-1}\left\{S_{DD}(w)-\frac{1}{2}\right\}(t)}\right\}(w),
\end{align}
where $\frac{1}{2}$ is substrated and added to account for the half a quanta of white noise which is lost in the symmetrization, and if not removed will cause appearance of a non-physical Dirac delta under the square root. While $S_{DD}(w)$ is found from simple scattering matrix calculations, all it takes now to find the measurable quantity $S_{BB}(w)$ is to take an inverse Fourier transform, followed by a square root and another Fourier transform. Similarly, one we may now define $\bar{S}_{BB}(w)=\frac{1}{2}[S_{BB}(w)+S_{BB}(-w)]$ as the symmetrized noise spectrum. The equation (\ref{T11-C6}) is the main key to recover the expected results from the higher-order operator algebra.

\subsection{Time-evolution of Operators}\label{S11-TimeDomain}

It is easy to obtain the explicit solution to the truncated system of Langevin equations (5)
\begin{align}
\label{T11-D1}
\left\{A(t)\right\}&=e^{i\left[\textbf{N}\right]2\Omega t}\left\{A(0)\right\}-2\Omega e^{i\left[\textbf{N}\right]2\Omega t} \int_0^t e^{-i\left[\textbf{N}\right]2\Omega \tau}\left(i\frac{\alpha}{2}\left\{A_\text{c}\right\}-\left[\sqrt{\Gamma}\right]\left\{A_\text{in}(\tau)\right\}\right)d\tau,
\end{align}
where $\exp(\cdot)$ represents the matrix exponentiation, and $[\textbf{N}]=[\textbf{M}]+i[\Gamma]$.

\subsection{Non-negative Integer Powers of Noise}\label{S11-HigherOrderNoise}

It is straightforward to see that any term involving a non-negative integer power of a noise such as $\hat{\alpha}_\text{in}^j (t); j\in\mathscr{N},j>1$ where $\hat{\alpha}_\text{in}^j (t)=\kappa^{\frac{1-j}{2}}\hat{a}_\text{in}^j (t)$ has identically zero contribution to the measured noise spectral density. In order to show this, let us the noise assume the normal autocorrelation
\begin{align}
\label{T11-E1}
\braket{\hat{\alpha}_\text{in}^{\dagger j}(\tau)\hat{\alpha}_\text{in}^j(t)}_\text{S}=\zeta \exp\left[-\pi \zeta^2 (t-\tau)^2\right].
\end{align}
In the limit of $\zeta\rightarrow +\infty$ this will settle back to the expected Dirac delta's function $\delta(t-\tau)$. The autocorrelation of the measurable optical field is connected to the operator $\hat{a}(t)$, which by means of the Isserlis-Wick theorem becomes 
\begin{align}
\label{T11-E2}
\braket{\hat{a}^{\dagger }(\tau)\hat{a}(t)}_\text{S}&=\left[\frac{\kappa^{j-1}}{j}\braket{\hat{a}_\text{in}^{\dagger j}(\tau)\hat{a}_\text{in}^j(t)}_\text{S}\right]^\frac{1}{j}=\left(\frac{\kappa^{j-1}\zeta}{j}\right)^{\frac{1}{j}}\exp\left[-\frac{\pi \zeta^2}{j} (t-\tau)^2\right].
\end{align}
The corresponding noise spectral density in frequency domain, where $w$ is the absolute optical frequency (and not the detuning referenced to a certain non-zero resonance frequency), is given by
\begin{align}
\label{T11-E3}
\mathscr{F}&\left\{\braket{\hat{a}^{\dagger }(\tau)\hat{a}(t)}_\text{S}\right\}(w)=\left(\frac{\kappa^{j-1}\zeta}{j}\right)^{\frac{1}{j}} \mathscr{F}\left\{\exp\left[-\frac{\pi \zeta^2}{j} (t-\tau)^2\right]\right\}(w)|_{\tau=0}
=\frac{\sqrt{j}}{j^\frac{1}{j}}\left(\frac{\kappa}{\zeta}\right)^\frac{j-1}{j}\exp\left[-\frac{j w^2}{4\pi \zeta^2}\right],
\end{align}
In the limit of $\zeta\rightarrow +\infty$ with $j>1$ the above expression is identically zero, and hence meeting the claim.

It is equally straightbackward to show that for any white Gaussian noise such as $\hat{a}_\text{in}$ satisfying 
\begin{equation}
\braket{\hat{a}_\text{in}^\dagger(t)\hat{a}_\text{in}(\tau)}_\text{S}=\frac{1}{2}\delta(t-\tau),
\end{equation}
the higher-power noise processes  $\hat{\alpha}_\text{in}^j (t)=\kappa^{\frac{1-j}{2}}\hat{a}_\text{in}^j (t)$ contribute only to the zero frequency of the noise spectral density. To show this, we assume
\begin{align}
\label{T11-E4}
\braket{\hat{a}_\text{in}^{\dagger}(\tau)\hat{a}_\text{in}(t)}_\text{S}=\frac{\zeta}{2} \exp\left[-\pi \zeta^2 (t-\tau)^2\right],
\end{align}
which again in the limit of $\zeta\rightarrow+\infty$ reproduces the Dirac's delta $\delta(t-\tau)$. Then the Isserlis-Wick theorem for such a Gaussian noise process could be exactly used to write
\begin{align}
\label{T11-E5}
\braket{\hat{\alpha}_\text{in}^{\dagger j}(\tau)\hat{\alpha}_\text{in}^j(t)}_\text{S}&=\frac{j}{2^j \kappa^{j-1}}\braket{\hat{a}_\text{in}^{\dagger}(\tau)\hat{a}_\text{in}(t)}_\text{S}^j=\frac{j\zeta^j}{2^j \kappa^{j-1}}\exp\left[-\pi j \zeta^2 (t-\tau)^2\right].
\end{align}
Taking the Fourier transform from both sides gives the resulting noise spectral density
\begin{align}
\label{T11-E6}
\mathscr{F}\left\{\braket{\hat{\alpha}_\text{in}^{\dagger j}(\tau)\hat{\alpha}_\text{in}^j(t)}_\text{S}\right\}(w)|_{\tau=0}=\frac{\sqrt{j}\zeta^{j-1}}{2^j \kappa^{j-1}}\exp\left[-\frac{w^2}{4\pi j \zeta^2}\right],
\end{align}
which in the limit of $\kappa=\zeta\rightarrow+\infty$ yields an upper bound to a constant number of quanta $\sqrt{j}/2^j$, being less than $\frac{1}{2}$ for $j>1$. This maximum bound to the background number of added noise quanta due to higher-power noise rapidly decays to zero with increasing $j$.

These limits are physically meaningful as long as cavity linewidth is much larger than the pump laser linewidth, which is quite accurately met in practice. So, when no squeezing is taking place and cavity resonances exhibit noise spectra corresponding to a much larger number of quanta than $\frac{1}{2}$, it should be safe to ignore the effect of square noise terms and higher powers. 

Ultimately, a numerical integration carried out on a nonlinear differential equation with exaggerated noise input amplitude could very visibly distinguish the zero contribution of the higher-power noise terms, quite expectedly, confirming the general above conclusions.

\subsection{Multiplicative Noise}

It is possible to make simple estimates for $b(w)$ to be used instead of $\bar{b}$ in (30), where a convolution such as (\ref{T11-C4}) would have been needed instead. This can be done by setting up the Langevin equations for operators $\{\hat{b},\hat{b}^\dagger\}$, which are coupled. Then these have to be Fourier-transformed and diagonalized to find $\hat{b}(w)$ explicitly, and the expectation of this expression $b(w)=\braket{\hat{b}(w)}$ can be now used as \cite{S2-Paper3}
\begin{align}
\label{T11-F1}
S_{DD}(w)&=\frac{1}{\gamma^2}|b(w)\ast[S_{11}(w)+S_{13}(w)]+b^\ast(w)\ast [S_{12}(w)+S_{13}(w)]|^2 S_{YY}(w).
\end{align}
This process is a lot more complicated and typically can be simplified by direct utilization of (30). Nevertheless, the corresponding Langevin equations are
\begin{align}
\label{T11-F2}
\frac{d}{d\tau}\left\{\begin{array}{c}
\hat{b} \\
\hat{b}^\dagger
\end{array}\right\}&=\frac{1}{2}
\left[\begin{array}{cc}
-i(2\beta\hat{n}+1)-\frac{1}{2}\gamma & -i4\alpha \\
i4\alpha & i(2\beta\hat{n}+1)-\frac{1}{2}\gamma
\end{array}\right]
\left\{\begin{array}{c}
\hat{b} \\
\hat{b}^\dagger
\end{array}\right\}-\sqrt{\gamma}\left\{\begin{array}{c}
\sqrt{\gamma}\bar{b}+\hat{y}_\text{in}\\
\sqrt{\gamma}\bar{b}^\ast+\hat{y}_\text{in}^\dagger
\end{array}
\right\}.
\end{align}
Replacing $\hat{n}$ with $\bar{n}$, taking the expectation and some simplification gives
\begin{align}
\label{T11-F3}
\left\{\begin{array}{c}
b(w) \\
b^\ast(w)
\end{array}\right\}&=\gamma\left([\textbf{W}]-iw[\textbf{I}]\right)^{-1}\left\{\begin{array}{c}
\bar{b} \\
\bar{b}^\ast
\end{array}\right\},\\
[\textbf{W}]&=\frac{1}{2}
\left[\begin{array}{cc}
-i(2\beta\bar{n}+1)-\frac{1}{2}\gamma & -i4\alpha \\
i4\alpha & i(2\beta\bar{n}+1)-\frac{1}{2}\gamma
\end{array}\right]. \nonumber
\end{align}
Here, the factor $\frac{1}{2}$ is included to take care of normalization with respect to $2\Omega$ rather than $\Omega$. Functions $b(w)$ and $b^\ast(w)$ from solution of (\ref{T11-F3}) can be plugged in the convolutions of (\ref{T11-F1}). Note the cancellation of $\gamma$ as it explicitly appears in (\ref{T11-F1}) and (\ref{T11-F3}).

To wrap up and summarize this section, we presented an exact diagonalization of the cross-Kerr nonlinear interaction with inclusion of parametric amplification. Cases of strong pump and classical pump were considered and also taken into account. It was shown that while it is expected that an infinite-dimensional basis could provide the mathematically exact solution, there exist an orthogonal transformation of infinite order, which can exactly reduce the problem into a finite-order $6\times 6$ formulation.

\section{Classical Nonlinearity} \label{Section-12}
This last section presents the application of higher-order algebra to other areas outside nonlinear quantum physics. Once we are concerned with classical scalar variables instead of operator quantities, setting up the governing equations and solutions are typically easier and more straightforward. Furthermore, the difficulty with the evaluation of spectral densities is no longer a major obstacle since operators are gone.

Once the evolution system of equations is obtained using the techniques discussed in the present study, then exact solutions normally exist, although sometimes in terms of infinite-dimensional matrices which may need cut-off in dimensions. However, convergence properties are noticeably good and even the first few orders suffice to obtain nonlinear corrections to fully-linearized solutions. If the system dynamics is both nonlinear and time-dependent containing system coefficients as functions of time, then combination with the previously developed differential transfer matrix method \cite{S4-59,S4-60} would be immediately helpful.

To this end, we present two very important examples with potentially huge impact on the further use and application of higher-order algebra. The first is an electronic circuit with exponential nonlinearity, which is even much more nonlinear than the previous systems. A simple p-n junction \cite{Gharekhanlou1} is as such. All bipolar circuits \cite{Wang,Gharekhanlou2,Gharekhanlou3} need to be described using exponentially nonlinear forms, while field-effect circuits \cite{Streetman} are second-order nonlinear and much easier to deal with. The second one is the Kuramoto model \cite{Kuramoto1,Kuramoto2}, which in the original Hamiltonian form \cite{Kuramoto3} is quadratic nonlinear and therefore can be well addressed using the higher-order operator algebra. However, the Kuramoto through an appropriate canonical transformation can exhibit sinusoidal nonlinearity, falling into the same class of exponential nonlinearity of electronic circuits. 

\subsection{Electronic Circuits}
We consider a simple capacitive/resistive p-n junction rectifier circuit with exponential nonlinearity and oscillatory decaying voltage drive, subject to noise. Even though this represents a very simple electronic circuit, it is known to be impossible to solve analytically. The general approach is to linearize the circuit around steady-state static point, and obtain the linear response to the driving voltage. The derivation of nonlinear response without numerical simulations is typically out of question.

At the quantized level the current passing through the resistor encompasses noise and generates noisy voltage across the rectifying junction. This can be well described using a stochastic nonlinear operator differential equation. It is possible to employ the expectation values within the mean field approximation to obtain the classical quantities, and demonstrate that the proposed analytical scheme actually works very well, is convergent, and uniformly converges to the accurate solution.  

First, consider the infinitely-ordered nonlinear operator equation
\begin{equation}
\label{T12-eqS1}
\tau \frac{d}{dt}\hat{u}\left(t\right)=-\mu \hat{u}\left(t\right)-\kappa \left[e^{\hat{u}\left(t\right)}-1\right]+v\left(t\right)-\hat{n}_{\rm in}\left(t\right),
\end{equation}
\noindent 
which models the voltage operator of an RC circuit shunted by a nonlinear ideal diode, driven by a sinusoidal voltage source $v\left(t\right)=V_0e^{-\alpha t}\sin\left(\omega t\right)$, and stochastic noise $\hat{n}_{\rm in}\left(t\right)$. We suppose that the noise $\hat{n}_{\rm in}\left(t\right)$ is governed by a Weiner process. Here, and without loss of generality, both $\kappa $ and $\mu $ are taken to be positive real parameters. Hence, this model does not include an oscillating part due to an imaginary $\mu $, which could have been otherwise absorbed into $\hat{u}\left(t\right)$ by a rotating frame transformation. This particular choice also eliminates the imaginary part of $\hat{u}\left(t\right)$. 

The reason for choosing this particular differential equation is that it is nonlinear to the infinite order, and also the extended basis of higher-order operators all commute and therefore trivially form a closed basis. Using the proposed method in this article, the above operator equation can be first put into the infinitely-ordered linear system of ordinary differential equations as
\begin{eqnarray}
\label{T12-eqS2}
\tau \frac{d}{dt}\left\{ \begin{array}{c}
\hat{u}\left(t\right) \\ 
{\hat{u}}^2\left(t\right) \\ 
{\hat{u}}^3\left(t\right) \\ 
{\hat{u}}^4\left(t\right) \\ 
{\hat{u}}^5\left(t\right) \\ 
\vdots 
\end{array}
\right\}=-\left[
\begin{array}{cccccc}
\kappa +1 & \frac{\kappa }{2!} & \frac{\kappa }{3!} & \frac{\kappa }{4!} & \frac{\kappa }{5!} & \cdots  \\ 
0 & 2\left(\kappa +1\right) & \frac{2\kappa }{2!} & \frac{2\kappa }{3!} & \frac{2\kappa }{4!} & \cdots  \\ 
0 & 0 & 3\left(\kappa +1\right) & \frac{3\kappa }{2!} & \frac{3\kappa }{3!} & \cdots  \\ 
0 & 0 & 0 & 4\left(\kappa +1\right) & \frac{4\kappa }{2!} & \cdots  \\ 
0 & 0 & 0 & 0 & 5\left(\kappa +1\right) & \cdots  \\ 
\vdots  & \vdots  & \vdots  & \vdots  & \vdots  & \ddots  \end{array}
\right]\left\{ \begin{array}{c}
\hat{u}\left(t\right) \\ 
{\hat{u}}^2\left(t\right) \\ 
{\hat{u}}^3\left(t\right) \\ 
{\hat{u}}^4\left(t\right) \\ 
{\hat{u}}^5\left(t\right) \\ 
\vdots  
\end{array}
\right\}\\ \nonumber 
+\left\{ \begin{array}{c}
v\left(t\right) \\ 
2\hat{u}\left(t\right)v\left(t\right) \\ 
3{\hat{u}}^2\left(t\right)v\left(t\right) \\ 
4{\hat{u}}^3\left(t\right)v\left(t\right) \\ 
5{\hat{u}}^4\left(t\right)v\left(t\right) \\ 
\vdots  \end{array}
\right\}-\left\{ \begin{array}{c}
\hat{n}_{\rm in}\left(t\right) \\ 
2\hat{u}\left(t\right)\hat{n}_{\rm in}\left(t\right) \\ 
3{\hat{u}}^2\left(t\right)\hat{n}_{\rm in}\left(t\right) \\ 
4{\hat{u}}^3\left(t\right)\hat{n}_{\rm in}\left(t\right) \\ 
5{\hat{u}}^4\left(t\right)\hat{n}_{\rm in}\left(t\right) \\ 
\vdots  \end{array}
\right\}.
\end{eqnarray}

Subsequently, the input terms can be linearized using a similar approximation used in the paper earlier for dealing with multiplicative noise terms. Doing this results in 
\begin{eqnarray}
\label{T12-eqS3}
\tau \frac{d}{dt}\left\{ \begin{array}{c}
\hat{u}\left(t\right) \\ 
{\hat{u}}^2\left(t\right) \\ 
{\hat{u}}^3\left(t\right) \\ 
{\hat{u}}^4\left(t\right) \\ 
{\hat{u}}^5\left(t\right) \\ 
\vdots  \end{array}
\right\}=-\left[
\begin{array}{cccccc}
\kappa +1 & \frac{\kappa }{2!} & \frac{\kappa }{3!} & \frac{\kappa }{4!} & \frac{\kappa }{5!} & \cdots  \\ 
0 & 2\left(\kappa +1\right) & \frac{2\kappa }{2!} & \frac{2\kappa }{3!} & \frac{2\kappa }{4!} & \cdots  \\ 
0 & 0 & 3\left(\kappa +1\right) & \frac{3\kappa }{2!} & \frac{3\kappa }{3!} & \cdots  \\ 
0 & 0 & 0 & 4\left(\kappa +1\right) & \frac{4\kappa }{2!} & \cdots  \\ 
0 & 0 & 0 & 0 & 5\left(\kappa +1\right) & \cdots  \\ 
\vdots  & \vdots  & \vdots  & \vdots  & \vdots  & \ddots  \end{array}
\right]\left\{ \begin{array}{c}
\hat{u}\left(t\right) \\ 
{\hat{u}}^2\left(t\right) \\ 
{\hat{u}}^3\left(t\right) \\ 
{\hat{u}}^4\left(t\right) \\ 
{\hat{u}}^5\left(t\right) \\ 
\vdots  \end{array}
\right\}\\ \nonumber
+\left\{ \begin{array}{c}
v\left(t\right) \\ 
2\bar{u}v\left(t\right) \\ 
3{\bar{u}}^2v\left(t\right) \\ 
4{\bar{u}}^3v\left(t\right) \\ 
5{\bar{u}}^4v\left(t\right) \\ 
\vdots  \end{array}
\right\}-\left\{ \begin{array}{c}
\hat{n}_{\rm in}\left(t\right) \\ 
2\bar{u}\hat{n}_{\rm in}\left(t\right) \\ 
3{\bar{u}}^2\hat{n}_{\rm in}\left(t\right) \\ 
4{\bar{u}}^3\hat{n}_{\rm in}\left(t\right) \\ 
5{\bar{u}}^4\hat{n}_{\rm in}\left(t\right) \\ 
\vdots  \end{array}
\right\},
\end{eqnarray}
\noindent in which $\bar{u}=\frac{1}{T}\int^T_0{\left\langle \hat{u}\left(t\right)\right\rangle dt}$ is the time-average of the input. Obviously, the system of equations (\ref{T12-eqS3}) is not a mathematically exact description of (\ref{T12-eqS1}). However, it allows far more accurate solutions than the fully linearized approximation, and even at a very small truncation number of 2, which is the lowest-order beyond full linearization, it can explain major non-linear corrections and distortions. 

Now, the above system of equations can be exactly integrated, after truncation to a finite-order. In the presence of noise, the above system of linear stochastic equations can be treated and integrated as an It\^{o} process, and the results typically show very rapid convergence for various orders of truncation between 2 and 6 versus the numerically exact solution. Furthermore, if the noise terms are dropped and noiseless approximation $\hat{n}_{\rm in}\to 0$ to the circuit is desired, then

It is still not quite clear that the method is convergent to the exact solution, since the It\^{o} integration of a Weiner process every time is carried over a different sequence of random numbers. This difficulty cannot be avoided in principle, since there is no way to reset the numerically random sequence.

Therefore, as a double check, we take the expectation values, which discards the noise term, and transform a mean-field approximation to reach a similar system of differential equations, however, expressed in terms of the expectation value function $\left\langle \hat{u}\left(t\right)\right\rangle $ and its higher orders. This is equivalent to solving the nonlinear differential equation

\begin{equation}
\label{T12-eqS4}
\tau \frac{d}{dt}\left\langle \hat{u}\left(t\right)\right\rangle =-\mu \left\langle \hat{u}\left(t\right)\right\rangle -\kappa \left[e^{\left\langle \hat{u}\left(t\right)\right\rangle }-1\right]+v\left(t\right),
\end{equation}
\noindent 
given the fact that $\left\langle \hat{n}_{\rm in}\left(t\right)\right\rangle =0$.

Doing this immediately reveals the convergence property of our proposed method. The numerical solutions are so rapidly and accurately converging to the exact solution, that they are practically indistinguishable beyond the two lowest truncation orders.

\subsection{Kuramoto Model}
For a network of $N-$particles with annihilators $\hat{a}_j,j\in\{1\dots N\}$ satisfying $[\hat{a}_i,\hat{a}_j^\dagger]=\delta_{ij}$ and $[\hat{a}_i,\hat{a}_j]=0$, and number operators $\hat{n}_j=\hat{a}_j^\dagger\hat{a}_j,j\in\{1\dots N\}$, the quantum analogue to the Kuramoto Hamiltonian \cite{Kuramoto3} can be written as 
\begin{eqnarray}
\label{T12-eqS5}
\mathbb{H}&=&\mathbb{H}_{\rm non}+\mathbb{H}_{\rm int}, \\ \nonumber
\mathbb{H}_{\rm non}&=&\sum_{j=1}^N\hbar\Omega_j\hat{a}_j^\dagger\hat{a}_j, \\ \nonumber
\mathbb{H}_{\rm int}&=&i\hbar\psi\sum_{l,m=1}^{N}(\hat{a}_j\hat{a}_m^\dagger-\hat{a}_m^\dagger\hat{a}_j)(\hat{n}_m-\hat{n}_l),
\end{eqnarray}
where $\psi$ is interaction strength. The classical counterpart may be written using the definition of classical dimensionless positions $x_j=\frac{1}{2}\braket{\hat{a}_j+\hat{a}_j^\dagger}$ and momenta $x_j=\frac{i}{2}\braket{\hat{a}_j-\hat{a}_j^\dagger}$ to obtain
\begin{eqnarray}
\label{T12-eqS6}
\mathcal{H}&=&\mathcal{H}_{\rm non}+\mathcal{H}_{\rm int}, \\ \nonumber
\mathcal{H}_{\rm non}&=&\frac{1}{2}\sum_{j=1}^N\Omega_j(x_j^2+p_j^2), \\ \nonumber
\mathcal{H}_{\rm int}&=&\frac{1}{4}\psi\sum_{l,m=1}^{N}(x_j p_m-x_m p_j)(x_j^2+p_j^2-x_m^2+p_m^2).
\end{eqnarray}
Either in the quantum (\ref{T12-eqS5}) or classical (\ref{T12-eqS6}) form, the Kuramoto model is quadratic nonlinear, and therefore well addressable using higher-order operator algebra.

Meanwhile, it can be put into an apparently different form using the canonical transformations \cite{Kuramoto3} 
\begin{eqnarray}
\label{T12-eq7}
I_l&=&\frac{1}{2}(x_l^2+p_l^2), \\ \nonumber
\phi_l&=&\arctan\left(\frac{x_l}{p_l}\right),
\end{eqnarray} 
to obtain after appropriate redefinition of $\omega$ and $\eta$
\begin{eqnarray}
\label{T12-eq8}
\mathcal{H}=\sum_{j=1}^N \omega_j I_j-\eta\sum_{l,m=1}\sqrt{I_l I_m}(I_m-I_l)\sin(\phi_m-\phi_l).
\end{eqnarray}
The corresponding equations of motion are exponentially nonlinear because of sinusoidal terms, and the same above approach of power-series expansion and truncation is expected to work well.

There is much room for deeper study of higher-order algebra in the broad field of electronic circuits, since it ultimately allows to study nonlinear response of circuits using relatively accurate closed-form expressions. Field-effect devices are accurately described by second-order nonlinearity and the resulting system of equations is finite-dimensional at the order of 2, and thus needing no truncation. In comparison, p-n junction and bipolar devices are exponentially nonlinear of infinite-order, however, truncation even at the lowest order of 2 should suffice to keep the accuracy reasonable. In case of Kuramoto model, which is gaining considerable attraction in diverse fields of science and technology, plausible applications are just countless. These shall remain and need to be set apart as topics of future dedicated and separate studies.
\section{Conclusions} \label{Section-13}
In conclusions, we presented a comprehensive overview and details of the higher-order algebra as applied to nonlinear quantum physics. Instances of studied systems using this approach so far are ranging from quantum and quadratic optomechanics, quantum read-out circuits and anharmonicity, all the way to side-band inequivalence and cross-Kerr interaction. All studied nonlinear systems gain lots of new physical insight using the explicit expressions found through this analysis, and yet there remains much unexplored domains proposed throughout. We also briefly discussed extensions of the higher-order algebra to deal with two very remarkable and classically nonlinear systems. These are exponentially nonlinear electronic circuits and quadratic nonlinear Kuramoto model. Electronic circuits are usually fully linearized at the study and design stages, while the infamous Kuramoto model with diverse applications in modern science and technology can barely be studied without relying on full numerical methods. These in separate deserve complete and focused studies in near future. It is hoped that the method of higher-order operators will soon be widely recognized and find its correct place within the community of scientists and engineers who are dealing with nonlinear problems on a daily basis.
\section*{Acknowledgments} \nonumber \label{Section-14}
This work has received no external funding, and has benefited from attending many lectures and discussions with countless interested students, researchers, and scientists. The author wishes to sincerely thank Professor Tobias J. Kippenberg at Laboratory of Photonics and Quantum Measurements at \'{E}cole Polytechnique F\'{e}d\'{e}rale de Lausanne (EPFL). In particular, the author is indebted to Professor Markus Aspelmeyer at Laboratory of Quantum Foundations and Quantum Information on the Nano- and Microscale at Vienna Center for Quantum Science and Technology (VCQ) in University of Vienna for encouragement and hosting this research during its initial development. Complete analysis of side-band inequivalence would have been impossible without brief visit and kind invitation to the group of Professor Eva M. Weig in the Laboratory of Nanomechanics at University of Konstanz, to whom the author wishes to extend much appreciations and thanks. This article is dedicated to the celebrated artist, Anastasia Huppmann.



\footnotesize
\begin{thebibliography}{99}

\bibitem{S2-1}
T. J. Kippenberg, K. J. Vahala, Cavity optomechanics, {Science} 321 (2008) 1172.
\bibitem{S2-2}
M. Aspelmeyer, T. Kippenberg, F. Marquardt, Cavity Optomechanics: Nano- and Micromechanical Resonators Interacting with Light, Springer, Berlin, 2014.
\bibitem{S2-3}
M. Aspelmeyer, T. J. Kippenberg, F. Marquardt, Cavity optomechanics, Rev. Mod. Phys. 86 (2014) 1391.
\bibitem{S2-4}
P. Meystre, A short walk through quantum optomechanics, Ann. Phys. 525 (2013) 215.
\bibitem{S2-5}
C. K. Law, Interaction between a moving mirror and radiation pressure: A Hamiltonian formulation, {Phys. Rev. A} 51 (1995) 2537.
\bibitem{S2-6}
W. P. Bowen, G. J. Milburn, {Quantum Optomechanics}, CRC Press, Boca Raton, 2016.
\bibitem{S2-Paper1}
S. Khorasani, Higher-order interactions in quantum optomechanics: Revisiting the theoretical foundations, {Appl. Sci.} 7 (2017) 656.
\bibitem{S2-Paper2}
S. Khorasani, Higher-order interactions in quantum optomechanics: Analytical solution of nonlinearity, {Photonics} 4 (2017) 48.
\bibitem{S2-Paper3}
S. Khorasani, Method of higher-order operators for quantum optomechanics, Sci. Rep. 8 (2018) 11566.
\bibitem{S2-Paper4}
S. Khorasani, Higher-order interactions in quantum optomechanics: Analysis of quadratic terms, {Sci. Rep.} 8 (2018) 16676.
\bibitem{S2-Paper5}
S. Khorasani, Solution of cross-Kerr interaction combined with parametric amplification, {Sci. Rep.} 9 (2019) 1830.
\bibitem{S2-Paper6}
S. Khorasani, Momentum-field interactions beyond standard quadratic optomechanics, in A. I. Arbab (ed.) {Quantum Mechanics: Theory, Analysis, and Applications}, Nova Science Publishers, 2018, pp. 1–17.
\bibitem{S2-Paper7}
S. Khorasani, Analysis of side-band inequivalence, {Sci. Rep.} 9 (2019) 9075.
\bibitem{S2-7}
L. Zhang, F. Ji, X. Zhang, W. Zhang, Photon-phonon parametric oscillation induced by quadratic coupling in an optomechanical resonator, {J. Phys. B} 50 ({2017}) 145501.
\bibitem{S2-8}
C. Doolin, B. D. Hauer, P. H. Kim, A. J. R. MacDonald, H. Ramp, J. P. Davis, Nonlinear optomechanics in the stationary regime, {Phys. Rev. A} 89 ({2014}) 053838.
\bibitem{S2-9}
M. Asjad, G. S. Agarwal, M. S. Kim, P. Tombesi, G. Di Guiseppe, D. Vitali, Robust stationary mechanical squeezing in a kicked quadratic optomechanical system, {Phys. Rev. A} 89 ({2014}) 023849.
\bibitem{S2-Bruschi}
D. E. Bruschi, A. Xuereb, Mechano-optics: An optomechanical quantum simulator, {New J. Phys.} 20 ({2018}) 065004.
\bibitem{S2-Quad}
S. Liu, W.-X. Yang, T. Shui, Z. Zhu, Z.-X. Chen, Tunable two-phonon higher-order sideband amplification in a quadratically coupled optomechanical system, {Sci. Rep.} 7 ({2017}) 17637.
\bibitem{S2-Noise0}
C. W. Gardiner, M. J. Collett, Input and output in damped quantum systems: Quantum stochastic differential equations and the master equation, {Phys. Rev. A} 31 ({1985}) 3761.
\bibitem{S2-Noise1} 
C. W. Gardiner, P. Zoller, {Quantum Noise}, Springer, Berlin, 2004.
\bibitem{S2-Noise2}
C. Gardiner, P. Zoller, {The Quantum World of Ultra-Cold Atoms and Light. Book I: Foundations of Quantum Optics}, Imperial College Press, London, 2014.
\bibitem{S2-Noise3}
J. Combesa, J. Kerckhoff, M. Sarovar, The SLH framework for modeling quantum input-output networks, {Adv. Phys.: X} 2 ({2017}) 784.
\bibitem{S2-Noise4}
K. J. Rubin, G. Pruessner, G. A. Pavliotis, Mapping multiplicative to additive noise, {J. Phys. A: Math. Theor.} 47 ({2014}) 195001.



\bibitem{S3-3}  F. Marquardt, S. M.  Girvin, Optomechanics Retarded radiation forces, Physics 2 (2009) 40.
\bibitem{S3-6}  Z. R. Gong, H. Ian, X. Y. Liu, C. P. Sun, F. Nori, Effective Hamiltonian approach to the Kerr nonlinearity in an optomechanical system, Phys. Rev. A 80 (2009) 065801.
\bibitem{S3-7}  X. Y. L\"{u}, W. M. Zhang, S. Ashhab, Y. Wu, F. Nori, Quantum-criticality-induced strong Kerr nonlinearities in optomechanical systems, Sci. Rep. 3 (2013) 2943.
\bibitem{S3-8}  I. Mahboob, M. Mounaix, K. Nishiguchi, A. Fujiwara, H. Yamaguchi, A multimode electromechanical parametric resonator array, Sci. Rep. 4 (2014) 4448.
\bibitem{S3-9}  G. Heinrich, M. Ludwig, J. Qian, B. Kubala, F. Marquardt, Collective dynamics in optomechanical arrays, Phys. Rev. Lett. 107 (2011) 043603.
\bibitem{S3-10}  J. H. Gan, H. Xiong, L. G. Si, X. Y. Lu, Y. Wu, Solitons in optomechanical arrays, Opt. Lett. 41 (2016) 2676.
\bibitem{S3-11}  W. Chen, A. A. Clerk, Photon propagation in a one-dimensional optomechanical lattice, Phys. Rev. A 89 (2014) 033854.
\bibitem{S3-12}  A. Xuereb, C. Genes, G. Pupillo, M. Paternostro, A. Dantan, Reconfigurable long-range phonon dynamics in optomechanical arrays, Phys. Rev. Lett. 112 (2014) 133604.
\bibitem{S3-13}  O. Houhou, H. Aissaoui, A. Ferraro, Generation of cluster states in optomechanical quantum systems, Phys. Rev. A 92 (2015) 063843.
\bibitem{S3-14}  V. Peano, C. Brendel, M. Schmidt, F. Marquardt, Topological phases of sound and light, Phys. Rev. X 5 (2015) 031011.
\bibitem{S3-15}  M. Ludwig, F. Marquardt, Quantum many-body dynamics in optomechanical arrays, Phys. Rev. Lett. 111 (2013) 073603.
\bibitem{S3-16}  J.-Q. Liao, Q.-Q. Wu, F. Nori, Entangling two macroscopic mechanical mirrors in a two-cavity optomechanical system, Phys. Rev. A 89 (2014) 014302.
\bibitem{S3-17}  J. R. Johansson, G. Johansson, F. Nori, Optomechanical-like coupling between superconducting resonators, Phys. Rev. A 90 (2014) 053833. 
\bibitem{S3-18}  A. Kronwald, F. Marquardt, A. A. Clerk, Arbitrarily large steady-state bosonic squeezing via dissipation, Phys. Rev. A 88 (2013) 063833.
\bibitem{S3-19}  X. Y. L\"{u}, J. Q. Liao, L. Tian, F. Nori, Steady-state mechanical squeezing in an optomechanical system via Duffing nonlinearity, Phys. Rev. A 91 (2015) 013834.
\bibitem{S3-20}  X. Y. L\"{u}, Y. Wu, J. R. Johansson, H.  Jing, J. Zhang, F. Nori, Squeezed optomechanics with phase-matched amplification and dissipation, Phys. Rev. Lett. 114 (2015) 093602.
\bibitem{S3-21}  R. chilling, H. Sch\"{u}tz, A. H. Ghadimi, V. Sudhir, D. J. Wilson, T. J. Kippenberg, Field integration of a SiN nanobeam and a SiO${}_{2}$ microcavity for Heisenberg-limited displacement sensing, Phys. Rev. Applied 5 (2016) 054019.
\bibitem{S3-22}  F. Ruesink, M. A. Miri, A. Alu, E. Verhagen, Nonreciprocity and magnetic-free isolation based on optomechanical interactions, Nature Commun. 7 (2016) 13662.
\bibitem{S3-23}  X.-W. Xu, Y. Li, A.-X. Chen, Y.-X. Liu, Nonreciprocal conversion between microwave and optical photons in electro-optomechanical systems, Phys. Rev. A 93 (2016) 023827.
\bibitem{S3-24}  K. Fang,  J. Luo,  A. Metelmann,  M. H. Matheny,  F. Marquardt, A. A. Clerk, O. Painter, Generalized nonreciprocity in an optomechanical circuit via synthetic magnetism and reservoir engineering, Nature Phys. 13 (2017) 465.
\bibitem{S3-26}  N. R. Bernier, L. D. T\'{o}th, A. Koottandavida, M. Ioannou, D. Malz, A. Nunnenkamp, A. K. Feofanov, T. J. Kippenberg, Nonreciprocal reconfigurable microwave optomechanical circuit,  Nature Commun. 8 (2017) 604.
\bibitem{S3-27}  G. A. Peterson, F. Lecocq, K. Cicak, R. W. Simmonds, J. Aumentado, J. D. Teufel, Demonstration of efficient nonreciprocity in a microwave optomechanical circuit, Phys. Rev. X 7 (2017) 031001.
\bibitem{S3-28}  D. Malz, L. D. T\'{o}th, N. R. Bernier, A. K. Feofanov, T. J. Kippenberg, A. Nunnenkamp, Quantum-limited directional amplifiers with optomechanics, Phys. Rev. Lett. 120 (2018) 023601.
\bibitem{S3-29}  S. Zippilli, J. Li, D. Vitali, Steady-state nested entanglement structures in harmonic chains with single-site squeezing manipulation. Phys. Rev. A 92 (2015) 032319.
\bibitem{S3-30}  S. Barzanjeh, S. Guha, C. Weedbrook, D. Vitali, J. H. Shapiro, S. Pirandola, Microwave quantum illumination. Phys. Rev. Lett. 114 (2015) 080503.
\bibitem{S3-31}  K. Zhang, F. Bariani, Y. Dong, W. Zhang, P. Meystre, Proposal for an optomechanical microwave sensor at the subphoton level, Phys. Rev. Lett. 114 (2015) 113601.
\bibitem{S3-32}  L. D. Toth, N. R. Bernier, A. Nunnenkamp, E. Glushkov, A. K. Feofanov, T. J. Kippenberg, A dissipative quantum reservoir for microwave light using a mechanical oscillator, Nature Physics 13 (2017) 787. 
\bibitem{S3-33}  C. Galland, N. Sangouard, N. Piro, N. Gisin, T. J. Kippenberg, Heralded single-phonon preparation, storage, and readout in cavity optomechanics. Phys. Rev. Lett. 112 (2014) 143602.
\bibitem{S3-34}  S. Khorasani, Coupled mode theory of optomechanical crystals. IEEE J. Quant. Electron. 52 (2016) 6100406.
\bibitem{S3-35}  M. H. Aram, S. Khorasani, Efficient analysis of confined guided modes in phoxonic crystal slabs, IEEE/OSA J. Lightwave Technol. 35 (2017) 3734-3742. 
\bibitem{S3-36}  M. H. Aram, S. Khorasani, Optical wave evolution due to interaction with elastic wave in a phoxonic crystal slab waveguide, Appl. Phys. B 123 (2017) 218. 
\bibitem{S3-37}  J. Bochmann, A. Vainsencher, D. D. Awschalom, A. N. Cleland, Nanomechanical coupling between microwave and optical photons, Nat. Phys. 9 (2013) 712.
\bibitem{S3-38}  R. W. Andrews, R. W. Peterson, T. P. Purdy, K. Cicak, R. W. Simmonds, C. A. Regal, K. W. Lehnert, Bidirectional and efficient conversion between microwave and optical light, Nat. Phys. 10 (2014) 321,
\bibitem{S3-39}  C. Javerzac-Galy, K. Plekhanov, N. R. Bernier, L. D. Toth, A. K. Feofanov, T. J. Kippenberg, On-chip microwave-to-optical quantum coherent converter based on a superconducting resonator coupled to an electro-optic microresonator, Phys. Rev. A 94 (2016) 053815.
\bibitem{S3-40}  L. Tian, Optoelectromechanical transducer: reversible conversion between microwave and optical photons, Ann. Phys. (Berlin) 527 (2015) 1.
\bibitem{S3-41}  V. Caprara Vivoli, T. Barnea, C. Galland, N.Sangouard, Proposal for an optomechanical Bell test, Phys. Rev. Lett. 116 (2016) 070405.
\bibitem{S3-42}  M. Tsang, Cavity quantum electro-optics. Phys. Rev. A 81 (2010) 063837.
\bibitem{S3-43}  M. Tsang, Cavity quantum electro-optics. II. Input-output relations between traveling optical and microwave fields, Phys. Rev. A 84 (2011) 043845.
\bibitem{S3-44}  Y. Chang, T. Shi, Y. X. Liu, C. P. Sun, F. Nori, Multistability of electromagnetically induced transparency in atom-assisted optomechanical cavities, Phys. Rev. A 83 (2011) 063826.
\bibitem{S3-45}  H. Jing, \c{S}. K. \"{O}zdemir, Z. Geng, J. Zhang, X. Y. L\"{u}, B. Peng, L. Yang, F. Nori, Optomechanically-induced transparency in parity-time-symmetric microresonators. Sci. Rep. 5 (2015) 9663.
\bibitem{S3-46}  H. Wang, X. Gu, Y. X. Liu, A. Miranowicz, F. Nori, Optomechanical analog of two-color electromagnetically-induced transparency: Photon transmission through an optomechanical device with a two-level system, Phys. Rev. A 90 (2014) 023817.
\bibitem{S3-47}  Y. L. Liu, R. Wu, J. Zhang, \c{S}. K. \"{O}zdemir, L. Yang, F. Nori, Y. X. Liu, Controllable optical response by modifying the gain and loss of a mechanical resonator and cavity mode in an optomechanical system, Phys. Rev. A 95 (2017) 013843.
\bibitem{S3-48}  S. Huang, Quantum state transfer in cavity electro-optic modulators, Phys. Rev. A 92 (2015) 043845.
\bibitem{S3-49}  J. Larson, M. Horsdal, Photonic Josephson effect, phase transitions, and chaos in optomechanical systems, Phys. Rev. A 84 (2011) 021804.
\bibitem{S3-51}  M. Alidoosty, S. Khorasani, M. H. Aram, Simulation of multipartite cavity quantum electrodynamics, IEEE J. Quant. Electron. 49 (2013) 1066.
\bibitem{S3-52}  M. Alidoosty, S. Khorasani, M. H. Aram, Theory and simulation of cavity quantum electro-dynamics in multi-partite quantum complex systems, Appl. Phys. A 115 (2014) 595.
\bibitem{S3-53}  F. Monifi, J. Zhang, \c{S}. K. \"{O}zdemir, B. Peng, Y.-X. Liu, F. Bo, F. Nori, L. Yang, Optomechanically induced stochastic resonance and chaos transfer between optical fields, Nat. Photon. 10 (2016) 399.
\bibitem{S3-54}  C. Neill, P. Roushan, M. Fang, Y. Chen, M. Kolodrubetz, Z. Chen, A. Megrant, R. Barends, B. Campbell, B. Chiaro, A. Dunsworth, E. Jeffrey, J.  Kelly, J. Mutus, P. J. J. O'Malley, C.  Quintana, D. Sank, A. Vainsencher, J. Wenner, T. C. White, A. Polkovnikov, J. M. Martinis, Ergodic dynamics and thermalization in an isolated quantum system, Nat. Phys. 12 (2016) 1037.
\bibitem{S3-55}  J. D. Thompson, B. M. Zwickl, A. M. Jayich, F. Marquardt, S. M. Girvin, J. G. E. Harris, Strong dispersive coupling of a high-finesse cavity to a micromechanical membrane, Nature 452 (2008) 72.
\bibitem{S3-56}  J. C. Sankey, C. Yang, B. M. Zwickl, A. M. Jayich, J. G. E. Harris,  Strong and tunable nonlinear optomechanical coupling in a low-loss system, Nat. Phys. 6 (2010) 707.
\bibitem{S3-57}  A. Nunnenkamp, K. B{\o}rkje, J. G. E. Harris, S. M. Girvin, Cooling and squeezing via quadratic optomechanical coupling, Phys. Rev. A 82 (2010) 021806.
\bibitem{S3-58}  C. U. Lei, A. J. Weinstein, J. Suh, E. E. Wollman, A. Kronwald, F. Marquardt, A. A. Clerk, K. C. Schwab, Quantum nondemolition measurement of a quantum squeezed state beyond the 3 dB limit. Phys. Rev. Lett. 117 (2016) 100801.
\bibitem{S3-59}  J. Q. Liao, F. Nori, Photon blockade in quadratically coupled optomechanical systems, Phys. Rev. A 88 (2013) 023853.
\bibitem{S3-60}  J. Q. Liao, F. Nori, Single-photon quadratic optomechanics, Sci. Rep. 4 (2014) 6302.
\bibitem{S3-61}  E. J. Kim, J. R. Johansson, F. Nori, Circuit analog of quadratic optomechanics, Phys. Rev. A 91 (2015) 033835.
\bibitem{S3-62}  T. P. Purdy, D. W. Brooks, T. Botter, N. Brahms, z. Y. Ma, D. M. Stamper-Kurn, Tunable cavity optomechanics with ultracold atoms, Phys. Rev. Lett. 105 (2010) 133602.
\bibitem{S3-63}  Z. Shen, Y.-L. Zhang, Y. Chen, C.-L. Zou, Y.-F. Xiao, X.-B. Zou, F.-W. Sun, G.-C. Guo, C.-H. Dong, Experimental realization of optomechanically induced non-reciprocity, Nat. Photon. 10 (2016) 657.
\bibitem{S3-64}  G. A. Brawley, M. R. Vanner, P. E. Larsen, S. Schmid, A. Boisen, W. P. Bowen, Nonlinear optomechanical measurement of mechanical motion, Nature Commun. 7 (2016) 10988.
\bibitem{S3-65}  R. Leijssen, G. La Gala, L. Freisem, J. T. Muhonen, E. Verhagen, Nonlinear cavity optomechanics with nanomechanical thermal fluctuations, Nature Commun. 8 (2017) 16024.
\bibitem{S3-66}  V. B. Branginskii, A. B. Manukn, Ponderomotive effects of electromagnetic radiation. Sov. Phys. JETP 25 (1967) 653.
\bibitem{S3-67}  H. K. Cheung, C. K. Law, Nonadiabatic optomechanical Hamiltonian of a moving dielectric membrane in a cavity, Phys. Rev. A 84 (2011) 023812.
\bibitem{S3-68}  L. O. Casta\~{n}os, R. Weder, Equations of a moving mirror and the electromagnetic field, Phys. Scr. 90 (2015) 068011.
\bibitem{S3-69}  C. M. Wilson, G. Johansson, A. Pourkabirian, M. Simoen, J. R. Johansson, T. Duty, F. Nori, P. Delsing, Observation of the dynamical Casimir effect in a superconducting circuit, Nature 479 (2011) 376.
\bibitem{S3-70}  J. Q. You, F. Nori, Atomic physics and quantum optics using superconducting circuits, Nature 474 (2011) 589.
\bibitem{S3-71}  W. Schleich, Quantum Optics in Phase Space, Wiley-VCH, Berlin, 2001.
\bibitem{S3-72}  S. Khorasani, A proof for Poisson bracket in non-commutative algebra of quantum mechanics. Elec. J. Th. Phys. 13 (2016) 57.
\bibitem{S3-73}  S. Khorasani, Applied Quantum Mechanics, Delarang, Tehran, 2010.
\bibitem{S3-74}  J. H. Hetherington, F. D. C. Willard, Two-, three-, and four-atom exchange effects in bcc ${}_{3}$He, Phys. Rev. Lett. 35 (1975) 1442.
\bibitem{S3-75}  M. Roger, J. H. Hetherington, J. M. Delrieu, Magnetism in solid ${}_{3}$He, Rev. Mod. Phys. 55 (1983) 1.
\bibitem{S3-76}  G. S. Agarwal, S. S. Jha, Theory of optomechanical interactions in superfluid He, Phys. Rev. A 90 (2014) 023812.
\bibitem{S3-76a}
A. B. Shkarin, A. B., A. D. Kashkanova, C. D. Brown, S. Garcia, K. Ott, J. Reichel, J. G. E. Harris, Quantum optomechanics in a liquid, Phys. Rev. Lett. 122 (2019) 153601.
\bibitem{S3-76b}
X. He, G. I. Harris, C. G. Baker, A. Sawadsky, Y. L. Sfendla, Y. P. Sachkou, S. Forstner, W. P. Bowen, Strong optical coupling through superfluid Brillouin lasing, arXiv (2019) 1907.06811.
\bibitem{S3-77}  J. Suh, A. J. Weinstein, K. C. Schwab, Optomechanical effects of two-level systems in a back-action evading measurement of micro-mechanical motion, Appl. Phys. Lett. 103 (2013) 052604.
\bibitem{S3-78}  P. Roelli, C. Galland, N. Piro, T. J. Kippenberg, Molecular cavity optomechanics as a theory of plasmon-enhanced Raman scattering. Nat. Nano. 11 (2016) 164.
\bibitem{S3-79}  T. Bagci, A. Simonsen, S. Schmid, L. G. Villanueva, E. Zeuthen, J. Appel, J. M. Taylor, A. S{\o}rensen, K. Usami, A. Schliesser, E. S. Polzik, Optical detection of radio waves through a nanomechanical transducer, Nature 507 (2014) 81.
\bibitem{S3-80}  S. Lannebere, M. G. Silveirinha, Wave instabilities and unidirectional light flow in a cavity with rotating walls, Phys. Rev. A 94 (2016) 033810.
\bibitem{S3-81}  LIGO Scientific Collaboration and Virgo Collaboration, Observation of gravitational waves from a binary black hole merger, Phys. Rev. Lett. 116 (2016) 061102.
\bibitem{S3-82}  T. K. Para\"{i}so, M. Kalaee, L. Zang, H. Pfeifer, F. Marquardt, O. Painter, Position-squared coupling in a tunable photonic crystal optomechanical cavity, Phys. Rev. X 5 (2015) 041024.
\bibitem{S3-83}  M. Kalaee, T. K. Paraiso, H. Pfeifer, O. Painter, Design of a quasi-2D photonic crystal optomechanical cavity with tunable, large x${}^{2}$-coupling, Opt. Express 24 (2016) 21308.
\bibitem{S3-84}  A. Metelmann, A. A. Clerk, Nonreciprocal photon transmission and amplification via reservoir engineering, Phys. Rev. X 5 (2015) 021025.
\bibitem{S3-85}  R. G\"{u}tig, C. Eberlein, Dynamical Casimir effect with Dirichlet and Neumann boundary conditions, J. Phys. A 31 (1998) 6819.
\bibitem{S3-87}  S. Khorasani, B. Rashidian, Guided light propagation in dielectric slab waveguide with conducting interfaces, J. Opt. A: Pure Appl. Opt. 3 (2001) 380.
\bibitem{S3-88}  S. Khorasani, B. Rashidian, Modified transfer matrix method for conducting interfaces, J. Opt. A: Pure Appl. Opt. 4 (2002) 251.
\bibitem{S3-89}  F. Karimi, S. Khorasani, Optical modulation by conducting interfaces, IEEE J. Quant. Electron. 49 (2013) 607. 




\bibitem{S4-7}
M. Ludwig, F. Marquardt, Quantum many-body dynamics in optomechanical arrays, {Phys. Rev. Lett.} 111 (2013) 073603.
\bibitem{S4-14a}
J. Q. Liao, C. K. Law, L. M. Kuang, F. Nori, Enhancement of mechanical effects of single photons in modulated two-mode optomechanics, {Phys. Rev. A} 92 (2015) 013822.
\bibitem{S4-17}
J.-H. Kim, S. Kim, G. Bahl, Complete linear optical isolation at the microscale with ultralow loss, {Sci. Rep.} 7 (2017) 1647.
\bibitem{S4-17a}
B. Peng, \c{S}. K. \"{O}zdemir, F. Lei, F. Monifi, M. Gianfreda, G. L. Long, S. Fan, F. Nori, C. M. Bender, L. Yang, Parity-time-symmetric whispering-gallery microcavities, {Nat. Phys.} 10 (2014) 394.
\bibitem{S4-23}
S. Khorasani, Coupled mode theory of optomechanical crystals, {IEEE J. Quant. Electron.} 52 (2016) 6100406.
\bibitem{S4-23.0}
H. Jing, \c{S}. K. \"{O}zdemir, H. L\"{u}, F. Nori, High-order exceptional points in optomechanics, {Sci. Rep.} 7 (2017) 3386.
\bibitem{S4-33.0} M. Bhattacharya, H. Uys, P. Meystre, Optomechanical trapping and cooling of partially reflective mirrors, {Phys. Rev. A} 77 (2008) 033819. 
\bibitem{S4-33.3} X.-G. Zhan, L.-G. Si, A.-S. Zheng, X. X. Yang, Tunable slow light in a quadratically coupled optomechanical system, {J. Phys. B} 46 (2013)  025501.
\bibitem{S4-33.4} L. F. Buchmann, L. Zhang, A. Chirivelli, P. Meystre, Macroscopic tunneling of a membrane in an optomechanical double-well potential, {Phys. Rev. Lett.} 108 (2012) 210403.
\bibitem{S4-33.5} H. Seok, L. F. Buchmann, E. M. Wright, P. Meystre, Multimode strong-coupling quantum optomechanics, {Phys. Rev. A} 88 (2013) 063850.
\bibitem{S4-33.6} H. Seok, E. M. Wright, P. Meystre, Dynamic stabilization of an optomechanical oscillator, {Phys. Rev. A} 90 (2014) 043840.
\bibitem{S4-33.6.0}
M. R. Vanner, Selective linear or quadratic optomechanical coupling via measurement, {Phys. Rev. X} 1 (2011) 021011.
\bibitem{S4-33.6.1}
H. Seok, E. M. Wright, Antibunching in an optomechanical oscillator, {Phys. Rev. A} 95 (2017) 053844. 
\bibitem{S4-33.6.3}
A. M. Jayich, J. C. Sankey, B. M. Zwickl, C. Yang, J. D. Thompson, S. M. Girvin, A. A. Clerk, F. Marquardt, J. G. E. Harris, Dispersive optomechanics: a membrane inside a cavity, {New J. Phys.} 10 (2008) 095008.
\bibitem{S4-33.6.4}
M. Kol\'{a}\v{r}, A. Ryabov, R. Filip, Optomechanical oscillator controlled by variation in its heat bath temperature, {Phys. Rev. A} 95 (2017)  042105.
\bibitem{S4-33.6.5}
B. Fan, M. Xie, Stochastic resonance in a tristable optomechanical system, {Phys. Rev. A} 95 (2017) 023808.
\bibitem{S4-33.6.6}
D. Lee, M. Underwood, D. Mason, A. B. Shkarin, S. W. Hoch, J. G. E. Harris, Multimode optomechanical dynamics in a cavity with avoided crossings. {Nature Commun.} 6 (2015) 6232.
\bibitem{S4-33.6.7}
J. H. Lee, H. Seok, Quantum reservoir engineering through quadratic optomechanical interaction in the reversed dissipation regime, Phys. Rev. A 97 (2018) 013805.
\bibitem{S4-33.8a}
I. Buluta, F. Nori, Quantum simulators, {Science} 326 (2009) 108.
\bibitem{S4-33.8b}
I. Georgescu, S. Ashhab, F. Nori, Quantum simulation, {Rev. Mod. Phys.} 86 (2014) 153.
\bibitem{S4-33.8c}
Y. Makhlin, G. Sch\"{o}n, A. Shnirman, Quantum-state engineering with Josephson-junction devices, {Rev. Mod. Phys.} 73 (2001) 357.
\bibitem{S4-33.9a} B. P. Venkatesh, D. H. J. O’Dell, J. Goldwin, An optomechanical elevator: Transport of a Bloch oscillating Bose–Einstein condensate up and down an optical lattice by cavity sideband amplification and cooling, {Atoms} 4 (2016) 2.
\bibitem{S4-33.10} N. Kiesel, F. Blaser, U. Deli\'{c}, R. Klatenbaek, M. Aspelmeyer, Cavity cooling of an optically levitated submicron particle, {Proc. Natl. Acad. Sci.} 110 (2013) 14180.
\bibitem{S4-33.10.0} M. J. Woolley, M. F. Emzir, G. J. Milburn, M. Jerger, M. Goryachev, M. E. Tobar, A. Fedorov, Quartz-superconductor quantum electromechanical system, {Phys. Rev. B} 93 (2016) 224518.
\bibitem{S4-33.11} P. Domokos, H. Ritsch, Mechanical effects of light in optical resonators, {J. Opt. Soc. Am. B} 20 (2003) 1098.
\bibitem{S4-33.12} M. J. Woolley, A. C. Doherty, G. J. Milburn, Continuous quantum nondemolition measurement of Fock states of a nanoresonator using feedback-controlled circuit QED, {Phys. Rev. B} 82 (2010) 094511.
\bibitem{S4-33.13} O. Romero-Isart, A. C.  Pflanzer, F. Blaser, R. Kaltenbaek, N. Kiesel, M. Aspelmeyer, J. I. Cirac, Large quantum superpositions and interference of massive nanometer-sized objects, {Phys. Rev. Lett.} 107 (2011) 020405.
\bibitem{S4-33.15}
H. Haug, Quantum-mechanical rate equations for semiconductor lasers, {Phys. Rev. A} 184 (1969) 338.
\bibitem{S4-33.16}
H. Haug, S. W. Koch, {Quantum Theory of the Optical and Electronic Properties of Semiconductors}, World Scientific, Singapore, 2009.
\bibitem{S4-33.19}
M. Lax, H. Yuen, Quantum noise. XIII. Six-classical-variable description of quantum laser fluctuations, {Phys. Rev.} 172 (1968) 362.
\bibitem{S4-33.20}
H. Risken, {The Fokker-Planck Equation: Methods of Solution and Applications}, Springer, Berlin, 1996.
\bibitem{S4-33a} M. Ludwig, B. Kubala, F. Marquardt, The optomechanical instability in the quantum regime, {New J. Phys.} 10 (2008) 095013.
\bibitem{S4-33b} R. Hamerly, H. Mabuchi, Quantum noise of free-carrier dispersion in semiconductor optical cavities, {Phys. Rev. A} 92 (2015)  023819.
\bibitem{S4-33c} A. Nunnenkamp, K. B{\o}rkje, S. M. Girvin, Single-photon optomechanics, {Phys. Rev. Lett.} 107 (2011) 063602.
\bibitem{S4-33d} S. Rips, M. Kiffner, I. Wilson-Rae, M. J. Hartmannnew, Steady-state negative Wigner functions of nonlinear nanomechanical oscillators, {New J. Phys.} 14 (2012) 023042.
\bibitem{S4-34}
W. Casteels, S. Finazzi, A. Le Boit\'{e}, F. Storme, C. Ciuti, Truncated correlation hierarchy schemes for driven-dissipative multimode quantum systems, {New J. Phys.} 18 (2016) 093007. 
\bibitem{S4-34a}
C. Jiang, Y. Cui, G. Chen, Dynamics of an optomechanical system with quadratic coupling: Effect of first order correction to adiabatic elimination, {Sci. Rep.} 6 (2016) 35583.
\bibitem{S4-34b}
D. R. Haaheim, F. M. Stein, Methods of solution of the Riccati differential equation, {Mathematics Magazine} 42 (1969) 233. 
\bibitem{S4-34c}
T. Schneider, M. Zannetti, R. Badii, H. R. Jauslin, Stochastic simulation of quantum systems and critical dynamics, {Phys. Rev. Lett.} 53 (1984) 2191.
\bibitem{S4-34.8}
E. B. Iversen, R. Juhl, J. K. M\o ller, J. Kleissl, H. Madsen, J. M. Morales, Spatio-temporal forecasting by coupled stochastic differential equations: Applications to solar power, {arXiv} (2017) 1706.04394.
\bibitem{S4-34.9}
G. Adomian, K. Malakian, Operator-theoretic solution of stochastic systems, {J. Math. Anal. Appl.} 76 (1980) 183.
\bibitem{S4-34.10}
G. Adomian, {Nonlinear stochastic operator equations}, Academic Press, Orlando, 1986. 
\bibitem{S4-34.11}
J.-P. Bouchaud, R.Cont, A Langevin approach to stock market fluctuations and crashes, {Eur. Phys. J. B} 6 (1998) 543.
\bibitem{S4-34.12}
J. J. H. Brouwers, Langevin and diffusion equation of turbulent fluid flow, {Phys. Fluids} 22 (2010) 085102.
\bibitem{S4-34.12a}
B. O. Heppe, Generalized Langevin equation for relative turbulent dispersion, {J. Fluid Mech.} 357 (1998) 167.
\bibitem{S4-34.13}
B. A. Bodo, M. E. Thompson, T. E. Unny, A review on stochastic differential equations for applications in hydrology, {Stochastic Hydrology and Hydraulics} 1 (1987) 81.
\bibitem{S4-34.14}
P. Wang, D. A. Barajas-Solano, E. Constantinescu, S. Abhyankar, D. Ghosh, B. F. Smith, Z. Huang, A. M. Tartakovsky, Probabilistic density function method for stochastic ODEs of power systems with uncertain power input, {SIAM/ASA J. Uncertainty Quantification} 3 (2015) 873.
\bibitem{S4-34.15}
A. V. Shapovalov, R. O. Rezaev, A. Y. Trifonov, Symmetry operators for the Fokker–Plank–Kolmogorov equation with nonlocal quadratic nonlinearity, {Sigma} 3 (2007) 005.
\bibitem{S4-34.16}
G. A. Pavliotis, {Stochastic Processes and Applications: Diffusion Processes, the Fokker-Planck and Langevin Equations}, Springer, New York, 2014.
\bibitem{S4-34.16a}
H. J. Carmichael, {Statistical Methods in Quantum Optics 1: Master Equations and Fokker-Planck Equations}, Springer, Berlin, 2002.
\bibitem{S4-34.16b}
H. J. Carmichael, {Statistical Methods in Quantum Optics 2: Nonclassical Fields}, Springer, Berlin, 2008.
\bibitem{S4-34.17}
K. I. Kim, Higher order bias correcting moment equation for M-estimation and its higher order efficiency, {Econometrics} 4 (2016) 48.
\bibitem{S4-35}
H. Xiong, L.-G. Si, X.-Y. Lu, Y. Wu, Optomechanically induced sum sideband generation, {Opt. Express} 24 (2016) 5773.
\bibitem{S4-35a}
F. Wang, W. Nie, C. H. Oh, Higher-order squeezing and entanglement of harmonic oscillators in superconducting circuits, {J. Opt. Soc. Am. B} 34 (2017)  130.
\bibitem{S4-36}
P. Ginzburg, Accelerating spontaneous emission in open resonators, {Ann. Phys.} 528 (2016) 571. 
\bibitem{S4-36.0}
P. D. Nation, J. R. Johansson, M. P. Blencowe, F. Nori, Stimulating uncertainty: Amplifying the quantum vacuum with superconducting circuits, {Rev. Mod. Phys.} 84 (2012) 1.
\bibitem{S4-36a} P. Rabl, Photon blockade effect in optomechanical systems, {Phys. Rev. Lett.} 107 (2011) 063601.
\bibitem{S4-36b} M. Dagenais, L. Mandel, Investigation of two-time correlations in photon emissions from a single atom, {Phys. Rev. A} 18 (1978) 2217.
\bibitem{S4-36c} S. Hong, R. Riedinger, I. Marinkovi\'{c}, A. Wallucks, S. G. Hofer, R. A. Norte, M. Aspelmeyer, S. Gr\"{o}blacher, Hanbury Brown and Twiss interferometry of single phonons from an optomechanical resonator, {Science} 358 (2017) 203-206.
\bibitem{S4-36d}
H. Wang, X. Gu, Y. X. Liu, A. Miranowicz, F. Nori, Tunable photon blockade in a hybrid system consisting of an optomechanical device coupled to a two-level system, {Phys. Rev. A} 92 (2015) 033806. 
\bibitem{S4-36e}
A. Roy, M. Devoret, Introduction to parametric amplification of quantum signals with Josephson circuits, {Comptes Rendus Physique} 17 (2016) 740.
\bibitem{S4-36e1}
C. A. Holmes, G. J. Milburn, Parametric self pulsing in a quantum opto-mechanical system, {Fortschr. Phys.} 57 (2009) 1052.
\bibitem{S4-36f}
Y. Yamamoto, K. Semba, {Principles and Methods of Quantum Information Technologies}, Springer, Tokyo, 2016.
\bibitem{S4-36g}
N. Imoto, H. A. Haus, Y. Yamamoto, Quantum nondemolition measurement of the photon number via the optical Kerr effect, {Phys. Rev. A} 32 (1985) 2287.
\bibitem{S4-36h}
R. H. Hadfield, G. Johansson, {Superconducting Devices in Quantum Optics}, Springer, Cham, 2016.
\bibitem{S4-36i} A. A. Gangat, T. M. Stace, G. J. Milburn, Phonon number quantum jumps in an optomechanical system, {New J. Phys.} 13 (2011)  043024.
\bibitem{S4-38}
M. Cirio, K. Debnath, N. Lambert, F. Nori, Amplified opto-mechanical transduction of virtual radiation pressure, {Phys. Rev. Lett.} 119 (2017) 053601.
\bibitem{S4-38a}
R. Khan, F. Massel, T. T. Heikkil\"{a}, Cross-Kerr nonlinearity in optomechanical systems, {Phys. Rev. A} 91 (2015) 043822.
\bibitem{S4-39}
N. R. Bernier, L. D. T\'{o}th, A. Koottandavida, M. Ioannou, D. Malz, A. Nunnenkamp, A. K. Feofanov, T. J. Kippenberg, Nonreciprocal reconfigurable microwave optomechanical circuit, {Nat. Comm.} 8 (2017) 604.
\bibitem{S4-39b}
A. A. Clerk, M. H. Devoret, S. M. Girvin, F. Marquardt, R. J. Schoelkopf, Introduction to quantum noise, measurement, and amplification, {Rev. Mod. Phys.} 82 (2010) 1155.
\bibitem{S4-40}
G. C. Wick, The evaluation of the collision matrix, {Phys. Rev.} 80 (1950) 268.
\bibitem{S4-40a}
P. W. Milonni, J. H. Eberly, {Lasers}, Wiley, New York, 1988.
\bibitem{S4-40b}
G. Bj\"{o}rk, A. Karlsson, Y. Yamamoto, Definition of a laser threshold, {Phys Rev A.} 50 (1994) 1675.
\bibitem{S4-40b1}
C. Z. Ning, What is Laser Threshold?, {IEEE J. Sel. Top. Quant. Elect.} 19 (2013) 1503604.
\bibitem{S4-40b2}
W. W. Chow, F. Jahnke, C. Gie, Emission properties of nanolasers during the transition to lasing, {Light: Sci. Appl.} 3 (2014) e201.
\bibitem{S4-40b3}
S. Strauf, F. Jahnke, Single quantum dot nanolaser, {Laser Photonics Rev.} 5 (2011) 607.
\bibitem{S4-40b4}
C. Gies, J. Wiersig, M. Lorke, F. Jahnke, Semiconductor model for quantum-dot-based microcavity lasers, {Phys. Rev. A} 75 (2007)  013803.
\bibitem{S4-40c}
H. Flayac, V. Savona, Non classical statistics in weakly nonlinear media. Presented at School on Recent Trends in Light-Matter Interaction, Lausanne, Switzerland, September 2017. 
\bibitem{S4-40d}
H. Flayac, V. Savona, Nonclassical statistics from a polaritonic Josephson junction, {Phys. Rev. A} 95 (2017) 043838.
\bibitem{S4-40e}
H. Flayac, V. Savona, Single photons from dissipation in coupled cavities, {Phys. Rev. A} 94 (2016) 013815.
\bibitem{S4-40f}
Y. Arakawa, S. Iwamoto, M. Nomura, A. Tandaechanurat, Y. Ota, Cavity quantum electrodynamics and lasing oscillation in single quantum dot-photonic crystal nanocavity coupled systems, {IEEE J. Sel. Top. Quant. Electron.} 18 (2012) 1818.
\bibitem{S4-40g}
M. Nomura, N. Kumagai, S. Iwamoto, Y. Ota, Y. Arakawa, Laser oscillation in a strongly coupled single-quantum-dot–nanocavity system, {Nat. Phys.} 6 (2010) 279.
\bibitem{S4-40h}
M. Mikkelsen, T. Fogarty, J. Twamley, T. Busch, Optomechanics with a Kerr-type nonlinear coupling, Phys. Rev. A 96 (2017) 043832.
\bibitem{S4-40i}
S. Shahidani, M. H. Naderi, M. Soltanolkotabi, S. Barzanjeh, Steady-state entanglement, cooling, and tristability in a nonlinear optomechanical cavity, {J. Opt. Soc. Am. B} 31 (2014) 1087.
\bibitem{S4-41}
M. Dykman, {Fluctuating Nonlinear Oscillators}, Oxford University Press, Oxford, 2012.
\bibitem{S4-42}
G. Wendin, V. S. Shumeiko, Quantum bits with Josephson junctions, {Low Temp. Phys.} 33 (2007) 724. 
\bibitem{S4-2DM} 
S. Khorasani, A. Koottandavida, Nonlinear graphene quantum capacitors for electro-optics, {npj 2D Mater. Appl.} 1 (2017) 7.
\bibitem{S4-43}
J. Clarke, F. K. Wilhelm, Superconducting quantum bits, {Nature} 453 (2008) 1031.
\bibitem{S4-45}
Y. Makhlin, G. Sch\"{o}n, A. Shnirman, Quantum-state engineering with Josephson-junction devices, {Rev. Mod. Phys.} 73 (2001) 357.
\bibitem{S4-46}
Y. A. Pashkin, O. Astafiev, T. Yamamoto, Y. Nakamura, J. S. Tsai, Josephson charge qubits: A brief review. {Quantum Inf. Process.} 8 (2009) 55.
\bibitem{S4-47}
J. M. Martinis, Superconducting phase qubits, {Quantum Inf. Process.} 8 (2009) 81.
\bibitem{S4-49}
S. M. Girvin, M. H. Devoret, R. J. Schoelkopf, Circuit QED and engineering charge based superconducting qubits, {Phys. Scr. T} 137 (2009) 014012.
\bibitem{S4-50}
Z. L. Xiang, S. Ashhab, J. Q. You, F. Nori, Hybrid quantum circuits: Superconducting circuits interacting with other quantum systems, {Rev. Mod. Phys.} 85 (2013) 623.
\bibitem{S4-51}
D. Malz, A. Nunnenkamp, Floquet approach to bichromatically driven cavity-optomechanical systems, {Phys. Rev. A} 94 (2016) 023803.
\bibitem{S4-56}
B. He, L. Yang, Q. Lin, M. Xiao, Radiation pressure cooling as a quantum dynamical process, {Phys. Rev. Lett.} 118 (2017) 233604.
\bibitem{S4-59}
S. Khorasani, A. Adibi, Analytical solution of linear ordinary differential equations by differential transfer matrix method, {Elect. J. Diff. Eq.} 2003 (2003) 1.
\bibitem{S4-60}
S. Khorasani, Differential transfer matrix solution of generalized eigenvalue problems, {Proc. Dynamic Sys. Appl.} 6 (2012) 213.





\bibitem{S8-Fokker5}
P. J. Colmenares, Fokker-Planck equation of the reduced Wigner function associated to an Ohmic quantum Langevin dynamics, {Phys. Rev. E} {97} (2018) 052126.
\bibitem{S8-Master5}
D. Boyanovsky, D. Jasnow, Heisenberg-Langevin vs. quantum master equation, {Phys. Rev. A} {96} (2017) 062108.
\bibitem{S8-Macri}
V. Macr\`{i}, A. Ridolfo, O. Di Stefano, A. F. Kockum, F. Nori, S. Savasta, Non-perturbative dynamical Casimir effect in optomechanical systems: Vacuum Casimir-Rabi splittings, {Phys. Rev. X} {8} (2018) 011031.
\bibitem{S8-Opto1}
Y. Chang, T. Shi, Y.-X. Liu, C. P. Sun, F. Nori, Multistability of electromagnetically induced transparency in atom-assisted optomechanical cavities, {Phys. Rev. A} {83} (2011) 063826. 
\bibitem{S8-Opto2}
Jing, H., \c{S}. K. \"{O}zdemir, Z. Geng, J. Zhang, X.-Y. L\"{u}, B. Peng, L. Yang, F. Nori, Optomechanically-induced transparency in parity-time-symmetric microresonators, {Sci. Rep.} {5} (2015) 9663. 
\bibitem{S8-Polariton}
H. Ian, Z. R. Gong, X.-Y. Liu, C. P. Sun, F. Nori, Cavity optomechanical coupling assisted by an atomic gas, {Phys. Rev. A} {78} (2008) 013824. 
\bibitem{S8-Multi1}
H. Jing, \c{S}. K. \"{O}zdemir, H. L\"{u}, F. Nori, High-order exceptional points in optomechanics, {Sci. Rep.} {7} (2017) 3386. 
\bibitem{S8-Multi2}
J.-Q. Liao, F. Nori, Spectrometric reconstruction of mechanical-motional states in optomechanics, {Phys. Rev. A} {90} (2014) 023851. 
\bibitem{S8-Multi3}
J.-Q. Liao, C. K. Law, L.-M. Kuang, F. Nori, Enhancement of mechanical effects of single photons in modulated two-mode optomechanics, {Phys. Rev. A} {92} (2015) 013822. 
\bibitem{S8-Kerr5}
H. Wang, X. Gu, Y.-X. Liu, A. Miranowicz, F. Nori, Optomechanical analog of two-color electromagnetically-induced transparency: Photon transmission through an optomechanical device with a two-level system, {Phys. Rev. A} {90} (2014) 023817. 
\bibitem{S8-Girvin}
K. B\o rkje, A. Nunnenkamp, J. D. Teufel, S. M. Girvin, Signatures of nonlinear cavity optomechanics in the weak coupling regime, {Phys. Rev. Lett.} {111} (2012) 053603.
\bibitem{S8-Lemonde2}
M.-A. Lemonde, N. Didier, A. A. Clerk, Nonlinear interaction effects in a strongly driven optomechanical cavity, {Phys. Rev. Lett.} {111} (2013) 053602.
\bibitem{S8-OMIT1}
S. Weis, R. Rivi\`{e}re, S. Del\'{e}glise, E. Gavartin, O. Arcizet, A. Schliesser, T. J. Kippenberg, Optomechanically induced transparency, {Science} {330} (2010) 1520.
\bibitem{S8-OMIT2}
J. D. Teufel, D. Li, M. S. Allman, K. Cicak, A. J. Sirois, J. D. Whittaker, R. W. Simmonds, Circuit cavity electromechanics in the strong-coupling regime, {Nature} {471} (2011) 204.
\bibitem{S8-Polaron}
E. Verhagen, S. Del\'{e}glise, S. Weis, A. Schliesser, T. J. Kippenberg, Quantum-coherent coupling of a mechanical oscillator to an optical cavity mode, {Nature} {482} (2012) 63.
\bibitem{S8-Lemonde1}
M.-A. Lemonde, N. Didier, A. A. Clerk, Enhanced nonlinear interactions in quantum optomechanics via mechanical amplification, {Nat. Commun.} {7} (2016) 11338. 
\bibitem{S8-Lemonde3}
M.-A. Lemonde, A. A. Clerk, Real photons from vacuum fluctuations in optomechanics: The role of polariton interactions, {Phys. Rev. A.} {91} (2015) 033836. 
\bibitem{S8-Self1}
T. J. Kippenberg, H. Rokhsari, T. Carmon, A. Scherer, K. J. Vahala, Analysis of radiation-pressure induced mechanical oscillation of an optical microcavity, {Phys. Rev. Lett.} {95} (2005) 033901.
\bibitem{S8-Self2}
F. Marquardt, J. G. E. Harris, S. M. Girvin, Dynamical multistability induced by radiation pressure in high-finesse micromechanical optical cavities, {Phys. Rev. Lett.} {96} (2006) 103901.
\bibitem{S8-Self3}
K. Vahala, M. Herrmann, S. Kn\"{u}nz, V. Batteiger, G. Saathoff, T. W. H\"{a}nsch, T. Udem, A phonon laser, {Nat. Phys.} {5} (2009) 682.
\bibitem{S8-Self4}
M. Bagheri, M. Poot, M. Li, W. P. H. Pernice, H. X. Tang, Dynamic manipulation of nanomechanical resonators in the high-amplitude regime and non-volatile mechanical memory operation, {Nat. Nanotechnol.} {6} (2011) 726.
\bibitem{S8-Self5}
A. G. Krause, J. T. Hill, M. Ludwig, A. H. Safavi-Naeini, J. Chan, F. Marquardt, O. Painter, Nonlinear radiation pressure dynamics in an optomechanical crystal, {Phys. Rev. Lett.} {115} (2015) 233601.
\bibitem{S8-Chaos1}
T. Carmon, M. Cross, K. Vahala, Chaotic quivering of micron-scaled on-chip resonators excited by centrifugal optical pressure, {Phys. Rev. Lett.} {98} (2007) 167203.
\bibitem{S8-Chaos2}
D. Navarro-Urrios, N. E. Capuj, M. F. Colombano, P. David Garc\i a, M. Sledzinska, F. Alzina, A. Griol, A. Mart\i nez, C. M. Sotomayor-Torres, Nonlinear dynamics and chaos in an optomechanical beam, {Nat. Commun.} {8} (2017) 14965.
\bibitem{S8-Stokes1}
R. Neuhaus, M. J. Sellars, S. J. Bingham, D. Suter, Breaking the Stokes–anti-Stokes symmetry in Raman heterodyne detection of magnetic-resonance transitions, {Phys. Rev. A} {58} (1998) 4961.
\bibitem{S8-Stokes2}
T. Goldstein, S.-Y. Chen, D. Xiao, A. Ramasubramaniam, J. Yan, Raman scattering and anomalous Stokes–anti-Stokes ratio in $\text{MoTe}_2$ atomic layers, {Sci. Rep.} {6} (2016) 28024.
\bibitem{S8-Stokes3}
E. A. Kittlaus, N. T. Otterstrom, P. T. Rakich, On-chip inter-modal Brillouin scattering, {Nat. Commun.} {8} (2017) 15819.
\bibitem{S8-Mika}
C. F. Ockeloen-Korppi, E. Damskagg, J.-M. Pirkkalainen, A. A. Clerk, F. Massel, M. J. Woolley, M. A. Sillanpaa, Entangled massive mechanical oscillators, {Nature} {556} (2018) 478.
\bibitem{S8-Weig1}
K. Gajo, S. Sch\"{u}z, E. M. Weig, Strong 4-mode coupling of nanomechanical string resonators, {Appl. Phys. Lett.} {111} (2017) 133109.
\bibitem{S8-Weig2}
M. J. Seitner, M. Abdi, A. Ridolfo, M. J. Hartmann, E. M. Weig, Parametric oscillation, frequency mixing, and injection locking of strongly coupled nanomechanical resonator modes, {Phys. Rev. Lett.} {118} (2017) 254301.
\bibitem{S8-Kip3}
A. Schliesser, R. Rivi\`{e}re, G. Anetsberger, O. Arcizet, T. J. Kippenberg, Resolved-sideband cooling of a micromechanical oscillator, {Nature Phys.} {4} (2008) 415.
\bibitem{S8-Teufel}
J. B. Clark, F. Lecocq, R. W. Simmonds, J. Aumentado, J. D. Teufel, Sideband cooling beyond the quantum backaction limit with squeezed light, {Nature} {541} (2017) 191.
\bibitem{S8-Vivishek}
V. Sudhir, R. Schilling, S. A. Fedorov, H. Schuetz, D. J. Wilson, T. J. Kippenberg, Quantum correlations of light from a room-temperature mechanical oscillator, {Phys. Rev. X} {7} (2017) 031055.
\bibitem{S8-Painter}
P. E. Barclay, K. Srinivasan, O. Painter, Nonlinear response of silicon photonic crystal microresonators excited via an integrated waveguide and fiber taper, {Opt. Express} {13} (2005) 801.
\bibitem{S8-Spring1}
J. D. Teufel, J. W. Harlow, C. A. Regal, K. W. Lehnert, Dynamical backaction of microwave fields on a nanomechanical oscillator, {Phys. Rev. Lett.} {101} (2008) 197203.
\bibitem{S8-Spring2}
M. Eichenfield, R. Camacho, J. Chan, K. J. Vahala, O. Painter, A picogram- and nanometre-scale photonic-crystal optomechanical cavity, {Nature} {459} (2009) 550.
\bibitem{S8-Spring3}
A. H. Safavi-Naeini, S. Groeblacher, J. T. Hill, J. Chan, M. Aspelmeyer, O. Painter, Squeezed light from a silicon micromechanical resonator, {Nature} {500} (2013) 185.
\bibitem{S8-Spring4}
P. B. Deotare, I. Bulu, I. W. Frank, Q. Quan, Y. Zhang, R. Ilic, M. Loncar, All optical reconfiuration of optomechanical fiters, {Nat. Commun.} {3} (2011) 846.
\bibitem{S8-Spring5}
C. J. Sarabalis, Y. D. Dahmani, R. N. Patel, J. T. Hill, A. H. Safavi-Naeini, Release-free silicon-on-insulator cavity optomechanics, {Optica} {4} (2017) 1147. 
\bibitem{S8-Transduction}
R. Leijssen, E. Verhagen, Strong optomechanical interactions in a sliced photonic crystal nanobeam, {Sci. Rep.} {5} (2015) 15974.
\bibitem{S8-SAspel4}
R. Riedinger, A. Wallucks, I. Marinkovic, C. L\"{o}schnauer, M. Aspelmeyer, S. Hong, S. Gr\"{o}blacher, Remote quantum entanglement between two micromechanical oscillators, {Nature} {556} (2018) 473.
\bibitem{S8-SKip2}
A. Schliesser, T. J. Kippenberg, Cavity optomechanics with whispering-gallery-mode optical micro-resonators, {Adv. At., Mol. Opt. Phys.} {58} (2010) 207. 
\bibitem{S8-SLoo}
S. G. Loo, Spectral density of random signals corrupted by multiplicative noise, {Electron. Lett.} {3} (1967) 238.
\bibitem{S8-SWick1}
R. H\"{u}bener, A. Mari, J. Eisert, Wick’s theorem for matrix product states, {Phys. Rev. Lett.} {110} (2013) 040401.
\bibitem{S8-SWick2}
L. Isserlis, On a formula for the product-moment coefficient of any order of a normal frequency distribution in any number of variables, {Biometrika} {12} (1918) 134.





\bibitem{S9-0f}
W. J. Gu, Z. Yi, L.-H. Sun, Y. Yan, Enhanced quadratic nonlinearity with parametric amplifications, {J. Opt. Soc. Am. B} {35} (2018) 652.
\bibitem{S9-0g}
J. S. Zhang, A.-X. Chen, Enhancing quadratic optomechanical coupling via nonlinear medium and lasers, {Phys. Rev. A} {99} (2019) 013843.
\bibitem{S9-0h}
M. Hossein-Zadeh, K. J. Vahala, Observation of optical spring effect in a microtoroidal optomechanical resonator, {Opt. Lett.} {32} (2007) 1611.
\bibitem{S9-0i}
K. Huang, M. Hossein-Zadeh, Direct stabilization of optomechanical oscillators, {Opt. Lett.} {42} (2017) 1946.
\bibitem{S9-Purdy1}
T. P. Purdy, P.-L. Yu, N. S. Kampel, R. W. Peterson, K. Cicak, R. W. Simmonds, C. A. Regal, Optomechanical Raman-ratio thermometry. {Phys. Rev. A} {92} (2015) 031802(R).
\bibitem{S9-Purdy2}
T. P. Purdy, K. E. Grutter, K. Srinivasan, J. M. Taylor, Quantum correlations from a room-temperature optomechanical cavity, {Science} {356} (2017) 1265.
\bibitem{S9-asym1}
F. Marquardt, J. P. Chen, A. A. Clerk, S. M. Girvin, Quantum theory of cavity-assisted sideband cooling of mechanical motion, {Phys. Rev. Lett.} {99} (2007) 093902.
\bibitem{S9-asym2}
I. Wilson-Rae, N. Nooshi, W. Zwerger, T. J. Kippenberg, Theory of ground state cooling of a mechanical oscillator using dynamical backaction, {Phys. Rev. Lett.} {99} (2007) 093901 (2007).
\bibitem{S9-Cabon}  
S. Khorasani, B. Cabon, Theory of Optimal Mixing in Directly Modulated Laser Diodes, {Scientia Iranica} {16} (2009) 157.
\bibitem{S9-Vivishek}
V. Sudhir, D. J. Wilson, R. Schilling, H. Sch\"{u}tz, S. A. Fedorov, A. H. Ghadimi, A. Nunnenkamp, T. J. Kippenberg, Appearance and disappearance of quantum correlations in measurement-based feedback control of a mechanical oscillator, {Phys. Rev. X} {7} (2017) 011001.
\bibitem{S9-Florian}
P. Rakich, F. Marquardt, Quantum theory of continuum optomechanics, {New J. Phys.} {20} (2018) 045005.
\bibitem{S9-Fullerene}
M. S. Dresselhaus, G. Dresselhaus, P. Eklund, Raman scattering in fullerenes, J. Raman Spect. {27} (1996) 351.
\bibitem{S9-CNT1}
A. G. Souza Filho, S. G. Chou, Ge. G. Samsonidze, G. Dresselhaus, M. S. Dresselhaus, Lei An, J. Liu, Anna K. Swan, M. S. \"{U}nl\"{u}, B. B. Goldberg, A. Jorio, A. Gr\"{u}neis, R. Saito, Stokes and anti-Stokes Raman spectra of small-diameter isolated carbon nanotubes, Phys. Rev. B {69} (2004) 115428.
\bibitem{S9-CNT2}
M. S. Dresselhaus, G. Dresselhaus, R. Saito, A. Jorio, Raman spectroscopy of carbon nanotubes, {Phys. Rep.} {409} (2005) 47.
\bibitem{S9-CNT3}
A. Jorio, A. G. Souza Filho, G. Dresselhaus, M. S. Dresselhaus, R. Saito, J. H. Hafner, C. M. Lieber, F. M. Matinaga, M. S. S. Dantas, M. A. Pimenta, Joint density of electronic states for one isolated single-wall carbon nanotube studied by resonant Raman scattering, {Phys. Rev. B} {63} (2001) 245416.
\bibitem{S9-CNT4}
Y. Chen, Z. Shen, Z. Xu, Y. Hu, H. Xu, S. Wang, X. Guo, Y. Zhang, L. Peng, F. Ding, Z. Liu, J. Zhang, Helicity-dependent single-walled carbon nanotube alignment on graphite for helical angle and handedness recognition, {Nat. Commun.} {4} (2013) 2205.
\bibitem{S9-Springer}
P.-H. Tan, ed. {Raman Spectroscopy of Two-dimensional Materials}, Springer, Singapore, 2019.
\bibitem{S9-Raman2D}
F. Shao, R. Zenobi, Tip-enhanced Raman spectroscopy: principles, practice, and applications to nanospectroscopic imaging of 2D materials, {Anal. Bioanal. Chem.} {411} (2019) 37.
\bibitem{S9-RamanGr}
A. Jorio, R. Saito, G. Dresselhaus, M. S. Dresselhaus, {Raman Spectroscopy in Graphene Related Systems}, Wiley-VCH, Weinheim, 2011.
\bibitem{S9-CTP}
S. Khorasani, Third-order optical nonlinearity in two-dimensional transition metal dichalcogenides, {Commun. Theor. Phys.} {70} (2018) 334.
\bibitem{S9-MoS2-1}
X. Zhang, X.-F. Qiao, W. Shi, J.-B. Wu, D.-S. Jiang, P.-H. Tan, Phonon and Raman scattering of two-dimensional transition metal dichalcogenides from monolayer, multilayer to bulk material, {Chem. Soc. Rev.} {44} (2015) 2757.
\bibitem{S9-MoS2-2}
D. Tuschel, Raman spectroscopy and imaging of low energy phonons, {Spectroscopy} {30} (2015) 18.
\bibitem{S9-CNT}
R. Saito, M. Hofmann, G. Dresselhaus, A. Jorio, M. S. Dresselhaus, Raman spectroscopy of graphene and carbon nanotubes, {Adv. Phys.} {60} (2011) 413.
\bibitem{S9-Graphene}
J.-B. Wu, M.-L. Lin, X. Cong, H.-N. Liua, P.-H. Tan, Raman spectroscopy of graphene-based materials and its applications in related devices, {Chem. Soc. Rev.} {47} (2018) 1822.
\bibitem{S9-Graphene2}
J.-B. Wu, X. Zhang, M. Ij\"{a}s, W.-P. Han, X.-F. Qiao, X.-L. Li, D.-S. Jiang, A. C. Ferrari, P.-H. Tan, Resonant Raman spectroscopy of twisted multilayer graphene, {Nat. Commun.} {5} (2014) 5309.
\bibitem{S9-Graphene3}
P. H. Tan, W. P. Han, W. J. Zhao, Z. H. Wu, K. Chang, H. Wang, Y. F. Wang, N. Bonini, N. Marzari, N. Pugno, G. Savini, A. Lombardo, A. C. Ferrari, The shear mode of multilayer graphene, {Nat. Mater.} {11} (2012) 294.
\bibitem{S9-NanoPhoton}
https://www.nanophoton.net/raman/raman-spectroscopy.html
\bibitem{S9-MOM}
S.-I. Zaitsu, H. Izaki, T. Tsuchiya, T. Imasaka, Continuous-wave phase-matched molecular optical modulator, {Sci. Rep.} {6} (2016) 20908.
\bibitem{S9-Tullio}
Ferrante, C. {et al.} Raman spectroscopy of graphene under ultrafast laser excitation. {Nat. Commun.} {9}, 308 (2018).
\bibitem{S9-Pico}
F. Benz, M. K. Schmidt, A. Dreismann, R. Chikkaraddy, Y. Zhang, A. Demetriadou, C. Carnegie, H. Ohadi, B. de Nijs, R. Esteban, J. Aizpurua, J. J. Baumberg, Single-molecule optomechanics in ``picocavities'', {Science} {354} (2016) 726.
\bibitem{S9-PicoData}
F. Benz, M. K. Schmidt, A. Dreismann, R. Chikkaraddy, Y. Zhang, A. Demetriadou, C. Carnegie, H. Ohadi, B. de Nijs, R. Esteban, J. Aizpurua, J. J. Baumberg, Single-molecule optomechanics in ``picocavities'', [Data set] {Apollo} (2016) doi:10.17863/CAM.1675.
\bibitem{S9-Fiber}
T. H. Runcorn, R. T. Murray, J. R. Taylor, Highly efficient nanosecond 560 nm source by SHG of a combined Yb-Raman fiber amplifier, {Opt. Express} {26} (2018) 4440.
\bibitem{S9-FiberData}
T. H. Runcorn, R. T. Murray, J. R. Taylor, Highly efficient nanosecond 560 nm source by SHG of a combined Yb-Raman fiber amplifier, [Data set]  {Zenodo} (2018) doi:10.5281/zenodo.1166082.
\bibitem{S9-Pair1}
O. Di Stefano, A. Settineri, V. Macr\i, A. Ridolfo, R. Stassi, A. F. Kockum, S. Savasta, F. Nori, Interaction of mechanical oscillators mediated by the exchange of virtual photon pairs, {Phys. Rev. Lett.} {122} (2019) 030402.
\bibitem{S9-Pair2}
E. Jansen, J. D. P. Machado, Y. M. Blanter, Realization of a degenerate parametric oscillator in electromechanical systems, {Phys. Rev. B} {99} (2019) 045401.
\bibitem{S9-Pair3}
U. Nabholz, F. Schatz, J. E. Mehner, P. Degenfeld-Schonburg, Spontaneous parametric down-conversion induced by non-degenerate phononic three-wave mixing in a scanning MEMS micro mirror, Sci. Rep. 9 (2019) 3997.
\bibitem{S9-EOV1}
J. Capmany, C. R. Fern\'{a}ndez-Pousa, Quantum model for electro-optical phase modulation, {J. Opt. Soc. Am. B} {27} (2010) A119.
\bibitem{S9-EOV2}
M. K. Schmidt, R. Esteban, A. Gonz\'{a}lez-Tudela, G. Giedke, J. Aizpurua, Quantum mechanical description of Raman scattering from molecules in plasmonic cavities, {ACS Nano} {10} (2016) 6291.
\bibitem{S9-EOV3}
A. Lombardi, M. K. Schmidt, L. Weller, W. M. Deacon, F. Benz, B. de Nijs, J. Aizpurua, J. J. Baumberg, Pulsed molecular optomechanics in plasmonic nanocavities: From nonlinear vibrational instabilities to bond-breaking, {Phys. Rev. X} {8} (2018) 011016.
\bibitem{S9-Breather1}
N. N. Akhmediev, V. M. Eleonskii, N. E. Kulagin, Exact first-order solutions of the nonlinear Schr\"{o}dinger equation, {Theor. Math. Phys.} {72} (1987) 809.
\bibitem{S9-Breather2}
H. Xiong, Y. Wu, Optomechanical Akhmediev breathers, {Laser Photon. Rev.} {12} (2018) 1700305.




\bibitem{S10-q1}
H. Xiong, L.-G. Si, A.-N. Zheng, X. Yang, Y. Wu, Higher-order sidebands in optomechanically induced transparency, {Phys. Rev. A} {86} (2012) 013815. 
\bibitem{S10-q12}
Y. F. Jiao, T. X. Lu, H. Jing, Optomechanical second-order sidebands and group delays in a Kerr resonator, {Phys. Rev. A} {97} (2018) 013843. 
\bibitem{S10-q3}
A. Kronwald, F. Marquardt, Optomechanically induced transparency in the nonlinear quantum regime, {Phys. Rev. Lett.} {111} (2013) 133601. 
\bibitem{S10-q4}
M. Karuza, C. Biancofiore, M. Bawaj, C. Molinelli, M. Galassi, R. Natali, P. Tombesi, G. Di Giuseppe, D. Vitali, Optomechanically induced transparency in a membrane-in-the-middle setup at room temperature. {Phys. Rev. A} {88} (2013) 013804. 
\bibitem{S10-q8}
Z. Shen, C.-H. Dong, Y. Chen, Y.-F. Xiao, F.-W. Sun, G.-C. Guo, Compensation of the Kerr effect for transient optomechanically induced transparency in a silica microsphere. {Opt. Lett.} {41} (2016) 1249.
\bibitem{S10-q5}
H. Jing, \c{S}. K. \"{O}zdemir, X.-Y. L\"{u}, J. Zhang, L. Yang, F. Nori, PT-symmetric phonon laser. {Phys. Rev. Lett.} {113} (2014) 053604.
\bibitem{S10-q13}
J. Zhang, B. Peng, \c{S}. K. \"{O}zdemir, K. Pichler, D. O. Krimer, G. Zhao, F. Nori, Y.-X. Liu, S. Rotter, L. Yang, A phonon laser operating at an exceptional point, {Nat. Photon.} {12} (2018) 479.
\bibitem{S10-q6}
J. Zhang, B. Peng, \c{S}. K. \"{O}zdemir, Y.-X. Liu, H. Jing, X.-Y. L\"{u}, Y.-L. Liu, L. Yang, F. Nori, Giant nonlinearity via breaking parity-time symmetry: A route to low-threshold phonon diodes, {Phys. Rev. B} {92} (2015) 115407.
\bibitem{S10-q7}
L. Fan, K. Y. Fong, M. Poot, H. X. Tang, Cascaded optical transparency in multimode-cavity optomechanical systems, {Nat. Commun.} {6} (2015) 5850. 
\bibitem{S10-q10}
Z.-P. Liu, J. Zhang, \c{S}. K. \"{O}zdemir, B. Peng, H. Jing, X.-Y. L\"{u}, C.-W. Li, L. Yang, F. Nori, Y.-X. Liu, Metrology with PT-symmetric cavities: Enhanced sensitivity near the PT-phase transition, {Phys. Rev. Lett.} {117} (2016) 110802.
\bibitem{S10-q14}
S. Qvarfort, A. Serafini, P. F. Barker, S. Bose, Gravimetry through non-linear optomechanics, {Nat. Commun.} {9} (2018) 3690.
\bibitem{S10-Quad4}
A. Dalafi, M. H. Naderi, A. Motazedifard, Effects of quadratic coupling and squeezed vacuum injection in an optomechanical cavity assisted with a Bose-Einstein condensate,	{Phys. Rev. A} {97} (2018) 043619.
\bibitem{S10-Sala}
K. Sala, T. Tufarelli, Exploring corrections to the Optomechanical Hamiltonian, {Sci. Rep.} {8} (2018) 9157.
\bibitem{S10-Thesis}
B. H. Pang, {Theoretical Foundations for Quantum Measurement in a General Relativistic Framework}, Ph.D. Dissertation, California Institute of Technology, Psadena, 2018, doi:10.7907/dfyy-y188. 





\bibitem{S11-Hoi}
I.-C. Hoi, A. F. Kockum, T. Palomaki, T. M. Stace, B. Fan, L. Tornberg, S. R. Sathyamoorthy, G. Johansson, P. Delsing, C. M. Wilson, Giant cross–Kerr effect for propagating microwaves induced by an artificial atom, {Phys. Rev. Lett.} 111 (2013) 053601.



\bibitem{Gharekhanlou1}
B. Gharekhanlou, S. Khorasani, Current-voltage characteristics of graphane p-n junctions, {IEEE Trans. Elect. Dev.} 57 (2010) 209-214.
\bibitem{Wang}
S. Wang, {Fundamentals of Semiconductor Theory and Device Physics}, Prentice‐Hall International, Englewood Cliffs, 1989. 
\bibitem{Gharekhanlou2}
B. Gharekhanlou, S. Khorasani, Generation and recombination in two-dimensional bipolar transistors, {Appl. Phys. A}  115 (2014) 737-740.
\bibitem{Gharekhanlou3}
B. Gharekhanlou, S. Khorasani, R. Sarvari, Two-dimensional bipolar junction transistors, {Mater. Res. Express} 1 (2014) 015604.
\bibitem{Streetman}
B. G. Streetman, S. Banerjee, {Solid State Electronic Devices}, 7th ed., Pearson Education, Harlow, 2019.
\bibitem{Kuramoto1}
F. A. Rodrigues, T. K. Peron, P. Ji, J. Kurths, The Kuramoto model in complex networks, Phys. Rep. 610 (2016) 1-98.
\bibitem{Kuramoto2}
C. C. Gong, C. Zheng, R. Toenjes, A. Pikovsky, Repulsively coupled Kuramoto-Sakaguchi phase oscillators ensemble subject to common noise, {Chaos} 29 (2019) 033127.
\bibitem{Kuramoto3}
D. Witthaut, M. Timme, Kuramoto dynamics in Hamiltonian systems, {Phys. Rev. E} 90 (2014) 032917.

\end{thebibliography}
\end{document}